\documentclass[preprint]{aastex631}
\usepackage{xspace}
\usepackage{booktabs}
\usepackage{longtable}
\usepackage{morefloats}
\usepackage{blindtext}
\usepackage{placeins}
\usepackage{gensymb}
\usepackage{makecell}
\usepackage{comment}
\usepackage{rotating}
\usepackage{listings} 
\usepackage{amsmath,amssymb}
\usepackage{multirow} 

\newcommand{\Fermi}{\emph{Fermi}\xspace}
\newcommand{\lat}{\emph{Fermi}-LAT\xspace}
\newcommand{\Swift}{\emph{Swift}\xspace}
\newcommand{\code}[1]{\lstinline!#1!\xspace}

\newcommand{\fermitools}{\code{fermitools}}

\newcommand{\fermipy}{\code{fermipy}}
\newcommand{\ThreeML}{\code{ThreeML}}

\newcommand{\gtburst}{\code{gtburst}}

\renewcommand{\deg}{\,$^{\circ}$}

\newcommand{\be}{\begin{equation}}
\newcommand{\ee}{\end{equation}}
\newcommand{\ba}{\begin{eqnarray}}
\newcommand{\ea}{\end{eqnarray}}

\def\p{\prime}

\newcommand{\ltsima} {$\; \buildrel < \over \sim \;$}
\newcommand{\gtsima} {$\; \buildrel > \over \sim \;$}
\newcommand{\lta} {\lower.5ex\hbox{\ltsima}}
\newcommand{\gta} {\lower.5ex\hbox{\gtsima}}

\newcommand{\g}{\gamma}
\newcommand{\G}{\Gamma}

\newcommand{\e}{\times 10^}

\newcommand{\tz}{$\rm T_{\rm \,LAT,0}$\xspace}
\newcommand{\tone}{$\rm T_{\rm \,LAT,1}$\xspace}
\newcommand{\toz}{$\rm T_{\rm \,LAT,100}$\xspace}

\newcommand{\grb}{GRB\,221009A\xspace}
\newcommand{\trig}{$\rm T_{0}$\xspace}

\newcommand{\fraz}[2]{\frac{\displaystyle #1}{\displaystyle #2}}

\newcommand{\nob}[1]{#1}

\newcommand{\nop}[1]{ #1}

\definecolor{royalpurple}{rgb}{0.47, 0.32, 0.66}

\setcitestyle{notesep={ }}

\received{\today}
\revised{\today}
\accepted{\today}
\submitjournal{ApJS}

\definecolor{blazeorange}{rgb}{1.0, 0.4, 0.0}
\definecolor{seagreen}{rgb}{0.18, 0.55, 0.34}
\definecolor{rufous}{rgb}{0.66, 0.11, 0.03}
\definecolor{royalfuchsia}{rgb}{0.79, 0.17, 0.57}
\definecolor{scarlet}{rgb}{1.0, 0.13, 0.0}
\definecolor{royalpurple}{rgb}{0.47, 0.32, 0.66}

\shorttitle{\grb}
\shortauthors{The Fermi LAT collaboration}
\begin{document}
\title{\grb: the B.O.A.T Burst that Shines in Gamma Rays}
\author[0000-0003-4378-8785]{M.~Axelsson }
\affiliation{Department of Physics, KTH Royal Institute of Technology, AlbaNova, SE-106 91 Stockholm, Sweden}
\affiliation{The Oskar Klein Centre for Cosmoparticle Physics, AlbaNova, SE-106 91 Stockholm, Sweden}
\author[0000-0002-6584-1703]{M.~Ajello}
\affiliation{Department of Physics and Astronomy, Clemson University, Kinard Lab of Physics, Clemson, SC 29634-0978, USA}
\author[0000-0003-1250-7872]{M.~Arimoto}
\affiliation{Faculty of Mathematics and Physics, Institute of Science and Engineering, Kanazawa University, Kakuma, Kanazawa, Ishikawa 920-1192}
\author[0000-0002-9785-7726]{L.~Baldini}
\affiliation{Universit\`a di Pisa and Istituto Nazionale di Fisica Nucleare, Sezione di Pisa I-56127 Pisa, Italy}
\author[0000-0002-8784-2977]{J.~Ballet}
\affiliation{Universit\'e Paris-Saclay, Universit\'e Paris Cit\'e, CEA, CNRS, AIM, F-91191 Gif-sur-Yvette Cedex, France}
\author[0000-0003-4433-1365]{M.~G.~Baring}
\affiliation{Rice University, Department of Physics and Astronomy, MS-108, P. O. Box 1892, Houston, TX 77251, USA}
\author[0000-0001-7233-9546]{C.~Bartolini}
\affiliation{Istituto Nazionale di Fisica Nucleare, Sezione di Bari, I-70126 Bari, Italy}
\affiliation{Universit\`a degli studi di Trento, via Calepina 14, 38122 Trento, Italy}
\author[0000-0002-6954-8862]{D.~Bastieri}
\affiliation{Istituto Nazionale di Fisica Nucleare, Sezione di Padova, I-35131 Padova, Italy}
\affiliation{Dipartimento di Fisica e Astronomia ``G. Galilei'', Universit\`a di Padova, Via F. Marzolo, 8, I-35131 Padova, Italy}
\affiliation{Center for Space Studies and Activities ``G. Colombo", University of Padova, Via Venezia 15, I-35131 Padova, Italy}
\author[0000-0002-6729-9022]{J.~Becerra~Gonzalez}
\affiliation{Instituto de Astrof\'isica de Canarias and Universidad de La Laguna, Dpto. Astrof\'isica, 38200 La Laguna, Tenerife, Spain}
\author[0000-0002-2469-7063]{R.~Bellazzini}
\affiliation{Istituto Nazionale di Fisica Nucleare, Sezione di Pisa, I-56127 Pisa, Italy}
% ...
\collaboration{1000}{(Fermi LAT collaboration)}


\author[0000-0001-8058-9684]{S.~Lesage}
\affiliation{Department of Space Science, University of Alabama in Huntsville, 320 Sparkman Drive, Huntsville, AL 35899, USA}
\affiliation{Center for Space Plasma and Aeronomic Research, University of Alabama in Huntsville, Huntsville, AL 35899, USA}

\author[0000-0003-2105-7711]{M.~S.~Briggs}
\affiliation{Center for Space Plasma and Aeronomic Research, University of Alabama in Huntsville, Huntsville, AL 35899, USA}
\affiliation{Department of Space Science, University of Alabama in Huntsville, 320 Sparkman Drive, Huntsville, AL 35899, USA}

\author[0000-0002-2942-3379]{E.~Burns} 
\affil{Department of Physics \& Astronomy, Louisiana State University, Baton Rouge, LA 70803, USA}


\author[0000-0002-6657-9022]{S.~Bala}
\affiliation{Science and Technology Institute, Universities Space Research Association, Huntsville, AL 35805, USA}

\author[0000-0001-7916-2923]{P.~N.~Bhat}
\affiliation{Center for Space Plasma and Aeronomic Research, University of Alabama in Huntsville, Huntsville, AL 35899, USA}

\author[0009-0003-3480-8251]{W.~H.~Cleveland}
\affiliation{Science and Technology Institute, Universities Space Research Association, Huntsville, AL 35805, USA}

\author[0000-0003-1835-570X]{S.~Dalessi}
\affiliation{Department of Space Science, University of Alabama in Huntsville, 320 Sparkman Drive, Huntsville, AL 35899, USA }
\affiliation{Center for Space Plasma and Aeronomic Research, University of Alabama in Huntsville, 320 Sparkman Drive, Huntsville, AL 35899, USA }

\author[0009-0003-7797-1243]{C.~de Barra}
\affiliation{School of Physics, University College Dublin, Belfield, Dublin, Ireland} 

\author{M.~Gibby}
\affiliation{Jacobs Space Exploration Group, Huntsville, AL 35806, USA}

\author{M.~M.~Giles}
\affiliation{Jacobs Space Exploration Group, Huntsville, AL 35806, USA}

\author[0000-0003-0761-6388]{R.~Hamburg}
\affiliation{Universit\'e Paris-Saclay, CNRS/IN2P3, IJCLab, 91405 Orsay, France}

\author[0000-0001-9556-7576]{B.~A.~Hristov}
\affiliation{Center for Space Plasma and Aeronomic Research, The University of Alabama in Huntsville, Huntsville, AL 35899}

\author[0000-0002-0468-6025]{C.~M.~Hui}
\affiliation{ST12 Astrophysics Branch, NASA Marshall Space Flight Center, Huntsville, AL 35812, USA}

\author[0000-0001-9201-4706]{D.~Kocevski} 
\affiliation{ST12 Astrophysics Branch, NASA Marshall Space Flight Center, Huntsville, AL 35812, USA}

\author[0000-0002-2531-3703]{B.~Mailyan}
\affiliation{Department of Aerospace, Physics and Space Sciences, Florida Institute of Technology, Melbourne, FL 32901, USA }

\author[0000-0002-0380-0041]{C.~Malacaria}
\affiliation{International Space Science Institute, Hallerstrasse 6, 3012 Bern, Switzerland}

\author[0000-0002-1477-618X]{S.~ McBreen}
\affiliation{School of Physics, University College Dublin, Belfield, Dublin, Ireland} 

\author[0000-0002-6269-0452]{S.~Poolakkil}
\affiliation{Department of Space Science, University of Alabama in Huntsville, Huntsville, AL 35899, USA}
\affiliation{Center for Space Plasma and Aeronomic Research, University of Alabama in Huntsville, Huntsville, AL 35899, USA}

\author[0000-0002-7150-9061]{O.J.~Roberts}
\affiliation{Science and Technology Institute, Universities Space Research Association, Huntsville, AL 35805, USA}

\author[0000-0002-0602-0235]{L.~ Scotton}
\affiliation{Department of Space Science, University of Alabama in Huntsville, 320 Sparkman Drive, Huntsville, AL 35899, USA}
\affiliation{Center for Space Plasma and Aeronomic Research, University of Alabama in Huntsville, Huntsville, AL 35899, USA}

\author[0000-0002-2149-9846]{P.~Veres}
\affiliation{Department of Space Science, University of Alabama in Huntsville, Huntsville, AL 35899, USA}
\affiliation{Center for Space Plasma and Aeronomic Research, University of Alabama in Huntsville, Huntsville, AL 35899, USA}

\author[0000-0002-0221-5916]{A.~von Kienlin}
\affil{Max-Planck-Institut f\"{u}r extraterrestrische Physik, Giessenbachstrasse 1, D-85748 Garching, Germany}

\author[ 0000-0002-8585-0084]{C.~ A.~Wilson-Hodge}
\affiliation{ST12 Astrophysics Branch, NASA Marshall Space Flight Center, Huntsville, AL 35812, USA}

\author[0000-0001-9012-2463]{J.~Wood}
\affiliation{ST12 Astrophysics Branch, NASA Marshall Space Flight Center, Huntsville, AL 35812, USA}

\collaboration{1000}{(Fermi GBM collaboration)}


\begin{abstract}
We present a complete analysis of {\it Fermi} Large Area Telescope (LAT) data of \grb, the brightest Gamma-Ray Burst (GRB) ever detected. 
The burst emission above 30\,MeV detected by the LAT preceded by 1\,s the low-energy ($<$\,10 MeV) pulse that triggered the Fermi Gamma-Ray Burst Monitor (GBM), as has been observed in other GRBs.
The prompt phase of GRB~221009A lasted a few hundred seconds. 
It was so bright that we identify a Bad Time Interval (BTI) of 64 seconds caused by the extremely high flux of hard X-rays and soft gamma rays, during which the event reconstruction efficiency was poor and the dead time fraction quite high.
The late-time emission decayed as a power law, but the extrapolation of the late-time emission during the first 450 seconds suggests that the afterglow started during the prompt emission. We also found that high-energy events observed by the LAT are incompatible with synchrotron origin, and, during the prompt emission, are more likely related to an extra component identified as synchrotron self-Compton (SSC). 
A remarkable 400\,GeV photon, detected by the LAT 33~ks after the GBM trigger and directionally consistent with the location of \grb, is hard to explain as a product of SSC or TeV electromagnetic cascades, and the process responsible for its origin is uncertain.
Because of its proximity and energetic nature, \grb is an extremely rare event.

\end{abstract}
%
%
\newpage
\section{Introduction} 
\label{sec:intro}
%
%
Gamma-ray bursts (GRBs) are extremely powerful events lighting up the gamma-ray sky approximately once per day, with the bulk of their emission extending from several keV to tens of MeV. They have been detected regularly since the late 1960s by many dedicated space missions, with a total of many thousands of events. 
The promptly emitted gamma-ray radiation on timescales of 
seconds is followed by a longer-lasting afterglow emission revealed at every wavelength (from radio to TeV gamma rays) on timescales extending to days, months, even years.

The Large Area Telescope \citep[LAT,][]{2009ApJ...697.1071A} onboard the {\it Fermi} Gamma-ray Space Telescope, operating since 2008, opened a new window to explore GRB emission in the GeV domain, establishing important spectral and temporal properties of the high-energy ($>$ 0.1\,GeV) burst emission \citep{2FLGC}.

For almost a decade, GRB\,130427A held many records among all LAT-detected GRBs \citep{2014Sci...343...42A}, including the highest fluence $(1.3 \pm 0.2 \times 10^{-4}$ erg $\rm{cm}^{-2}$ in the 0.1--100 GeV energy range), the highest energy photon ($94$\,GeV at \trig+ 243\,s; with \trig referring to the burst trigger), the largest number of $>$ 0.1\,GeV-photons recorded from a GRB ($>$  600, with 17 events above 10\,GeV) and the longest high-energy duration $34.4 \pm 0.3$\,ks ($\sim$10 hrs). Despite its intensity and proximity at $z=0.34$, no very-high-energy ($>$ 0.1\,TeV, VHE) emission was detected by any ground-based experiment at the time. 
The VERITAS array reported bright moonlight conditions precluding observations at the time of trigger \citep{2014ApJ...795L...3A} and follow-up could not be initiated until $\sim 71$\,ks ($\sim$20\,hrs) after the onset of the burst, leading to upper limit derivations and constraints. 

VHE emission was finally detected from a GRB a few years later, both at very early times (from 1 to 40 min after the GRB onset) as in the case of GRB\,190114C ($z=0.42$) observed by the MAGIC telescopes at energies $0.3-1$\,TeV \citep{2019Natur.575..455M} as well as in the afterglow regime ($\sim 4-10$\,hrs after the GRB onset) for GRB\,180720B ($z=0.65$) and for GRB\,190829A ($z=0.078$), both observed by the H.E.S.S. telescopes at energies $0.1-0.44$\,TeV \citep{2019Natur.575..464A} and $0.18-3.3$\,TeV \citep{2021Sci...372.1081H}, respectively\nob{, or GRB~201216C observed by MAGIC at redshift $z = 1.1$ \citep{2024MNRAS.527.5856A}}. 
All these remarkable afterglow detections at TeV energies have been favored by the proximity of the bursts, preventing such high-energy emission from being severely attenuated by pair production against the the extragalactic background light \citep[EBL,][]{2010ApJ...712..238F}. Interestingly, none of these three breakthrough detections was accompanied by extraordinary MeV/GeV emission. In fact, the highest-energy photons detected by the LAT were 21 and 5 GeV for GRB~190114C and GRB~180720B, respectively, while the LAT detected no prompt or afterglow high-energy emission in the case of GRB~190829A \citep{2019GCN.25574....1P}.
The MeV emission is thought to arise mainly from synchrotron radiation of relativistic electrons in the jet plasma due to ultra-relativistic bulk motion. The same synchrotron component also likely dominates the LAT emission; however an additional high-energy spectral component is typically needed to explain emission at energies $\gtrsim10$\,GeV that exceeds the maximum energy of synchrotron photons at late times \citep{2014Sci...343...42A, 2020ApJ...890....9A}. Thus, \lat covers the key energy band in which the transition of the GRB emission mechanism occurs.

Here we report the \Fermi-LAT observation and interpretation of the highest-energy emission ever measured from a GRB. \grb is so bright that it allowed a complete spectral and temporal analysis to be performed from keV to TeV energy from the triggering pulse to the late afterglow. 
For the first time, rapid spectral variability in the LAT during the prompt emission clearly revealed the simultaneous presence of the afterglow, which extended up to several TeV, with the prompt emission. The spectrum shows clear evidence of synchrotron self-Compton (SSC) and its evolution with time is presented in this paper. \grb broke many previous records, not only at high energies, but over many energy bands, quickly deserving the title of the ``Brightest Of All Time", or the ``B.O.A.T." GRB \citep{2023ApJ...946L..31B}. 

This event was so strong and long-lived that it was detected by dozens of space- and ground-based observatories, caused even a sudden disturbance of the Earth's ionosphere \citep{2022GCN.32744....1S}, and resulted in more than 120 rapid observation reports and communications on the \nob{General} Coordinate Network (GCN) and via The Astronomer’s Telegram service. 
Over \nob{100 publications, including the detection of gamma-rays from AGILE~\citep{BOAT_AGILE},} discussions about X-ray polarization~\citep{IXPEpapaer}, radio observations \citep{BOAT_MW}, and optical observations \citep{Fulton_2023}, as well as several follow-up campaigns carried out by e.g. VHE telescopes \citep{Aharonian_2023} and neutrino observatories \citep{Abbasi_2023}\nob{, have been published to date}. 

The Fermi Gamma-Ray Burst Monitor (GBM) onboard {\it Fermi} was the first detector to trigger on \grb on 2022 October 9, at 13:16:59.988 UT (trigger 221009553 / 687014224)~\citep{2023ApJ...952L..42L}. We will refer to this time as \trig of the burst throughout the text. Unfortunately, prompt notices containing classification and localization information were not immediately distributed to the community through standard procedure due to issues in the ground segment~\citep{2023ApJ...952L..42L}.
Therefore, the LAT GRB pipeline, the so-called ``LAT Transient Factory" \citep[LTF,][]{2015arXiv150203122V} did not automatically receive and process this event. 
About an hour after the trigger, at 14:10:17 UT (\trig+ 3197\,s), the {\it Swift} Burst Alert Telescope (BAT) reported the detection of a very bright hard X-ray, soft X-ray, and UV/optical transient. BAT tentatively classified it as a new source called {\it Swift $J1913.1+1946$} \citep{2022GCN.32632....1D} with possible Galactic origin given the proximity ($4$ deg) to the Galactic plane. This trigger was successfully received by the LTF and the resulting preliminary analysis immediately revealed a high-energy detection with high significance~\citep{2022GCN.32637....1B}, with the highest-energy event being a 7.7 GeV photon detected 766 s after the BAT trigger. Meanwhile, the initial GBM trigger had been analyzed and found to be spatially consistent with the BAT and LAT localizations~\citep{2022GCN.32642....1L}. GBM data revealed a first emission episode with duration of $\sim 40$~s, followed by a second much longer and extremely bright emission episode, affected by strong pulse pile-up and other systematic effects\footnote{\url{https://fermi.gsfc.nasa.gov/ssc/data/analysis/grb221009a.html}}. 
The preliminary GBM-reported peak flux and fluence clearly indicated that \grb was the most intense and fluent burst detected throughout the lifetime of the \Fermi mission.
Following the GBM notice, a new analysis was immediately initiated covering the first 3200\,s. 
This revealed an extremely bright and structured emission episode temporally coincident with the GBM main emission episode starting at \trig+ 200\,s and a highest-energy photon of $99.3$\,GeV at \trig+ 240\,s, which represented the highest photon energy ever detected by the LAT~\citep{2022GCN.32658....1P} associated with a GRB .
However, as first reported in \citet{2022GCN.32748....1X}, LAT detected an even more energetic photon with an energy of 400\,GeV $\sim 9$\,hrs after the GBM trigger \citep{2024NatCo..15.4280X}.
Over the following days, more than 100 different teams reported follow-up observations and different preliminary analyses of \grb.
The first redshift measurement of $z = 0.151$ was provided by the X-shooter/VLT Team $11.5$\,hrs after the GBM trigger~\citep{2022GCN.32648....1D}, and a spectroscopic redshift confirmation ($z=0.1505$) came from the $10.4$\,m Gran Telescopio de Canarias \citep[GTC,][]{2022GCN.32686....1C}.
Another remarkable feature of the burst afterglow was detected in the {\it Swift} X-ray Telescope (XRT) analysis, which produced a stacked image in the $ 0.3-10$\,keV energy band, showing a complex system of expanding dust-scattering rings with radii from about $2.5$ to $7.5$ arcmin~\citep{2023ApJ...946L..30T, 2023MNRAS.521.1590V}, a system that was also observed by the Imaging X-ray Polarimetry Explorer \citep[IXPE,][]{IXPEpapaer}, and X-ray Multi-Mirror Mission \citep[XMM-Newton,][]{BOAT_XMM}. 

Searches for neutrino and gravitational-wave counterparts around the time of the burst did not result in detections \citep{2022GCN.32741....1K, 2023ApJ...944..115A, 2023ApJ...946L..26A}. 
Additionally, the full moon on the night of October 9, 2022 prevented all major IACT telescopes (MAGIC, H.E.S.S. and VERITAS) from immediately following up the event.
The High-Altitude Water Cherenkov Observatory (HAWC) reported no detection in the burst region $\sim8$ hours after the GBM trigger \citep{2022GCN.32683....1A}. 

Two unexpected discoveries were announced by ground-based experiments a couple of days after the trigger. 
First, the Large High Altitude Air Shower Observatory (LHAASO) collaboration reported the detection of $> 500$\,GeV photons within 2000 seconds of the GBM trigger, with a significance greater than 100 standard deviations by the WDCA instrument. Moreover, LHAASO-KM2A observed a record-breaking 18\,TeV photon coming from the direction of \grb with a significance of 10 standard deviations \citep{2022GCN.32677....1H}.
Following the LHAASO announcement, an even more remarkable claim was made by the Carpet-2 air-shower array collaboration, reporting the unprecedented detection of a 250 TeV photon-like air shower at \trig+ 4536\,s \citep{2022ATel15669....1D}, but, to date, no confirmation of it has been published in the literature.
Finally, the discovery of a MeV emission line in the spectrum of \grb was announced by \citet{BOAT_LINE}, and will be also discussed later in this paper.

This paper is organized as follows: In Sec.~\ref{sec:analysis} 
we present the \nob{LAT} analysis, including the special treatment of the data during the Bad Time Interval (BTI). 
In Sec.~\ref{sec:results} we collect the results of the analysis of the triggering pulse, of the prompt emission and on the temporally extended emission.
In Sec.~\ref{sec:discussion} we model the high-energy emission from prompt to afterglow, disentangling the contribution from internal shocks from that of the external shock, that was responsible for the late-time emission. In this section we also present a broad-band modeling of the prompt emission combining LAT and GBM data, which indicates the presence of an extra high-energy component needed to describe the data. Finally, we compare \grb with the other LAT-detected GRBs. Summary and conclusions are drawn in Sec.~\ref{sec:conclusion}. In this paper two appendices are provided: App.~\ref{sec:BTI} presents the details of the analysis performed to measure the flux during the brightest part of the GRB; App.~\ref{sec:time_resolved_spectral_analysis} collects additional information on the GBM-LAT joint fit.


\section{Analysis Methods and Procedures}
\label{sec:analysis}
The LAT instrument~\citep{2009ApJ...697.1071A} comprises three subsystems: a tracker with 18 $(x,y)$ tracking planes of silicon strip detectors interleaved with tungsten layers, a segmented calorimeter made of CsI(Tl) crystals and an anticoincidence detector (ACD) made of plastic scintillator tiles. The tracker induces $e^+e^-$ pair conversion of the incident gamma ray and measures its direction. The energy of the electromagnetic shower induced by the $e^+e^-$ pair is measured in the calorimeter. The energy of the incident gamma-ray is estimated using information from both the tracker and calorimeter. The role of the ACD is to reject the charged cosmic-ray background.
Due to the peculiar brightness of this GRB, the X-ray and soft gamma-ray ($\mathrm{E} \lesssim 30~\mathrm{MeV}$) flux was so intense during the brightest part of the emission that it produced a very high level of \nob{additional hits and energy deposits in all three subsystems}~\citep{2022GCN.32760....1O}. Because the instrument and the event reconstruction were not designed to deal with such conditions, we have to carefully identify the time interval \nob{(flagged as BTI)} during which standard analysis methods cannot be used, either \texttt{fermitools}\footnote{\url{https://fermi.gsfc.nasa.gov/ssc/data/analysis/software/}} or the LAT Low Energy Event technique, LLE \citep{PelassaLLE, LLESolarFlare}, which require that the \nob{standard} instrument response functions (IRFs) correctly model the instrument response.


\subsection{Dedicated data analysis of the brightest part of the GRB prompt emission} 
\label{sec:BTI_prescriptions}
Shortly after launch, it was realized that a significant fraction of the recorded events\footnote{Each time the LAT instrument triggers on an incident particle and the corresponding signals in the subdetectors are measured, all the related information is recorded and it is named an event.} contained, in addition to the signals due to the triggering particle, the remnants of the signals due to a particle that passed through the instrument a few $\mu \mathrm{s}$ before the trigger was issued~\citep{2012ApJS..203....4A}. Part of the Pass~8 effort~\citep{2013arXiv1303.3514A} to improve the event reconstruction was dedicated to mitigate the effect of these so-called ghost signals. 
It required modifying the instrument simulation to correctly take into account the presence of these spurious out-of-time signals. Together with the electronic noise, they constitute the background noise with respect to the signals of the triggering particle. Hence we refer to these spurious out-of-time signals as an additional noise. This noise was actually captured with events that are recorded independently of the activity in the instrument thanks to a 2~Hz {\tt PERIODIC} trigger (PT). Almost all of these ghosts correspond to only one out-of-time particle. Therefore, the event reconstruction and selection were designed to be as insensitive as possible to this additional noise.

\begin{figure*}[t]
    \centering    
    \includegraphics[width=0.9\linewidth]{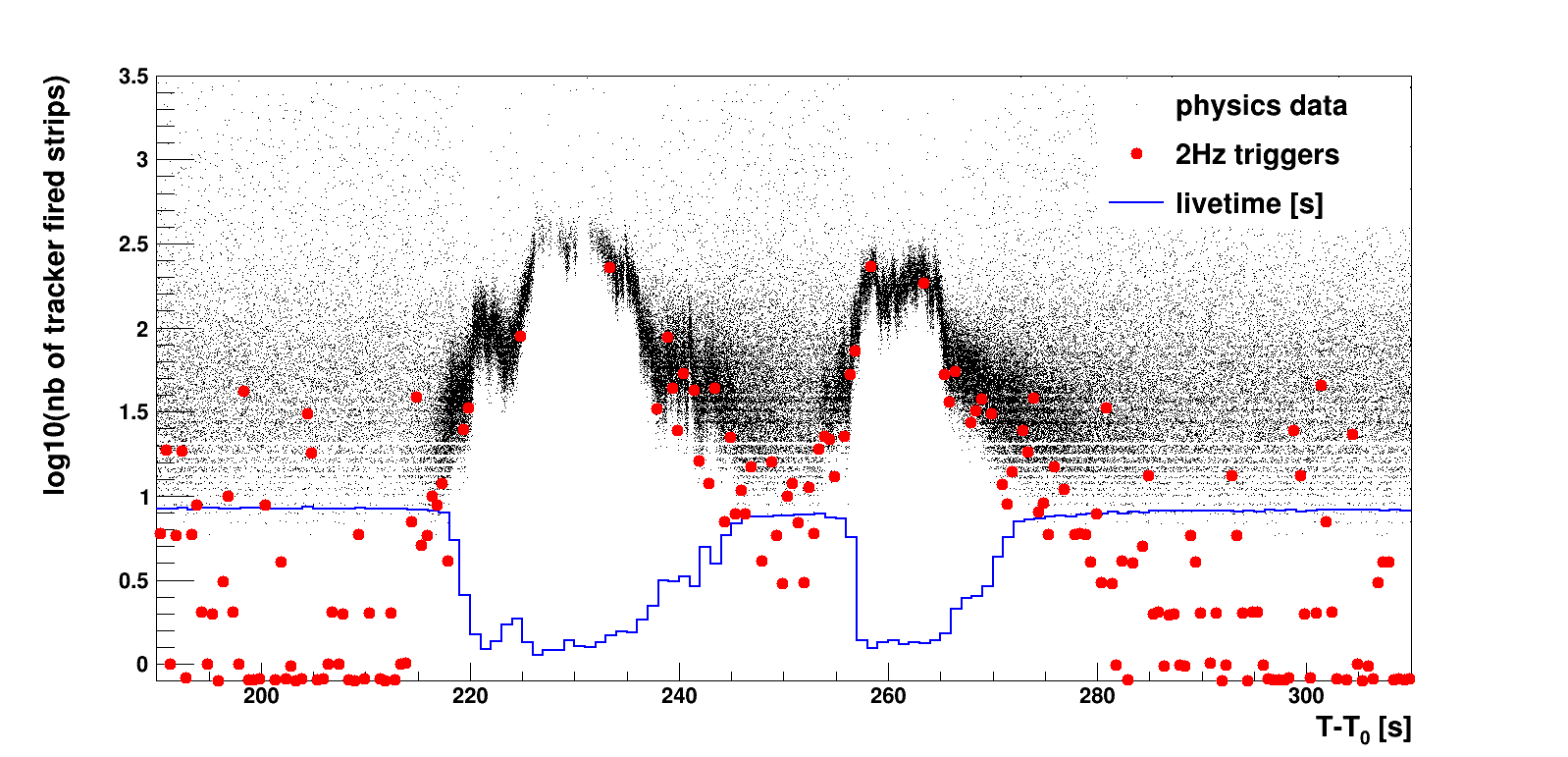}
    \caption{Number of hit strips in the tracker in the events with at least one track that pass the gamma-ray main trigger and on-board filter as a function of time since the GBM trigger (black dots). The red dots correspond to events triggered with a 2~Hz cadence to monitor the noise (both electronic and physical) in the instrument. For clarity's sake, events with no hit strips are plotted with $y=-0.1$. The blue histogram shows the live time recorded in the 1~second spacecraft file. During normal data taking conditions (before $\mathrm{T}_0 \sim 217~\mathrm{s}$ and after $\mathrm{T}_0 + \sim 280~\mathrm{s}$) the noise in the tracker can be very low (at most one strip), while during the brightest part of the GRB prompt emission it is never empty and can reach more than 100 strips. It can be noted that, because of the very high deadtime during the burst, which is a consequence of the extremely high GRB flux, the periodic trigger 2~Hz cadence could not be sustained.}
    \label{fig:bti_fssc}
\end{figure*}

Fig.~\ref{fig:bti_fssc} shows the number of fired strips in the tracker as a function of time for events with at least one track that passed the main gamma-ray trigger and on-board filter, as well as for the PT events\footnote{See also \url{https://fermi.gsfc.nasa.gov/ssc/data/analysis/grb221009a.html}}. 
Both before (\trig + $\sim 217~\mathrm{s}$) and after (\trig + $\sim 280~\mathrm{s}$) the main episode of the GRB, there are a few PT events with more than 10 strips that correspond to the remnants of out-of-time cosmic rays, but most of the PT events contain less than 10 strips (the majority of them having no signal in the tracker), which characterizes normal conditions of data taking.
On the contrary, during most of the time interval [\trig + $\sim 217~\mathrm{s}$, \trig + $\sim 280~\mathrm{s}]$ the number of fired strips in PT events is greater than 10 and it can even exceed 100. Due to the very high deadtime, only a few PT events could be recorded during the brightest parts of the burst. However, we can use the bottom edge of the main-trigger event distribution (black dots in Fig.~\ref{fig:bti_fssc}) to estimate the noise in the tracker as a function of time: it exceeds 100 during two peaks of activity, for T$-$\trig between 225 and 236~s and between 257 and 265~s. The intense X-ray and soft gamma-ray flux of the GRB also affected the calorimeter and the anti-coincidence detector. As a consequence, the event reconstruction and selection are strongly impacted and the standard Instrument Response Functions (IRFs) cannot be expected to be valid during all the prompt emission.

A detailed assessment of the effect of the X-ray and soft gamma-ray flux on the LAT is presented in App.~\ref{sec:BTI}, as well as the dedicated analysis that we develop to be able to measure the flux of the GRB during its brightest period, which includes a special event reconstruction and selection. Here we summarize the main findings. \nob{In the following, because we need to follow the 1~s binning of the information stored in the LAT pointing and livetime history file and because the GRB happened to reach the LAT out of phase by 0.4~s with respect to this 1~s binning, the times with respect to the GRB trigger are given with a decimal part of 0.6~s.}

Regarding the standard energy measurement, we find that the noise contamination is greater than 10~MeV during almost all the time interval [\trig + $ 217.6~\mathrm{s},\mathrm{T}_0+273.6~\mathrm{s}]$ and that it is beyond 100~MeV during [\trig + $ 225.6~\mathrm{s},\mathrm{T}_0+235.6~\mathrm{s}]$ and [\trig + $ 257.6~\mathrm{s},\mathrm{T}_0+264.6~\mathrm{s}]$. We use the fraction of events with some fired strips in the tracker top corner facing the GRB direction to monitor the level of noise. It is $~2$\% during normal conditions and we use the times when it becomes $>3.5$\% and then $<3.5$\% to define the start and end of the BTI: [\trig + $ 216.6~\mathrm{s},\mathrm{T}_0+280.6~\mathrm{s}]$.
    
As explained in App.~\ref{sec:BTI}, to overcome the impact of the extra noise during this time interval, we use an energy estimator based only on the tracker information around the track. We select events passing the trigger and the on-board filter, with at least one track. Thanks to Earth-limb data and multi-photon simulations, we estimate the selection efficiency as a function of the noise level, and thus as a function of time. We require the event energy to be greater than 160~MeV so that the false positive rate is low enough. We calculate the GRB flux in 5~s intervals by fitting the distribution of the angular separation to the GRB with templates based on data and simulation.

\subsection{The fitting procedure} 
\label{sec:fitting}
To perform all the fits in this analysis, except during the BTI, we used \ThreeML\citep{3ML}, a python-based software package for parameter inference using the maximum likelihood formalism (with Bayesian posterior sampling supported as well). 
The likelihood calculation, including an interface to instrument response files and reading in data, is encapsulated in plugins, each corresponding to a certain instrument or data format, allowing joint likelihood fits of datasets recorded by different instruments. 
GBM and LLE data can be imported in \ThreeML using the \texttt{OGIPLike} plugin, which supports any type of file format based on an OGIP\footnote{OGIP stands for the Office of Guest Investigators Program, which established conventions for FITS files for high-energy astrophysics projects.} standard, while for the LAT dataset we used two different plugins, \texttt{FermiLATLike} for un-binned analysis (which interfaces with \gtburst\footnote{\url{https://fermi.gsfc.nasa.gov/ssc/data/analysis/scitools/gtburst.html}}, part of the \fermitools), and \texttt{FermipyLike} for binned likelihood analysis via and interface to \fermipy \citep{2017ICRC...35..824W}. 
Depending on the instrumental conditions and on the viewing angle of \grb, we combine different datasets and use different plugins. In all our analyses we use the best-known localization R.A. = 288\degr.264587, Dec. (J2000) = 19\degree.773397 reported by \citet{BOAT_RADIOPOSITION}.
%
%
%
\section{Results}
\label{sec:results}
\begin{figure*}[t]
    \centering    
    \includegraphics[width=5cm,height=13cm]{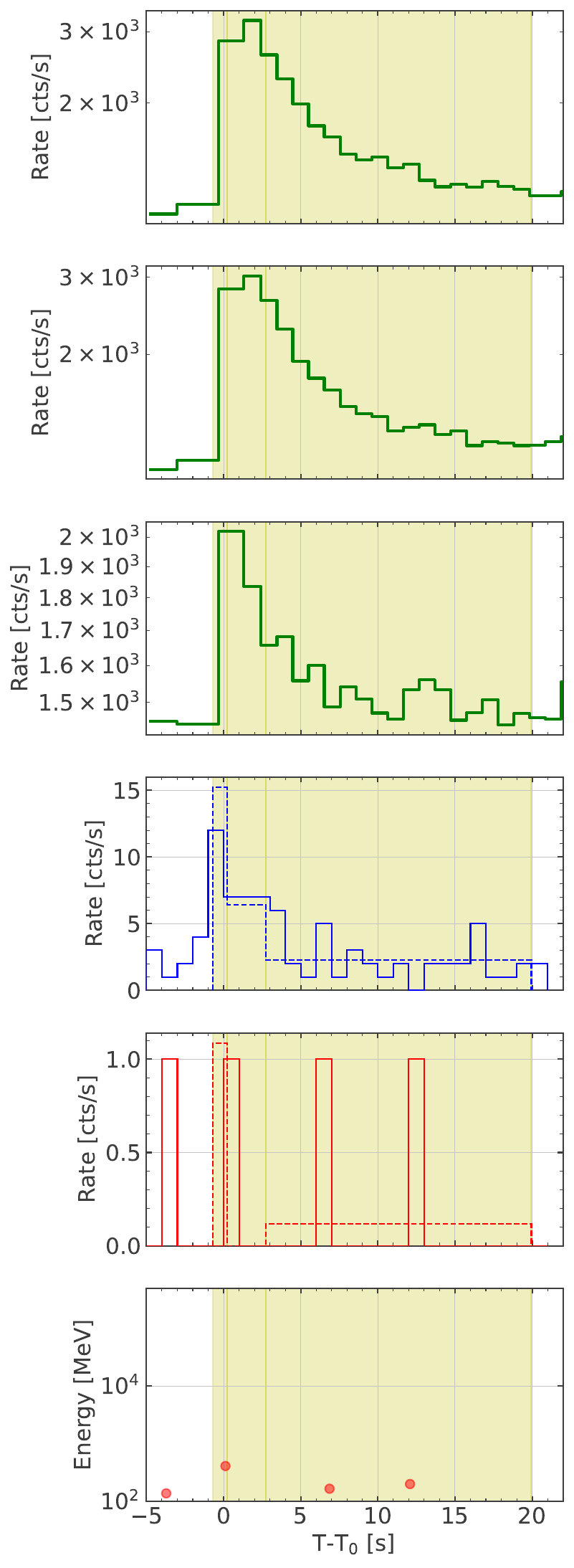}
    \includegraphics[width=12cm,height=13cm]{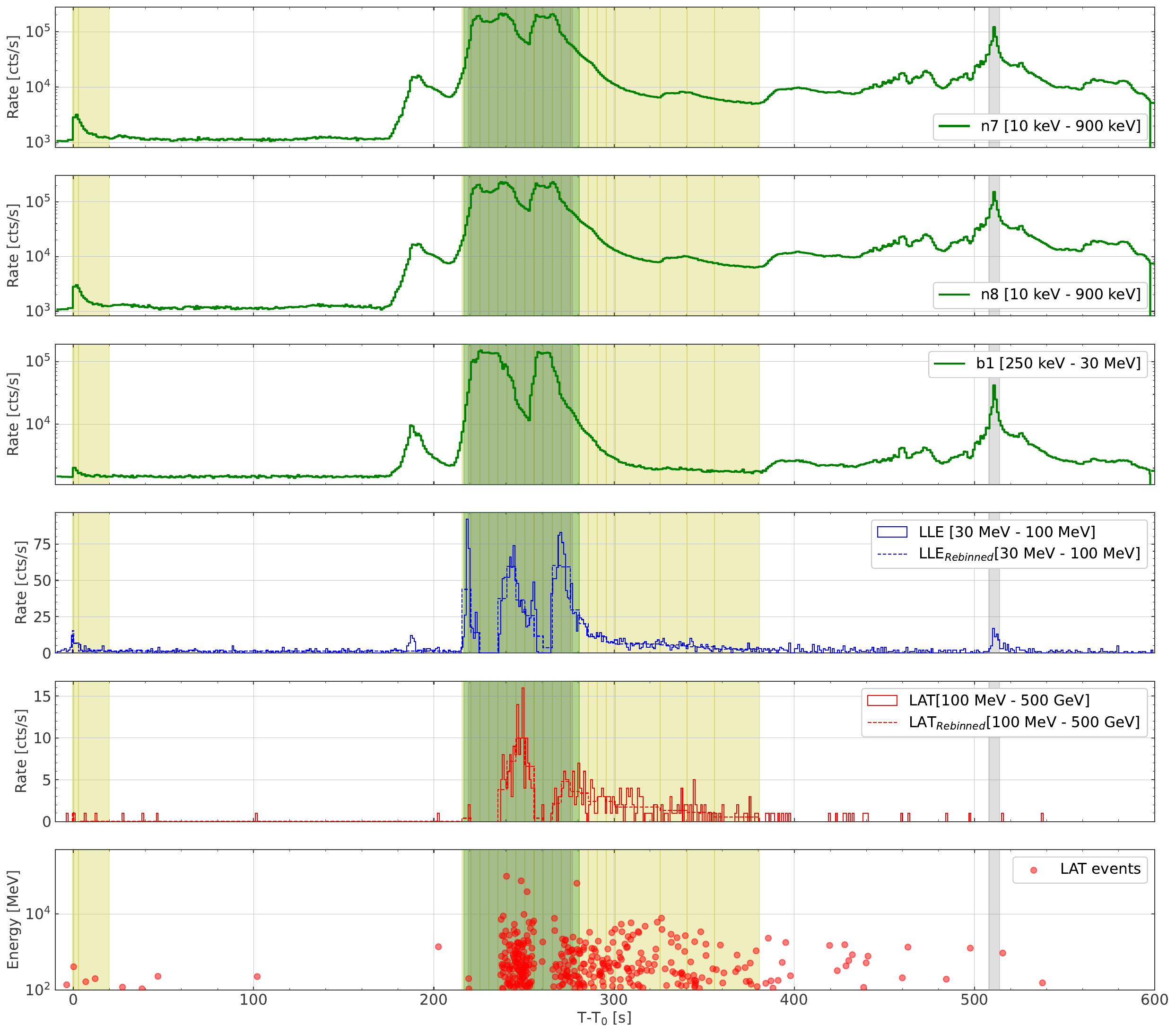}
    \caption{Left: light curve of the first 20 seconds after the GBM trigger (\trig) demonstrating the bright triggering pulse; 
    Right: light curve of the entire 600 s interval after \trig showing the main emission across the broad gamma-ray band including the shaded regions indicating the BTI: green indicates the LAT BTI, and the two gray areas, one partially overlapping the green area, are the GBM BTIs \citep{2023ApJ...952L..42L}. Yellow shows the intervals selected for joint spectral analysis of the triggering pulse, and of the prompt emission (numerical values in Tab.\,\ref{tab:lat_flux}). In both figures, the LAT and the LLE light curves are also shown re-binned using these intervals (blue and red dashed line, respectively) while, in the last panel, the reconstructed event energy is displayed as a function of the arrival time. The GBM light curves (first three panels) are not corrected for the pulse pile-up effects.}
    \label{fig:prompt_lc}
\end{figure*}

Fig.~\ref{fig:prompt_lc} shows the \Fermi broad band light curve, with two NaI detectors (n4, n7), a BGO detector (b1), the LLE and LAT light curves, as well as the LAT \texttt{TRANSIENT\_010E} events with energies greater than 100\,MeV. We identify as the triggering pulse the emission episode from \trig to \trig + 20 s (left panel), while the entire prompt emission lasting approximately 10 minutes is shown in the right panel.

\subsection{The triggering pulse}
\label{sec:precursor}
 We first subdivide this interval by performing a Bayesian Blocks analysis \citep{2013ApJ...764..167S} on LLE data \citep{PelassaLLE,LLESolarFlare}, obtaining 3 time intervals. 
In order to perform a spectral analysis over a wider energy range, we included also low-energy ($<10$\,MeV) data collected by the GBM and followed the prescriptions of the GBM Team \citep{2023ApJ...952L..42L} in terms of detectors and data format selection, avoiding data types possibly affected by pulse pile-up effects. We used data from detectors n4, n7 and b1, which have the best viewing angles with respect to the source, and are not affected by any blockage by the spacecraft or solar panels, in their nominal energy ranges. In particular, we analyzed TTE \citep[Time-Tagged Event,][]{2009arXiv0908.0450M} data in the time interval prior to the main emission episode (from \trig $-$ 0.69\,s to \trig + 19.95\,s, the triggering pulse analysis). We used detector response files (RSP files) specifically produced by the GBM Team  \nop{for \grb} \citep{2023ApJ...952L..42L}.
For the spectral analysis of the triggering pulse we used LLE data from 30 MeV to 100 MeV, and \texttt{TRANSIENT\_010E} ($>100$\,MeV) event class LAT data\footnote{We use the \texttt{P8R3\_TRANSIENT010E\_V3} instrument response functions.}.

First, we tested different phenomenological spectral models, including a simple power law (\texttt{PL}), \texttt{Band}, Comptonized (\texttt{COMP}) -- as defined in the GBM GRB spectral catalog \citep{Poolakkil_2021} -- and a double smoothly broken power law \citep[\texttt{2SBPL},][]{2018A&A...613A..16R}. 
We also tested for the presence of an extra \texttt{PL} for the \texttt{COMP}, \texttt{Band} and \texttt{2SBPL} models. To compare different models, we used the Bayesian Information Criterion, defined as: BIC = $k\ln(N)-2\ln(\mathcal{L})$, where $k$ is the number of free parameters of the model, $N$ is the number of data points (equivalent to the number of spectral bins in a binned analysis or to the number of events in an un-binned analysis), and $\mathcal{L}$ is the value of the likelihood. In Tab.~\ref{tab:bic} we summarize the values for the BIC of the different models tested. The models that best describe the triggering pulse with the lowest BIC are \texttt{COMP} and \texttt{Band}, but the parameters of the Band model are not well constrained in all time intervals. 
In particular, the first two time bins result in a Band parameter $\beta$ value of $-4.0$, which is extremely soft, and an unconstrained $E_{\rm peak}$. In the last interval, the value of $\beta$ is compatible with the results for the \texttt{COMP} model. As a result of this comparison, we conclude that \texttt{COMP} is the preferred model for describing the triggering pulse.

\begin{figure}[t]
    \centering    
    \includegraphics[height=6.5cm]{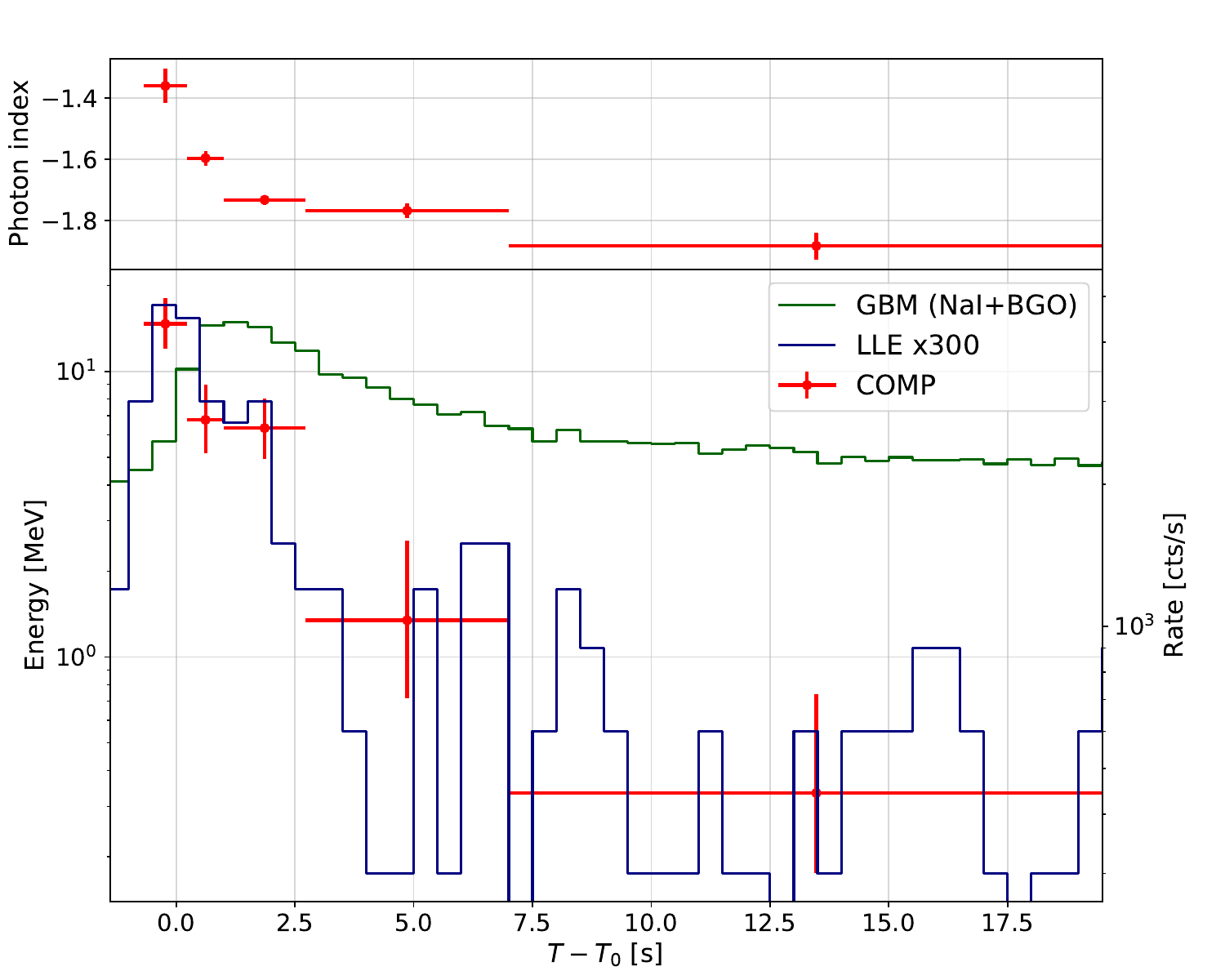}
    \includegraphics[height=6.3cm]{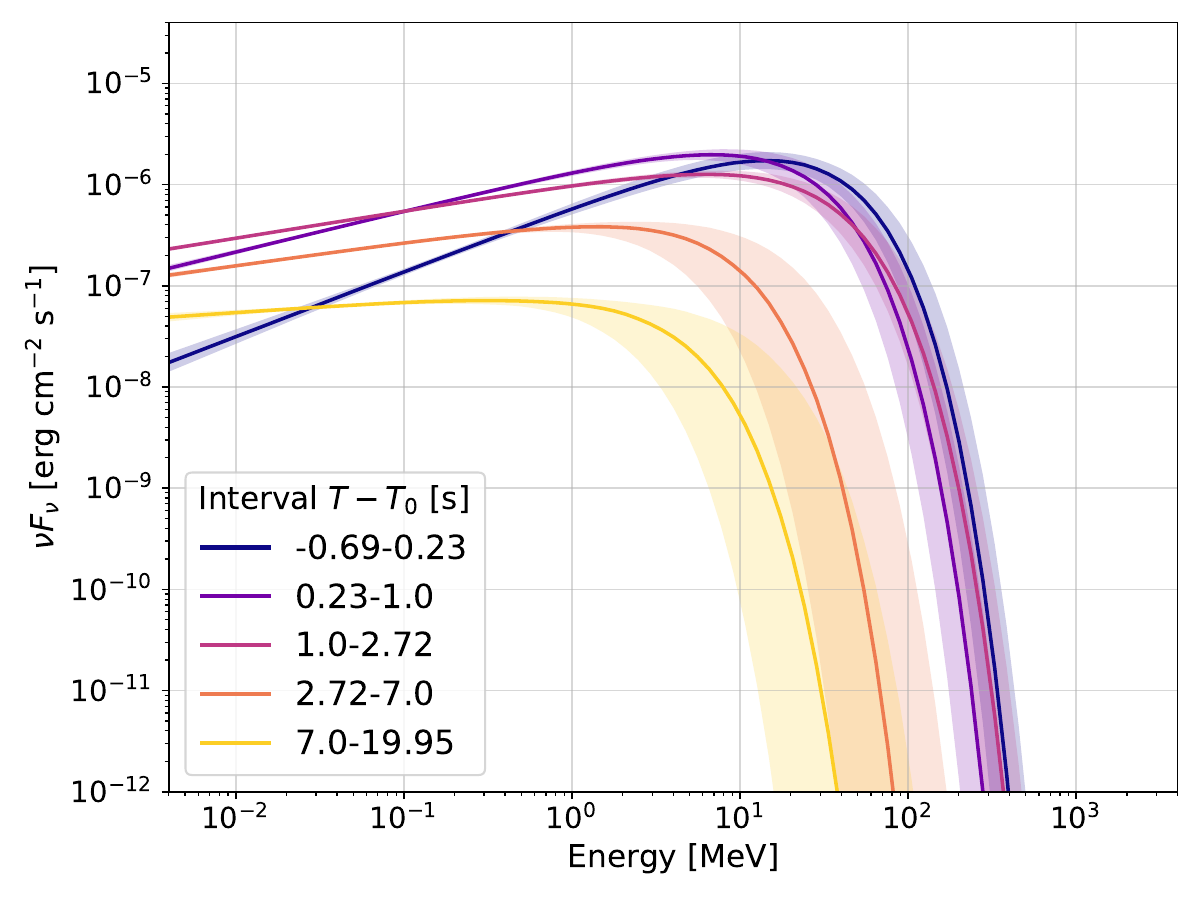}
    \caption{Left: Temporal evolution of the spectral fit parameters during the triggering pulse. Top panel: low-energy power law index; bottom panel: $E_{\rm peak}$. GBM and LLE light curve are also displayed and referenced to the right-hand y-axis.  Right: Time resolved $\nu$F($\nu$) spectrum from $-0.69$ to 19.95 seconds with shaded regions representing the 68\% confidence levels. \nop{The emission is extremely energetic in the first temporal bin, featuring a very high $E_{\rm peak}$ and a hard spectral index, and it softens over time.}}
    \label{fig:precursor}
\end{figure}

We further divided the last two intervals based on the GBM count rate in order to track the parameters of the \texttt{COMP} model on a finer timescale. In the left panel of Fig.~\ref{fig:precursor} we show the temporal evolution of the photon index and $E_{\rm peak}$ over the GBM and LAT LLE light curves. The values are listed in Tab.~\ref{tab:cparam}. In the first figure
of Sec.~\ref{sec:time_resolved_spectral_analysis} we show the spectral fits for all time bins. The emission is very energetic in the first temporal bin with a very high $E_{\rm peak}$ and a hard spectral index, and becomes softer at later times. The right panel of Fig.~\ref{fig:precursor} shows the spectral evolution of the model. From \trig+ 19.95 to \trig+ 175\,s, there is no detection of the GRB in either standard LAT nor LLE data, and therefore we derive a flux upper limit using standard LAT \texttt{TRANSIENT\_010E} event class (Tab.~\ref{tab:lat_flux}).

\begin{table*}[]
    \centering
    \begin{tabular}{r | c c c c c c c }
             Interval (s)   &  \multicolumn{7}{c}{BIC values}\\
              (s from \trig)   & \texttt{PL} & \texttt{COMP} & \texttt{Band}  & \texttt{2SBPL} & \texttt{COMP+PL} & \texttt{Band+PL}  & \texttt{2SBPL+PL} \\
        \hline
        \hline
        --0.69 -- 0.23 &   798 & \textbf{742} &  749 & 767  & 754	  & 761	  & 781   \\
        0.23 -- 2.72  &  2182 & \textbf{1927} & 1928 & -- & 1937 & 1939  & --\\
        2.72 -- 19.95 &  4032 & \textbf{3913} & 3918 & -- & 3922 & 3928  & -- \\

        \hline        
    \end{tabular}
    \caption{Values for the BIC model selection estimator for the three time intervals of the triggering pulse obtained with Bayesian Blocks analysis. In bold we highlight the values for the lowest BIC in each interval. The \texttt{COMP} model is preferred in all intervals.}
    \label{tab:bic}
\end{table*}

\begin{table}[]
    \centering
    \begin{tabular}{r c c }
        Interval  & index & $E_{\rm peak}$ [MeV]\\
        \hline
        \hline
        --0.69 -- 0.23 & $-1.36 \pm 0.06 $  & $15 \pm 3 $ \\ 
        0.23 -- 1.00 & $-1.60 \pm 0.02 $ & $7 \pm  2 $ \\ 
        1.00 -- 2.72 & $-1.73 \pm 0.02 $ & $6 \pm 2 $\\ 
        2.72 -- 7.00 & $-1.77^{+0.03}_{-0.02}$ & $1.3^{+0.6}_{-1.2}$ \\ 
        7.00 -- 19.95 & $-1.88^{+0.04}_{-0.05}$ & $0.33^{+0.02}_{-0.04}$ \\ 

        \hline
    \end{tabular}
    \caption{Parameter values for the \texttt{COMP} model in each interval of the triggering pulse.}
    \label{tab:cparam}
\end{table}

\subsection{The Prompt Phase}
\label{sec:prompt}
\begin{figure*}[t]
    \centering    
    \includegraphics[width=\textwidth]{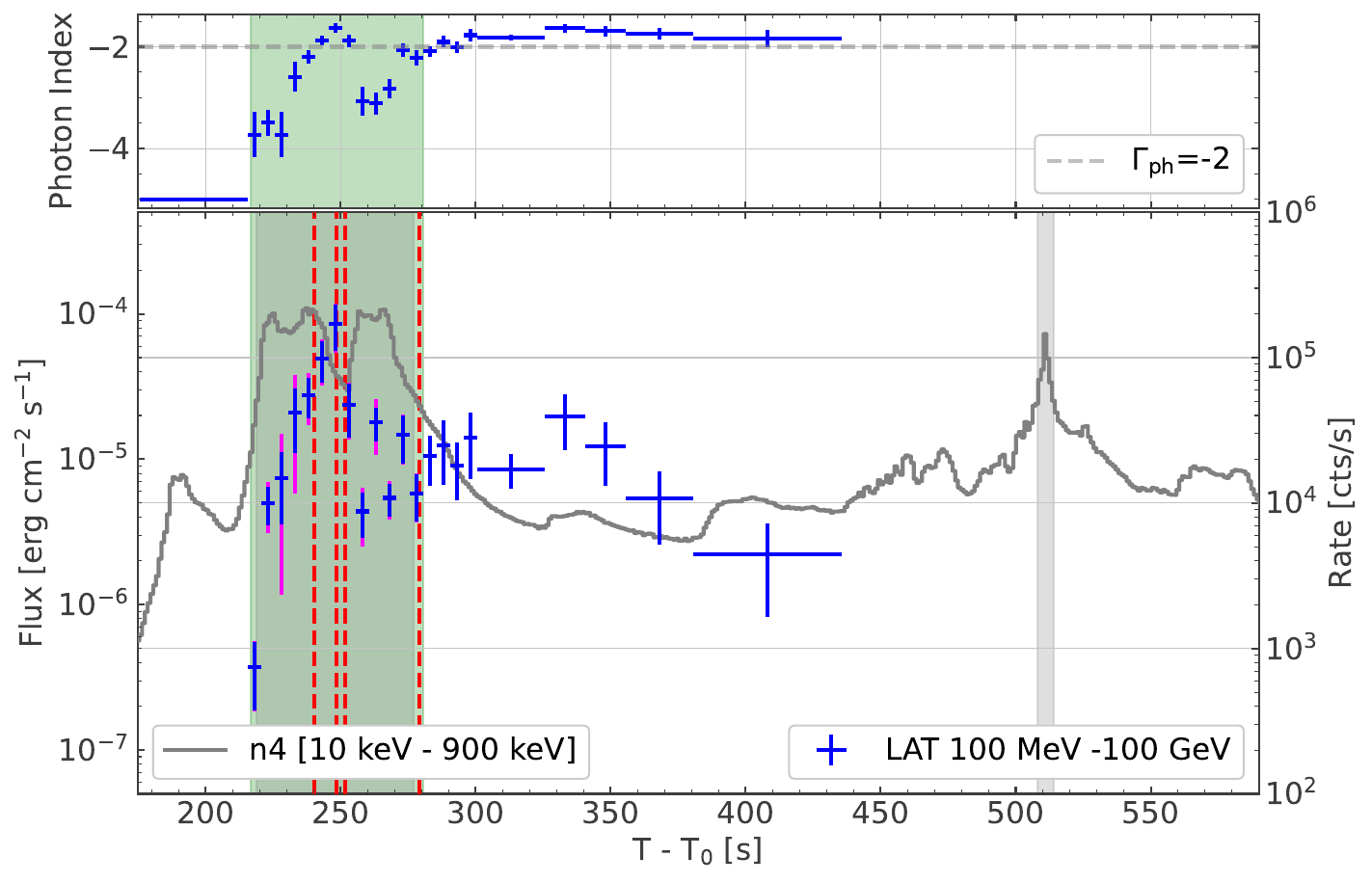}
    \caption{Fermi high-energy emission between 175\,s and 590\,s displaying the photon index (top panel) and the flux between 100~MeV and 100~GeV measured by the \Fermi LAT (bottom panel). 
    Magenta error bars indicate the sum of the statistical and systematic uncertainties estimated during the BTI. The gray curve is the rate of events in the GBM detector n4 \nob{(not corrected for pulse pile-up effects)} with numerical values on the right-hand vertical axis. The vertical red dashed lines correspond to the arrival times of the $>$ 10 GeV photons while the shaded regions indicate the BTI: green indicates the LAT BTI, and the two gray areas are the GBM BTIs \citep{2023ApJ...952L..42L}.}
    \label{fig:flux_prompt}
\end{figure*}

During the prompt phase (from \trig+ 175\,s to \trig+ 600\,s) we used data from different detectors for different time intervals. 
From \trig+ 175\,s to \trig+ 215.6\,s and from \trig+ 280.6\,s to \trig+ 435\,s we used LLE data from 30~MeV to 100~MeV and \texttt{TRANSIENT\_010E} above 100 MeV. We jointly fitted them using unbinned likelihood for the standard LAT event class with \ThreeML. During [\trig + $ 215.6~\mathrm{s},\mathrm{T}_0+280.6~\mathrm{s}]$, corresponding to the BTI, we apply the analysis described in App.~\ref{sec:BTI}. At \trig+ 435\,s the GRB has a large incidence angle ($\sim$80\deg), and, even if the pulse at \trig+ 510\,s is clearly visible in the LLE event selection, the instrument responses are not well modeled, so we did not perform any spectral analysis. 
In each of these time windows, we divided and analyzed the data into several intervals. Fig.\,\ref{fig:flux_prompt} shows the flux light curve in the $0.1-100$\,GeV energy band and Tab.\,\ref{tab:lat_flux} lists the best model used (\texttt{COMP} or \texttt{PL}), the type of data used (GBM, LLE and standard LAT analysis), the numerical values for the energy flux, the estimated photon index $\Gamma_{\rm ph}$, and the value of the Test Statistic (TS). 
Since the TS is defined as $-2\times(\log(\mathcal{L}_{0}) - \log(\mathcal{L}_{1}))$, with $\mathcal{L}_{1}$ the value of the likelihood and $\mathcal{L}_{0}$ the value of the likelihood when the source (in this case the GRB) is removed, it is only available when the null hypothesis is defined.
This is the case for LAT-only analysis, for which we modeled the background components, and not in case of GBM- or LLE-only analysis, where background-subtracted data were analyzed.

%

The first GBM BTI ([\trig + $  219\,\mathrm{s},\mathrm{T}_0+ 277\,\mathrm{s}]$) is similar to the LAT BTI. During this main part of the GRB prompt phase, the emission in the GBM energy range exhibits two very bright peaks at about \trig+ 230\,s and \trig+ 260\,s~\citep{2023ApJ...952L..42L}, which is confirmed by the noise monitoring in the LAT instrument reported in App.~\ref{sec:LEspectrum}. 
The emission above $\sim100$\,MeV, on the other hand, has only one peak, whose maximum occurred during [\trig + $  245\,\mathrm{s},\mathrm{T}_0+ 250\,\mathrm{s}$], with a spectral index $\Gamma_{\rm ph} = -1.6 \pm 0.1$. 
After this peak, the emission became softer ($\Gamma_{\rm ph} \sim -3$) for 15\,s and then, between \trig+ 270\,s and \trig+ 435\,s, the power law became quite hard again, compared to other \lat GRBs, peaking at $\Gamma_{\rm ph} =-1.6 \pm 0.1$ between \trig+ 325\,s to \trig+ 355\,s. This rapid variation of the spectral index during the prompt emission and the striking variability observed in the LAT flux, with multiple episodes up to 400~s after \trig, is a new feature not observed in other LAT-detected GRBs, and clearly indicates the detection of the prompt emission at high energies.

The probability that an event is \nob{spatially and spectrally} associated with the GRB was calculated following \citet{2FLGC}, by using \texttt{gtsrcprob} after running the un-binned likelihood analysis in each time bin. During the BTI, we only used events with energies $>$ 1\,GeV to limit the effect of the noise induced by low-energy events. Four events with reconstructed energy greater than 10\,GeV and probability $\sim$ 1 were recorded by the LAT and are summarized in the first part of Tab.\,\ref{tab:he_photons}. 
The highest-energy event had an energy of 99\,GeV, the highest photon energy recorded so far by the LAT during the prompt emission of a GRB.
Measurement of the energies of these events, each of which was greater than 10\,GeV, was unaffected by the extra noise in the calorimeter, for which the average energy deposition per event was $<$\,100 MeV (see App.~\ref{sec:energy_estimation} and Fig.\ref{fig:cal_cluster_energy}).

\subsection{Late Time Emission}
\label{sec:afterglow}
\begin{figure*}[t]
    \centering    
    \includegraphics[width=\textwidth]
    {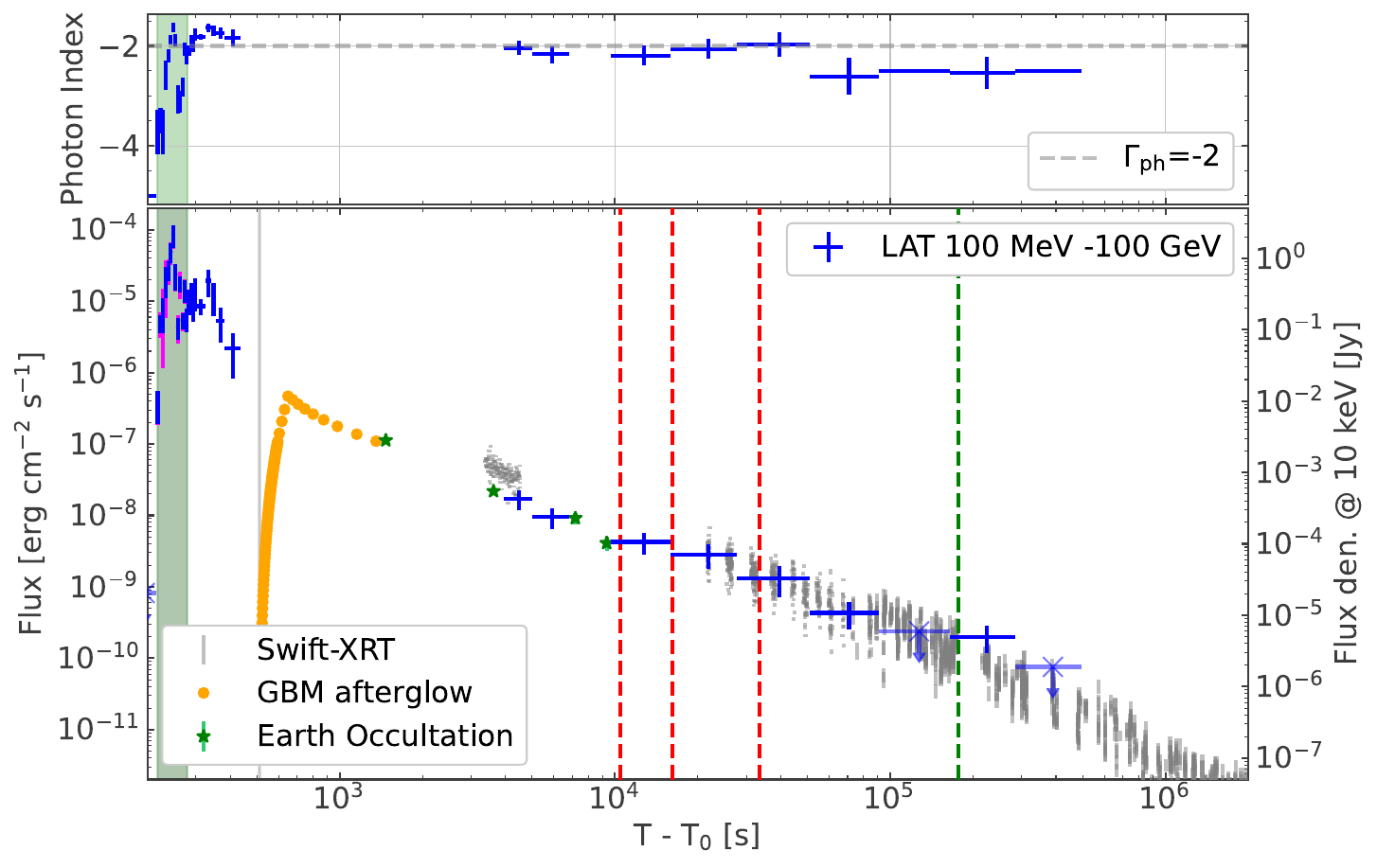}
    \caption{Late-time emission observed by \Fermi and \Swift. Top panel: photon index \nob{of the LAT emission, in blue}. 
    Bottom panel: flux between 100\,MeV and 100\,GeV (blue) with, in magenta, the sum of the statistical and systematic uncertainties during the BTI. 
    Orange points, green stars and gray dots are the \Fermi GBM and \Swift XRT afterglow measurements (flux density at 10\,keV) obtained by \citet{2023ApJ...952L..42L} with values reported on the right-hand vertical axis. The vertical red dashed lines correspond to the arrival times of the $> 10$\,GeV photons while the green dashed line is the arrival time of the last event with probability $>$ 0.9 to be associated with \grb. As in the previous figure, the shaded regions indicate the BTIs.}
    \label{fig:flux_extended}
\end{figure*}

For the late-time emission, we started our analysis at \trig + 3917.6\,s, when the GRB reentered the LAT field of view. 
First, we computed the exposure as a function of time. Considering that the typical decay time for GRB gamma-ray emission goes as T$^{-1}$, we multiplied the exposure by $\rm (T-T_0)^{-1}$, providing us with a proxy for defining a temporal binning with an approximately constant number of expected events. 
Until \trig + 40\,ks we performed unbinned likelihood analysis with \texttt{P8\_TRANSIENT010E} class events, while between \trig + 40\,ks and \trig + 10$^6$\,s, given the longer exposures, we switched to binned likelihood analysis using \fermipy in \ThreeML with \texttt{P8\_SOURCE} class events, which has a lower contamination from misclassified cosmic rays.
In each time bin, we included all the 4FGL \citep{4FGL} sources with a predicted number of events $>$ 1, estimated by using the photon flux above 100 MeV in the 4FGL catalog multiplied by the bin exposure (in cm$^{2}$\,s). We left their normalizations free to vary\footnote{In our analysis we used the 4FGL-DR3 release of the 4FGL catalog.}.
Three events with energies greater than 10\,GeV arrived during this phase, as summarized in Tab.\,\ref{tab:he_photons} and in Fig.\,\ref{fig:flux_extended}. In particular, the LAT recorded an event with an energy of 400\,GeV \citep{2022GCN.32748....1X} arriving 33\,ks after the GBM trigger. Its origin and association with \grb will be discussed in Sec.\,\ref{sec:400GeV}. The light curve for \grb is displayed in logarithmic scale in Fig.\,\ref{fig:flux_extended} with numerical values in the last part of Tab.\,\ref{tab:lat_flux}.

To compute the onset time and the duration of the 100\,MeV--100\,GeV flux light curve, we used the same algorithm developed in \citet{2FLGC}: First, we calculated the probability for each photon to be associated with the GRB using the result of the likelihood analysis (using \texttt{gtsrcprob}). 
The onset time \tz corresponds to the time when the first photon with probability $p > 0.9$ to be associated with the GRB is detected, while \tone corresponds to the last photon with $p > 0.9$. \toz of the signal is simply \tone$- $\tz. 
To estimate the uncertainty on \tone ($\delta$\tone) for an event with $n$ detected photons with probability $p > 0.9$, we define $\Delta t_{n-1,n}$ as the time interval between the second to last and the last event. Assuming Poisson statistics, the probability of measuring an event between $t$ and $t+dt$ is $P(t,t+dt) = \lambda\,dt$, where $\lambda$ is the rate: in our case $\lambda = 2/\Delta t_{n-1,n}$. Therefore, we conservatively compute the uncertainty as $\delta$\tone$ = 1/\lambda = \Delta t_{n-1,n}/2$. Similarly, considering the first two events with probability $p > 0.9$, we define the uncertainty on \tz as $\delta$\tz$= \Delta t_{1,2}/2$. The error on \toz follows using standard error propagation.
The arrival times of the first and last events of \grb with $p > 0.9$ are \tz=\trig+ 6.86~s and \tone=\trig+ 176673.012~s. 
The uncertainties estimated with the procedure described above, are $\delta$\tz= 106.22\,s and $\delta$\tone= 2822.44\,s, which lead to \toz$=176 \pm 3$\,ks. The first event arrived during the triggering pulse. The second event associated with \grb arrived $\sim$200 seconds later, emphasizing the separate nature of the triggering pulse with the bulk of the GRB emission. The last events arrived more than two days after the GBM trigger, setting a new record on the duration of LAT-detected GRBs.

\newpage
\begin{table}[ht]
\begin{small}
    \centering
    \begin{tabular}{r | c | c @{\hspace{-3pt}} c @{\hspace{-3pt}} c | c | r | r}
        Interval    &  Best Model & & Data Used &  & Flux [erg cm$^{-2}$ s$^{-1}$] & $\Gamma_{\rm ph}$ & TS \\
    (T-\trig) [s] &             &  GBM  & LLE   & LAT            &  0.1--100.0 GeV               &   \,      &    \\
        \hline
        \hline
        \multicolumn{8}{c}{Triggering pulse}\\
        \hline
  $-$0.69 -- 0.23  &       \texttt{COMP} & X& X& X&                       $<$ 1.0$\times$10$^{-8}$ & - & 0.0 \\
    0.23 -- 2.72  &       \texttt{COMP} & X& X& X&                       $<$2.0$\times$10$^{-9}$ & - & 0.0 \\
    2.72 -- 19.95  &       \texttt{COMP} & X& X& X&                       $<$ 1.0$\times$10$^{-9}$ & - & 5.0 \\
    \hline
    19.95 -- 175.00  &         \texttt{PL} &  &  & X&                       $<$ 3.2$\times$10$^{-8}$ &                               $-$2.5 (fixed) & 7.8 \\
    \hline
\multicolumn{8}{c}{Prompt}\\
\hline
  175.6 -- 215.6  &         \texttt{PL} &  & X& X&                 $<$ 8.0$\times$10$^{-10}$ &                                     $<$ $-$5 & 0.0 \\
  215.6 -- 220.6  &         \texttt{PL} &  &  & X$^{\dagger}$ & (4 $\pm$ 2 ($^{+0.06}_{-0.04}$ sys))$\times$10$^{-7}$ &                           $-$3.7 $\pm$ 0.4 & 83.4 \\
  220.6 -- 225.6  &         \texttt{PL} &  &  & X$^{\dagger}$ & (5 $\pm$ 1 ($^{+0.5}_{-0.4}$ sys))$\times$10$^{-6}$ &                           $-$3.5 $\pm$ 0.3 & 260.8 \\
  225.6 -- 230.6  &         \texttt{PL} &  &  & X$^{\dagger}$ & (7 $\pm$ 4 ($^{+4}_{-2}$ sys))$\times$10$^{-6}$ &                           $-$3.7 $\pm$ 0.4 & 59.6 \\
  230.6 -- 235.6  &         \texttt{PL} &  &  & X$^{\dagger}$ & (2 $\pm$ 1 ($^{+0.7}_{-0.5}$ sys))$\times$10$^{-5}$ &                           $-$2.6 $\pm$ 0.3 & 181.1 \\
  235.6 -- 240.6  &         \texttt{PL} &  &  & X$^{\dagger}$ & (2.8 $\pm$ 0.9 ($^{+0.3}_{-0.2}$ sys))$\times$10$^{-5}$ &                           $-$2.2 $\pm$ 0.1 & 965.4 \\
  240.6 -- 245.6  &         \texttt{PL} &  &  & X$^{\dagger}$ & (5 $\pm$ 2 ($\pm$ 0.2 sys))$\times$10$^{-5}$ &                           $-$1.9 $\pm$ 0.1 & 1346.3 \\
  245.6 -- 250.6  &         \texttt{PL} &  &  & X$^{\dagger}$ & (9 $\pm$ 3 ($\pm$ 0.1 sys))$\times$10$^{-5}$ &                         $-$1.63 $\pm$ 0.09 & 1539.8 \\
  250.6 -- 255.6  &         \texttt{PL} &  &  & X$^{\dagger}$ & (2.4 $\pm$ 0.9 ($^{+0.06}_{-0.04}$ sys))$\times$10$^{-5}$ &                           $-$1.9 $\pm$ 0.1 & 828.1 \\
  255.6 -- 260.6  &         \texttt{PL} &  &  & X$^{\dagger}$ & (4 $\pm$ 2 ($^{+0.4}_{-0.3}$ sys))$\times$10$^{-6}$ &                           $-$3.1 $\pm$ 0.3 & 220.7 \\
  260.6 -- 265.6  &         \texttt{PL} &  &  & X$^{\dagger}$ & (1.8 $\pm$ 0.5 ($\pm$ 0.3 sys))$\times$10$^{-5}$ &                           $-$3.1 $\pm$ 0.2 & 405.4 \\
  265.6 -- 270.6  &         \texttt{PL} &  &  & X$^{\dagger}$ & (5 $\pm$ 1 ($^{+0.3}_{-0.2}$ sys))$\times$10$^{-6}$ &                           $-$2.8 $\pm$ 0.2 & 485.3 \\
  270.6 -- 275.6  &         \texttt{PL} &  &  & X$^{\dagger}$ & (1.5 $\pm$ 0.5 ($^{+0.03}_{-0.02}$ sys))$\times$10$^{-5}$ &                           $-$2.1 $\pm$ 0.1 & 807.8 \\
  275.6 -- 280.6  &         \texttt{PL} &  &  & X$^{\dagger}$ & (6 $\pm$ 2 ($^{+0.09}_{-0.07}$ sys))$\times$10$^{-6}$ &                           $-$2.2 $\pm$ 0.2 & 466.3 \\
  280.6 -- 285.6  &         \texttt{PL} &  & X& X&        (1.0 $\pm$ 0.4)$\times$10$^{-5}$ &                           $-$2.1 $\pm$ 0.1 & 204.0 \\
  285.6 -- 290.6  &         \texttt{PL} &  & X& X&        (1.2 $\pm$ 0.6)$\times$10$^{-5}$ &                           $-$1.9 $\pm$ 0.1 & 138.0 \\
  290.6 -- 295.6  &         \texttt{PL} &  & X& X&            (9 $\pm$ 4)$\times$10$^{-6}$ &                           $-$2.0 $\pm$ 0.1 & 178.0 \\
  295.6 -- 300.6  &         \texttt{PL} &  & X& X&        (1.4 $\pm$ 0.7)$\times$10$^{-5}$ &                           $-$1.8 $\pm$ 0.1 & 148.0 \\
  300.6 -- 325.6  &         \texttt{PL} &  & X& X&            (9 $\pm$ 2)$\times$10$^{-6}$ &                         $-$1.82 $\pm$ 0.06 & 505.0 \\
  325.6 -- 340.6  &         \texttt{PL} &  & X& X&        (2.0 $\pm$ 0.8)$\times$10$^{-5}$ &                  $-$1.63$^{+0.08}_{-0.09}$ & 263.0 \\
  340.6 -- 355.6  &         \texttt{PL} &  & X& X&        (1.2 $\pm$ 0.6)$\times$10$^{-5}$ &                           $-$1.7 $\pm$ 0.1 & 172.0 \\
  355.6 -- 380.6  &         \texttt{PL} &  & X& X&            (5 $\pm$ 3)$\times$10$^{-6}$ &                           $-$1.7 $\pm$ 0.1 & 139.0 \\
  380.6 -- 435.6  &         \texttt{PL} &  & X& X&            (2 $\pm$ 1)$\times$10$^{-6}$ &                           $-$1.8 $\pm$ 0.2 & 42.0 \\

\hline
  \multicolumn{8}{c}{Extended}\\
\hline
 3930 -- 5010  &         \texttt{PL} &  &  & X&        (1.7 $\pm$ 0.6)$\times$10$^{-8}$ &                     $-$2.1$^{+0.2}_{-0.1}$ & 116.3 \\
 5010 -- 6810  &         \texttt{PL} &  &  & X&        (1.0 $\pm$ 0.3)$\times$10$^{-8}$ &                     $-$2.2$^{+0.1}_{-0.2}$ & 100.0 \\
 9630 -- 15870  &         \texttt{PL} &  &  & X&            (4 $\pm$ 1)$\times$10$^{-9}$ &                           $-$2.2 $\pm$ 0.2 & 75.8 \\
 15.87k -- 27.81k &         \texttt{PL} &  &  & X&            (3 $\pm$ 1)$\times$10$^{-9}$ &                           $-$2.1 $\pm$ 0.2 & 67.6 \\
 27.81k -- 51.09k &         \texttt{PL} &  &  & X&        (1.3 $\pm$ 0.6)$\times$10$^{-9}$ &                           $-$2.0 $\pm$ 0.2 & 33.4 \\
 51.09k -- 90.57k &         \texttt{PL} &  &  & X$^{*}$&           (4 $\pm$ 2)$\times$10$^{-10}$ &                           $-$2.6 $\pm$ 0.4 & 13.4 \\
 90.57k --164.13k &         \texttt{PL} &  &  & X$^{*}$&                 $<$ 2.0$\times$10$^{-10}$ &                               $-$2.0 (fixed) & 5.2 \\
164.13k -- 284.49k &         \texttt{PL} &  &  & X$^{*}$&       (2.0 $\pm$ 0.8)$\times$10$^{-10}$ &                           $-$2.5 $\pm$ 0.3 & 10.0 \\
284.49k -- 495.33k &         \texttt{PL} &  &  & X$^{*}$&                 $<$ 8.0$\times$10$^{-11}$ &                               $-$2.0 (fixed) & 3.1 \\
    \hline
    \end{tabular}
    \caption{Results of the time-resolved spectral analysis from the GRB onset to \trig$+500$\,ks, including the best model (\texttt{COMP} or \texttt{PL}), the data used in the fit, the energy flux, the photon index $\Gamma_{\rm ph}$ (only for the \texttt{PL} model), and the Test Statistic (TS).  UL are at 95\% confidence. For the entries marked with $^\dagger$ the results were obtained with the special analysis described in App.~\ref{sec:BTI}. For the entries marked with $^{*}$ we used the \ThreeML \fermipy plugin with \texttt{SOURCE} class.}
    \label{tab:lat_flux}
\end{small}
\end{table}

\begin{table}[]
    \centering
    \begin{tabular}{r c c c c}
        T-\trig &  Energy & Prob. & Conv. type. & Ang. Sep.\\
        (s) &     (GeV) &            &  & (\deg)\\
        \hline
        \hline
  \multicolumn{5}{c}{Prompt}\\
  \hline
   240.336 &       99 &      -- & Back & 0.70$^\dagger$ \\
   248.427 &       75 &      1.000 & Back & 0.05 \\
   251.724 &       39 &      1.000 & Back & 0.25 \\
   279.342 &       65 &      1.000 & Front & 0.19 \\
\hline
  \multicolumn{5}{c}{Extended}\\
  \hline
 10475.104 &       24 &      0.998 & Front & 0.10 \\
 16176.428 &       15 &      0.993 & Front & 0.16 \\
 33552.966 &      398 &      1.000 & Back & 0.02 \\
        \hline
    \end{tabular}
    \caption{\Fermi LAT events with energies $>$\,10 GeV that arrived within one degree ROI centered at the position of the GRB optical counterpart. For the 99\,GeV event, the angular separation from the GRB localization is marked with a $^\dagger$ because the high level of noise in the tracker affected the direction reconstruction. We checked this event carefully and we concluded that its origin is compatible with the GRB location.}
    \label{tab:he_photons}
\end{table}

\section{Discussion}
\label{sec:discussion}
The results presented in the previous section allow us to put some constraints on the modeling and to discuss possible scenarios responsible for the emission at high energies.

\subsection{Modeling the high-energy light curve}
\label{sec:models}

\begin{figure*}[t]
    \centering    
    \includegraphics[width=0.49\linewidth]{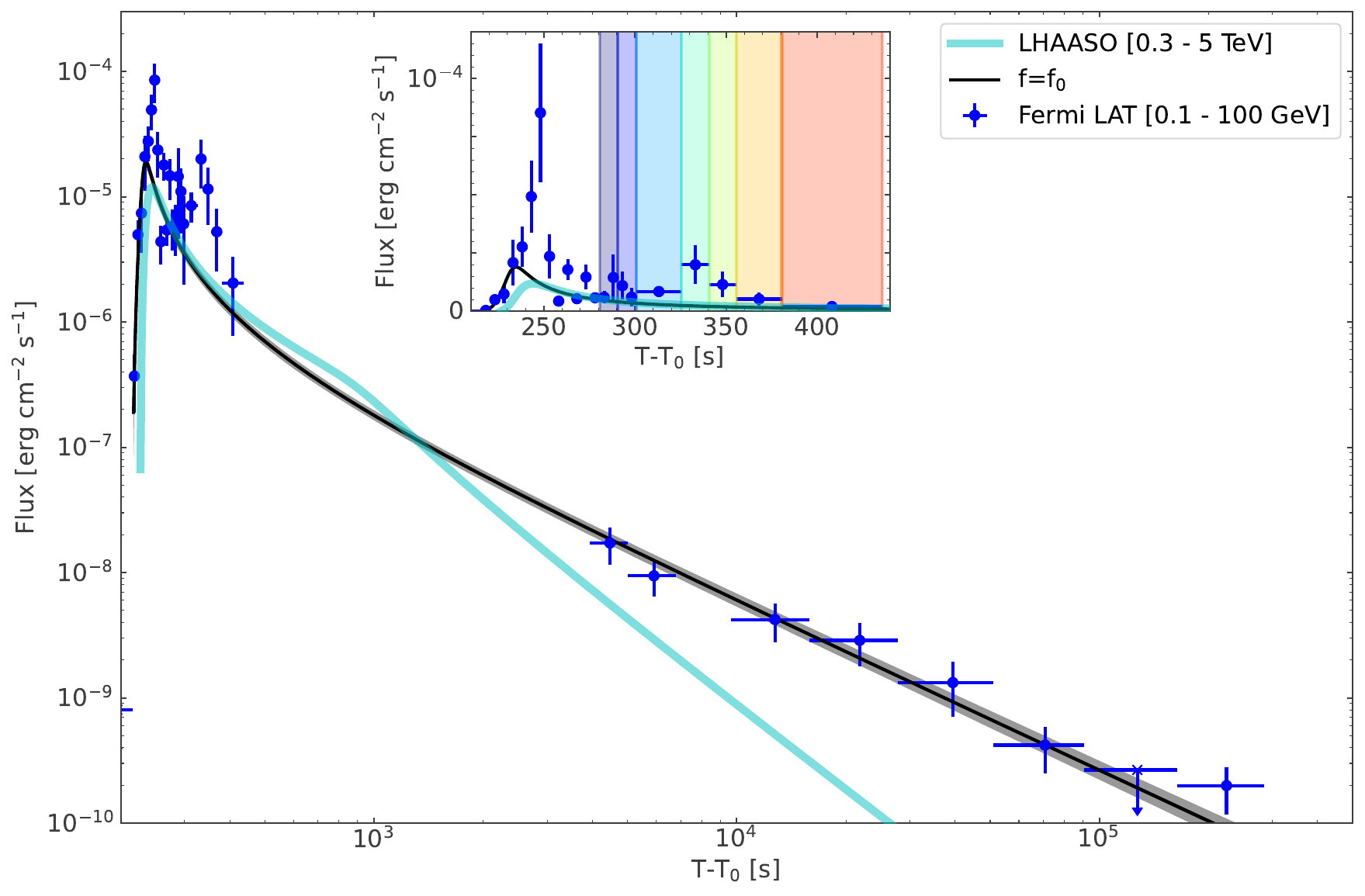}
    \includegraphics[width=0.49\linewidth]{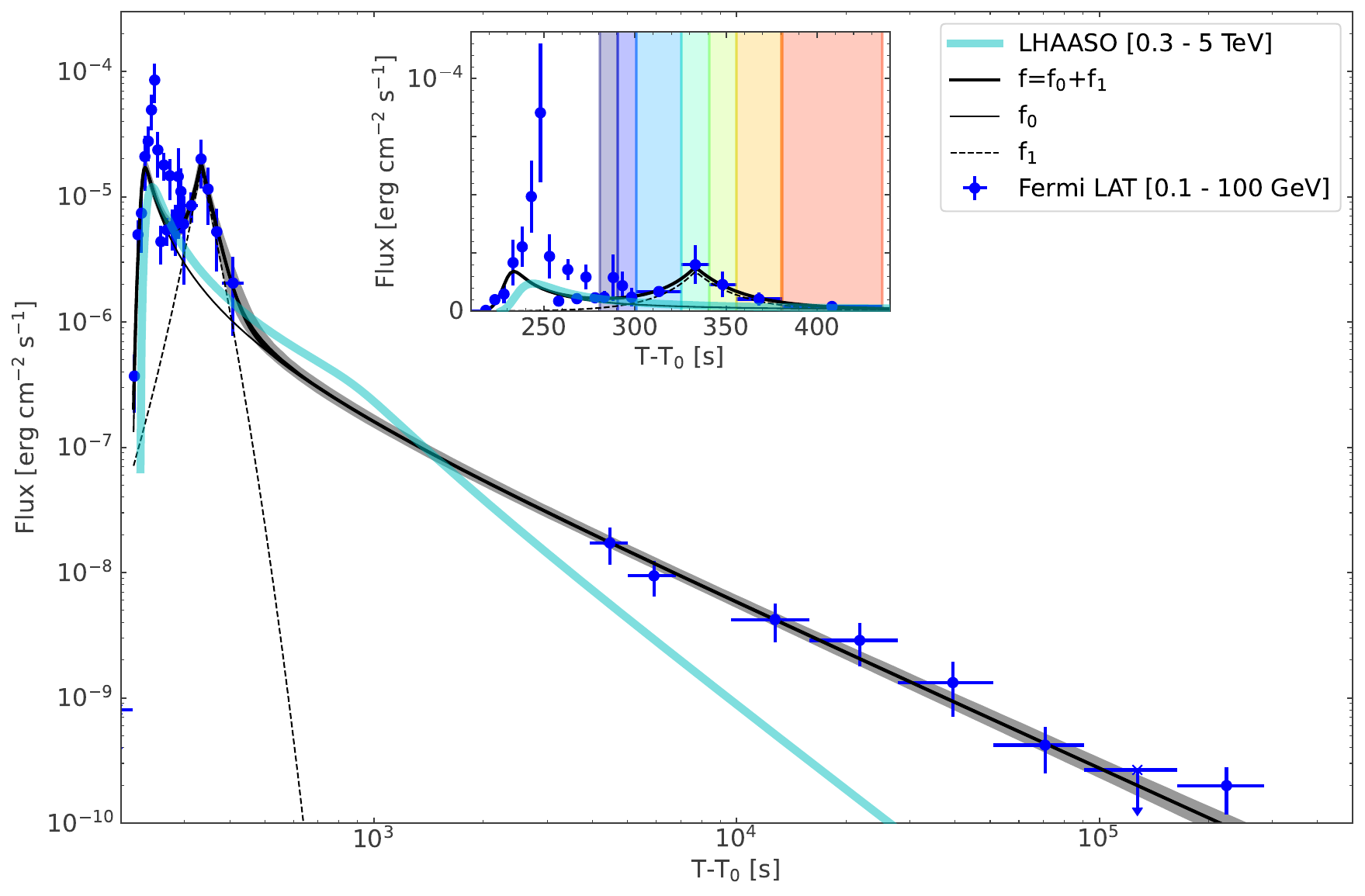}
    \includegraphics[width=0.49\linewidth]{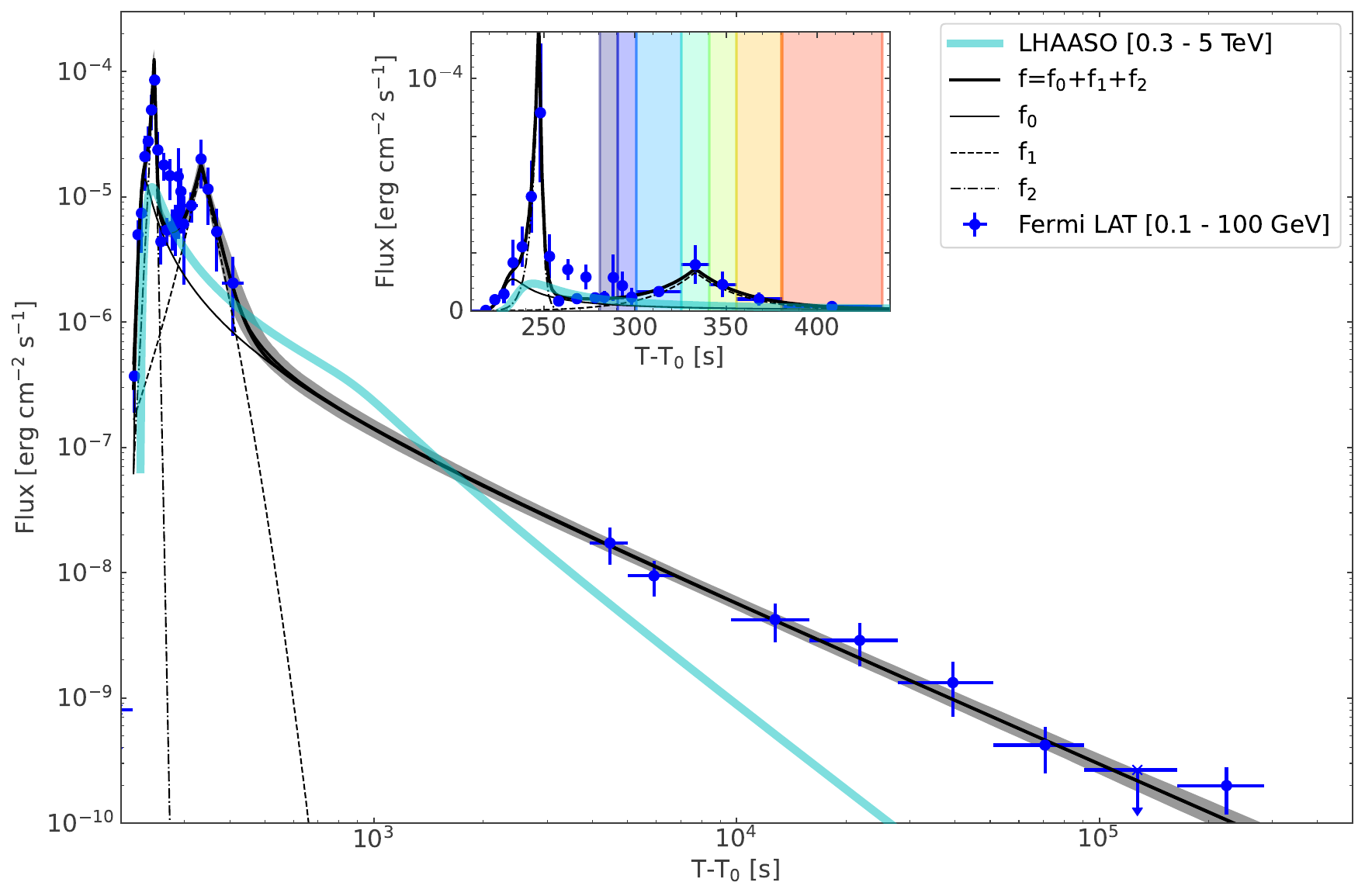}
    \includegraphics[width=0.49\linewidth]{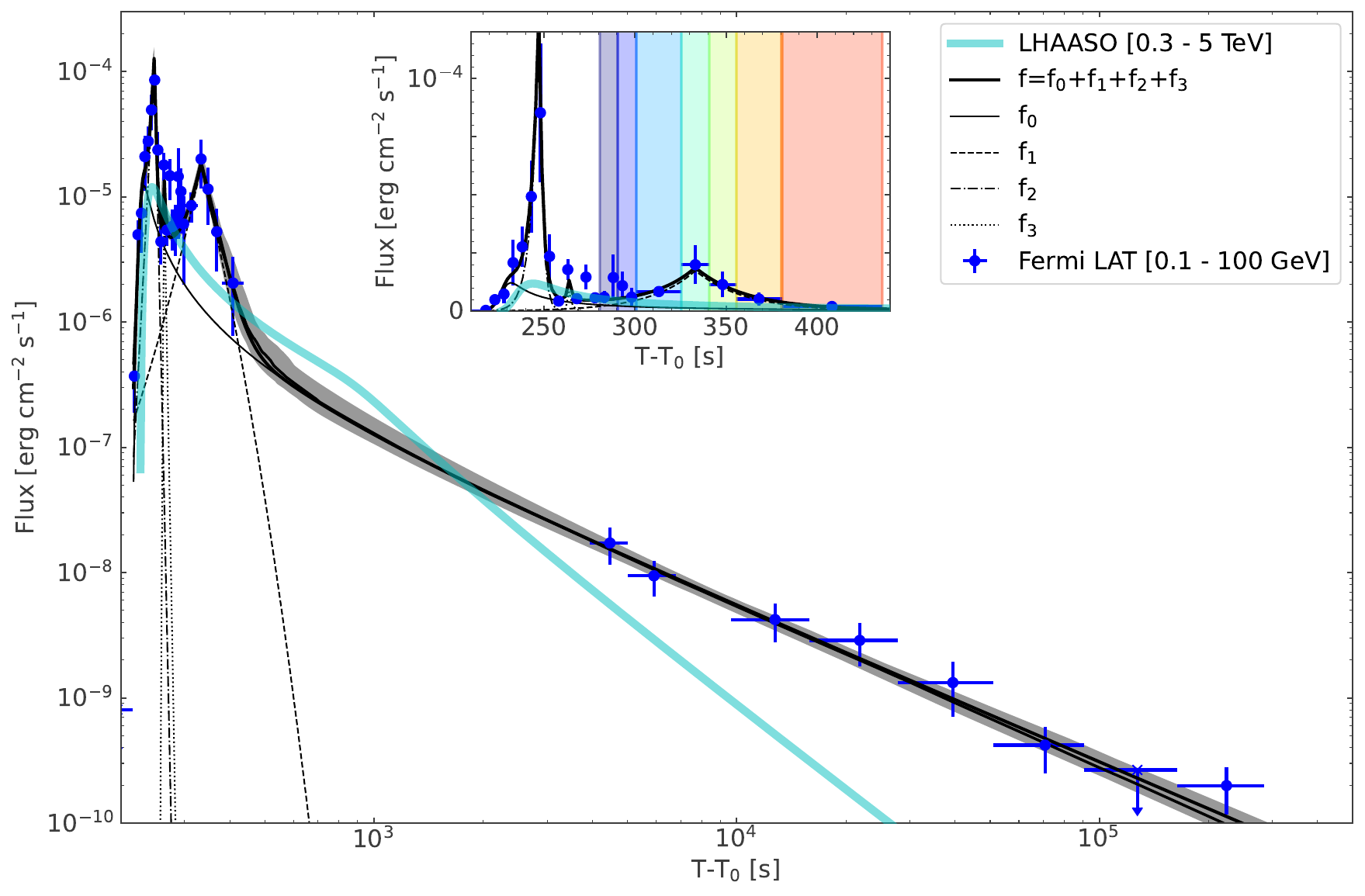}

    \caption{Model (and the components) used to describe the $>$ 100 MeV flux light curve as a function of time. Top-left: model with one component describing a pure afterglow emission (f$_{0}$). Top-right: f$_{0}$+f$_{1}$, bottom-left: \nob{The best fit model}: f$_{0}$+f$_{1}$+f$_{2}$, bottom-right: f$_{0}$+f$_{1}$+f$_{2}$+f$_{3}$ models including both afterglow emission and one, two or three pulses. In each panel, the gray band displays the 68\% contour levels, obtained by sampling the posterior distribution. The cyan curve is the LHAASO-WCDA 300~GeV -- 5~TeV light curve \citep{BOAT_LHAASO}. In the inset plots, the prompt emission (in linear $y$ scale) is shown. The seven colored bands highlight the time intervals used in the combined \Fermi GBM, \Fermi LAT and LHAASO fit displayed in Fig.~\ref{fig:joint_fit}.}
    \label{fig:lc_models}
\end{figure*}
We modeled the 100 MeV--100 GeV energy flux light curve as the sum of multiple functions. First, we modeled the entire light curve with a smoothly broken power law ($\rm f_{0}$)  representing the typical forward shock emission for which we adopted the definition presented in \citet{2004MNRAS.353..647N}:
\begin{equation}
\label{eq:f0}
    {\rm f_0(T)} =\mathrm{K \left(\frac {1}{2} \left( \frac{T-T_{on}}{\tau} \right) ^{-\xi \alpha_1}
    +  \frac {1}{2} \left(\frac{T-T_{on}}{\tau} \right) ^{-\xi\alpha_2} \right)
    ^{-\frac{1}{\xi}}} ,
\end{equation}
where $\alpha_1$ and $\alpha_2$ are the values of the early and late time indexes, $\tau$ is a scale parameter and $\xi$ defines how smooth the transition is. For our analysis, to limit the number of parameters, we fixed $\xi=2$ and $\alpha_1=3$ as predicted by the standard afterglow model \citep{2004MNRAS.353..647N}. The time T$_{\rm p}$ of the peak of the afterglow relative to onset time T$_{\rm on}$ is:
\begin{equation}
\label{eq:tp}
\rm T_{\rm p} = \tau \left (-\frac{\alpha_2}{\alpha_1}\right )^{\frac{1}{\xi (\alpha_2-\alpha_1)}} \;.
\end{equation}
After performing the fit with ${\rm f_0(T)}$, we repeated the fit adding one, two or three pulses using the pulse shape described in \citet{norris:2005} and \citet{Hakkila_2014} with the following functional form:
\begin{equation}
  {\rm f_{i}(T)} = K_{i} \begin{cases}
    \rm \exp{\left(\frac{T-T_{p,i}}{\tau_{i}}\right)}, & \text{for T $<$ T$_{\rm p,i}$},\\
    \rm \exp{\left(\frac{T_{p,i}-T}{\xi_{i}\tau_{i}}\right)}, & \text{for T $>$ T$_{\rm p,i}$},
  \end{cases}
\label{eq:fc}
\end{equation}
where $\tau_{i}$ is the rise time \nob{of the i-th pulse} and $\xi_{i}\ne1$ for asymmetric pulses. 

The best-fit parameters are summarized in Tab.~\ref{tab:model} and the best-fit models are displayed in Fig.~\ref{fig:lc_models}. The LHAASO-WCDA light curve \citep{BOAT_LHAASO} is also displayed (but is not used in the fitting procedure). 
The LAT light curve clearly shows multiple emission episodes and the preferred model fit (with the lowest value of the BIC) is the one with three components. 
A single emission episode described by $\rm f_{0}$ fails to reproduce the complex behavior of the flux during the prompt emission, even if it is consistent with the flux observed at high energy by LHAASO-WCDA. In order to obtain a significantly better fit of the LAT data, two additional pulses are required, one describing the pulse at 350\,s after the trigger, and another one describing the narrow feature observed during the BTI, at $\sim$250 s. Adding more pulses increases the BIC value (indicating that the increased complexity of the model is not justified).

\subsection{Internal Shocks vs External Shock}
\label{sec:internal_shocks}
GRB~221009A is a complex burst in which multiple emission mechanisms from multiple radii from the central engine could be at play simultaneously. The nature of the late-time LAT emission during $\sim \rm{T}_0+4\times10^3$~s -- $\sim \rm{T}_0+ 3\times 10^5$~s is similar to the extended emission detected by the LAT from other GRBs and the emission is due to the synchrotron process from an external forward shock \citep[see, e.g.,][]{2009MNRAS.400L..75K, 2010MNRAS.403..926G, 2010ApJ...716.1178A, 2010ApJ...724L.109R, 2010ApJ...720.1008C}. The temporal decay of the flux during this time is $F_\nu\propto T^{-1.27\pm 0.05}$, or $T^{-1.35\pm 0.03}$, depending on the model, which is between the flux decay $F_\nu\propto t^{-1}$ and $F_\nu\propto t^{-10/7}$ expected for an adiabatic blastwave and a radiative blastwave, respectively. 
However, an adiabatic blastwave is favored by previous LAT detections \citep{2FLGC}. 
The onset time $\rm{T}_0$ ranges from $213\pm1$ s to $214\pm1$ s, slightly before the onset time $\rm T^{*}$ measured by LHAASO, while the peak T$_{\rm p}$ of the afterglow from the onset time ranges from $23 \pm 4$ s to $17 \pm 4$ s, with the value for the best fit model being $18^{+4}_{-3}$ s being consistent with the value derived by LHAASO for the 0.3--5 TeV light curve.
The two emission episodes at $\rm{T}_0 + 330$~s and $\rm{T}_0 + 250$~s described by $\rm f_{1}$ and $\rm f_2$, respectively, are very unlikely to have originated from the afterglow forward shock emission. 
Although we could not perform a detailed joint spectral analysis of the first pulse, basically due to the presence of BTIs in both the \Fermi LAT and GBM, a detailed joint fit of GBM and LAT data is presented in Sec.~\ref{sec:jointLATGBM}, including information from LHAASO-WCDA, too.

Synchrotron emission from a reverse shock, with temporal flux decay $F_\nu \propto T^{-(73p+21)/96}$ for a thick shell and $F_\nu \propto T^{-(27p+7)/35}$ for a thin shell \citep{2000ApJ...545..807K}, can explain flux decaying as $F_\nu\propto T^{-2.1}$ only for $p = 2.5$ and is also very unlikely to produce exponentially decaying emission. Finally, even the high-latitude emission from a shell, a so called ``naked GRB'' \citep{2000ApJ...541L..51K} with $F_\nu\propto T^{-(2-\beta)} \propto T^{-3}$ for the spectral index $\beta = -1$, is unable to produce this steep flux decay detected by the LAT. 
Therefore, the most natural explanation is that the LAT detected prompt emission from internal shocks until at least $\sim \rm{T}_0 + 380$~s. 
Our model suggests that during the prompt emission, the afterglow is subdominant the prompt emission, at least in the energy range below 100\,GeV, while it becomes dominant at higher energies, in agreement with the LHAASO-WCDA light curve.
A combination of shells ejected earlier in the burst with higher bulk Lorentz factors could initiate an afterglow even while shells with lower bulk Lorentz factors continue to produce prompt emission. 
Therefore, the afterglow onset time could lie between $\sim \rm{T}_0 + 215$~s, i.e. the beginning of the bright emission detected by the LAT, and $\sim$  $\rm{T}_0 + 600$~s, when the afterglow-like feature was detected in X-rays by Konus-Wind \citep{BOAT_KW}. 
Using the values of the best fit model (f$_{0}$+f$_{1}$+f$_{2}$), the onset of f$_{0}$ (T$_{\rm on} = 215 \pm 2$~s) and its peak (T$_{\rm p}$ = 18$^{+4}_{-3}$~s) do indeed suggest that the afterglow peaks at early times.
By identifying the peak time with the deceleration time of the afterglow, and using the kinetic energy ${E_K}$ from \citet{2023ApJ...952L..42L}, we can derive the Lorentz factor, as in e.g. \citet{GammaNappo,Ghirlanda+18Lorentz}:
\begin{equation}\label{eq:N14}
\Gamma 
= \left[\left(\frac{17-4s}{16\pi(4-s)}\right)
\left(\frac{E_K}{n_0 m_{\rm p}c^{5-s}}\right)\right]^{\frac{1}{8-2s}}\left(\fraz{T_{\rm p}}{1+z}\right)^{-\frac{3-s}{8-2s}},
\end{equation}
where T$_{\rm p}$ is the peak time in seconds, $m_{\rm p}$ is the mass of a proton and $c$ is the speed of light. We considered both a constant density Inter Stellar Medium (ISM) ($s=0$) and a wind density profile ($s=2$). For the first case,  we assumed density $n_{0}=1$ cm$^{-3}$, while for the second case $n(r)=n_{0}r^{-2}$ with $n_{0}$ = 3.2$\times 10^{35}$\,cm$^{-1}$. 

The derived values for $\Gamma$ (wind) and $\Gamma$ (ISM) are summarized in Tab.\ref{tab:model}. For the best-fit model, using T$_{\rm p}$ = 18$^{+4}_{-3}$, we obtained $\Gamma = 250 \pm 10$ for the wind scenario ($s=2$) and $\Gamma = 520 \pm 40$ for the constant density ISM ($s=0$). 
Assuming the derivation from \citet{SP99} instead, also applied by \citet{BOAT_LHAASO}, we obtain their exact same value of $\Gamma  = 440$.
These values represent lower limits on the outflow Lorentz factor, which can be significantly higher for a relativistic reverse shock (i.e. a ``thick shell'', which is likely the case for this GRB).

\begin{table}[h!]
    \centering
    \begin{tabular}{l| r | r | r |  r }
         & f$_{0}$                                & f$_{0}$+f$_{1}$ & f$_{0}$+f$_{1}$+f$_{2}$ & f$_{0}$+f$_{1}$+f$_{2}$+f$_{3}$ \\
    \hline
    \hline
K$_{0}$ [erg cm$^{-2}$ s$^{-1}$]    & (1.8 $\pm$ 0.3)$\times$10$^{-5}$ & (1.6 $\pm$ 0.3)$\times$10$^{-5}$     & (1.3 $\pm$ 0.3)$\times$10$^{-5}$         & (1.1$^{+0.4}_{-0.2}$)$\times$10$^{-5}$    \\
$\alpha_{1}$                        &        3.0 (fixed)                  & 3.0 (fixed)                             & 3.0 (fixed)                                 & 3.0 (fixed)                                 \\
$\alpha_{2}$                        &     -1.35 $\pm$ 0.03              & -1.31 $\pm$ 0.04                      & -1.27 $\pm$ 0.05                          & -1.25$^{+0.06}_{-0.05}$                   \\
$\xi$                               &            2 (fixed)                 &  2 (fixed)                               &  2(fixed)                                   &  2 (fixed)                                   \\
$\tau$ [s]                          &       20 $\pm$ 2                  & 18$^{+3}_{-2}$                        & 17$^{+4}_{-3}$                            &  15$^{+4}_{-3}$                           \\
T$_{\rm on}$ [s]                    &          213 $\pm$ 1              & 214 $\pm$ 1                           & 215 $\pm$ 2                               & 215 $\pm$ 2                               \\
\hline
K$_{1}$ [erg cm$^{-2}$ s$^{-1}$]    &              -                    & (1.7$^{+1}_{-0.6}$)$\times$10$^{-5}$  & (1.7$^{+0.6}_{-0.5}$)$\times$10$^{-5}$      & (1.7$^{+0.7}_{-0.4}$)$\times$10$^{-5}$  \\
T$_{\rm p,1}$  [s]                  &              -                    &     330 $\pm$ 10.0                    & 333 $\pm$ 2                               & 333 $\pm$ 2                               \\
$\tau_1$       [s]                  &              -                    &           22$^{+10}_{-8}$             & 25$^{+9}_{-6}$                            & 25$^{+8}_{-5}$                            \\
$\xi_1$                             &              -                    &  1.1$^{+1}_{-0.6}$                    & 1.1 $\pm$ 0.4                             & 1.0$^{+0.4}_{-0.3}$                       \\
\hline
K$_{2}$ [erg cm$^{-2}$ s$^{-1}$]    &              -                    &                                    -  & (1.2$^{+0.7}_{-0.4}$)$\times$10$^{-4}$    & (1.2$^{+0.6}_{-0.4}$)$\times$10$^{-4}$    \\
T$_{\rm p,2}$  [s]                  &              -                    &                                    -  & 247.4 $\pm$ 0.9                           & 247.3 $\pm$ 0.9                           \\
$\tau_2$       [s]                  &              -                    &                                    -  &  4$^{+2}_{-1}$                            & 4$^{+2}_{-1}$                             \\
$\xi_2$                             &              -                    &                                    -  & 0.5$^{+0.3}_{-0.2}$                       & 0.5$^{+0.3}_{-0.2}$                       \\
\hline  
K$_{3}$ [erg cm$^{-2}$ s$^{-1}$]    &              -                    &                                    -  &                                       -   & (1.1$^{+4}_{-0.9}$)$\times$10$^{-5}$      \\
T$_{\rm p,3}$  [s]                  &              -                    &                                   -   &                                       -   & 263$^{+3}_{-2}$                           \\
$\tau_3$       [s]                  &              -                    &                                   -   &                                       -   & 0.6$^{+0.2}_{-0.1}$                       \\
$\xi_3$                             &              -                    &                                    -  &                                       -   & 3$^{+3}_{-1}$                             \\
\hline  
BIC                                 & 69.2                              & 64.9                                  & 62.6                                      & 71.2                                      \\
$\Delta_{BIC}$                      & 0                                 & -4.3                                  & -2.3                                      & 8.6                                       \\
\hline  
\hline  
T$_{\rm p}$                         & 21$^{+3}_{-2}$                    & 23 $\pm$ 4                            & 18$^{+4}_{-3}$                            & 17 $\pm$ 4                                \\
$\Gamma$ (wind)                     & 240 $\pm$ 7                       & 237 $\pm$ 10                          & 250 $\pm$ 10                              & 250.0$^{+20.0}_{-10.0}$                   \\
$\Gamma$ (ISM)                      & 490 $\pm$ 20                      & 480 $\pm$ 30                          & 520 $\pm$ 40                              & 530.0$^{+50.0}_{-40.0}$                   \\
Fluence [erg cm$^{-2}$]             & (1.4 $\pm$ 0.1)$\times$10$^{-3}$ & (2.2 $\pm$ 0.3)$\times$10$^{-3}$     & (2.6 $\pm$ 0.4)$\times$10$^{-3}$         & (2.7 $\pm$ 0.4)$\times$10$^{-3}$         \\
E$_\mathrm{iso}$ [erg]                          & (9 $\pm$ 0.8)$\times$10$^{52}$   & (1 $\pm$ 0.2)$\times$10$^{53}$       & (2 $\pm$ 0.3)$\times$10$^{53}$           & (2 $\pm$ 0.3)$\times$10$^{53}$           \\

 \hline
    \end{tabular}
    \caption{Parameters of the best-fit models for f$_{0}$, f$_{1}$, f$_{2}$ and f$_{3}$ and derived quantities T$_{\rm p}$, $\Gamma$, fluence and E$_{\rm iso}$ (both calculated between 100 MeV and 100 GeV from the onset time T$_{\rm on}$ and 300 ks). 
    The value of the BIC and its increase (or decrease) with respect to that with one less component is also shown.}
    \label{tab:model}
    \end{table}

To estimate the released energy of \grb, we integrated the best-fit model in time between the onset time T$_{\rm on}$ and 300~ks. The last two rows of Tab.~\ref{tab:model} show the values for each model. For the best-fit model, we obtained a fluence of (2.6 $\pm$ 0.4)$\times$10$^{-3}$ erg cm$^{-2}$ which corresponds to an isotropic energy E$_{\rm iso}$ = (2 $\pm$ 0.3)$\times$10$^{53}$ erg, \nob{both calculated between 100~MeV and 100~ GeV}.

\subsection{Estimation of the bulk Lorentz factor from pair opacity}
\label{sec:gamma}
The bulk Lorentz factor can be directly estimated by measuring the spectral cutoff in the high-energy band due to the pair opacity \citep{2018ApJ...864..163V,2020ApJ...890....9A,2020ApJ...891..106A}. 
We tested the presence of a spectral cutoff during the prompt phase by comparing the fit with a simple power law with one that has an additional exponential cutoff at high energies. We found that the model with a cutoff was not preferred in any of the time intervals; thus we can set a lower limit on the bulk Lorentz factor, $\Gamma_{\rm bulk,min}$:

\begin{equation}\label{eq:Gamma-min}
\Gamma_{\rm bulk,min} = 500\left[N \left(\frac{1+z}{1.151}\right)^{-\Gamma_{\rm ph}}\left(\frac{\rm L_0}{6\times10^{52}\,{\rm erg~s}^{-1}}\right)\left(\frac{99.3\,{\rm GeV}}{\rm E_{\rm max, obs}}\right)^{1+\Gamma_{\rm ph}} \left(\frac{-\Gamma_{\rm ph}}{1.9}\right)^{-5/3}\left(\frac{0.1\,{\rm s}}{\rm t_v}\right)\right]^{1/(2-2\Gamma_{\rm ph})}
\end{equation}
where the luminosity $L_0 = 4 \pi d_{\rm L}^{2} (1+z)^{-\Gamma_{\rm ph}-2}$ F$_0$, with $d_{\rm L}$ the luminosity distance (in cm) and F$_0$ the energy flux in $\nu F_\nu$ at 511\,keV.
$N$ is a numerical parameter of the order of unity obtained from previous studies \citep{2018ApJ...864..163V}, while $\rm E_{\rm max, obs}$ is the energy of the highest-energy photon, observed at \trig + 240.3~s by the LAT (Tab.~\ref{tab:he_photons}). 
In the interval containing this event [\trig + 244\,s, \trig + 255\,s] the LAT emission above 100 MeV can be well fitted by a single power-law function with a photon index of $\Gamma_{\rm ph} = -1.9 \pm 0.1$ (see Tab.~\ref{tab:lat_flux}).  
For the minimum variability time scale $\rm t_v$, we cannot use LAT data because the intense flux of hard X-rays constitutes a major source of noise in every LAT subsystem (see App.~\ref{sec:BTI} for details). Instead, we used $\rm t_v \sim\,0.1$ ms obtained from the GBM data \citep{2023ApJ...952L..42L}.
To estimate the density of target photons at 511\,keV, we first assumed that low-energy events and high-energy events come from the same physical place (so that the $\sim$MeV photons that we see are the same that could absorb the GeV events), and, using the results of the analysis presented in \citet{2023ApJ...952L..42L}, we obtained a value of $\nu F_\nu\sim 10^{-3}$\,erg\,cm$^{-2}$\,s$^{-1}$, which implies $\Gamma_{\rm bulk,min} \sim 550$.
If instead we assume that the high-energy spectral component is separated from the low-energy spectral component, we can use the value of F$_0$ estimated by extrapolating the LAT power law to 511\,keV, obtaining F$_0\sim 10^{-6}$\,erg\,cm$^{-2}$\,s$^{-1}$ and a value for $\Gamma_{\rm bulk,min} = 170$. 
The estimates of the bulk $\Gamma$ obtained in this analysis are consistent with those obtained using the deceleration time scale (Sec.~\ref{sec:internal_shocks}, Eq.~\ref{eq:N14}, and Tab.~\ref{tab:model}) and, considering the uncertainties of the model, we cannot firmly establish if the preferred density is a constant ISM as opposed to a wind profile.

\begin{figure}[t]
    \centering    
    \includegraphics[width=0.7\linewidth]{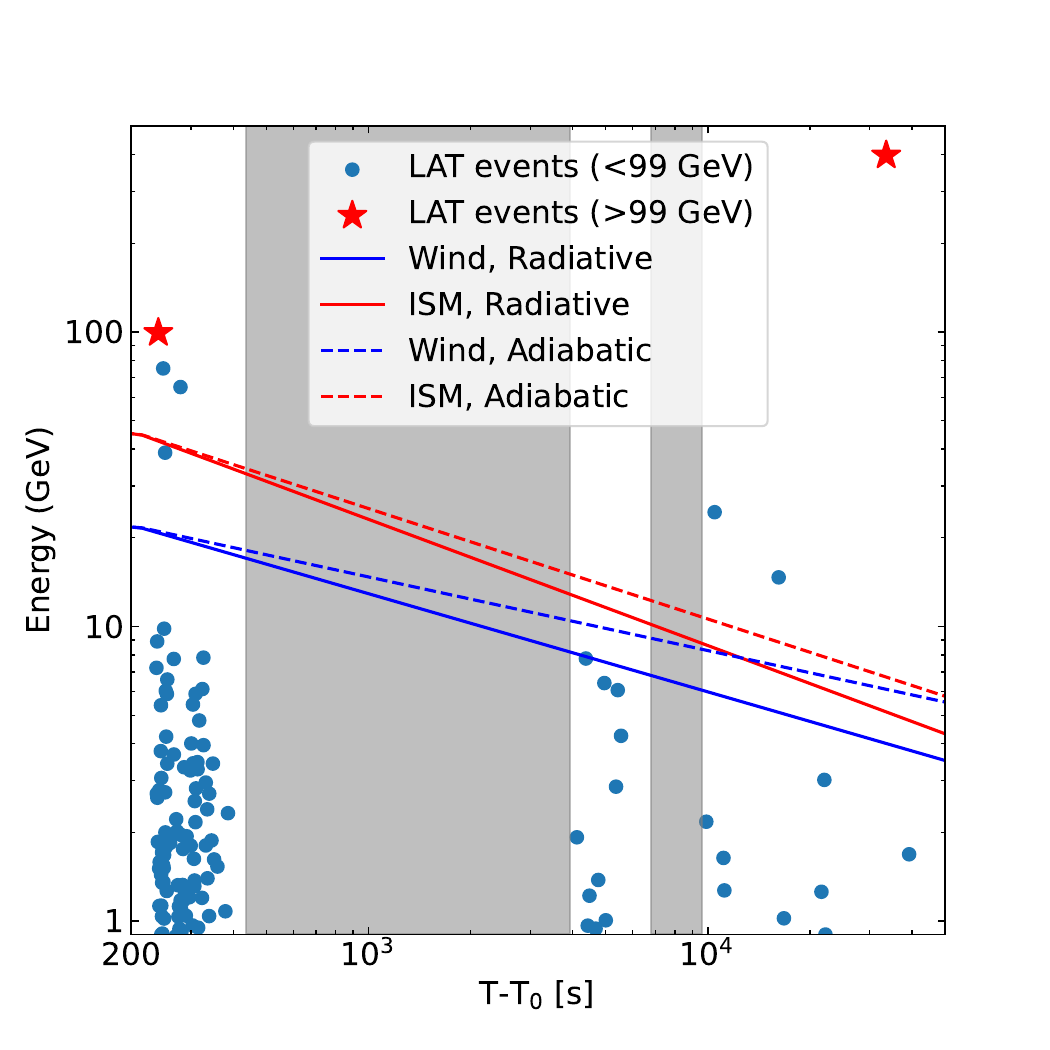}
    \caption{Observed LAT photons with $>$ 90\% probabilities of being associated with \grb. Red stars highlight events above 99 GeV. The lines show the highest possible energies for synchrotron photons based on the synchrotron shock model of the adiabatic/radiative expansion of a fireball in a circumburst medium with an ISM or wind profile, using the initial bulk Lorentz factors calculated from Eq.~\ref{eq:N14} and displayed in Tab.~\ref{tab:model}.
    Photons with energies above these lines exceed the maximum energy allowed by synchrotron emission.
    The gray shaded areas represent intervals in which \grb was not in the LAT field of view. Note that such intervals are not displayed for times greater than \trig + 10$^4$ s.}
    \label{fig:maximum_syn_photon}
\end{figure}

However, from the estimated bulk Lorentz factor, we can constrain the maximum synchrotron photon energy (with electrons shock accelerating over the Larmor time and then cooling over the synchrotron cooling time) under the assumption that the extended gamma-ray emission originates from the external forward shock. 
Fig.~\ref{fig:maximum_syn_photon} shows the gamma-ray photons detected by the LAT compared with the predicted maximum synchrotron photon energy when using $\Gamma=250$ for the wind case scenario and $\Gamma = 520$ for the ISM case, for both radiative and adiabatic expansion of the fireball. 
High-energy gamma rays detected by the LAT are in tension with the maximum synchrotron energy during both the prompt and afterglow phases, suggesting that their energies are boosted by inverse Compton scattering.

\subsection{Synchrotron, synchrotron self Compton?}
\label{sec:jointLATGBM}
To study the spectral evolution of \grb after the BTI, where the high-energy emission is prominent, we jointly fitted GBM\footnote{We used CSPEC (Continuous Spectroscopy) GBM data}, LLE and LAT data. 
We analyzed the time intervals after the first GBM BTI [\trig + 280\,s, \trig + 435\,s], using the same binning as in Tab.~\ref{tab:lat_flux}. 
In our analysis, we included detectors n7, n8 and b1, which all have a good viewing angle during the interval. 
In addition, we added LLE and LAT (\texttt{TRANSIENT\_010E}) data using the \ThreeML \texttt{OGIPLike} and  \texttt{FermiLATLike} plugins.
We also calculated the predicted intrinsic spectrum during the various time intervals from the LHAASO analysis to constrain the best-fit model. 
To this end, we used the best-fit model for the light curve N(t) described in \citet{BOAT_LHAASO} (Fit [0,+$\infty$), Table S3 in the supplementary material), combined with equation S3 for the evolution of the spectral index $\Gamma_{\rm ph}$(t).  
For a set of 100 energy points from 300 GeV to 5 TeV, we computed the fluxes by randomly sampling the model parameters within their errors and integrating in time. The 15.9, and the 84.1 percentiles of the flux distribution define the 68\% contour level.
Finally, we used the \ThreeML plugin \texttt{XYLike} that allows flux errors (assumed to be Gaussian) to be taken into account in the fit by computing the $\chi^2$ relative to the assumed model.
To find the model that best fits the data, we started from the same model adopted in \citet{BOAT_LINE} for a similar time interval (\texttt{SBPL}). Similarly to \citet{BOAT_LINE}, we modeled the spectrum at low energy by adding a Gaussian line (\texttt{G}) with a peak free to vary between 1\,MeV and 20\,MeV, and a width between 1\,MeV and 10\,MeV. 
The decision whether to add the Gaussian feature was based on minimization of the residuals. 
At high energy, we also added an additional \texttt{PL} or a second smoothly broken power law \texttt{SBPL} to mimic the typical ``two-hump'' SSC spectrum. Given the increase in the number of parameters, we used the Bayesian inference analysis available in \ThreeML to explore the parameter space, selecting \texttt{multinest} \citep{Multinest} as a sampler.
The combination of these components resulted in six different models, and we used the Bayesian evidence $\log\,Z$ to compare models\footnote{In the statistical literature this term, which is the normalization integral on the right-hand-side of Bayes' theorem, is often called ``marginal likelihood''.}. 
The decrease of $\log\,Z$ with respect to the best-fit model is shown in Tab.~\ref{tab:logZ}. 
The model with the highest value of $\log\,Z$ is \texttt{SBPL+G+SBPL} for every time interval. 
Nevertheless, according to Jeffreys' scale~\citep{Jeffreys}, the Gaussian line is strongly favored ($\Delta \log\,Z > 5$) only in the first five intervals, while there is only moderate evidence (2.5$< \Delta \log\,Z < 5$) in the sixth interval, and no evidence in the last interval. 
With the same criterion, but now looking in Tab.~\ref{tab:logZ} at the values of $\Delta \log\,Z$ in the \texttt{SBPL+G+PL} column, a break at high energies is always strongly favored.

\begin{figure}[t]
     \centering    
     \includegraphics[width=1\linewidth]{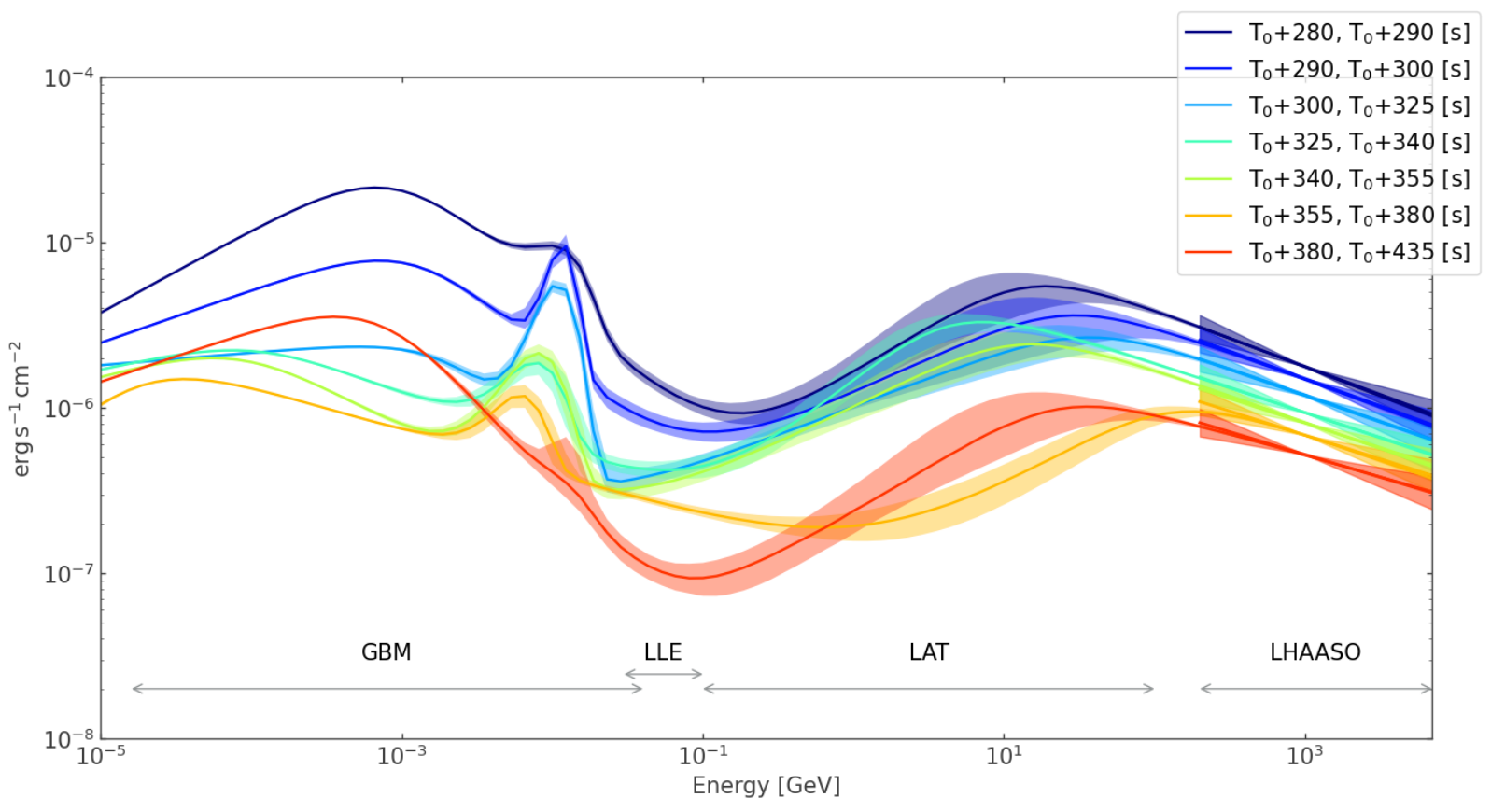}    
     \caption{Best fit model \nob{(\texttt{SBPL+G+SBPL})} for the time resolved joint analysis with 68\% c.l. \nob{The arrows at the bottom of the plot indicate} the energy band covered by each of the detectors. \nob{The overall spectrum exhibits a characteristic two-bump structure typical of synchrotron and SSC, where the first bump corresponds to the synchrotron emission, and the second bump corresponds to the SSC emission. In addition, a Gaussian line is statistically significant in the first five intervals, and only marginally significant (or not significant) in the last two.}
     }
     \label{fig:joint_fit}
 \end{figure}

Fig.~\ref{fig:joint_fit} shows the temporal evolution of the model. The Gaussian peak progressively shifts to lower energies with time, except in the last interval, although the Gaussian line is less significant. 
A break in the spectrum at $\sim$ 10~GeV is required to jointly fit LAT and LHAASO-WCDA data in each time bin.
The extra component at high energy could be responsible for producing TeV photons, the detection of which within the first 500 seconds after the trigger was indeed reported by the LHAASO experiment \citep{2022GCN.32677....1H}.
The required presence of the extra power law suggests that the high-energy gamma-ray emission is either SSC during the internal shocks or SSC during the external shock in the early phase of the afterglow.
In App.~\ref{sec:time_resolved_spectral_analysis} we summarize the results in every time bin for the LAT+GBM analysis, showing numerical values of the best-fit models (Tab.~\ref{tab:joint_best}) and the residuals for the \texttt{SBPL}, \texttt{SBPL+G} and \texttt{SBPL+G+PL} models (figures in App.~\ref{sec:time_resolved_spectral_analysis}).

  \begin{table}[h!]
 \centering
 \begin{tabular}{c|rrrrrr}
 \hline
 Interval (s from \trig)&       \texttt{SBPL} &     \texttt{SBPL+G} &    \texttt{SBPL+PL} &  \texttt{SBPL+G+PL} & \texttt{SBPL+SBPL} & \texttt{SBPL+G+SBPL} \\
\hline 
\hline 
 280.6--290.6  & 413.0 & 413.1 & 155.4 & 143.8 & 10.9 & 0.0\\
 290.6--300.6  & 156.0 & 155.9 & 91.4 &  73.2  & 18.9 & 0.0\\
 300.6--325.6  & 161.5 & 144.4 & 146.1 & 118.7 & 26.5 & 0.0\\
 325.6--340.6  & 137.8 & 137.9 & 88.9 &  89.0  & 30.8 & 0.0\\
 340.6--355.6  & 203.4 & 203.5 & 72.9 &  69.8  & 5.6 &  0.0\\
 355.6--380.6  & 186.2 & 185.9 & 63.9 &  61.1  & 2.7 &  0.0\\
 380.6--435.6  & 472.6 & 472.7 & 173.7 & 173.9 & 0.4 &  0.0\\
\hline
\end{tabular}
\caption{Decrement of the Bayesian evidence ($\Delta \log\,Z$) with respect the best fit model. For example, the decrease of $\Delta \log\,Z$ obtained by removing a Gaussian line from the \texttt{SBPL+G+SBPL} model (passing from \texttt{SBPL+G+SBPL} to \texttt{SBPL+SBPL}) can be read in the second-to-last column, while the decrease obtained by removing a second break in the high-energy component (from \texttt{SBPL+G+SBPL} to \texttt{SBPL+G+PL}) can be read in the third-to-last column.}
\label{tab:logZ}
\end{table}

\subsection{Is \grb really the B.O.A.T?}
\label{catalog}
\begin{figure}[t]
    \centering    
    \includegraphics[width=0.49\textwidth]{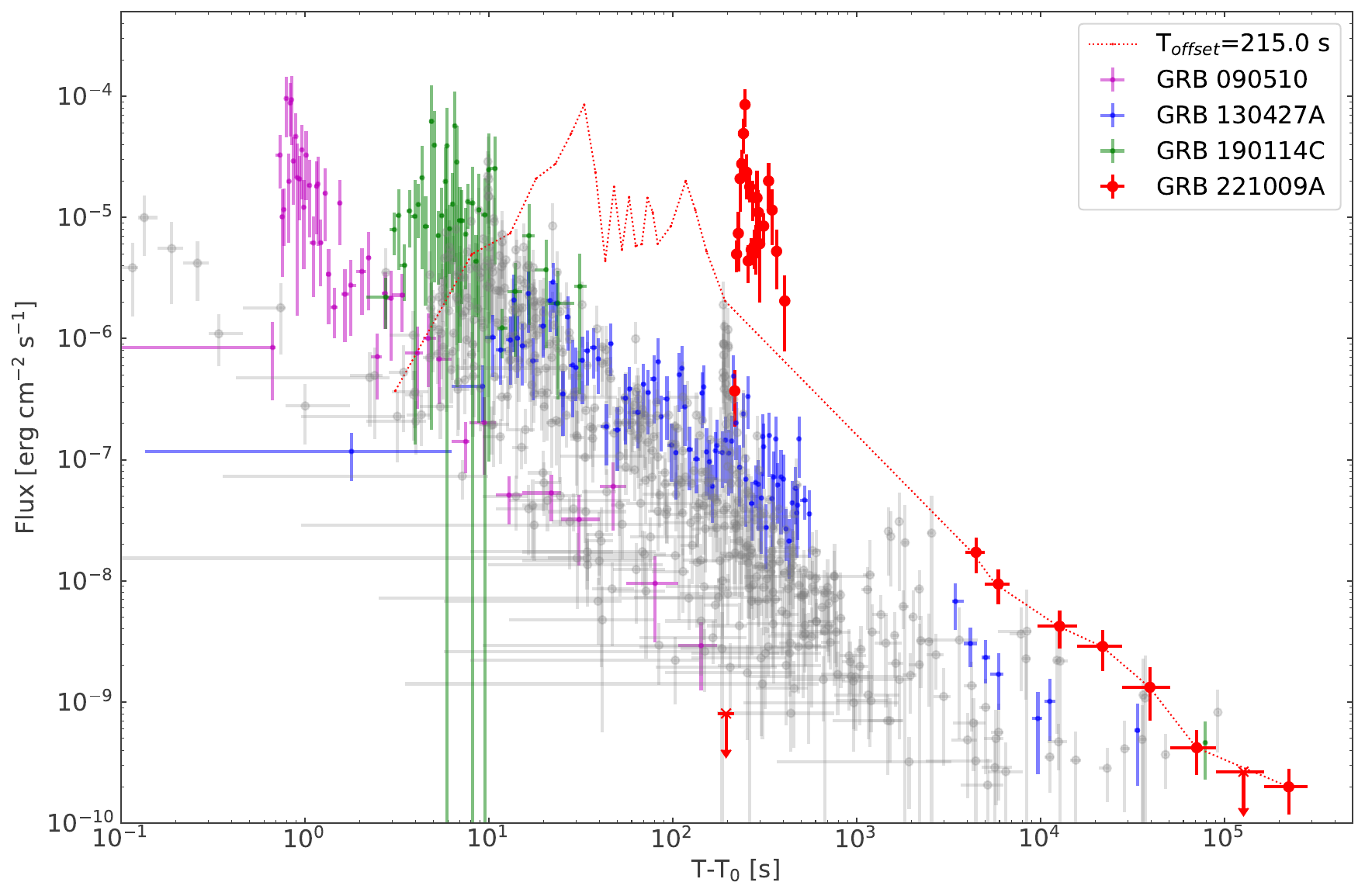}
    \includegraphics[width=0.49\textwidth]{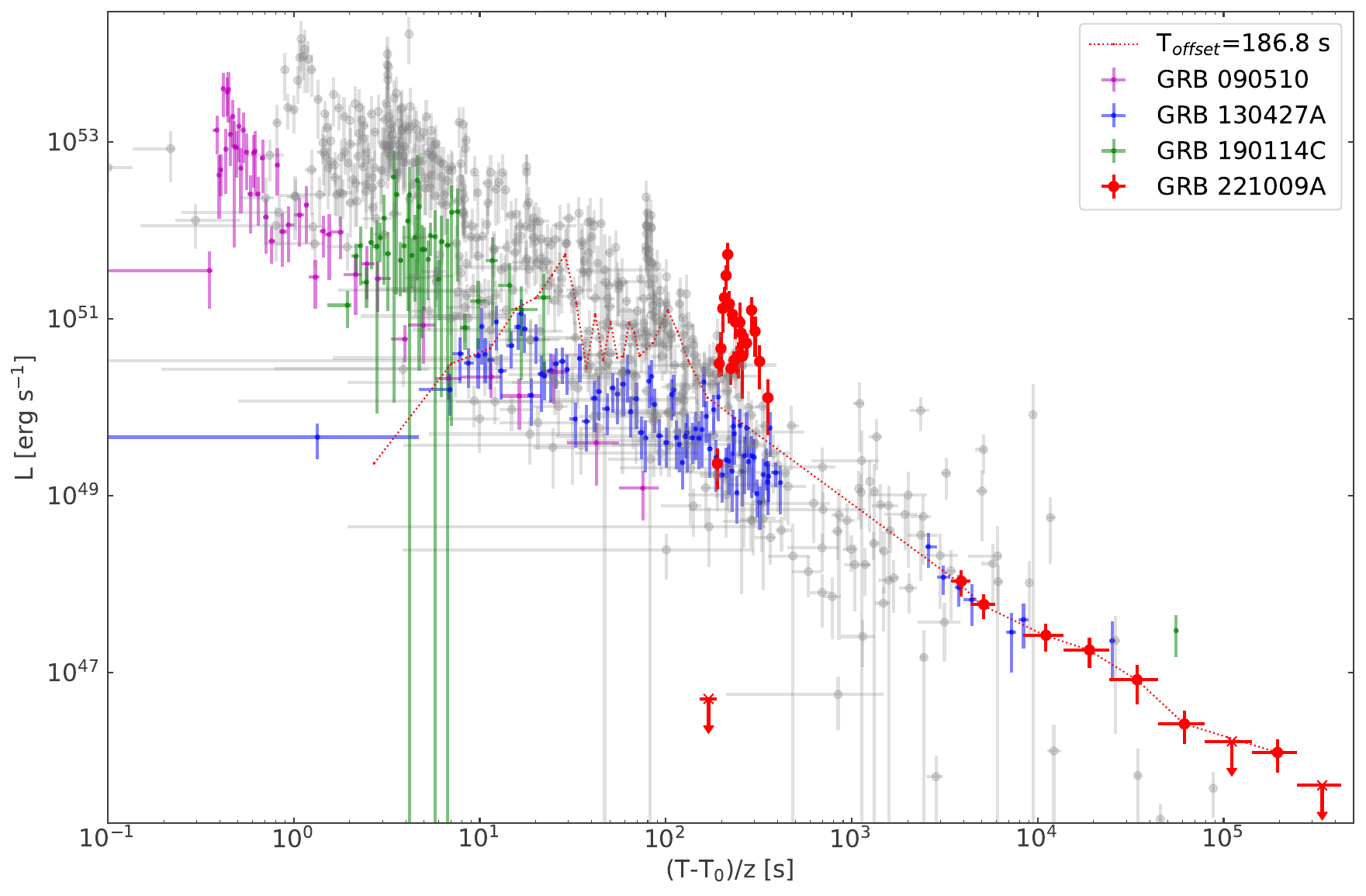}
    \caption{Flux light curves (left panel) and rest-frame luminosity light curves (right panel) \nob{in the 100~MeV-100~GeV rest frame} energy range for \grb (red) and for other LAT-detected GRBs with redshifts included in the 2FLGC (gray). Three previous record-holder GRBs are highlighted in different marker colors. The red dashed line, which indicates the light curve of \grb shifted by 215~s (i.e. the estimated onset time T$_{\rm on}$), shows how the light curve would look like if the GBM had triggered on the beginning of the bright pulse, and it is shown only for comparison with the other GRBs for which the triggering pulse could have been below threshold. \nob{\grb is particularly bright thanks to its proximity.}}
    \label{fig:fluence_luminosity}
\end{figure}

To answer this question, we need to put \grb in the context of other LAT GRBs, comparing some of its characteristics which those in the Second \Fermi-LAT GRB Catalog 2FLGC \citep{2FLGC}. Fig.~\ref{fig:fluence_luminosity} shows the fluence and luminosity as a function of time for \grb and for all the GRBs detected by the LAT that have known redshifts. 
We also highlight a few previous record holders: GRB\,090510 is still the brightest short\footnote{``Short'' GRBs have a duration in the MeV energy range less than two seconds; the other ones are classified as ``long''.} GRB ever detected by the LAT, and still the one whose fluence peaks earliest. GRB\,130427A was the previous record holder in terms of both fluence and highest-energy event, while GRB\,190114C, also very bright, is the first detection announced at TeV energies.
When comparing the two panels of Fig.~\ref{fig:fluence_luminosity}, \grb is an outlier if we consider the fluence light curve, which has an extremely intense pulse lasting several hundred seconds during the prompt phase, as well as late-time emission being on the brightest side of the GRB distribution. On the other hand, if we take into account its distance the luminosity of \grb is comparable with other bright LAT-detected GRBs. 

\begin{figure}[t]
    \centering    
    \includegraphics[width=0.45\textwidth]{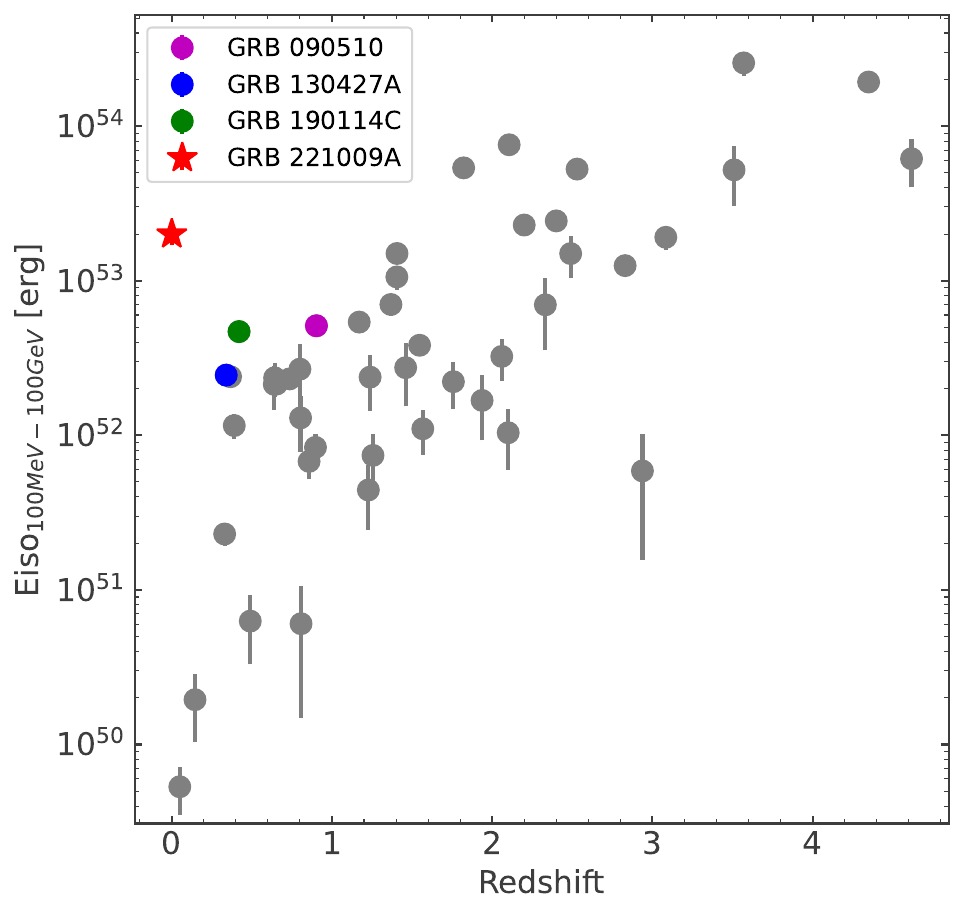}
    \includegraphics[width=0.45\textwidth]{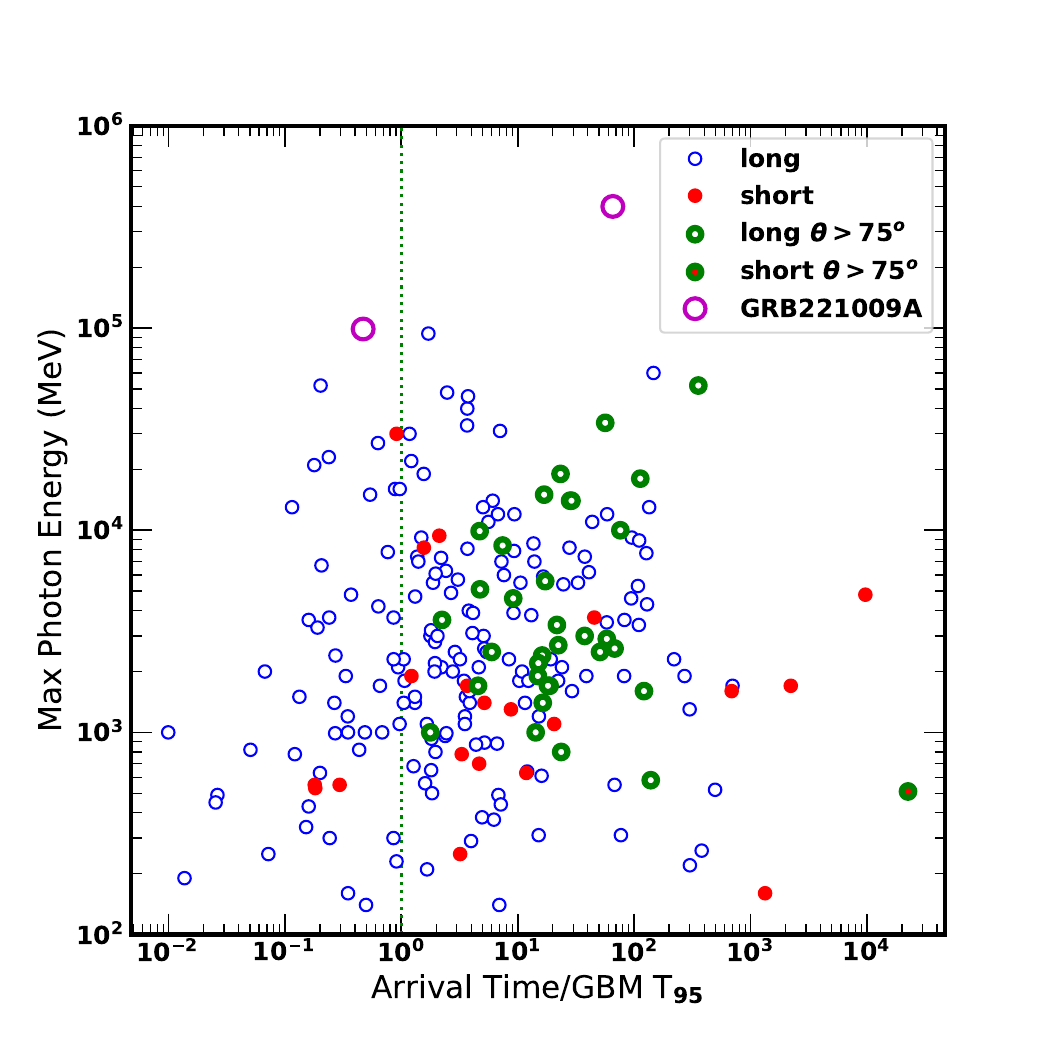}    
    \caption{Left: Distribution of the isotropic equivalent energies (E$_{\rm iso}$) as a function of redshift for the sample of \lat detected GRBs. 
    The red star highlights \grb. 
    As in Fig.~\ref{fig:fluence_luminosity}, we have highlighted the previous record holders with different marker colors. 
    In terms of E$_{\rm iso}$, \grb is clearly an outlier among GRBs at z $<$ 2.
    Right: Maximum photon energy vs. arrival time for LAT GRB photons normalized to the GRB\,T$_{95}$, the total duration of the burst (dashed vertical line indicates unity). Blue circles/red filled circles are long/short GRBs, and green circles highlight LAT-detected GRBs that were outside the field of view at the time of the GBM trigger.
    For \grb we display both the highest-energy event during the prompt emission (at 99\,GeV) and the highest-energy event during the temporally extended emission (at 400 GeV). }
    \label{fig:emax}
\end{figure}

The estimated isotropic energy E$_{\rm iso}$ can also be compared with that reported in the 2FLGC (Fig.~\ref{fig:emax}, left). 
Given its proximity, \grb is surprisingly more energetic than other GRBs at similar redshifts, and it is as energetic as other GRBs at redshift $\gta$ 2. 
Considering the difference in comoving volume confined within $z=0.151$ and $z\gta 2$, this unequivocally demonstrates that \grb is an extremely rare event and will likely remain unparalleled in our lifetimes. 
The right panel of Fig.~\ref{fig:emax} shows the maximum photon energy recorded by the LAT for each GRB and its arrival time. We have indicated both the highest energy events from \grb during the prompt emission as well as the 400 GeV event arriving 33\,ks after \trig. 
 An interesting coincidence: for the previous record holder, GRB\,130427A \nob{\citep[redshift $z=0.34$,][]{2013GCN.14455....1L}}, the highest energy event was a 94 GeV photon arriving at 243.13 seconds after the trigger, almost exactly the same numbers as for the 99 GeV event from \grb. With respect to the population of LAT GRBs, the 400\,GeV photon possibly associated with \grb clearly stands out. The association of this event with \grb will be discussed in detail in Sec.~\ref{sec:400GeV}.

\begin{figure}[t]
    \centering    
    \includegraphics[width=0.45\textwidth]{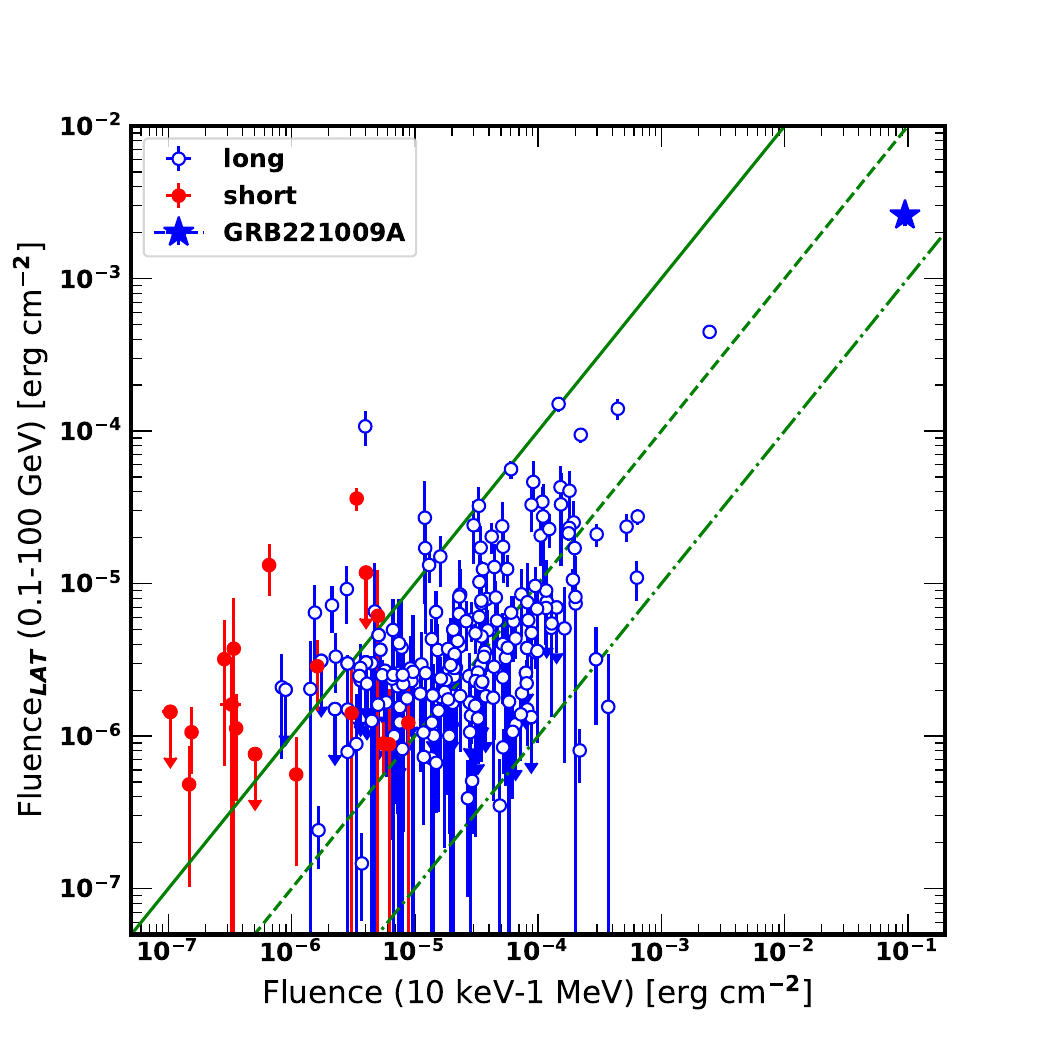}
    \includegraphics[width=0.45\textwidth]{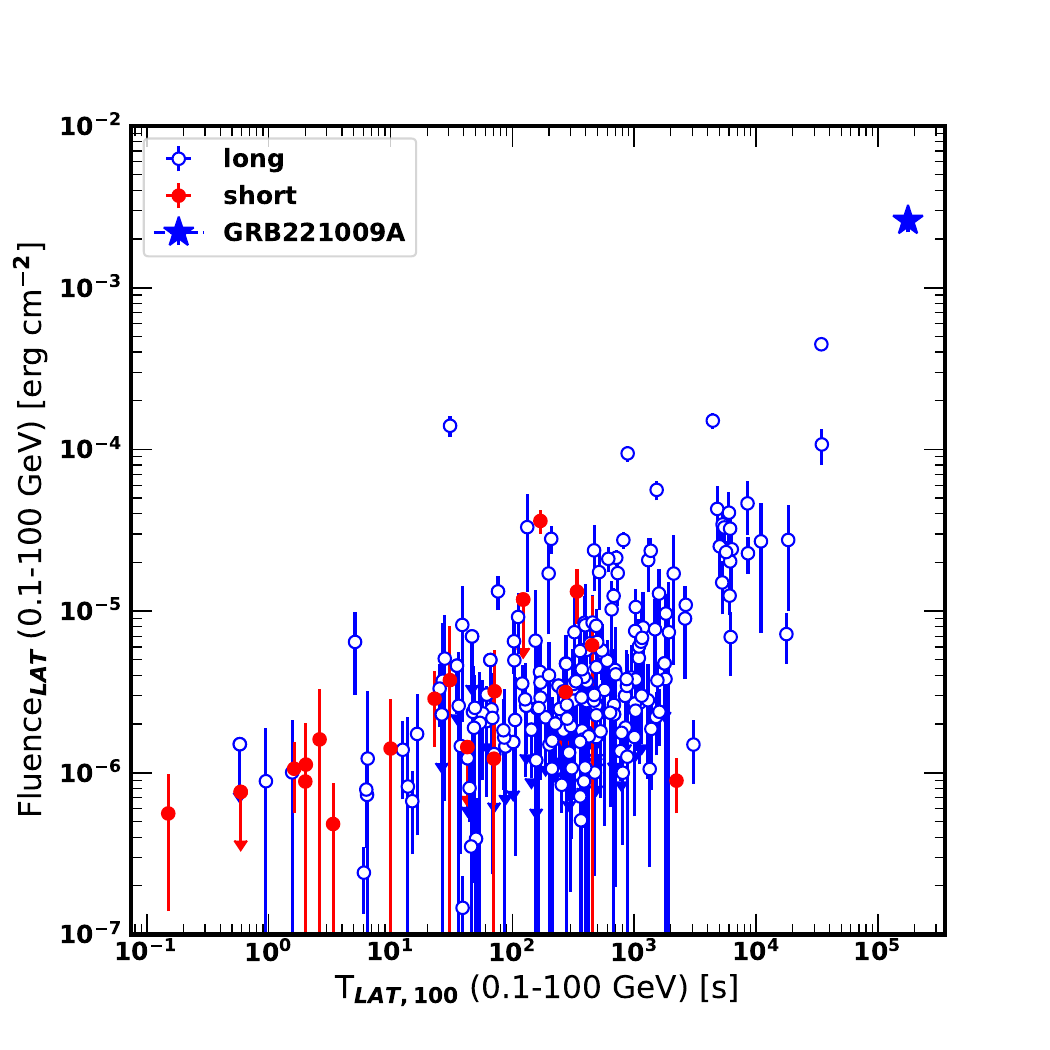}
    \caption{Left: Fluence in the 0.1--100 GeV energy range integrated over the entire duration of the high-energy emission detected by the LAT compared to the 10~keV--1 MeV fluence, as measured by the GBM \citep{2023ApJ...952L..42L}. 
    The green lines correspond to the fluence in the 0.1--100 GeV energy range being 1 (solid), 0.1 (dashed), and 0.01 (dot-dashed) times the fluence in the lower energy band.
    Right: Fluence in the 0.1--100 GeV energy band as a function of the GRB duration estimated using the \toz parameter as in \citet{2FLGC}. In both panels, \grb is marked with a blue star.}
    \label{fig:fluence_duration}
\end{figure}

By two other measures \grb can be classified as the B.O.A.T: the fluence integrated over the entire emission and the estimated duration \toz.
Fig.~\ref{fig:fluence_duration} shows the fluence from 0.1--100~GeV measured by the LAT as a function of the 10~keV--1~MeV fluence measured by the GBM (left panel), and as a function of \toz (right panel). 
The ratio of the 0.1--100\,GeV to the 10\,keV--1\,MeV fluence is close to 0.1, compatible with the rest of the long GRB population. 
In the fluence-duration scatter plot, \grb clearly stands out as the longest (with \toz$\sim 1.8 \times10^{5}$\,s) and the brightest GRB yet observed by the LAT.

\subsection{Closure relation and density profile}
\label{sec:closure}
During the temporally extended emission, the observed LAT photon index is $\Gamma_{\rm ph} \sim$ $-$2.2, while the observed temporal index $\alpha_{\rm LAT}$ = $-1.2 \pm 0.1$. Considering the spectral index $\beta$ = $-\Gamma_{\rm ph}-1$, we can test the closure relations \citep{CLOSURE_RELATIONS_TAK}. In particular, for $\nu_{\rm obs} > \nu_{\rm c}$, with $\nu_{\rm c}$ the synchrotron cooling frequency, $\alpha = (3/2) \: \Gamma_{\rm ph} + 2 \sim -1.3$ for both the ISM case and the wind profile, which is compatible with $\alpha_{\rm LAT}$. For $\nu_{\rm obs} < \nu_{\rm c}$, $\alpha = (3/2) \: (\Gamma_{\rm ph} + 1) \sim -1.8$ for the ISM case while $\alpha = (3/2) \: \Gamma_{\rm ph} + 1 \sim -2.3$ for the wind case, where both are in tension with the observed LAT photon index.

The derived spectral index from \nob{{\it Swift}-}XRT in the 0.3--10\,keV band from 1\,ks to 6.3\,ks from \trig is $\Gamma_{\rm ph, XRT}$ = $-1.91 \pm 0.09$, while the combined {\it Swift}, {\it MAXI} and {\it NICER} light curve can be fitted with a broken power law with indices $\alpha_{1,X} = -1.498 \pm 0.004$ and  $\alpha_{2,X} = -1.672 \pm 0.008$, with a break at T = \trig + $(7.9 \pm 1.1) \times 10^{4}$\,s \citep{BOAT_SWIFT}. 
For $\nu_{\rm obs} > \nu_{\rm c}$, both the ISM and wind scenario predict an expected decay index $\alpha = -0.85$, and for $\nu_{\rm obs} < \nu_{\rm c}$ $\alpha = -1.35$ for the ISM case scenario and $\alpha = -1.85$ for the wind case. As also noted by \citet{BOAT_SWIFT}, standard closure relations \citep{Racusin_2009} fail to reproduce the behavior of the X-ray afterglow emission, and the density profile cannot be firmly constrained.



\subsection{The 400 GeV event}
\label{sec:400GeV}
\begin{figure}[t]
    \centering    
    \includegraphics[width=0.8\linewidth]{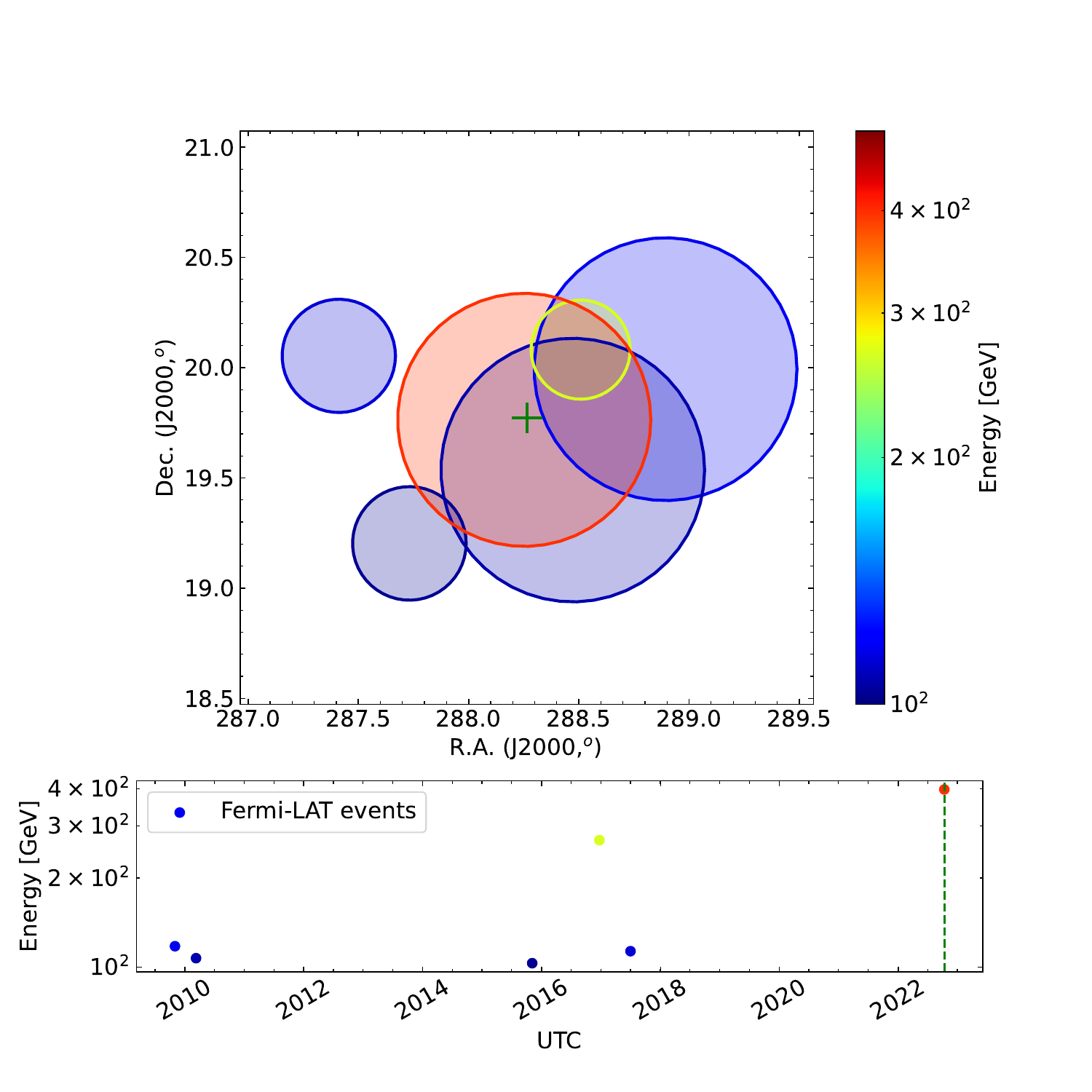}    
    \caption{Top: Arrival directions of all events with energies $>$ 100 GeV within 1$^\circ$ of the GRB location (green marker). The circles' radii are equal to the 95\% containment PSF radius and their colors represent their energies. Bottom: Arrival times of these events (for precise arrival times see Tab.~\ref{tab:he_photons_mission}). The green dashed line corresponds to the \Fermi-GBM trigger time.}
    \label{fig:he_photons_mission}
\end{figure}

The highest-energy event detected with a direction compatible with the location of \grb is a 400\,GeV photon, which converted in the lower part of the tracker, where the thicker tungsten converter layers cause a slightly worst point-spread function (PSF). It arrived approximately 33\,ks after the GBM trigger time. 
We performed a series of statistical tests of significance of this event\footnote{With ``event'' we mean the classification of the trigger as a gamma ray}.  First, we examined the LAT mission dataset from the beginning of the mission until the end of 2022, extracting all \texttt{SOURCE} events within a Region Of Interest (ROI) of 1 degree radius centered at the optical location of the GRB with energies $>$100\,GeV. Tab.~\ref{tab:he_photons_mission} shows the arrival time, the energy, the reconstructed direction, and the conversion type for each of these events, and Fig.~\ref{fig:he_photons_mission} shows their arrival directions and times. The highest-energy event ever detected by LAT in this ROI is in fact the 400\,GeV event arriving 9.3 hours after the GBM trigger. 
Comparing T$_{\rm off}$, the accumulated time that the ROI was in the LAT field of view (approximately 4.2 years) and T$_{\rm in}$, the accumulated time from \trig to the detection of the 400\,GeV (approximately 3.8 hours), the probability $p$ of observing $\rm N_{sig}=1$ event above 100\,GeV during the T$_{\rm in}$ when $\rm N_{bkg}=5$ events have been detected during T$_{\rm off}$ is calculated using the method described in \citet{LiMa} and corresponds to $p=1.6\times10^{-4}$ (3.6\,$\sigma$). This probability is slightly lower ($p=4.0\times10^{-5}$, or 3.9\,$\sigma$) if we consider that only one event ($N_{bkg}=1$) is compatible with the 95\% containment radius and it becomes $p = 9.3 \times 10^{-6}$ (4.3\,$\sigma$) considering that no other events at 400\,GeV or above are consistent with \grb ($N_{bkg}\rightarrow 0$).

\begin{table}[]
    \centering
    \begin{tabular}{r c c c c c c}
        UTC & Energy & R.A. & Dec. & Conv. type. & Ang. Sep. \\
        (s) & (GeV) & (J2000,\deg) & (J2000,\deg) &  & \deg\\
        \hline
        \hline
        2009-11-01 06:21:03.90 & 118 & 288.9 & 20.0 & Back & 0.63\\
        2010-03-10 17:30:57.05 & 107 & 288.5 & 19.5 & Back & 0.31\\
        2015-11-04 12:22:50.10 & 103 & 287.7 & 19.2 & Front & 0.76\\
        2016-12-21 08:42:19.47 & 268 & 288.5 & 20.1 & Front & 0.38\\
        2017-06-30 08:24:03.47 & 113 & 287.4 & 20.0 & Front & 0.85\\
        2022-10-09 22:36:17.96 & 398 & 288.2 & 19.8 & Back & 0.02\\
        \hline
    \end{tabular}
    \caption{\Fermi LAT events with energies $>$ 100 GeV that arrived since the beginning of the mission within 1 deg of the position of the GRB optical counterpart. ``Back'' stands for events converting in the lower part of the tracker as opposed to ``Front'' events converting in the upper part of the tracker. At these energises $\Delta$E/E $\lesssim$ 10\%.}
    \label{tab:he_photons_mission}
\end{table}

On the assumption that the 400~GeV event indeed came from the GRB, we performed a second test to check if this event is compatible with the low-energy part of the spectrum. To do so, we repeated the spectral analysis between 27.81\,ks and 51.09\,ks, selecting this time only the events with energies $<$\,100 GeV. 
The derived photon index is $\Gamma_{\rm ph}=-2.5\pm0.3$. 
Then we computed the exposure above 400 GeV during the same time interval and assumed the spectral index $\Gamma_{\rm ph}$. The number of expected events N$_{\rm exp}$ is given by the product of the integrated flux and the exposure. 

The number of expected events above 400 GeV varies from N$_{\rm exp}=4.3\times10^{-6}$ to N$_{\rm exp}=5.8\times10^{-4}$ (at 68\% confidence level) which corresponds to a probability ranging from 4.4\,$\sigma$ to 3.2\,$\sigma$ to observe at least one event.

If we lower the threshold to 360 GeV, which results from a 10\% energy resolution at 400\,GeV, the number of expected events above this threshold now ranges from N$_{\rm exp}=7.3\times10^{-6}$ to N$_{\rm exp}=1.0\times10^{-3}$ or probabilities from 4.3\,$\sigma$ to 3.1\,$\sigma$.

Finally, a similar approach that allows us to take into account the uncertainty on the flux with its probability density function is to compute the upper bound of the flux, by increasing the confidence level c.l. until N$_{\rm exp}=1$. The corresponding c.l. is 0.999970, or a probability of 3.1$\times10^{-5}$ (4.0\,$\sigma$) or a probability of 4.7$\times10^{-5}$ (3.9\,$\sigma$) for the lower energy threshold. 

In summary, considering the entire \Fermi mission elapsed time, the chance probability that an event above 100\,GeV is detected within 33~ks after the GBM trigger, in spatial coincidence with the GRB is approximately 4\,$\sigma$, indicating that we cannot firmly exclude that the photon is a background event. If the photon was indeed a background event, given the proximity to the Galactic plane and the absence of nearby point sources, it would likely be associated with diffuse emission from the Milky Way or from the unresolved population of blazars that constitute the majority of the extragalactic gamma-ray background~\citep{FermiEBL}.

On the other hand, if we assume that the 400\,GeV event is indeed associated with the GRB, the probability that the event is generated from the high-energy extrapolation of the low-energy part of the spectrum is relatively low (between 3.1 and 4.4\,$\sigma$), suggesting that this is unlikely to be the case.
A preliminary HAWC upper limit~\citep{2022GCN.32683....1A} was calculated during a time interval that included the high-energy event, concluding that no bright emission was detected at TeV energies. A more in-depth analysis is currently underway and will help to assess the origin of this event.

Electromagnetic cascades from high-energy cosmic-ray interactions with the magnetic fields of the host galaxies or TeV photons interacting with the EBL could, in principle, explain such a delayed emission. In the latter scenario, the resulting electron-positron pairs would in turn interact with the Cosmic Microwave Background (CMB) via the inverse Compton (IC) process, leading to a secondary cascade of multi-GeV photons, which could be delayed with respect to the GRB depending on the intergalactic magnetic field strength \citep[see, e.g.][]{Neronov09}. As a first estimate, producing a 400 GeV cascade photon would require a $\sim 23$ TeV primary photon assuming an equal distribution of energy among the pairs, while a 33 ks delay implies a $\sim 2\times10^{-18}$~G magnetic field. The exceptionally high energies reported by LHAASO would make this scenario feasible, but incompatible with intergalactic magnetic field limits derived from blazar observations \citep[$B>3\times10^{-16}$~G for conservative blazar duty cycles;][]{Fermi-LATextended18}. This tension could be alleviated if plasma instabilities in voids dominate particle cascade losses instead of IC, preventing persistent pair-beams induced by blazars from radiating at GeV energies \citep{Broderick12}. However, having a cascade implies the presence of an additional component besides the afterglow. Estimates of these cascade components as produced by GRBs are highly uncertain and require a complete multi-wavelength characterization \citep{DaVela23}.

\section{Summary and conclusions}
\label{sec:conclusion}
\grb is by far the brightest GRB ever detected by the LAT. In this analysis, we presented a comprehensive examination of this unique and intense burst.

The \Fermi-GBM instrument was triggered by an initial pulse, which was also observable in LAT data. This triggering pulse was well fitted by a \texttt{COMP} model with a particularly high peak energy of $\sim$15~MeV, which decreased to roughly 300~keV within the first 20 seconds. \nob{The high-energy emission observed by the LAT during the triggering pulse preceded the low-energy emission observed by the GBM, which is an uncommon feature in LAT-detected GRBs~\citep{2FLGC}}.

During the prompt phase, we identified a 64-second-long BTI, [\trig + $ 216.6~\mathrm{s},\mathrm{T}_0+280.6~\mathrm{s}]$, during which extremely high fluxes of X-rays and soft gamma rays created an  additional noise in all the LAT subsystems. Thanks to a special analysis, including a dedicated event reconstruction and selection, we were able to successfully analyze the LAT data and derive the GRB light curve during the BTI. 

During the first 450 seconds, the $>$ 100~MeV flux was rapidly variable with several pulses closely spaced in time. 
We were able to clearly characterize the two brightest ones, the first peaking at \trig + $247.4 \pm 0.9$\,s and the second at \trig + $330 \pm 10$\,s, although other less-significant pulses are visible in the LAT light curve. The peak of the LAT emission is not synchronized with either of the two bright pulses at $\sim$~230\,s and $\sim$~270\,s in the GBM light curve. This can be explained by the rapid variability of the spectral index, a typical signature of internal shock emission.
We performed a time-resolved spectral analysis after the LAT BTI, combining data from GBM, LAT and incorporating the results of the LHAASO-WCDA analysis. 
This study showed that a broken smoothly joined power law, which we interpreted as the synchrotron self-Compton component, is required to describe the high-energy part of the emission.
The presence of a $\sim$10~MeV line was also statistically favored and, if associated with a blue-shifted electron-positron annihilation line \citep{BOAT_LINE}, its evolution with time indicates a rapid deceleration of the jet, as suggested in~\citet{2024arXiv240312851Z}.
However, the relatively slow bulk Lorentz factor ($\Gamma \sim$~5--10) needed to boost the 511~keV line to MeV energies, compared to the much larger value obtained from GeV and TeV observations, suggests that the emission line is probably related to a slower component of the jet or a slower-evolving shell, as also proposed by \citet{BOAT_LINE}.

Such a slow region could be envisaged as connecting to the familiar spine-sheath morphology of extragalactic radio jets emanating from supermassive black holes (SMBHs).  
Evidence for such lateral structure comes from limb-brightening of
radio galaxy jets revealed in parsec scale VLBI observations \citep[e.g.,][]{2002MNRAS.336..328L,Giroletti-2004-ApJ}.  
This has driven the paradigm \citep[e.g.,][]{2005A&A...432..401G} of a mildly relativistic, outer jet, a dense sheath that carries most its kinetic energy and that entrains the external medium, surrounding a highly relativistic, lower density inner jet (spine) that possesses most of the jet's angular momentum.  
Both decelerate rapidly through the drag imposed by entrained circum-jet material. When such a configuration is employed in hydrodynamic simulations, with density ratios in the 10--100 range, the spine-sheath structure is preserved uninterrupted despite the generation of extensive turbulence in the shear layer between the two zones \citep[e.g.,][]{Meliani-2008-AandA}. 
In the GRB context, bulk motions are on average much faster than for SMBH jets. The rate of pair annihilation would be much greater in a denser and slower periphery than in a more tenuous and faster spine, which is consistent with the disparity in bulk motions inferred for the line and from the pair creation transparency arguments in Sec.~\ref{sec:gamma}.

The late-time emission decayed as a power law, but extrapolation of the late-time emission during the first 450~s showed that this component, which is commonly associated with the afterglow, was only a sub-dominant component. This suggests that the afterglow began during the prompt emission. A fit of the light curve required multiple components, two of which were necessary to reproduce the general behavior during the prompt phase, and the third was needed to reproduce the late-time emission. 
This last component could peak as early as \trig$+230$ s (rest frame), which corresponds to a deceleration time of 24 s. According to the standard fireball model, this corresponds to a bulk Lorentz factor of 230 (400) in the wind (ISM) case scenario.
Under the assumption of radiative or adiabatic expansion, high-energy events observed by the LAT are incompatible with a synchrotron origin, as seen also in GRB\,130427A \citep{2014Sci...343...42A} and in GRB\,190114C \citep{2020ApJ...890....9A}. 
In addition, the joint analysis of GBM and LAT data showed the presence of an additional component at high energies with a photon index $\Gamma_{\rm ph}$ ranging from $-1.7$ to $-1.4$. This suggests that high-energy events were more likely to be related to this extra component identified as SSC, as also suggested  by \citet{BOAT_AGILE} from the analysis of AGILE data. 
In this scenario, the keV-MeV afterglow observed by the GBM up to $\sim 10^{4}$\,s would be the synchrotron emission, although whether this emission could extend up to $\sim 33$~ks to explain the 400 GeV event detection is not clear.
However, the absence of MeV synchrotron emission disfavors an interpretation of the late-time high-energy events as due to SSC. 
Additionally, interpreting these events as TeV electromagnetic cascades from TeV events interacting with the EBL is difficult because of the particularly low magnetic fields that would be required to explain the 33~ks delay. 
We cannot definitively say whether these events originate from external Compton emission or from another source.

Overall, \grb stands out compared to other LAT GRBs, and its energetic nature, given its proximity, confirms that it is extremely rare \citep{BOAT}.
\begin{acknowledgments}
\nob{We dedicate this paper to the memory of our colleague Magnus Axelsson, who passed away prematurely in July 2022. Magnus's kindness, competence, and helpfulness made him one of our most valued colleagues. He played a central role in Fermi GRB science, and his presence on the Fermi team is greatly missed. More than a colleague, he was a friend.}

The {\lat} Collaboration acknowledges generous ongoing support from a number of agencies and institutes that have supported both the
development and the operation of the LAT as well as scientific data analysis.
These include the National Aeronautics and Space Administration and the
Department of Energy in the United States, the Commissariat \`a l'Energie Atomique and the Centre National de la Recherche Scientifique / Institut National de Physique Nucl\'eaire et de Physique des Particules in France, the Agenzia Spaziale Italiana and the Istituto Nazionale di Fisica Nucleare in Italy, the Ministry of Education, Culture, Sports, Science and Technology (MEXT), High Energy Accelerator Research Organization (KEK) and Japan Aerospace Exploration Agency (JAXA) in Japan, and the K.~A.~Wallenberg Foundation, the Swedish Research Council and the Swedish National Space Board in Sweden.

Additional support for science analysis during the operations phase is gratefully acknowledged from the Istituto Nazionale di Astrofisica in Italy and the Centre National d'\'Etudes Spatiales in France. This work performed in part under DOE Contract DE-AC02-76SF00515.

\nob{W. M. is funded by National Research Council of Thailand (NRCT): High-Potential Research Team Grant Program (N42A650868).}

This work made use of data supplied by the UK Swift Science Data Centre at the University of Leicester.
\end{acknowledgments}

\clearpage

\appendix

\section{Flux measurement during the BTI}
\label{sec:BTI}
The LAT energy range defines the part of the GRB emission that is considered as a signal for the LAT. In contrast, the intense emission of GRB221009A GRB below $\sim$30~MeV is actually a source of noise in the instrument. For clarity's sake, we name the GRB emission below and above $\sim$30~MeV the low-energy (LE) and high-energy (HE) emissions, respectively, keeping in mind that this denomination does not imply that these two parts of the GRB emission are not due to the same physical process.

\nob{During the brightest part of the prompt emission of GRB221009A, the extra noise produced by the LE emission is so important that it severely impacts the standard event reconstruction, preventing us from performing a standard data analysis. To overcome this limitation and to be able to measure the GRB HE emission, we have modified the event reconstruction and devised a new event selection, as well as original analysis and validation methods.}

In this appendix we first quantify the effect of the extra noise on the instrument and then present our dedicated data analysis to characterize the HE emission of GRB221009A. To do so, we make extensive use of simulations of photons in the LAT based on the Geant4 Monte-Carlo toolkit~\citep{AGOSTINELLI2003250,1610988}, \nob{including} the simulation of photons between 10~MeV and 10~GeV produced for the LLE analysis.
All time intervals are indicated in seconds with respect to the GRB trigger $\mathrm{T}_0$.

\subsection{Effect of the intense X-ray and soft gamma-ray flux on the LAT instrument}

\subsubsection{Energy estimation}
\label{sec:energy_estimation}

Below $\sim10$~GeV, the standard Pass~8 energy reconstruction~\citep{2013arXiv1303.3514A} makes use of the tracker and the calorimeter information, the former being critical for gamma rays below $\sim1$~GeV. We estimate the energy deposited in the tracker from the number of strips fired in the thin and thick sections of the tracker. For the calorimeter, we perform a clustering of the energy deposits and use the highest-energy cluster to compute the event energy.

Because the numbers of fired strips are computed in the whole tracker, they include the strips due to the extra noise. In order to quantify this noise contribution to the energy estimation, we design an alternative way to count the fired strips, namely within a cone around the event track axis ($d<20+0.1 \times l$, where $d$ and $l$ are the distance in mm to and along the track axis, respectively, with $l=0$ at the conversion point and only $l>0$ is considered). Using simulations, we find that the cone selection reduces the number of fired strips included in the tally by $\sim$~40\% and the resulting energy estimation is thus affected by a negative bias that ranges from 40\% at 60~MeV to 10\% at 1~GeV. We correct for this bias to compute the alternative energy estimator $E_\mathrm{cor}$.

\begin{figure*}[t]
    \centering    
    \includegraphics[width=0.7\linewidth]{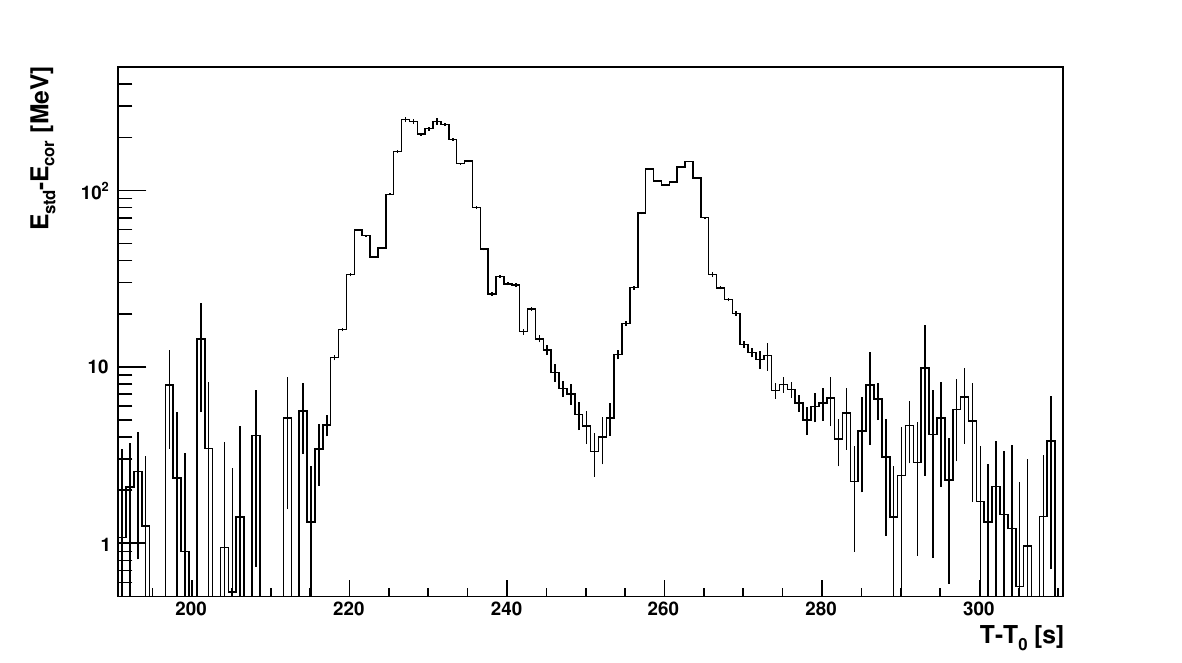}
    \includegraphics[width=0.7\linewidth]{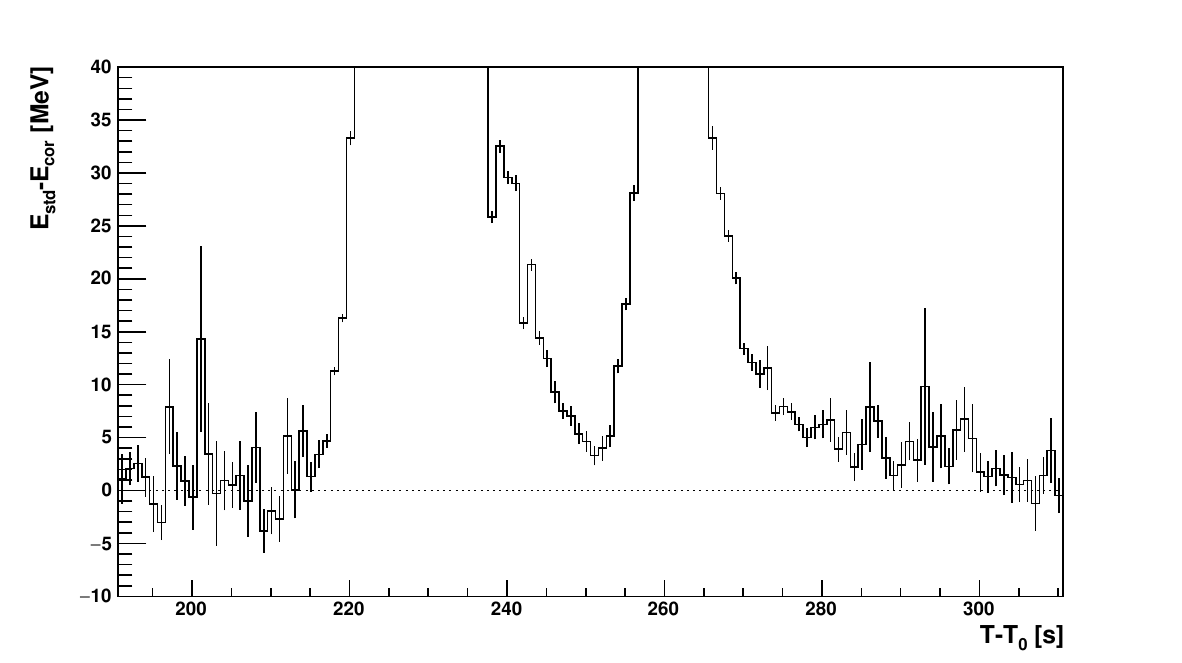}
    \caption{Average difference as a function of time since the GBM trigger between the reconstructed energy using the standard algorithm and the new modified energy estimator in log scale (top) and linear scale (bottom). \nob{This difference corresponds to the additional equivalent energy due to the extra noise in the tracker induced by the GRB LE flux.}}
    \label{fig:diff_energy_corr}
\end{figure*}

The difference between the standard and modified energies corresponds to the additional equivalent energy due to the extra noise in the tracker induced by the GRB LE flux. Fig.~\ref{fig:diff_energy_corr} shows the average of this difference as a function of time for events with at least one track passing the trigger and gamma filter. Before 218~s and after 290~s the extra energy is close to 0, as during normal conditions, but during most of the time interval [218~s,\,290~s], the extra energy is clearly not negligible. As a consequence, the standard energy evaluation is strongly biased and we need to develop a new event energy estimation. One can see two peaks of activity, P1 around 230~s and P2 around 260~s, that match the double-peak structure of the GBM light curve reported in \cite{2023ApJ...952L..42L}. In the interpeak (IP) period, the extra energy reduces almost to 0, which means that the impact of the GRB LE flux during the IP period is very much reduced compared to P1 and P2.

\begin{figure*}[t]
    \centering
    \includegraphics[width=0.9\linewidth]{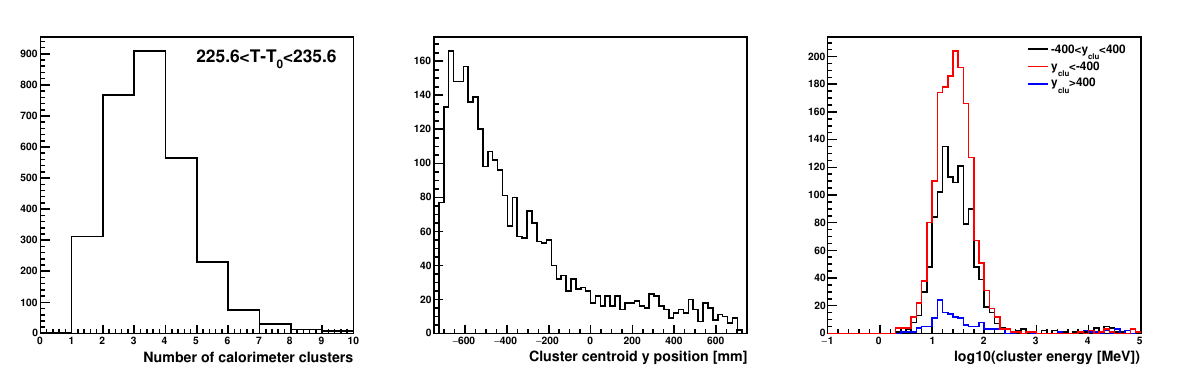}
    \includegraphics[width=0.9\linewidth]{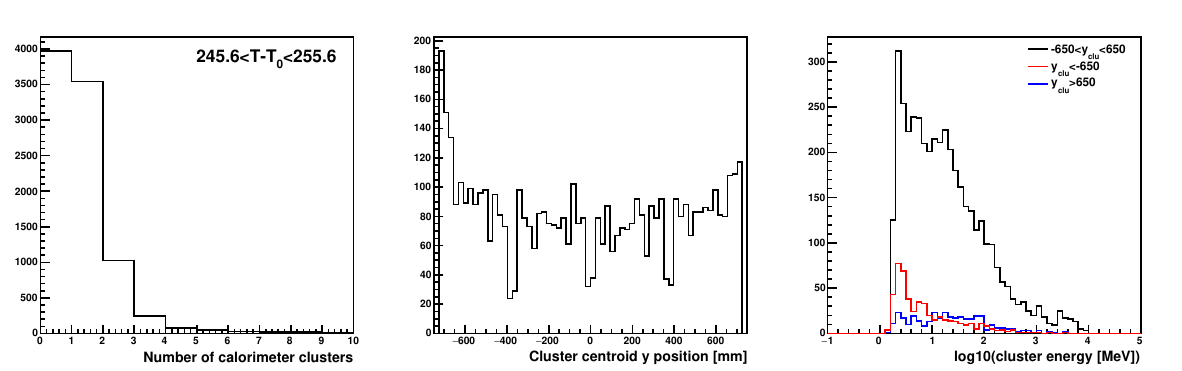}
    \includegraphics[width=0.9\linewidth]{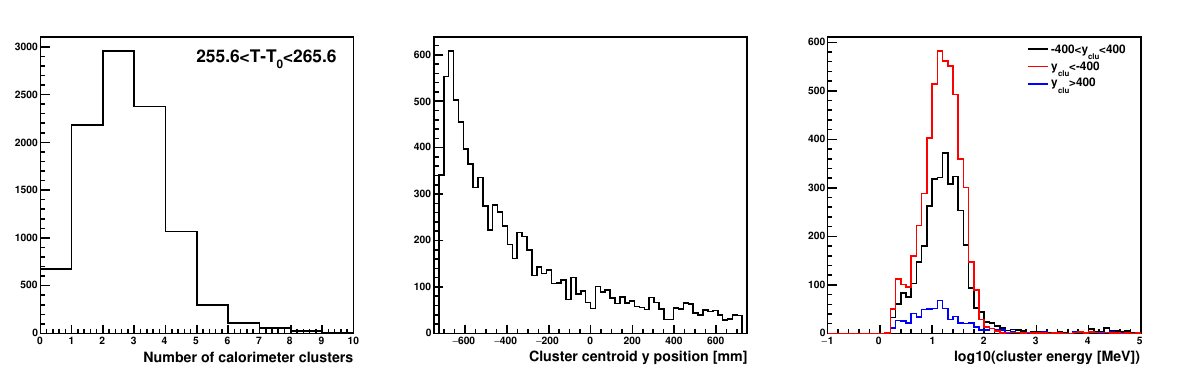}
    \caption{The number of calorimeter clusters (left), the y-position of the highest-energy cluster (center) and its energy depending on whether or not it is close to the $-$Y face (right) for events with at least one track passing the trigger and gamma filter in the time interval [225.6~s,\,235.6~s] (top), [245.6~s,\,255.6~s] (middle) and [255.6~s,\,265.6~s] (bottom). \nob{The distributions of the y-position of the highest-energy cluster show that the extra noise induced in the calorimeter by the GRB LE flux is located in the $-$Y face of the calorimeter. Comparing the energy distribution of the clusters close to the $-$Y and +Y faces (red and blue histograms, respectively) allows an estimation of the equivalent additional energy due to the extra noise in the calorimeter.}}
    \label{fig:cal_cluster_energy}
\end{figure*}

Regarding the impact of the extra noise on the calorimeter reconstruction, it must be noted first that this impact is naturally mitigated by the clustering introduced in Pass~8. The calorimeter is much denser than the tracker, so the X rays and soft gamma rays deposit their energy over shorter distances. During the prompt phase, the $-$Y-face of the LAT\footnote{The Z-axis of the instrument reference frame corresponds to the LAT boresight.} faced the GRB and the noise-induced clusters are expected to lie close to the $-$Y~edge of the calorimeter. This is what we observe, as can be seen in Fig.~\ref{fig:cal_cluster_energy}, for events with at least one track passing the trigger and gamma filter. During P1 and P2, most events have at least one cluster in the calorimeter and its position lies in the $-$Y~half of the instrument. Comparing the energies of the clusters close to the $-$Y~edge and the +Y~edge allows us to quantify the additional energy due to the extra noise in the calorimeter: it is most of the time above 10~MeV and can reach 100~MeV. The situation is very different for the IP interval, during which the extra energy is less than 10~MeV.

\begin{figure*}[t]
    \centering
    \includegraphics[width=0.45\linewidth]{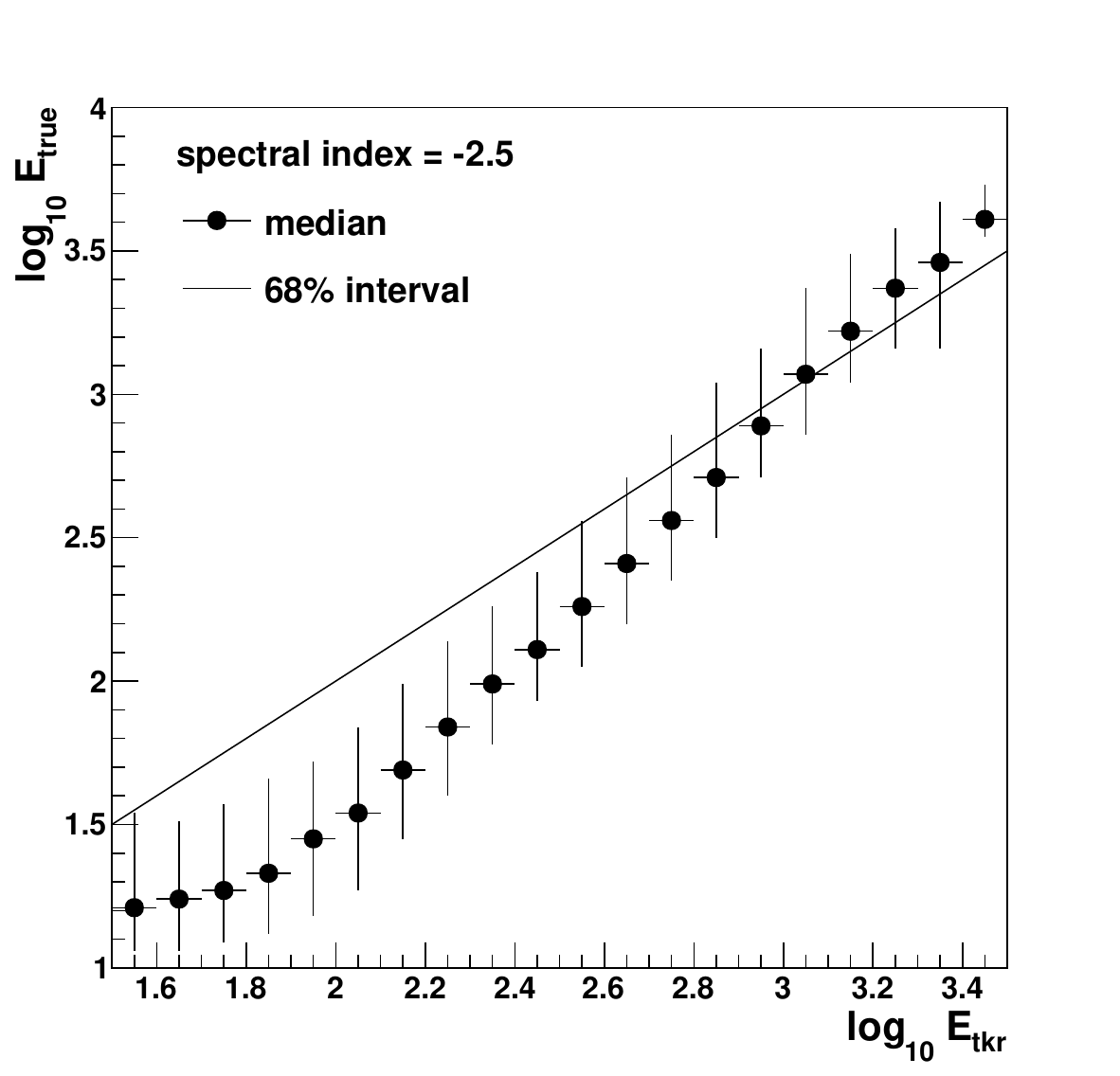}
    \includegraphics[width=0.45\linewidth]{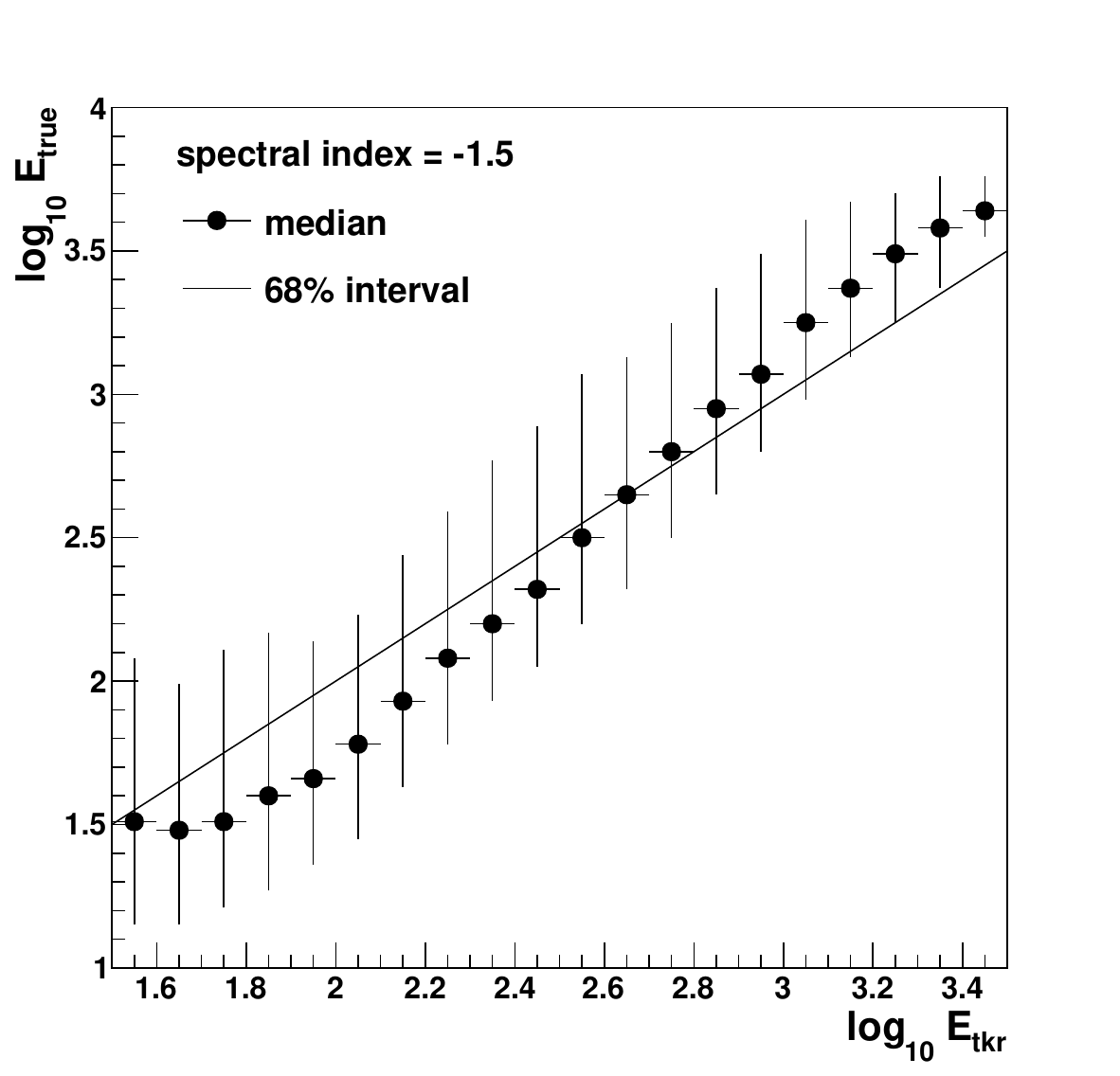}    \caption{Characterization of the energy estimator $E_\mathrm{tkr}$. Left: \nob{true energy median (dots) and 68\% interval (vertical bars) as a function of $\log_{10} E_\mathrm{tkr}$ for a simulation of a gamma-ray source with a spectral index of $-$2.5.} The dashed line corresponds to equality between the true energy median and $E_\mathrm{tkr}$. Right: same but with a spectral index of $-$1.5. For $E_\mathrm{tkr}<100~\mathrm{MeV}$, the true energy median is almost constant and therefore $E_\mathrm{tkr}$ can only be used above 100~MeV.}
    \label{fig:MClogEtkr}
\end{figure*}

Because the first cluster energy information cannot be used during P1 and P2, we choose to rely only on the tracker information to estimate the event energies during the BTI. As described above, we use the number of fired strips within a cone around the event track axis to compute $E_\mathrm{tkr}$. In order to characterize the performance of this energy estimator and especially to check the energy range in which it is usable, we look at the true energy median and the true energy 68\% containment interval as a function of $E_\mathrm{tkr}$. They are shown in Fig.~\ref{fig:MClogEtkr} for a simulation of a power-law spectrum, for two spectral indices ($-$2.5 and $-$1.5). One can see that the $E_\mathrm{tkr}$ energy estimator does rather poorly at low energies: the true energy median is almost constant for $E_\mathrm{tkr}<100~\mathrm{MeV}$ and therefore $E_\mathrm{tkr}$ can only be used above 100~MeV.

\subsubsection{Estimation of the trigger and on-board filter efficiency}
\label{sec:Limb_efficiency}

In addition to creating extra noise in the tracker and the calorimeter, the GRB LE flux also produced signals in the ACD. Since the ACD is instrumental to the trigger and the on-board filter selection, we must check whether the extra noise in the ACD significantly changed the trigger and filter efficiency.

To be recorded, an incoming photon must trigger the LAT and pass the on-board gamma-ray filter. The LAT trigger system defines several sets of conditions which are called trigger engines~\citep{2012ApJS..203....4A}. Two trigger engines are designed to accept gamma rays. Trigger engine~6 requires at least one calorimeter CsI crystal with more than 1~GeV, which corresponds to relatively high-energy gamma rays ($\sim 5$~GeV). Lower-energy gamma rays, which do not pass this condition, are accepted by trigger engine~7, which requires a trigger signal from one tracker tower and no signal above the veto threshold ($\sim 0.45$ of the signal of a minimum ionizing particle) in any of the ACD tiles above the tracker tower. The on-board gamma-ray filter performs a sequence of veto tests that depend on the event topology. An important case is trigger engine~7 events without any calorimeter CsI crystal collecting more than 100~MeV: one ACD tile above the veto threshold is enough to reject these events. 

Verifying a selection efficiency implies comparing it with a reference. Very fortunately, the Earth limb came into the LAT field of view at about the time of the GBM trigger, providing a reference signal that we use to check the trigger and gamma filter efficiencies.
The Earth-limb signal in the LAT depends on the position and orientation of the spacecraft. We have selected a reference data set including about 60~h~\footnote{Between 2020 September 1, at 04:09:09.000 UTC and 2022 November 20, at 14:35:11.000 UTC.} in which the satellite has a position/orientation very close to that between 100 and 1000~s after the GRB trigger. Using the reference data set, we prepared the live-time corrected zenith angle $\theta_\mathrm{zen}$ distribution for a set of bins in geocentric longitude and latitude (the longitude and latitude bin widths are $10^\circ$ and $0.5^\circ$, respectively). Then, for a given time interval during the GRB emission, we construct the reference $\theta_\mathrm{zen}$ template by adding the $\theta_\mathrm{zen}$ distributions scaled by the live time in each of the geocentric longitude and latitude bins during the time interval.

For the verification of trigger and gamma filter efficiencies, we use trigger engine~7 events that passed the gamma filter, with at least one track and without any calorimeter CsI crystal collecting more than 100~MeV. The latter condition allows the selection to be sensitive to the presence of any ACD tile veto. To increase the significance of the Earth-limb signal, we also require that $E_\mathrm{tkr}>125$~MeV.

\begin{figure*}[h]
    \centering    
    \includegraphics[width=0.45\linewidth]{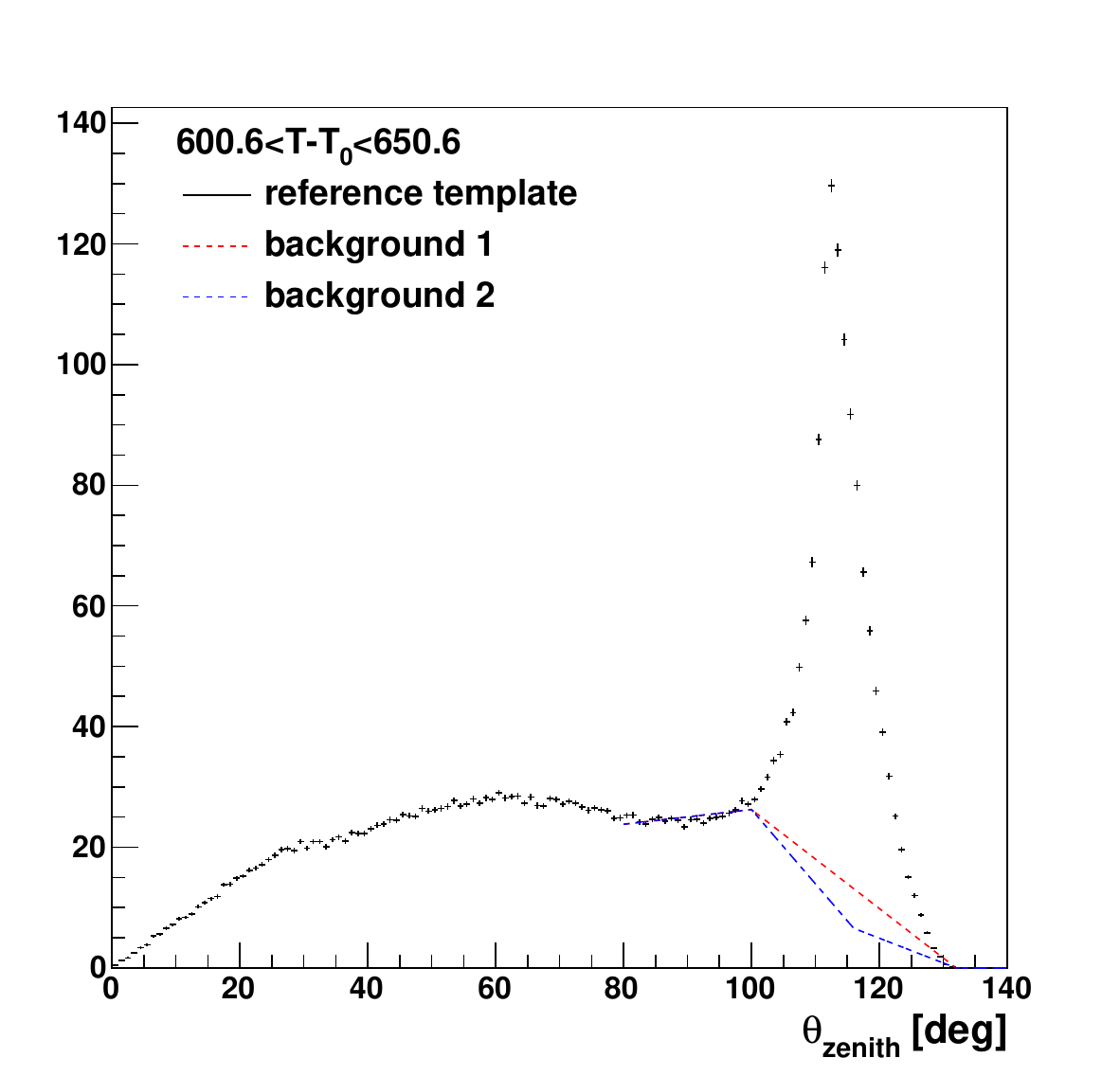}
    \includegraphics[width=0.45\linewidth]{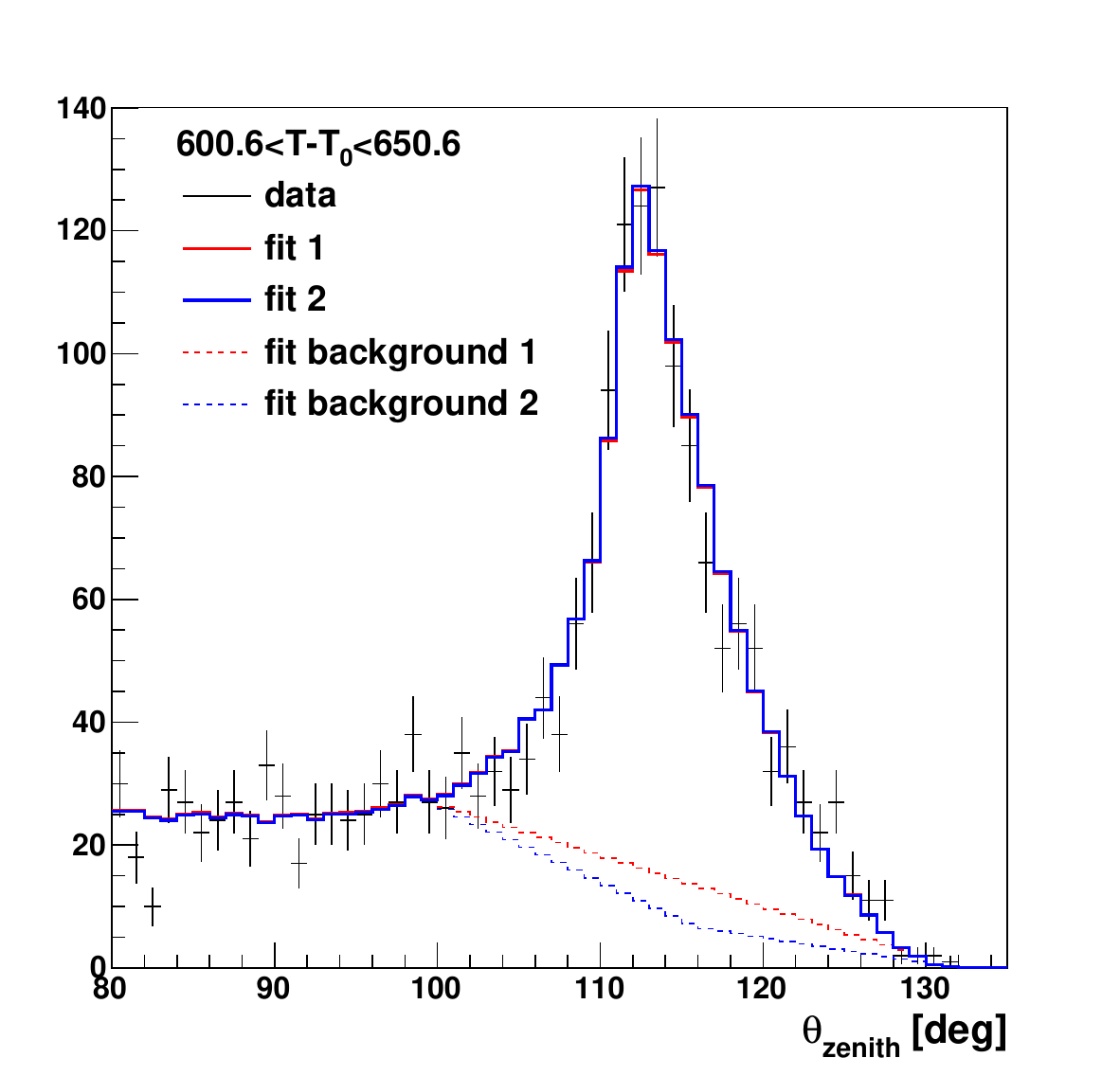}
    \caption{Left: reference $\theta_\mathrm{zen}$ template for the  time interval [600.6,\,650.6]. \nob{The Earth limb peak is centered at about $113^\circ$.} The red and blue dashed curves correspond to the background template for the two background hypotheses, $b_{116} = b_{100}/2$ (background 1) and $b_{100}/4$ (background 2), respectively. Right: $\theta_\mathrm{zen}$ distributions for the [600.6,\,650.6] time interval (black). The solid red and blue curves (barely distinguishable) correspond to the signal+background distribution predicted by the fits for the two background hypotheses, while the dashed red and blue curves show the corresponding background contributions.}
    \label{fig:trigger_filter_efficiency_template}
\end{figure*}

Fig.~\ref{fig:trigger_filter_efficiency_template}~(left) shows the reference distribution for the time interval [600.6,\,650.6]. The Earth-limb peak is clearly visible on top of a broader background distribution due to charged cosmic rays as well as celestial gamma rays. Above $80^\circ$ zenith angle the background distribution is assumed to be a piece-wise linear function defined by four parameters $b_{80}, b_{100}, b_{116}$ and $b_{132}$, the background levels at $\theta_\mathrm{zen} = 80$, 100, 116 and $132^\circ$, respectively, with $b_{132}$ fixed to 0. The Earth-limb peak is modeled as a sum of two gaussians centered on $\sim 113^\circ$. Fitting several reference templates of 50~s time intervals above $80^\circ$, we find that the background ratio $b_{116}/b_{100}$ fluctuates between 0.25 and 0.5. As a consequence, we assume two hypotheses for the background distribution: $b_{116} = b_{100}/4$ and $b_{100}/2$. For each background hypothesis, the fit of the reference distribution allows us to derive the reference background template, as shown in Fig.~\ref{fig:trigger_filter_efficiency_template}~(left). The signal template is simply defined by subtracting the background template from the reference distribution.

For any time interval during the GRB emission, we fit the $\theta_\mathrm{zen}$ distribution with the sum of the background and signal templates, weighted by the free parameters of the fit, $f_b$ and $f_s$, respectively. The relative efficiency with respect to normal conditions is directly given by $f_s$. The result of the fit for the time interval [600.6,\,650.6] is shown in Fig.~\ref{fig:trigger_filter_efficiency_template}~(right). For each background hypothesis, the fit gives $f_s$ and $\delta f_s$, its uncertainty. To combine the results of the two background hypotheses, we use the minimum of $f_s - \delta f_s$ and the maximum of $f_s + \delta f_s$ of the two background hypotheses to define a 68\% confidence interval, from which we derive the combined relative efficiency and its uncertainty.
 
\begin{figure*}[t]
    \centering    
    \includegraphics[width=0.9\linewidth]{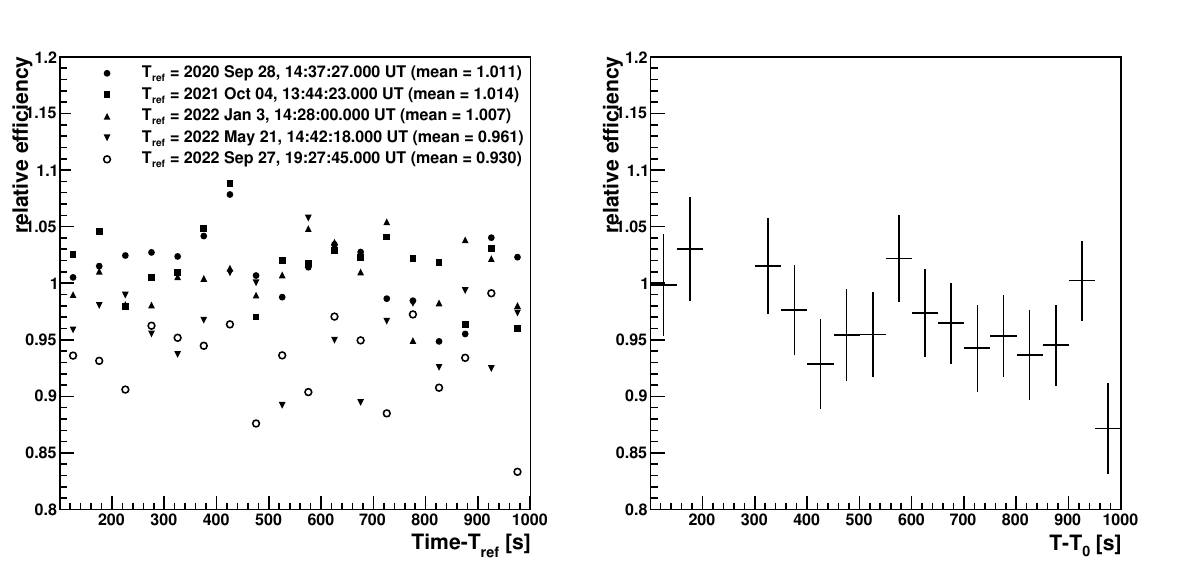}
    \caption{Left: relative efficiency in 50~s bins for five data subsets from 100~s to 1000~s after $\mathrm{T}_\mathrm{ref}$, the time when the satellite is at the same position/orientation as at the time of the GRB trigger. All points have a $\sim$5\% uncertainty which is not displayed for clarity's sake. \nob{The mean relative efficiency of each subset is indicated between parentheses.} Right: relative efficiency in 50~s bins from $\mathrm{T}_0+100$ to $\mathrm{T}_0+1000~\mathrm{s}$., except for the two bins corresponding to the BTI.} 
    \label{fig:trigger_filter_efficiency_scan}
\end{figure*}

To verify the method, we compute the relative efficiency for 50~s time intervals using several 900~s subsets of the reference data set. For each subset, we do not use the subset to build the reference template and the times are computed with respect to $T_\mathrm{ref}$, the time for which the satellite is at the same position/orientation as at the time of the GRB trigger. The results are shown in Fig.~\ref{fig:trigger_filter_efficiency_scan}~(left). The uncertainties, not shown in the figure for the sake of clarity, are of the order of 0.05. The average efficiencies of the subsets vary between 0.93 and 1.05 and the reduced $\chi^2$ of a fit by a constant is $\sim 1$, which proves that, when taking into account the average shift of each subset, the uncertainty on the relative efficiency is correctly estimated.

The relative efficiency for 50~s time intervals from 100 to 1000~s after the GBM trigger, except during the BTI, is shown in Fig.~\ref{fig:trigger_filter_efficiency_scan}~(right). A fit by a constant gives an average of 0.97 and a reduced $\chi^2$ of 1.

\begin{figure*}[t]
    \centering    
    \includegraphics[width=0.45\linewidth]{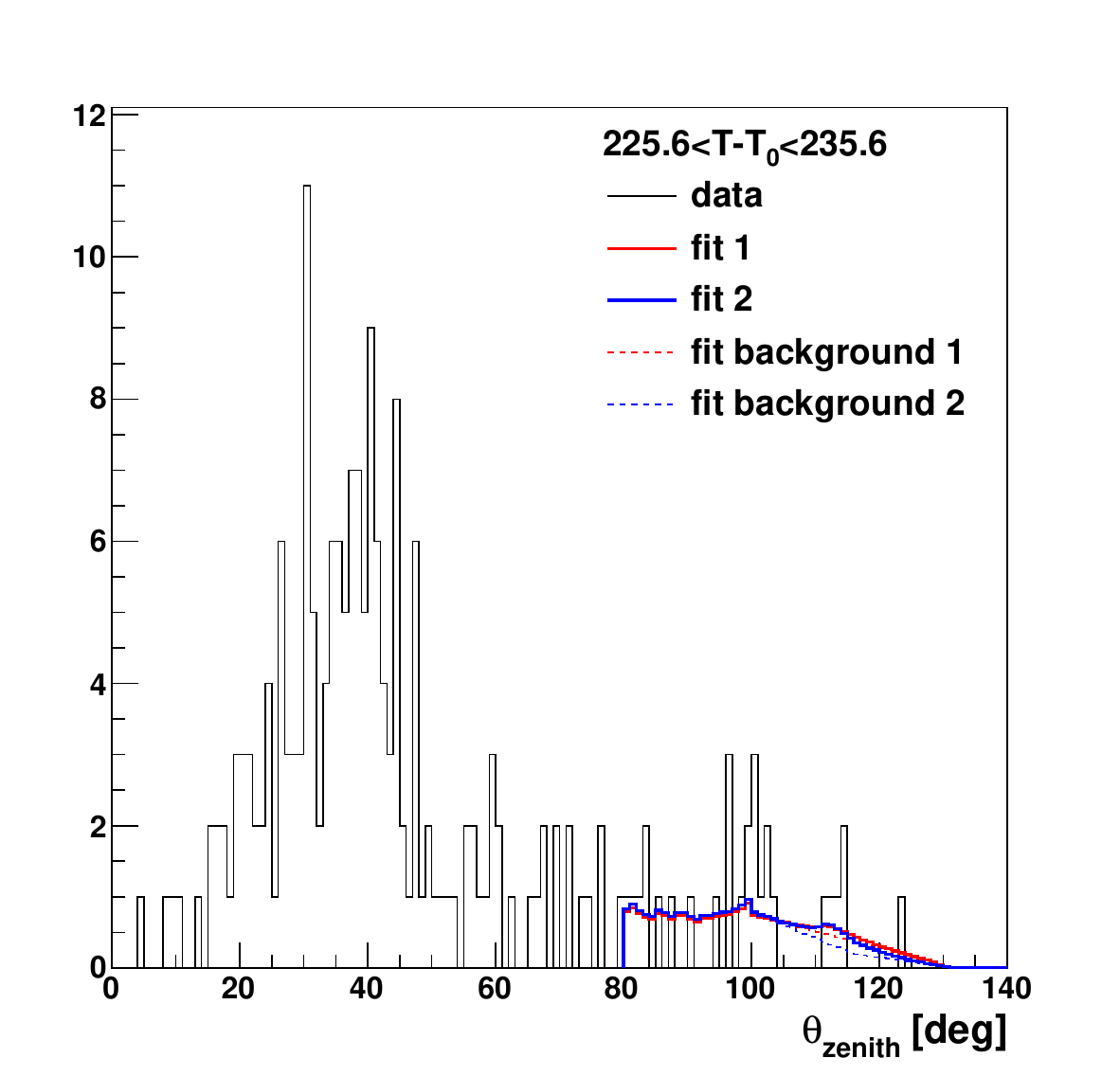}
    \includegraphics[width=0.45\linewidth]{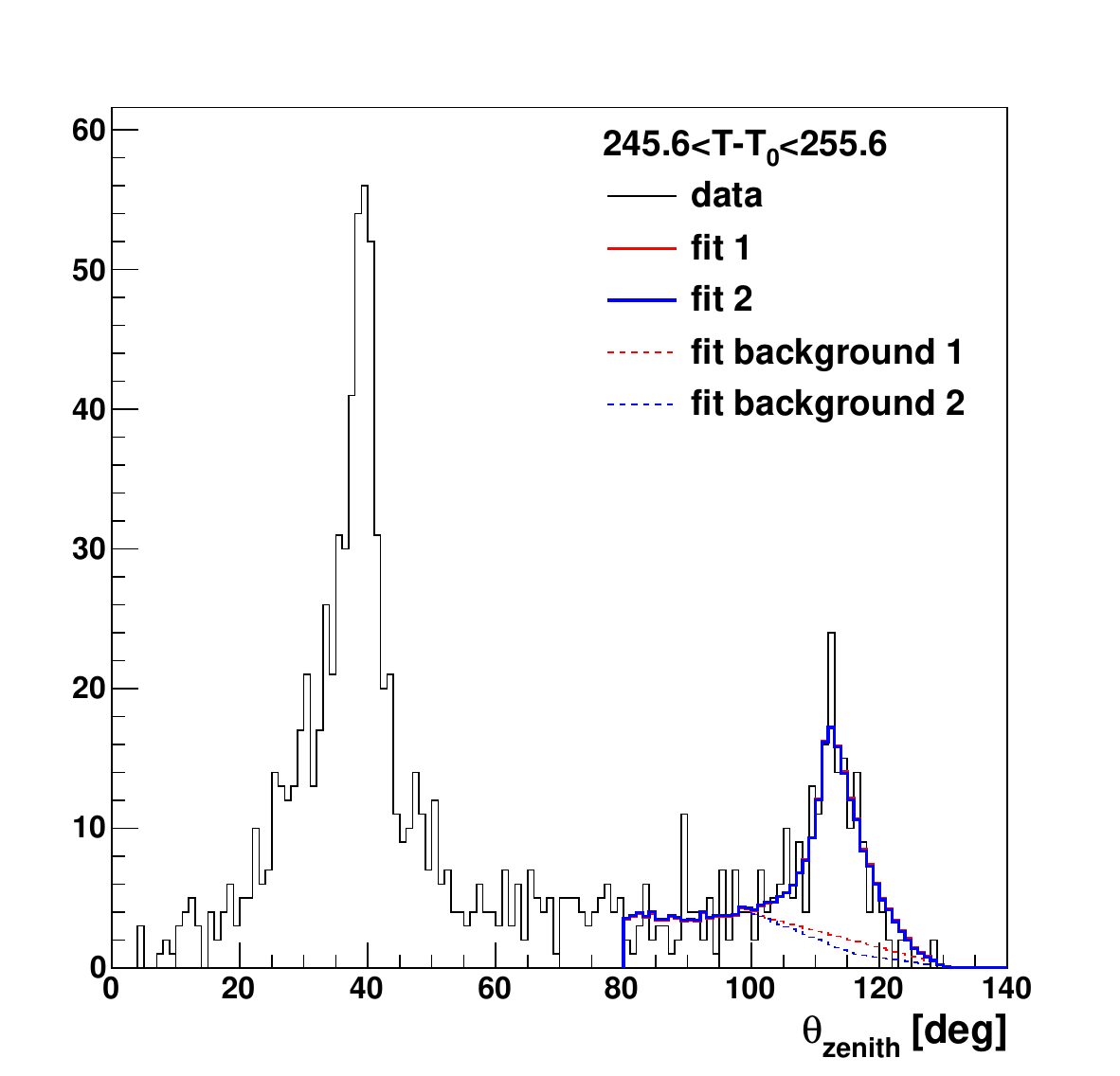}
    \caption{Left: $\theta_\mathrm{zen}$ distribution for the [225.6~s,\,235.6~s] time interval. Right: same for the [245.6~s,\,255.6~s] time interval. The solid red and blue curves correspond to the signal+background distribution predicted by the fits for the two background hypotheses, whereas the dashed red and blue curves show the corresponding background contributions. \nob{The Earth limb peak at about $113^\circ$ is clearly visible in the [245.6~s,\,255.6~s] time interval, whereas it seems absent in the [225.6~s,\,235.6~s] time interval.}}
    \label{fig:trigger_filter_efficiency_P1IPP2}
\end{figure*}

The results of the fit of the $\theta_\mathrm{zen}$ distributions for the time intervals [225.6~s,\,235.6~s] and [245.6~s,\,255.6~s], roughly corresponding to P1 and IP, are shown in Fig.~\ref{fig:trigger_filter_efficiency_P1IPP2}. The Earth-limb peak is clearly visible for IP but not for P1, which is consistent with the derived relative efficiencies: $0.13 \pm 0.12$, $0.92 \pm 0.10$ for P1 and IP, respectively. We also note that the two $\theta_\mathrm{zen}$ distributions both peak at about $40^\circ$, which corresponds to events coming from the GRB direction.

\begin{figure*}[t]
    \centering    
    \includegraphics[width=0.45\linewidth]{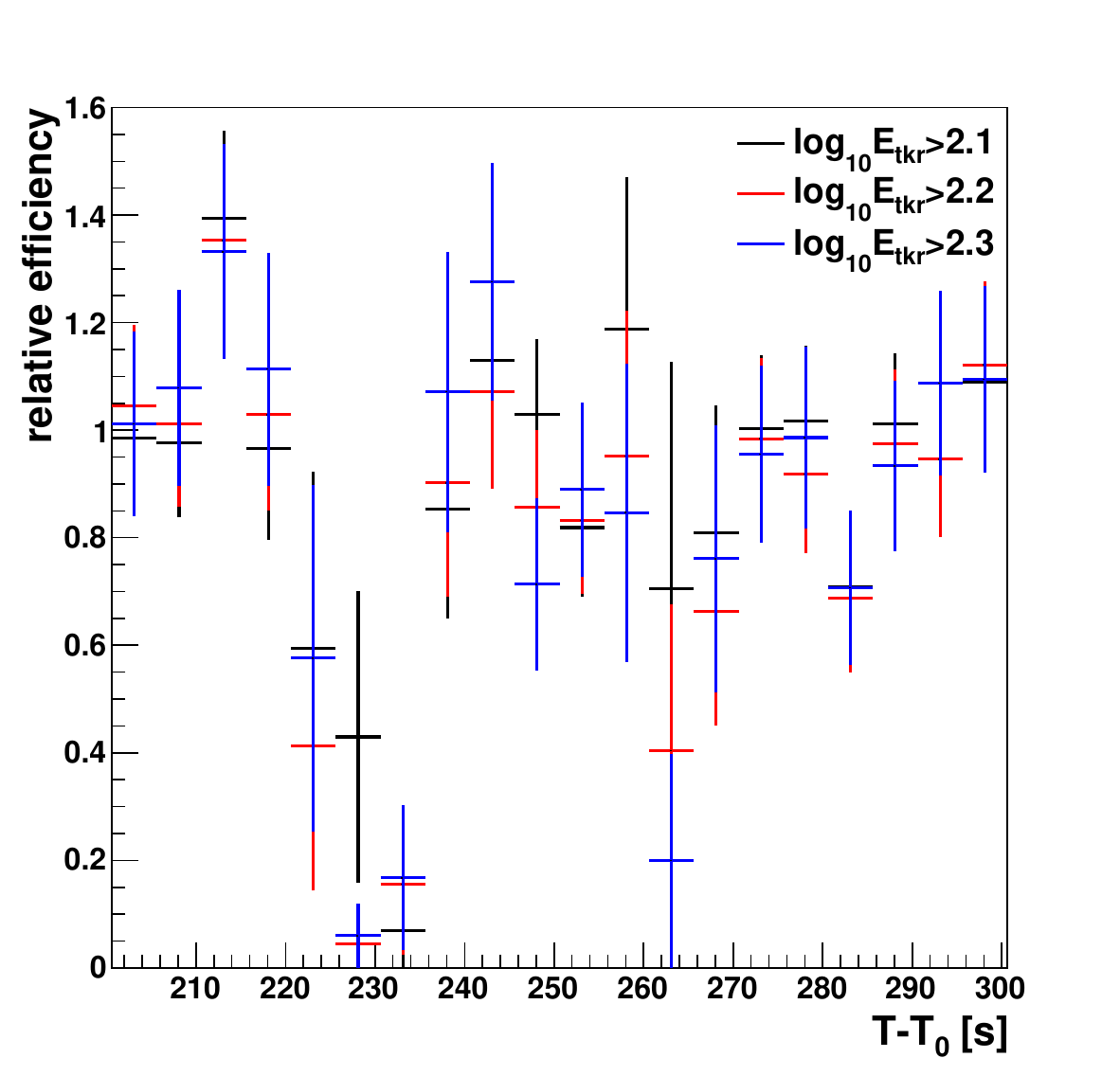}
    \includegraphics[width=0.45\linewidth]{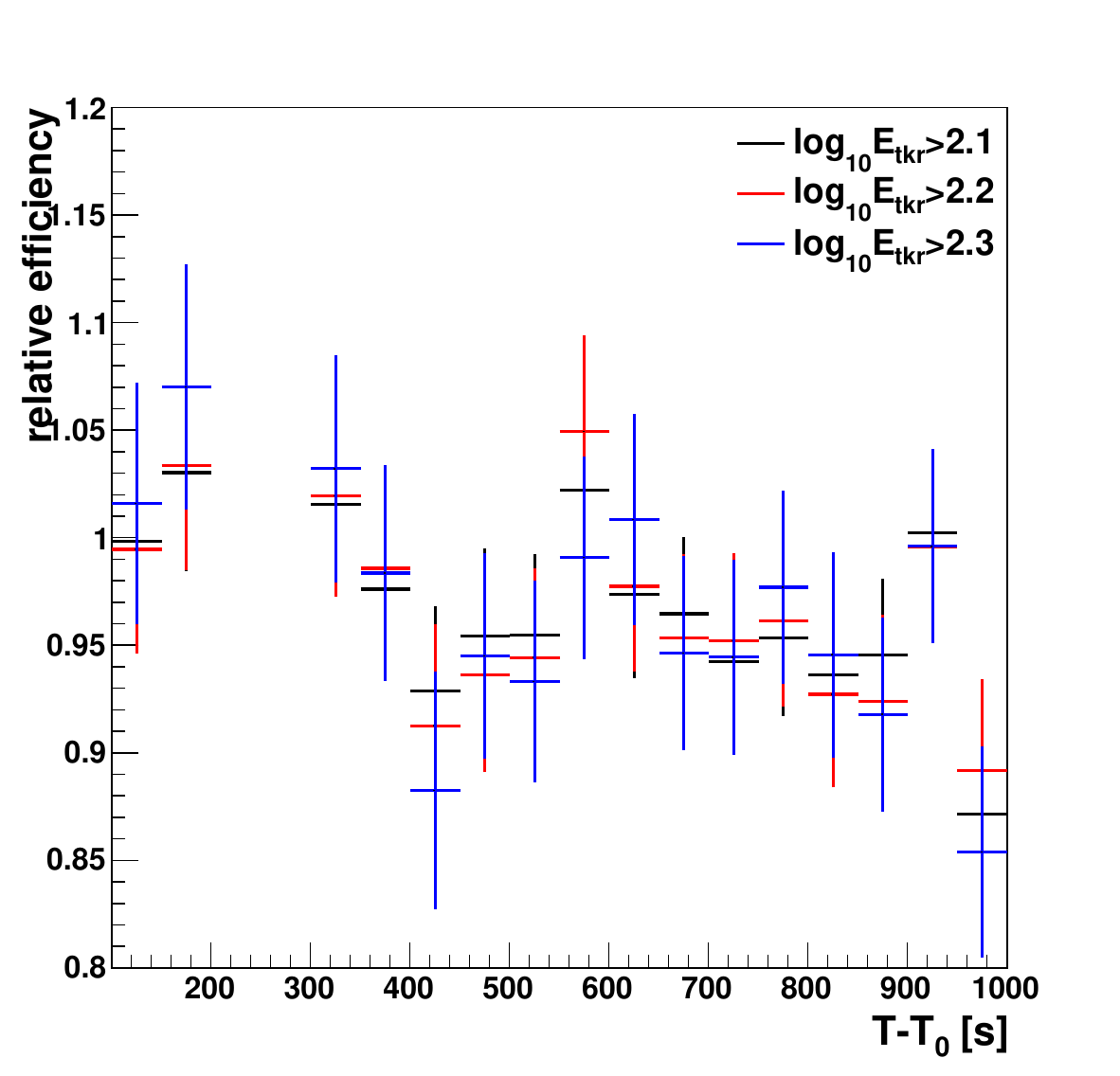}
    \caption{Left: relative efficiencies in 5~s bins from $\mathrm{T}_0+200$ to $\mathrm{T}_0+300~\mathrm{s}$ for three energy thresholds. \nob{The efficiencies decrease significantly during P1 and P2.} Right: same for 50~s bins from $\mathrm{T}_0+100$ to $\mathrm{T}_0+1000~\mathrm{s}$.}
    \label{fig:trigger_filter_efficiency_comp}
\end{figure*}

To derive the GRB light curve, we apply the \nob{same} method to 5~s time intervals between 200~s and 300~s. The results are shown in Fig.~\ref{fig:trigger_filter_efficiency_comp}~(left), for three $\log_{10}E_\mathrm{tkr}$ thresholds. These results allow us to characterize the relative efficiency variation with time (especially the efficiency drop during P1), but the derived efficiencies are not very precise. Furthermore, they change significantly when increasing the energy threshold, which is not the case for the 50~s time intervals, as demonstrated by Fig.~\ref{fig:trigger_filter_efficiency_comp}~(right). Another issue is that the efficiency during P1 is compatible with 0. Since there seems to be a GRB signal during P1, as suggested by the peak at approximately $40^\circ$ zenith angle in Fig.~\ref{fig:trigger_filter_efficiency_P1IPP2}~(left), an efficiency compatible with 0 would not allow us to derive an upper limit of the GRB flux. To overcome this limitation, we develop an alternative method presented in the next section.

\subsection{GRB flux measurement during the BTI}

Measuring the GRB flux during the BTI requires an event selection and the estimation of its effective area. In the case of fast transient signals like GRBs, good sensitivity can be achieved with a minimal event selection. For the prompt emission of GRB~221009A, using a minimal event selection is actually recommended because it helps us to mitigate the systematics produced by the extra noise induced by the GRB LE emission. We choose the most simple event selection: Engine~7 events passing the gamma filter, with at least one track and above a given $E_\mathrm{tkr}$ threshold, which will be defined later.

As explained in the previous Section, it is not possible to precisely estimate the event selection effective area with LAT data only. The only possibility is thus to use realistic simulations of individual HE photons on top of a background of LE photons. In order to do so, we need to know the GRB LE spectrum. This information is not provided by the GBM results due to the pulse pile-up effect during P1 and P2~\citep{2023ApJ...952L..42L}. In this section, we first describe how we constrain the GRB LE flux and we derive the event selection effective area as a function of time during the BTI. We then present the GRB light curve analysis. 

\subsubsection{Characterization of the extra noise in the instrument}
\label{sec:LEspectrum}

\begin{figure*}[t]
    \centering    
    \includegraphics[width=0.5\linewidth]{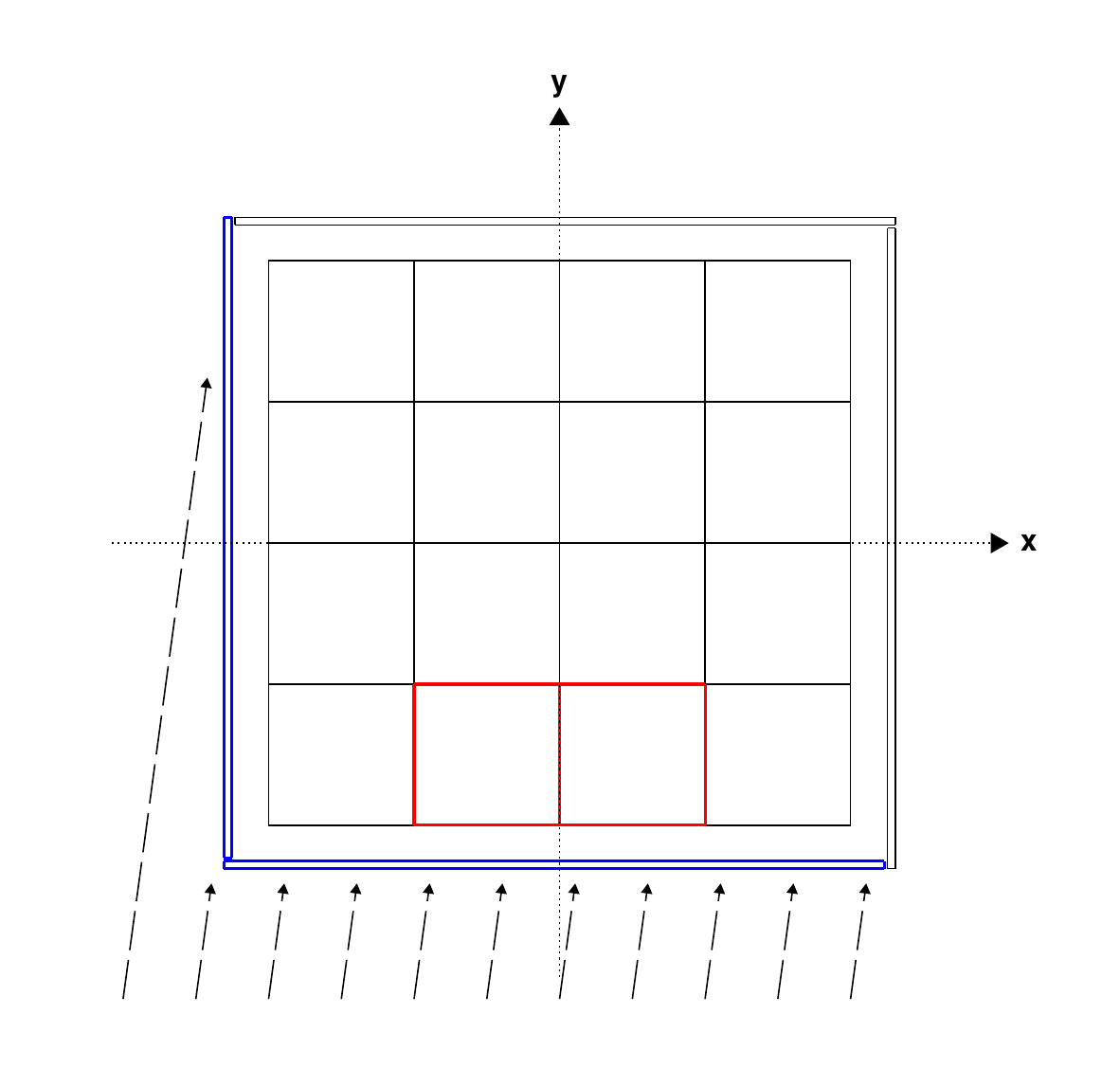}
    \caption{Top view of the LAT. The $4\times4$ square matrix corresponds to the $4\times4$ tracker towers. The two towers used to measure the corner occupancy and mean number of strips are highlighted in red. The two ACD $-$X and $-$Y bottom tiles are highlighted in blue. \nob{The dashed arrows illustrate the GRB photon direction, showing that the GRB photons first arrive at the $-$Y face of the LAT.}}
    \label{fig:LATgeometry}
\end{figure*}

The photon absorption cross section strongly decreases with energy until it reaches a minimum at about 3 MeV, above which it is almost constant. So, the interaction of the GRB LE photons occurs very soon after the photons enter the LAT, while the interaction of HE photons occurs deeper in the instrument. Since the GRB is very far off-axis during the BTI ($\sim 74^\circ$ with respect to boresight) and its direction is close to the $y$-axis, as shown in Fig.~\ref{fig:LATgeometry}, the most exposed part of the tracker is the top planes of the $-$Y-face towers. That is the reason why we use the information of the two top planes of the two central towers of the $-$Y face of the LAT to monitor the extra noise level. We refer to this part of the tracker as the tracker corner. These two towers are highlighted in red in Fig.~\ref{fig:LATgeometry}.

\begin{figure*}[t]
    \centering
    \includegraphics[width=0.9\linewidth]{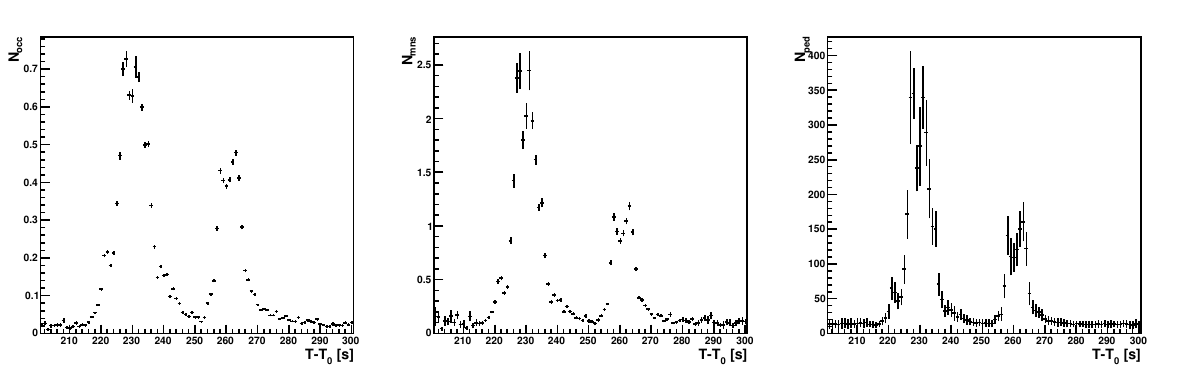}
    \caption{The time variation of the tracker-based quantities used to monitor the noise induced by the GRB LE emission. Left: the corner occupancy. Center: the corner mean number of strips. Right: the total number of strips pedestal. \nob{All quantities show the P1/IP/P2 structure.}}
    \label{fig:TkrNoise}
\end{figure*}

We compute two quantities: $\mathrm{N}_\mathrm{occ}$, the corner occupancy (the average of the fractions of events with at least one fired strip in the corner planes), and $\mathrm{N}_\mathrm{mns}$ the corner mean number of fired strips. Fig.~\ref{fig:TkrNoise} shows their variation, which makes apparent the features already seen in Fig.~\ref{fig:diff_energy_corr}, namely P1, IP and P2. We note that, for individual HE photons, the expected levels of the corner occupancy and the mean number of fired strips are about 1\% and 0.1, which is much lower than the levels reached during P1 and P2.

The average number of fired strips in the whole tracker could help us to characterize the extra noise, but it is likely biased by the signal induced by HE photons. To avoid this bias, we estimate the pedestal, that is to say the bottom edge, of this quantity, for any given time interval. To do so, we compute $N_{1\%}$ and $N_{20\%}$, the 1\% and 20\% quantiles of the distribution of the total number of fired strips, and define $\mathrm{N}_\mathrm{ped} = (N_{1\%}+N_{20\%})/2$ and its uncertainty $(N_{20\%}-N_{1\%})/2$. The P1, IP and P2 features are also clearly apparent in the variation of $\mathrm{N}_\mathrm{ped}$, as can be seen in Fig.~\ref{fig:TkrNoise}.

\begin{figure*}[t]
    \centering
    \includegraphics[width=0.9\linewidth]{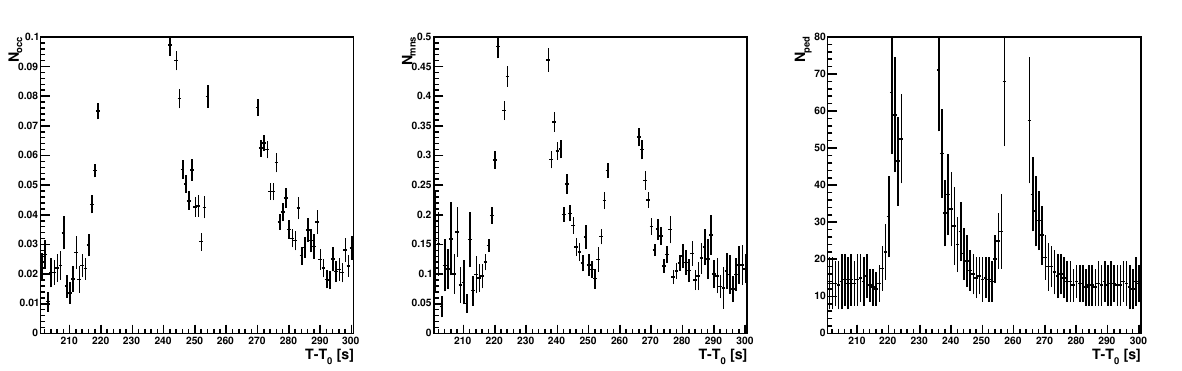}
    \caption{Same as in Fig.~\ref{fig:TkrNoise} but with the $y$-axis ranges set to show the baseline levels.}
    \label{fig:TkrNoiseZoom}
\end{figure*}

Figure~\ref{fig:TkrNoiseZoom} shows the same information as in Fig.~\ref{fig:TkrNoise}, but with a $y$-axis range set to show the baselines of $\mathrm{N}_\mathrm{occ}$, $\mathrm{N}_\mathrm{mns}$ and $\mathrm{N}_\mathrm{ped}$.

\begin{figure*}[t]
    \centering
    \includegraphics[width=0.45\linewidth]{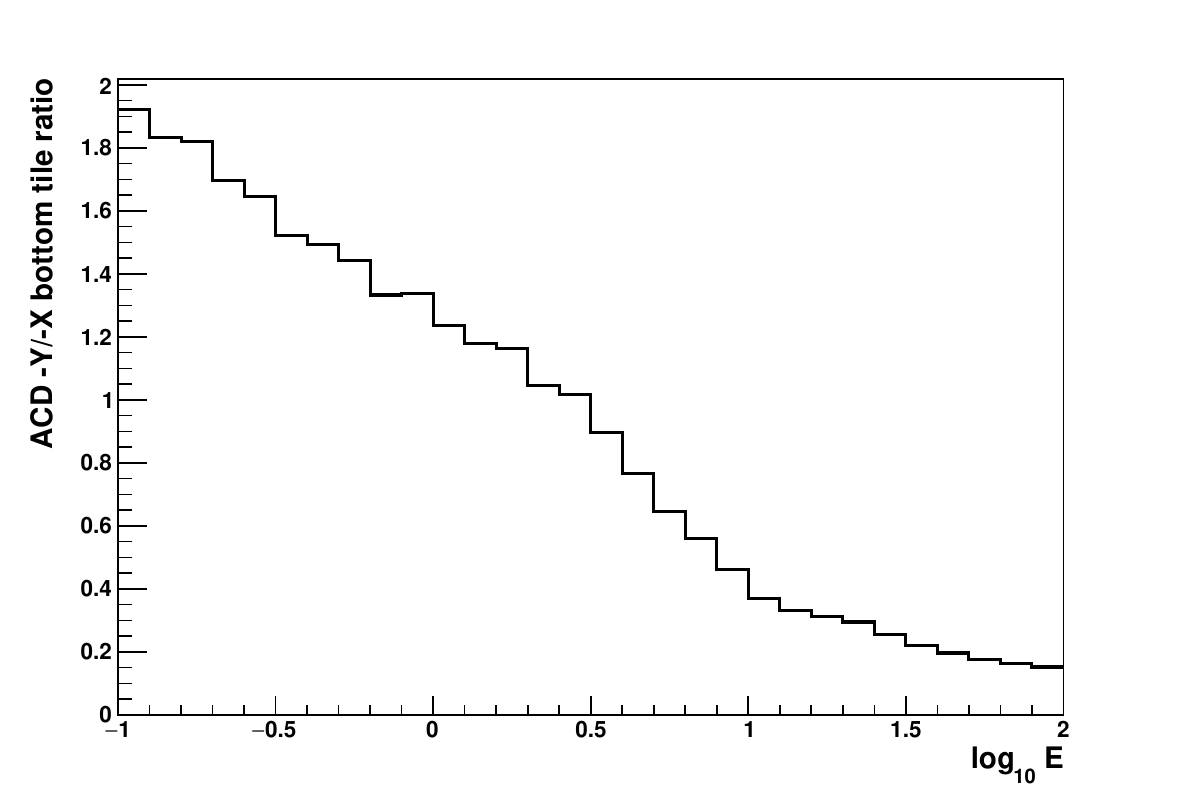}
    \includegraphics[width=0.45\linewidth]{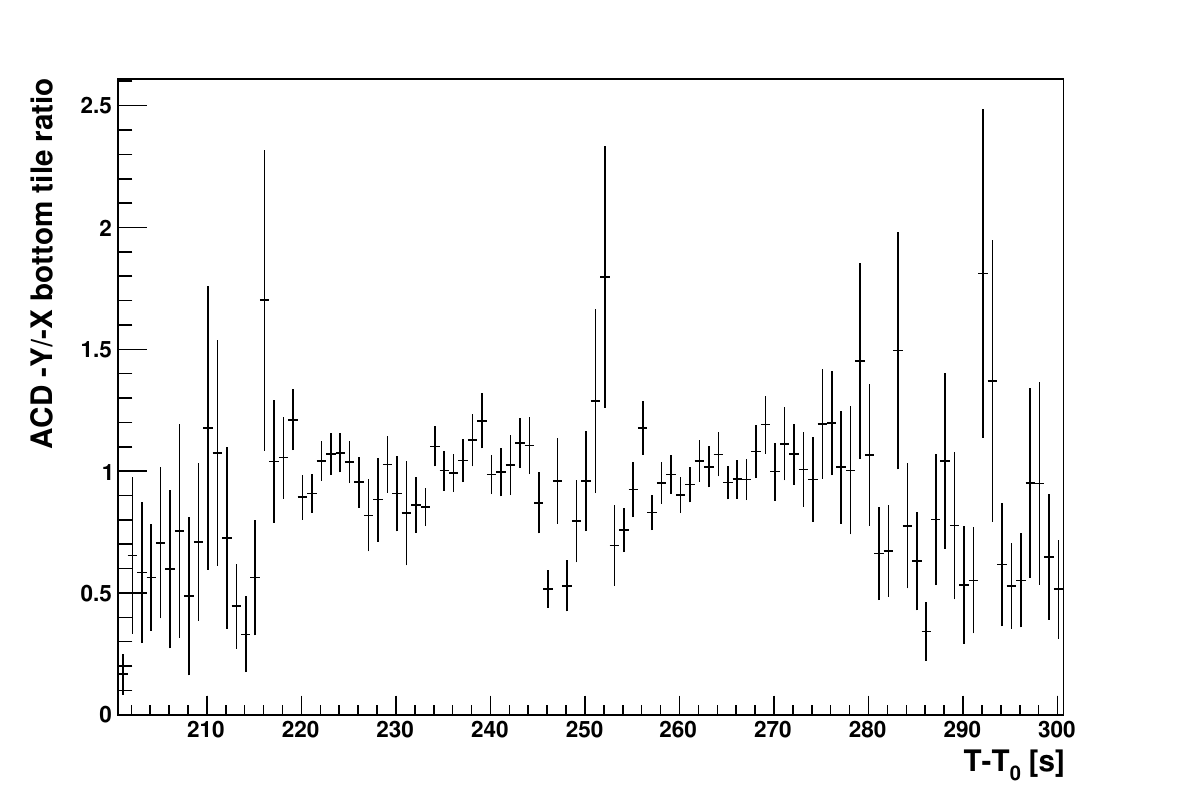}
    \caption{Ratio of the ACD $-$Y/$-$X bottom tile signals. Left: dependence with energy in MeV derived from simulations. Right: observed time variation during the BTI. \nob{Because the ratio varies with energy, it is sensitive to a spectral change. The fact that it is almost constant during the BTI, especially during P1 and P2, shows that the GRB LE spectrum did not change much during the BTI.}}
    \label{fig:ACDbottomtileratio}
\end{figure*}

The ACD information is less straightforward to use because its key role in the trigger and the on-board gamma-ray filter can very easily create a bias. We mitigate this problem by only using the information from the bottom tile of the $-$X face and the bottom tile of the $-$Y face. The four ACD bottom tiles (1.7~m long and 15~cm high, whose top parts cover the two bottom planes of the tracker) do not participate in the trigger decision and are not part of the most sensitive test of the gamma-ray filter. We use the $-$X and $-$Y bottom tiles and not the +X and +Y because the former face the GRB direction and their signals are thus not biased by the GRB HE emission. As in the tracker, the interaction of LE photons in the ACD depends on their energy. Using dedicated simulations, we find that the ratio of the signals in the $-$Y and $-$X bottom tiles is sensitive to the photon energy, as shown in Fig.~\ref{fig:ACDbottomtileratio}~(left). So, this ratio is sensitive to the GRB LE spectrum. Fig.~\ref{fig:ACDbottomtileratio}~(right) shows that this ratio does not vary much during the BTI, especially during P1 and P2. As a consequence, we assume that the shape of the GRB LE spectrum did not change significantly during the BTI.

\begin{figure*}[t]
    \centering
    \includegraphics[width=0.9\linewidth]{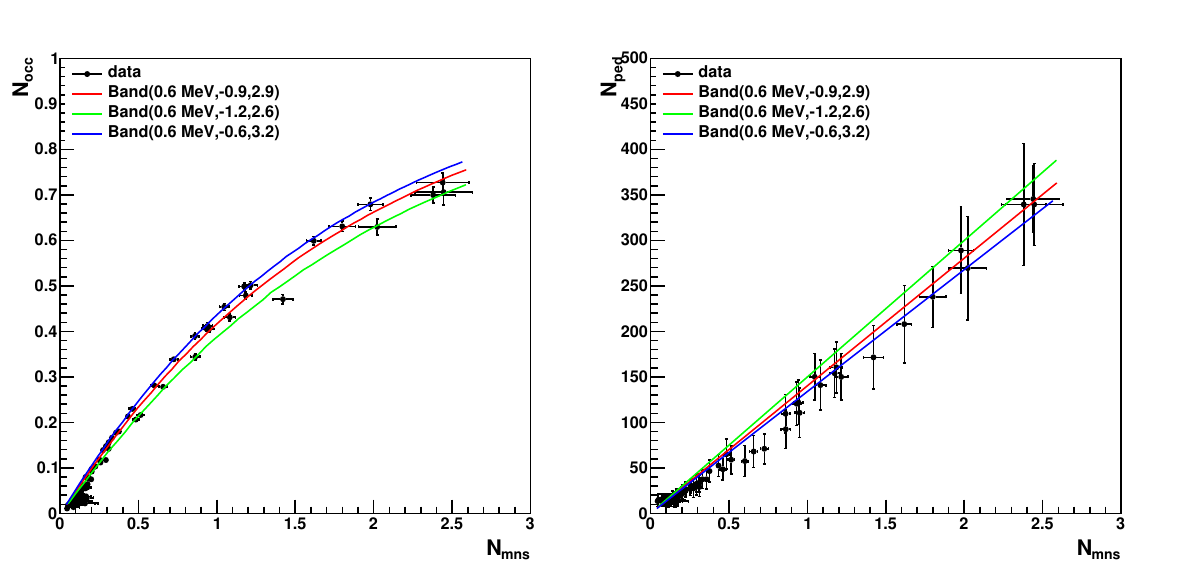}
    \caption{Correlations of the tracker-based quantities used to monitor the noise induced by the GRB LE emission. Left: corner occupancy vs. corner mean number of strips. Right: total number of strips pedestal vs. corner mean number of strips. The black dots correspond to the 1~s intervals in [200.6~s,\,300.6~s]. The red, green, and blue curves correspond to simulations produced with Band models with $E_\mathrm{p} = 0.6$~MeV.}
    \label{fig:TkrNoise_correlationB0.6}
\end{figure*}

This assumption allows us to use the tracker 1~s information during the BTI all together in order to constrain the LE spectrum. This is done by looking at the correlation of $\mathrm{N}_\mathrm{occ}$ with $\mathrm{N}_\mathrm{mns}$ and the correlation of $\mathrm{N}_\mathrm{ped}$ with $\mathrm{N}_\mathrm{mns}$. These correlations, shown in Fig.~\ref{fig:TkrNoise_correlationB0.6}, are both relatively tight, strengthening the assumption of a constant shape of the LE spectrum. As expected, the $\mathrm{N}_\mathrm{occ}$ correlation exhibits an inflection, while the $\mathrm{N}_\mathrm{ped}$ correlation is linear.

\begin{figure*}[t]
    \centering
    \includegraphics[width=0.9\linewidth]{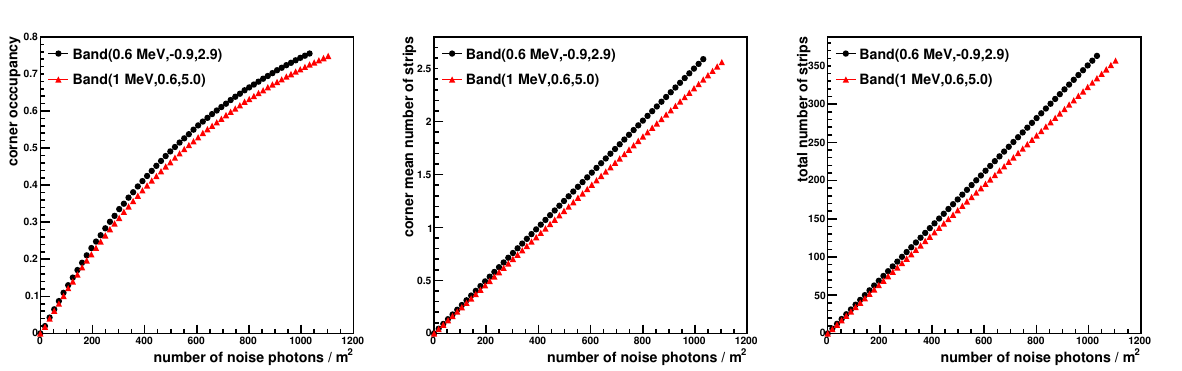}
    \caption{Tracker-based quantities as a function of the LE noise flux, as predicted with simulations. Left: the corner occupancy. Center: the corner mean number of strips. Right: the total number of strips. The black points correspond to $(E_\mathrm{p},\alpha,\beta)$=$(0.6~\mathrm{MeV},-0.9,2.9)$ and the red points correspond to $(1~\mathrm{MeV},0.6,5.0)$. \nob{For a given LE spectrum hypothesis, these graphs can be used to estimate the LE noise flux ($x$-axis) from a measured quantity ($y$-axis).}}
    \label{fig:SimTkrNoise}
\end{figure*}

We model the GRB LE emission with a Band function~\citep{1993ApJ...413..281B}. It corresponds to two smoothly connected power laws and, in addition to its normalization, it is defined by $E_\mathrm{p}$, the SED peak energy, and $\alpha$ and $\beta$, the spectral indices below and above $E_\mathrm{p}$, respectively. For each set of parameters $(E_\mathrm{p},\alpha,\beta)$, we perform multi-photon simulations, with photons between 0.1 and 30~MeV coming from the GRB direction, randomly distributed over a 2.8~$\mathrm{m}^2$ surface, large enough to cover the entire LAT sensitive detectors. We can then predict $\mathrm{N}_\mathrm{occ}$, $\mathrm{N}_\mathrm{mns}$ and $\mathrm{N}_\mathrm{ped}$ as a function of the incoming LE flux, as shown in Fig.~\ref{fig:SimTkrNoise} for $(E_\mathrm{p},\alpha,\beta) = (0.6~\mathrm{MeV},-0.9,2.9)$ and $(1~\mathrm{MeV},0.6,5.0)$. These results are then used to predict the same correlations as measured in the data.

For a given value of $E_\mathrm{p}$, we can perform a fit to the data to find the optimal values of $(\alpha,\beta)$. The fit for $E_\mathrm{p}=0.6$~MeV gives $(\alpha,\beta) \sim (-0.9,2.9)$, providing a good data/prediction agreement as can be seen in Fig.~\ref{fig:TkrNoise_correlationB0.6}. One can also see that $\alpha$ and $\beta$ have to be changed by 0.3 to clearly see a deviation from the data.

\begin{figure*}[t]
    \centering
    \includegraphics[width=0.9\linewidth]{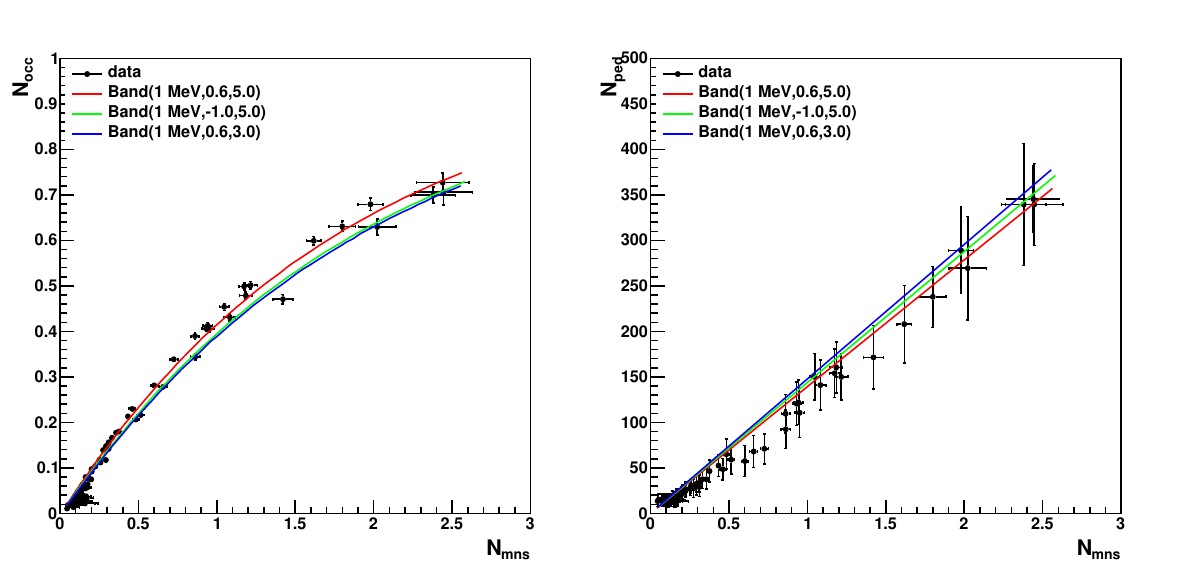}
    \caption{Same as Fig.~\ref{fig:TkrNoise_correlationB0.6}, except that the red, green and blue curves correspond to simulations produced with Band models with $E_\mathrm{p} = 1$~MeV.}
    \label{fig:TkrNoise_correlationB1.0}
\end{figure*}

We find a good data/prediction agreement for $0.4~\mathrm{MeV} \lesssim E_\mathrm{p} \lesssim 1~\mathrm{MeV}$, with the following trend: $\beta$ decreases when $E_\mathrm{p}$ decreases. As will be shown in Sec.~\ref{sec:BTIlightcurve}, the HE power-law index during P1 and P2 is found to be $\lesssim 3$. This constrains $\beta$ to also be $\lesssim 3$, which is reached for $E_\mathrm{p} \sim 0.6$~MeV. The interval for $E_\mathrm{p}$ thus shrinks to [0.6,\,1~MeV]. The fit for $E_\mathrm{p}=1.0$~MeV leads to $(\alpha,\beta) \sim (0.6,5.0)$, which also provides a good data/prediction agreement, as shown in Fig.~\ref{fig:TkrNoise_correlationB1.0}.

In order to assess the impact of the GRB LE flux, we will use in the following sections two hypotheses for the LE spectrum: $\mathrm{B}_{0.6} = \mathrm{Band}(0.6~\mathrm{MeV},-0.9,2.9)$ and $\mathrm{B}_{1.0} = \mathrm{Band}(1~\mathrm{MeV},0.6,5.0)$.

\subsubsection{Event selection effective area}

The event selection requires that a photon triggers the instrument, passes the gamma-ray filter, that at least one track is reconstructed, and that the reconstructed energy is above a given threshold. The corresponding effective area is estimated thanks to simulations of individual photons in the LAT.

Trigger engine~7 requires that the incoming photon triggers at least one tracker tower without vetoing any of the surrounding ACD tiles. Almost all photons below $\sim 1$~GeV do not have more than 100~MeV in one calorimeter crystal and are thus rejected at the gamma-ray filter level by any vetoing ACD tile, which is a more stringent requirement than the Engine~7 ACD veto requirement. As a consequence we can divide the selection into the following three steps:
\begin{itemize}
    \item the tracker tower trigger;
    \item the ACD veto by any tile above threshold;
    \item the track and energy requirements based on information provided by the event reconstruction.
\end{itemize}

The GRB LE flux impacts the event selection and we need to estimate the resulting correction factor, that is to say the ratio of the effective area in the presence of LE photon background with respect to the effective area without background. This correction factor depends on time (because of the time variation of the LE background flux) and we can factorize it into three parts:
\begin{equation} \label{eq:efficiency_correction}
\mathrm{C}_\mathrm{sel} (t) = \mathrm{C}_\mathrm{trig} (t) \times \mathrm{C}_\mathrm{veto} (t) \times \mathrm{C}_\mathrm{rec} (t)
\end{equation}
where $\mathrm{C}_\mathrm{trig}$, $\mathrm{C}_\mathrm{veto}$ and $\mathrm{C}_\mathrm{rec}$ account for the three steps of the event selection. Since the LE background only adds activity in the instrument, it cannot prevent a HE photon from triggering a tracker tower, from which we conclude that $\mathrm{C}_\mathrm{trig} = 1$ and that it can be ignored when estimating $\mathrm{C}_\mathrm{sel}$. However, this trigger factor must be taken into account when estimating the false positive rate, as will be discussed in Sec.~\ref{sec:FalsePositive}.

In the standard simulation of an individual photon in the LAT (like the one produced for the LLE analysis), all the energy deposits in the various parts of the instrument are assumed to occur at the same time. The situation is more complex when simulating one individual photon on top of a background of many photons because the background photons are not in-time with the signal photon. As a consequence, a realistic simulation has to take into account the characteristic readout times of the various detectors. For the tracker, the width of the trigger coincidence window is $\sim 700$~ns, while the average time over threshold of the shaped signal is $\tau_\mathrm{tkr}\sim 10~ \mu\mathrm{s}$. For the ACD, our analysis depends only on the trigger fast signal, whose shaping time is $\tau_\mathrm{ACD}\sim 0.4~\mu\mathrm{s}$. Rather than modifying the standard simulation to take into account these various characteristic times, we make use of the factorization of the effective area correction. Each part can be studied with a simulation performed with the number of LE photons that corresponds to the relevant characteristic time: $\tau_\mathrm{ACD}$ for $\mathrm{C}_\mathrm{veto}$ and $\tau_\mathrm{tkr}$ for $\mathrm{C}_\mathrm{rec}$.

For a given time interval, the background LE flux is estimated using the tracker corner occupancy. Assuming a certain LE spectrum (either $\mathrm{B}_{0.6}$ or $\mathrm{B}_{1.0}$), we use the information of Fig.~\ref{fig:SimTkrNoise}-(left) to derive $\mathrm{n}_\mathrm{tkr}$, the number of LE photons per $\mathrm{m}^2$ in the simulation needed to reproduce the observed noise. Since the corner occupancy is based on the information of the fired strips recorded in the tracker, $\mathrm{n}_\mathrm{tkr}$ actually corresponds to the time window $\tau_\mathrm{tkr}$. It means that we use a simulation with $\mathrm{n}_\mathrm{tkr}$ LE photons to assess $\mathrm{C}_\mathrm{rec}$. To assess $\mathrm{C}_\mathrm{veto}$ for the same LE flux level, we use a simulation with a number of LE photons set to $n_\mathrm{ACD} = r_\tau \mathrm{n}_\mathrm{tkr}$, where $r_\tau = \tau_\mathrm{ACD}/\tau_\mathrm{tkr}$, the ratio between the characteristic times of the ACD fast signal and the tracker readout signal, which is expected to be $\sim 0.04$. 

Because of the importance of $r_\tau$ in the estimation of $\mathrm{C}_\mathrm{veto}$, we want to constrain it with the help of the Earth-limb relative efficiency presented in Sec.~\ref{sec:Limb_efficiency}. This requires that we compute $\mathrm{C}_\mathrm{veto}$ and $\mathrm{C}_\mathrm{rec}$ both for HE photons coming from the GRB direction and for HE photons coming from the Earth-limb direction, which corresponds more or less to the mirror opposite of the GRB direction with respect to the plane $y=0$.

\begin{figure*}[t]
    \centering
    \includegraphics[width=0.9\linewidth]{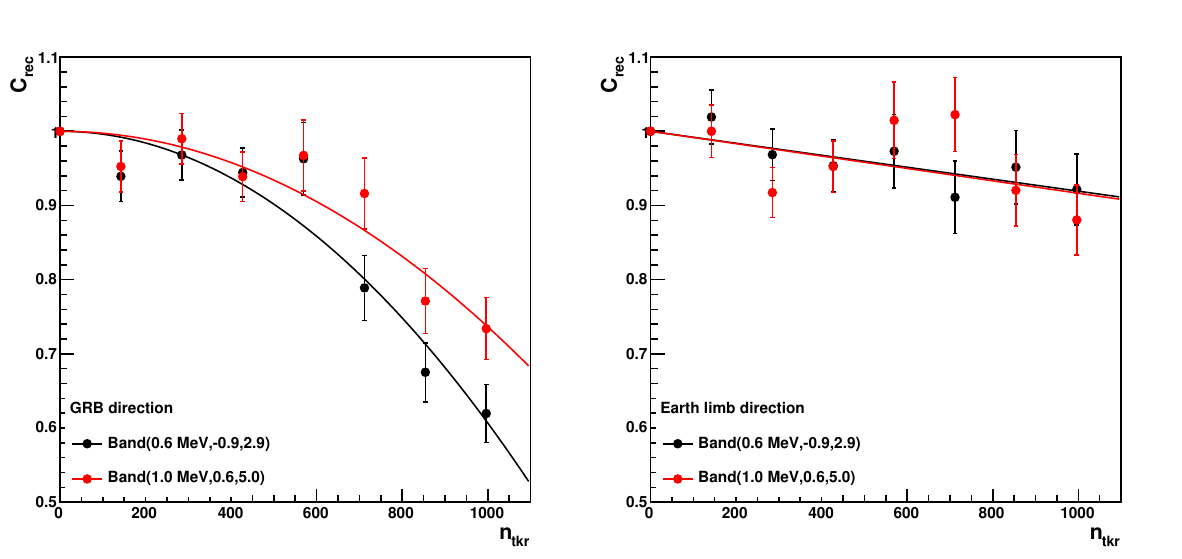}
    \caption{The effective area correction factor $\mathrm{C}_\mathrm{rec}$ corresponding to the final part of the event selection (at least one track and above a given energy threshold) as a function of $\mathrm{n}_\mathrm{tkr}$ the number of LE photons per $\mathrm{m}^2$ within $\tau_\mathrm{tkr}$, the tracker characteristic readout time. Left: for HE photons coming from the GRB direction and with $E_\mathrm{tkr}>100$~MeV. Right: for HE photons coming from the Earth limb direction and with $E_\mathrm{tkr}>125$~MeV. For both cases, the black and red points correspond to $\mathrm{B}_{0.6}$ and $\mathrm{B}_{1.0}$ respectively.}
    \label{fig:EffCorRec}
\end{figure*}

When estimating $\mathrm{C}_\mathrm{rec}$, we are only interested in the effect of the LE background on the tracker reconstruction capability. Therefore, we perform simulations in which we ignore the energy deposited by the background LE photons in the ACD. $\mathrm{C}_\mathrm{rec}$ is given by the ratio of the selection efficiency with background compared to the efficiency without background. The selection threshold $E_\mathrm{tkr}$ is set to 100~MeV for the GRB direction but we use 125~MeV for the Earth-limb direction in order to match the threshold used in Sec.~\ref{sec:Limb_efficiency}. We find that $\mathrm{C}_\mathrm{rec}$ does not vary significantly with energy for HE photons between 100~MeV and 1~GeV. The variations of $\mathrm{C}_\mathrm{rec}$ with $\mathrm{n}_\mathrm{tkr}$ are shown in Fig.~\ref{fig:EffCorRec}. One can see that the GRB LE background impacts differently the HE photons of the GRB and Earth limb: the loss of efficiency at large $\mathrm{n}_\mathrm{tkr}$ is greater for the former.

\begin{figure*}[t]
    \centering
    \includegraphics[width=0.9\linewidth]{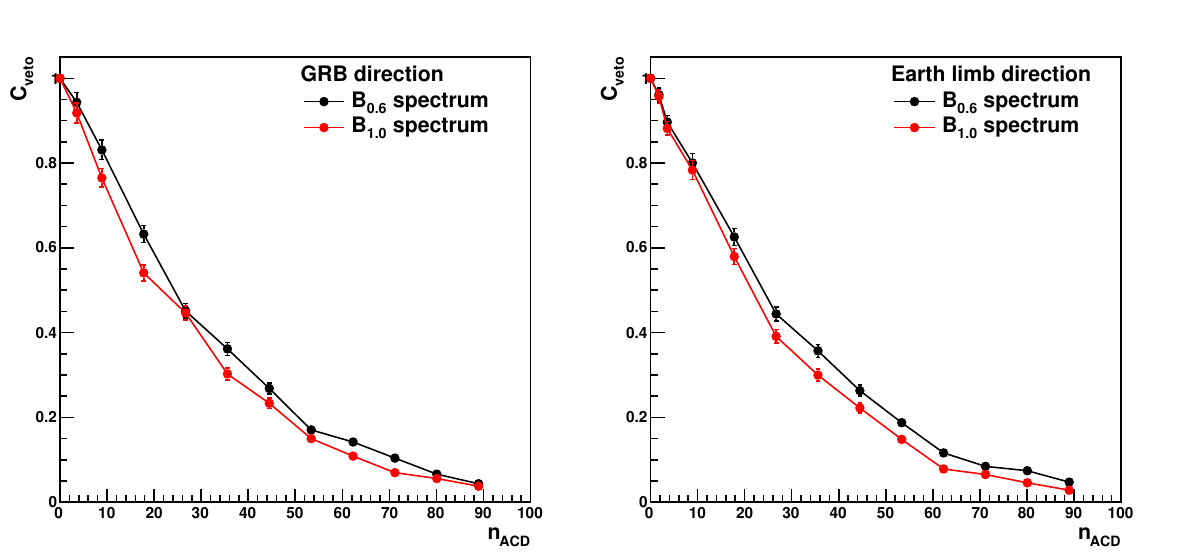}
    \caption{The effective area correction factor $\mathrm{C}_\mathrm{veto}$ corresponding to the ACD veto part of the event selection as a function of $\mathrm{n}_\mathrm{ACD}$, the number of LE photons per $\mathrm{m}^2$ within $\tau_\mathrm{ACD}$, the ACD fast signal characteristic time. Left: for HE photons coming from the GRB direction and with $E_\mathrm{tkr}>100$~MeV. Right: for HE photons coming from the Earth limb direction and with $E_\mathrm{tkr}>125$~MeV. For both cases, the black and red points correspond to $\mathrm{B}_{0.6}$ and $\mathrm{B}_{1.0}$ respectively.}
    \label{fig:EffCorVeto}
\end{figure*}

We also use simulations to estimate $\mathrm{C}_\mathrm{veto}$ as a function of $\mathrm{n}_\mathrm{ACD}$, the number of LE photons per $\mathrm{m}^2$ within $\tau_\mathrm{ACD}$. But in that case, the energy deposited by the background LE photons in the ACD is not ignored. As for $\mathrm{C}_\mathrm{rec}$, we do not see any significant dependence on the HE photon energy. We neither see any significant difference between the GRB and Earth limb directions, as can be seen in Fig.~\ref{fig:EffCorVeto}.

For each hypothesis of the LE spectrum, we can estimate $\mathrm{n}_\mathrm{tkr}$ as a function of time from the corner occupancy shown in Fig.~\ref{fig:TkrNoise}. In normal conditions, the corner occupancy is about $0.02$ and we consider that no selection correction is needed when the corner occupancy is less than 0.035. $\mathrm{n}_\mathrm{tkr}$ allows us to know $\mathrm{C}_\mathrm{rec} (\mathrm{n}_\mathrm{tkr})$. We then use $r_\tau$ to compute $\mathrm{C}_\mathrm{veto} (r_\tau \mathrm{n}_\mathrm{trk})$ and eventually get the selection correction factor as a function of time:
\begin{equation} \label{eq:efficiency_correction2}
\mathrm{C}_\mathrm{sel} (t) = \mathrm{C}_\mathrm{veto} \left(r_\tau \mathrm{n}_\mathrm{trk}(t) \right) \times \mathrm{C}_\mathrm{rec} \left (\mathrm{n}_\mathrm{trk}(t) \right)
\end{equation}

\begin{figure*}[t]
    \centering
    \includegraphics[width=0.6\linewidth]{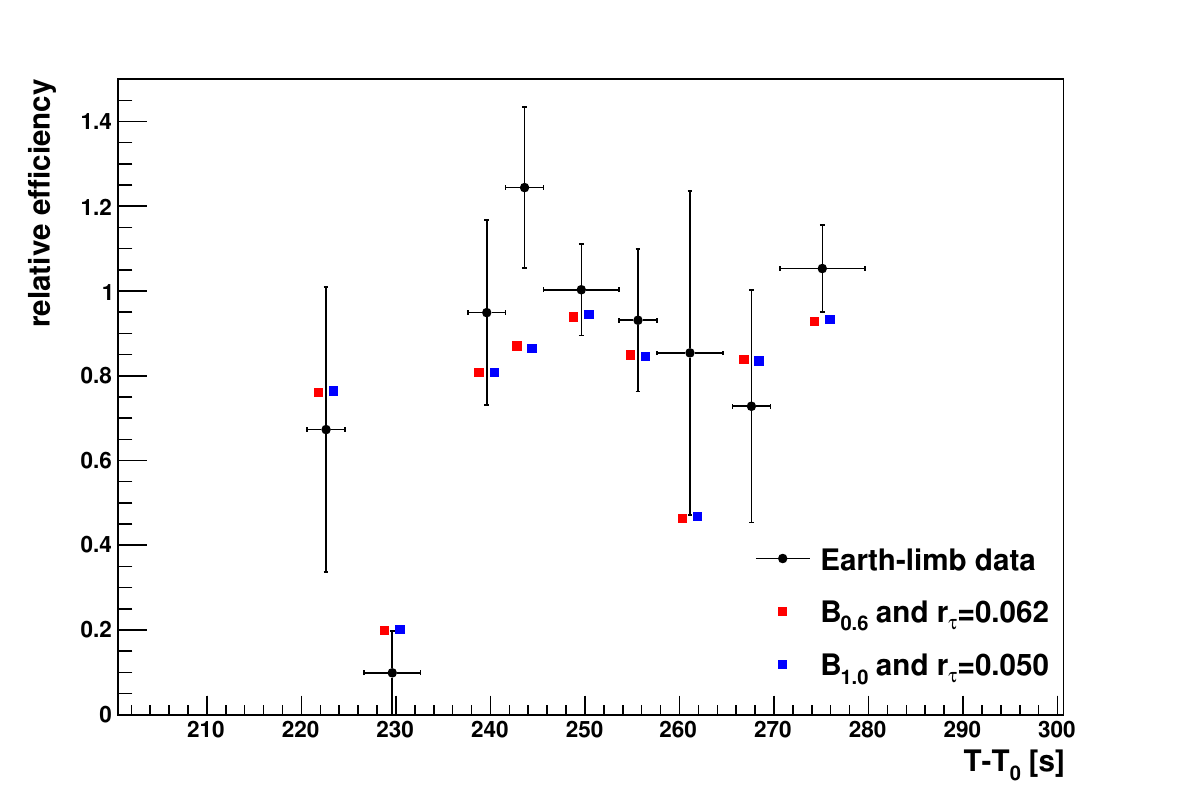}
    \caption{Result of the fit of the relative efficiency to constrain the time ratio $\mathrm{r}_\tau$. The black points correspond to the measurement obtained with Earth-limb data. The red points correspond to the relative efficiency predicted by the simulation with $\mathrm{B}_{0.6}$ spectrum and $\mathrm{r}_\tau=0.062$. The blue points correspond to $\mathrm{B}_{1.0}$ spectrum and $\mathrm{r}_\tau=0.050$. The red and blue points have been slightly shifted along the $x$-axis for clarity's sake.}
    \label{fig:Fit_rtau}
\end{figure*}

We constrain $r_\tau$ using the information of the Earth-limb relative efficiency. For each LE spectrum hypothesis, we perform a $\chi^2$ fit to find the value of $r_\tau$ that provides the best prediction of the relative efficiency in nine time intervals chosen such that the noise variation is minimal within each of these intervals. For each time interval, we estimate $\mathrm{n}_\mathrm{tkr}$ from the corner occupancy and the predicted efficiency can be computed for any $r_\tau$ using Eq.~\ref{eq:efficiency_correction2}. We find $r_\tau \pm \mathrm{d}r_\tau = 0.062 \pm 0.013$ for the $\mathrm{B}_{0.6}$ spectrum and $r_\tau \pm \mathrm{d}r_\tau = 0.050 \pm 0.012$ for the $\mathrm{B}_{1.0}$ spectrum. The data/prediction agreement is good, with a $\chi^2 = 8.6$, as shown in Fig.~\ref{fig:Fit_rtau}. One can see that the predicted efficiencies of the two LE spectrum hypotheses are almost identical. We note that the estimated $r_\tau$ values are not far from the expected value of 0.04.

\begin{figure*}[t]
    \centering
    \includegraphics[width=0.6\linewidth]{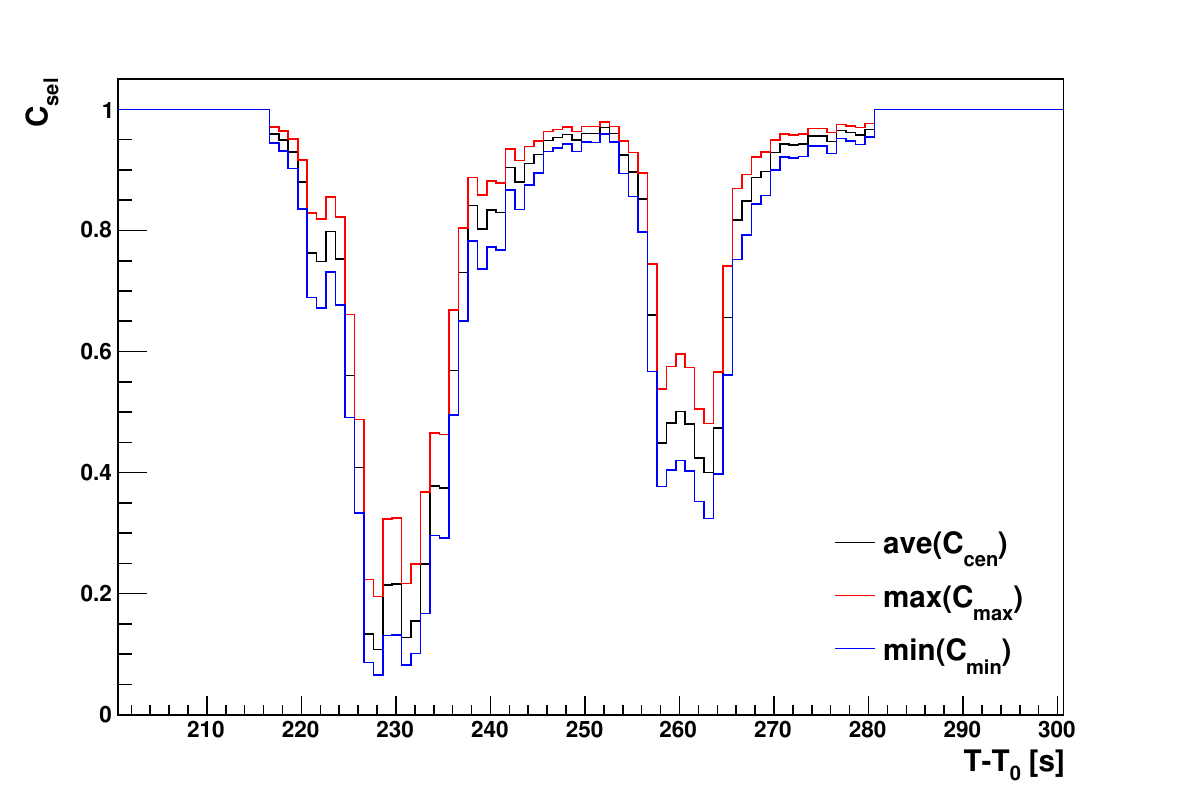}
    \caption{Effective area correction factor $\mathrm{C}_\mathrm{rec}$ as a function of time during the BTI after combining the corrections of the $\mathrm{B}_{0.6}$ and $\mathrm{B}_{1.0}$ spectrum hypotheses. The black curve corresponds to the average of $\mathrm{C}_\mathrm{cen}$. The red and blue curves correspond to the maximum of $\mathrm{C}_\mathrm{max}$ and to the minimum of $\mathrm{C}_\mathrm{min}$, respectively.}
    \label{fig:EffCorSel}
\end{figure*}

For each LE spectrum hypothesis, we compute $\mathrm{C}_\mathrm{sel}$ in three flavors: $\mathrm{C}_\mathrm{cen}$ with $r_\tau$, $\mathrm{C}_\mathrm{min}$ with $r_\tau + \mathrm{d}r_\tau$, and $\mathrm{C}_\mathrm{max}$ with $r_\tau - \mathrm{d}r_\tau$. We combine the results of the two LE spectrum hypotheses by taking the average of $\mathrm{C}_\mathrm{cen}$, as well as the minimum of $\mathrm{C}_\mathrm{min}$ and the maximum of $\mathrm{C}_\mathrm{max}$. The resulting effective area correction during the BTI is shown in Fig.~\ref{fig:EffCorSel}. 

\subsubsection{False positive}
\label{sec:FalsePositive}

The previous sections were devoted to the impact of the GRB LE flux on our ability to correctly detect HE photons. But the GRB LE flux alone can create a fake HE signal. In order to quantify the false positive rate induced by the GRB LE flux, we perform simulations with only background LE photons, between 0.1 and 30~MeV. In that situation, the background alone has to trigger a tracker tower and we thus have to take into account $\tau_\mathrm{trig}$, the $\sim 700$~ns tracker trigger coincidence window, to estimate the rate of background events that trigger a tracker tower. As for the efficiency correction factor, we factorize the problem by assuming that the false positive rate is:
$$\rho = \frac{1}{\tau_\mathrm{trig}} \times \epsilon_\mathrm{trig} \times \epsilon_\mathrm{veto} \times \epsilon_\mathrm{rec} $$
where $\epsilon_\mathrm{trig}$ is the fraction of events that trigger a tracker tower, $\epsilon_\mathrm{veto}$ the fraction of events with a tracker tower trigger that are not vetoed by the ACD and $\epsilon_\mathrm{rec}$ the fraction of events for which a track is found and that pass the chosen energy threshold. The $1/\tau_\mathrm{trig}$ term approximates the number of independent time intervals during which the background can trigger a tracker tower.

For an LE background flux given by $\mathrm{n}_\mathrm{trk}$, we use simulations with $\tau_\mathrm{trig}/\tau_\mathrm{tkr} \times \mathrm{n}_\mathrm{trk}$ to estimate $\epsilon_\mathrm{trig}$. Regarding $\epsilon_\mathrm{veto}$, we can assume that it is equal to $\mathrm{C}_\mathrm{veto}(r_\tau \mathrm{n}_\mathrm{trk})$ because the tracker energy deposits and the ACD energy deposits are not correlated. $\epsilon_\mathrm{rec}$ is derived from simulations with $\mathrm{n}_\mathrm{trk}$ LE photons. We use two values for $\tau_\mathrm{trig}$, 700~ns and 1500~ns, and find no difference in false positive rate.

\begin{figure*}[t]
    \centering
    \includegraphics[width=0.6\linewidth]{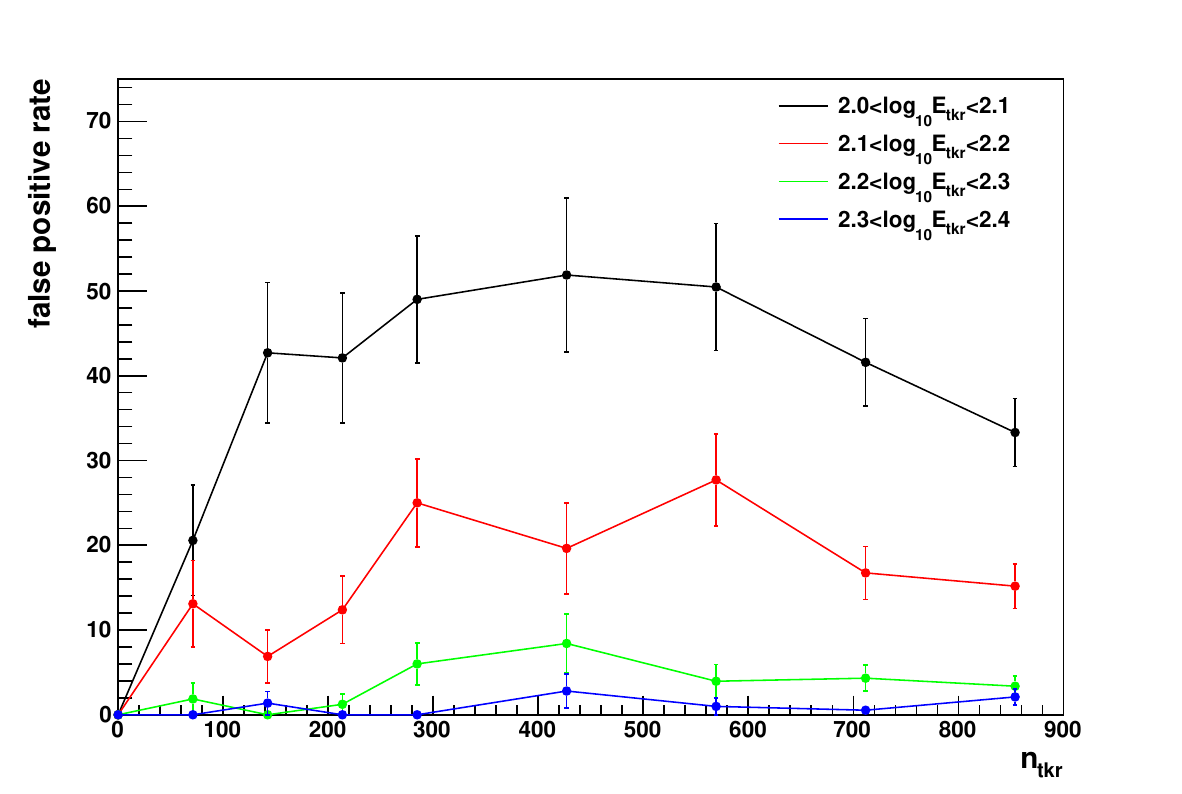}
    \caption{False positive rate in Hz within $20^\circ$ of the GRB for several $\log_{10}E_\mathrm{tkr}$ bins as a function of the number of LE photons per $\mathrm{m}^2$ within $\sim 10~\mu\mathrm{s}$, derived with the $\mathrm{B}_{0.6}$ spectrum hypothesis.}
    \label{fig:FalsePositeRate}
\end{figure*}

As expected, the false positive rate depends very much on $\beta$, the second spectral index of the LE spectrum. We actually find that for $\mathrm{LE}_{1.0}$ with $\beta=5.0$, it is always less than 1~Hz. On the contrary, the $\mathrm{B}_{0.6}$ spectrum with $\beta=2.9$ predicts a significant false positive rate, which can be considered as an upper limit. Fig.~\ref{fig:FalsePositeRate} breaks down the false positive rate in $\log_{10}E_\mathrm{tkr}$ bins. One can see that it decreases very much with energy. As a consequence, we set the selection $E_\mathrm{trk}$ threshold to 160~MeV to ensure that the false positive rate is less than 10~Hz. We note that, as can be seen in Fig.~\ref{fig:MClogEtkr}, a 160~MeV $E_\mathrm{trk}$ threshold corresponds to about a 100~MeV threshold in true energy.

\subsubsection{BTI light curve}
\label{sec:BTIlightcurve}

In order to determine the GRB HE energy flux for a given time interval, we perform a fit to the data with templates of the expected distribution of $\theta_\mathrm{GRB}$, the angular separation with respect to the GRB direction. We used two kinds of templates: one to model the GRB HE signal and the other to model the HE background induced by the gamma-ray sky (including the Galactic diffuse emission and the isotropic emission). The former is predicted by the LLE simulation while the latter is estimated with data recorded between 0 and 180~s, that is to say before the GRB HE emission hits the LAT. The GRB HE emission is modeled with a power law with a spectral index $\Gamma_\mathrm{ph}$. Since the LLE simulation was produced with a spectral index of $-$1, it is weighted by $E^{\Gamma_\mathrm{ph}+1}$ to predict the HE signal templates.

\begin{figure*}[t]
    \centering
    \includegraphics[width=0.9\linewidth]{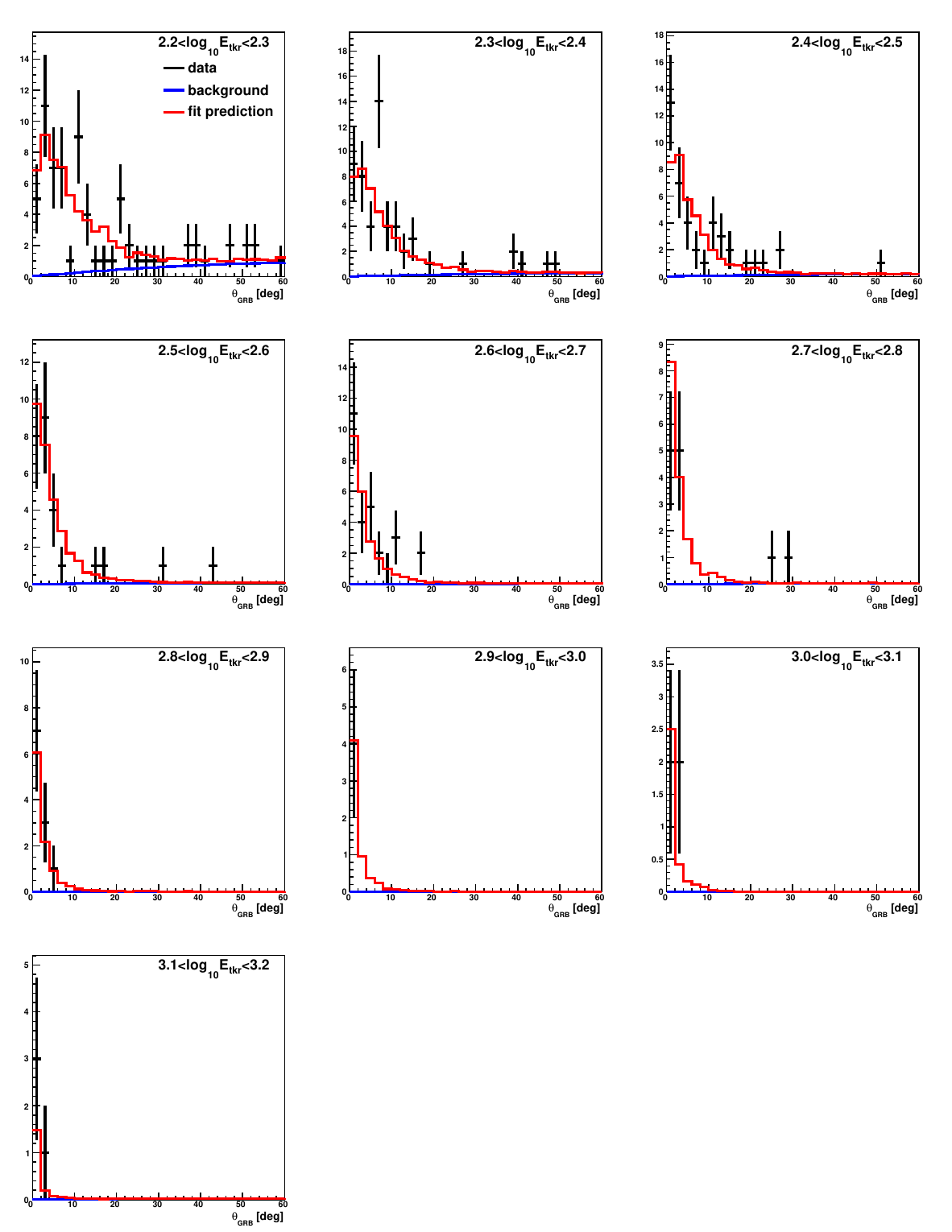}
    \caption{Distribution of $\theta_\mathrm{GRB}$, the angular separation to the GRB direction for the time interval [245.6~s,\,250.6~s]. The black points correspond to the data while the blue (background) and the red (background + GRB signal) histograms correspond to the result of the template fit.}
    \label{fig:TemplateFit}
\end{figure*}

The fit is performed over 10~bins in $\log_{10}E_\mathrm{tkr}$ between 2.2 and 3.2. There are two parameters for the HE emission (normalization and $\Gamma_\mathrm{ph}$) and one background scaling parameter per energy bin. The results of the fit for the time interval [245.6~s,\,250.6~s] are displayed in Fig.~\ref{fig:TemplateFit}. For each time interval, we take into account the variation of the selection effective area during the BTI by computing the average $\mathrm{C}_\mathrm{sel}$. We use the uncertainty on $\mathrm{C}_\mathrm{sel}$ to estimate the systematic uncertainty on the HE emission.

To estimate $n_\mathrm{fp}$, the expected upper limit on the number of false positive events, for each time interval, we multiply the live time by the false positive rate of the first fit energy bin ($2.2<\log_{10}E_\mathrm{tkr}<2.3$). We then compare this number with $n_\mathrm{obs}$, the number of observed events within $20^\circ$ in the same energy bin. The correction factor that we use to account for the false positive rate (by multiplying the normalization of the HE signal) is therefore assumed to be in the interval $[1-n_\mathrm{fp}/n_\mathrm{obs},1]$.

\begin{figure*}[t]
    \centering
    \includegraphics[width=0.6\linewidth]{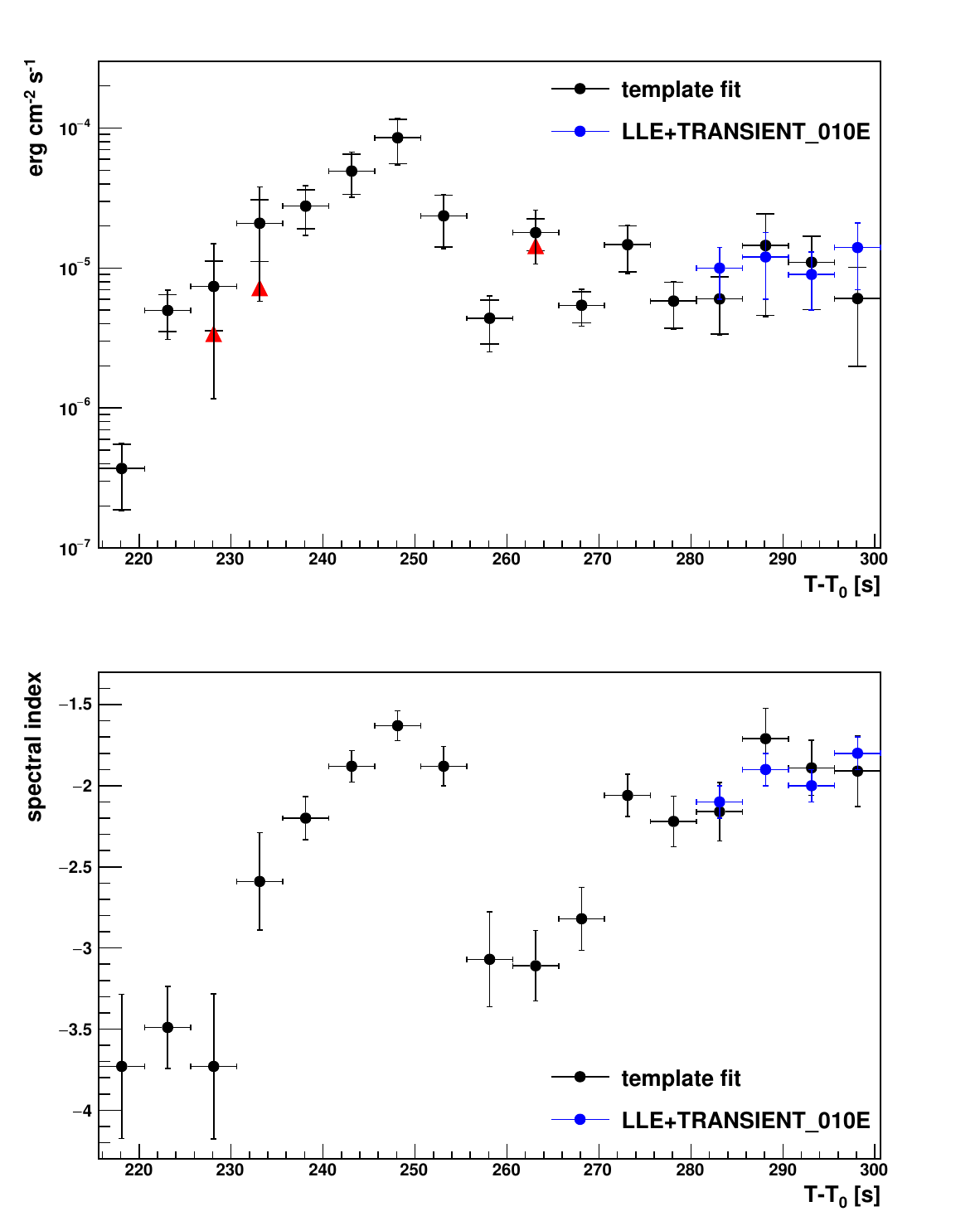}
    \caption{Light curve of the GRB HE emission during the BTI. Top: energy flux between 100~MeV and 10~GeV. The inner and the outer error bars correspond to the statistical uncertainty and to the sum of the statistical and systematic uncertainties, respectively. The red triangles show the energy flux after subtraction of the contribution of the false positive maximum rate. For clarity's sake, the red triangles are only shown when the induced difference is greater than half the statistical uncertainty. Bottom: spectral index of the HE power-law spectrum. The blue points correspond to the results of the LLE+{\tt TRANSIENT\_010E} joint fit with \ThreeML reported in Tab.~\ref{tab:lat_flux}.}
    \label{fig:LightCurveBTI}
\end{figure*}

We compute the light curve of the GRB HE emission during the BTI in 5~s time intervals. We find a significant HE emission after 215.6~s. The energy flux as well as the spectral index are shown in Fig.~\ref{fig:LightCurveBTI} and included in Tab.~\ref{tab:lat_flux}. The most striking result is that the HE emission exhibits a unique peak, occurring during IP, \nob{as discussed in Sec.~\ref{sec:prompt}}. The spectrum during this peak is rather hard, with $\Gamma_\mathrm{ph} = -1.63 \pm 0.09$, whereas the spectral index during P1 and P2 is $\sim -3$. After $\mathrm{T}_0+275.6~\mathrm{s}$, the spectral index is about $-$2. We note that the template fit results after $\mathrm{T}_0+280.6~\mathrm{s}$ are compatible with the results obtained with the LLE+{\tt TRANSIENT\_010E} joint fit with {\tt ThreeML} and reported in Tab.~\ref{tab:lat_flux}.

\subsubsection{Conclusion}

\nob{The GRB LE emission is so bright during a time interval of 64\,s (BTI) that it is unfortunately not possible to perform a standard analysis of the LAT data. In this appendix, we have presented all the solutions and methods that we needed to develop in order to be able to analyse the LAT data and to eventually characterize the GRB HE emission during this BTI. Here is a summary of the main items:
\begin{itemize}
    \item we replace the standard reconstructed energy with a new energy estimator, $\mathrm{E}_\mathrm{tkr}$, based on the number of fired strips in a cone along the track axis;
    \item we use Earth-limb data to constrain the selection efficiency;
    \item we use the information in the top corner of the tracker and the information in the bottom ACD tiles to constrain the LE spectrum and precisely define the boundaries of the BTI;
    \item we perform simulations of individual HE photons on top of a background of LE photons to estimate the selection effective area as a function of time, as well as the false positive rate due to the LE background;
    \item we apply a template fit approach to derive the HE emission light curve.
\end{itemize}
}

\clearpage

\newpage

\section{Time resolved joint analysis between GBM and LAT data} 
\label{sec:time_resolved_spectral_analysis}

\begin{figure*}[t]
    \centering    
    \includegraphics[width=\linewidth]{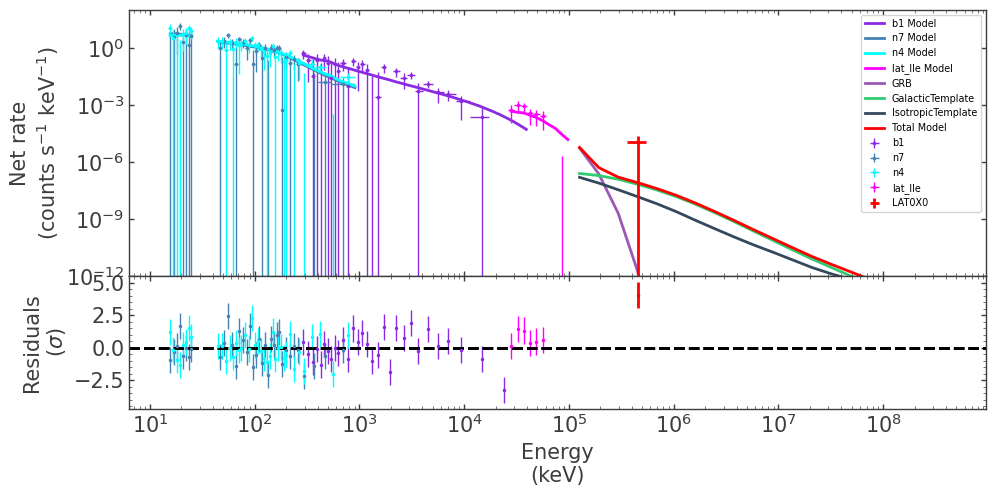}
    \includegraphics[width=\linewidth]{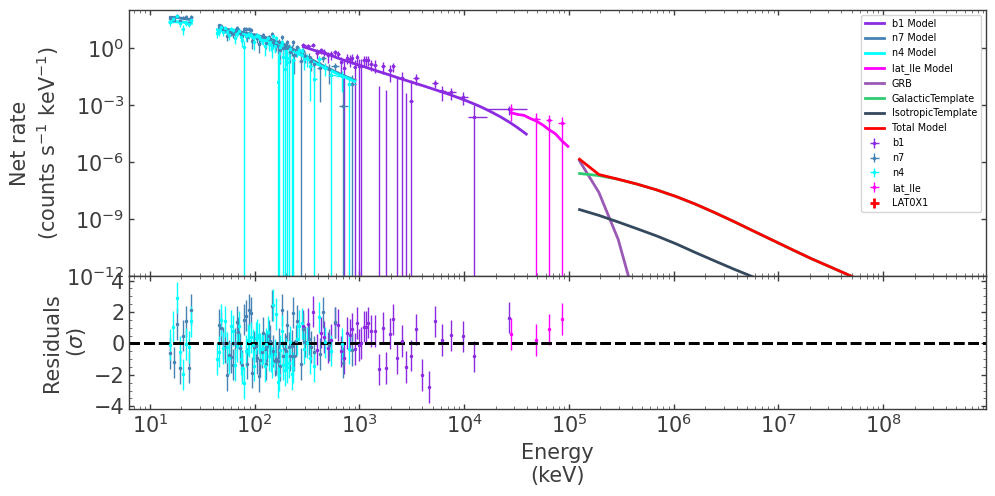}
\caption{
Count spectra \nob{(in counts s$^{-1}$ keV$^{-1}$)} and residuals (in sigma units) for the \texttt{COMP} model for the \nob{[\trig-0.69,\,\trig+ 0.23] (top) and [\trig+ 0.23,\,\trig+ 1.0] (bottom)} time intervals.}
    \label{fig:prec_spectra}
\end{figure*}

\nob{In Sec.~\ref{sec:results} we presented the results of the time-resolved spectral analysis of the GBM and LAT data to analyze the triggering pulse (Sec.~\ref{sec:precursor}) and the prompt emission (Sec.~\ref{sec:prompt}). Here we provide some more details about the analysis, clarifying some technical aspects related to the joint spectral analysis. 
All the models used in this paper are available in the \texttt{astromodels} package of \ThreeML and we refer to its documentation for their specific implementation.
We performed a time-resolved spectral fit of the triggering pulse in five time intervals, using the \texttt{COMP} model (\texttt{Cutoff\_powerlaw\_Ep} model in \texttt{astromodels}). 
Fig.~\ref{fig:prec_spectra} shows the count spectra and residual for \nob{the first two time intervals, which displayed the highest value of E$_{\rm peak}$}.}


\nob{For the prompt phase,} a time-resolved spectral fit was performed in seven time intervals, using Bayesian analysis available in \ThreeML, selecting \texttt{multinest} as the sampler. We compared the models using the Bayesian evidence estimator ($\log\,Z$). 
In Tab.~\ref{tab:priors} we summarize the priors defined in the analysis.

\begin{figure*}[t]
    \centering    
    \includegraphics[width=0.4\linewidth]{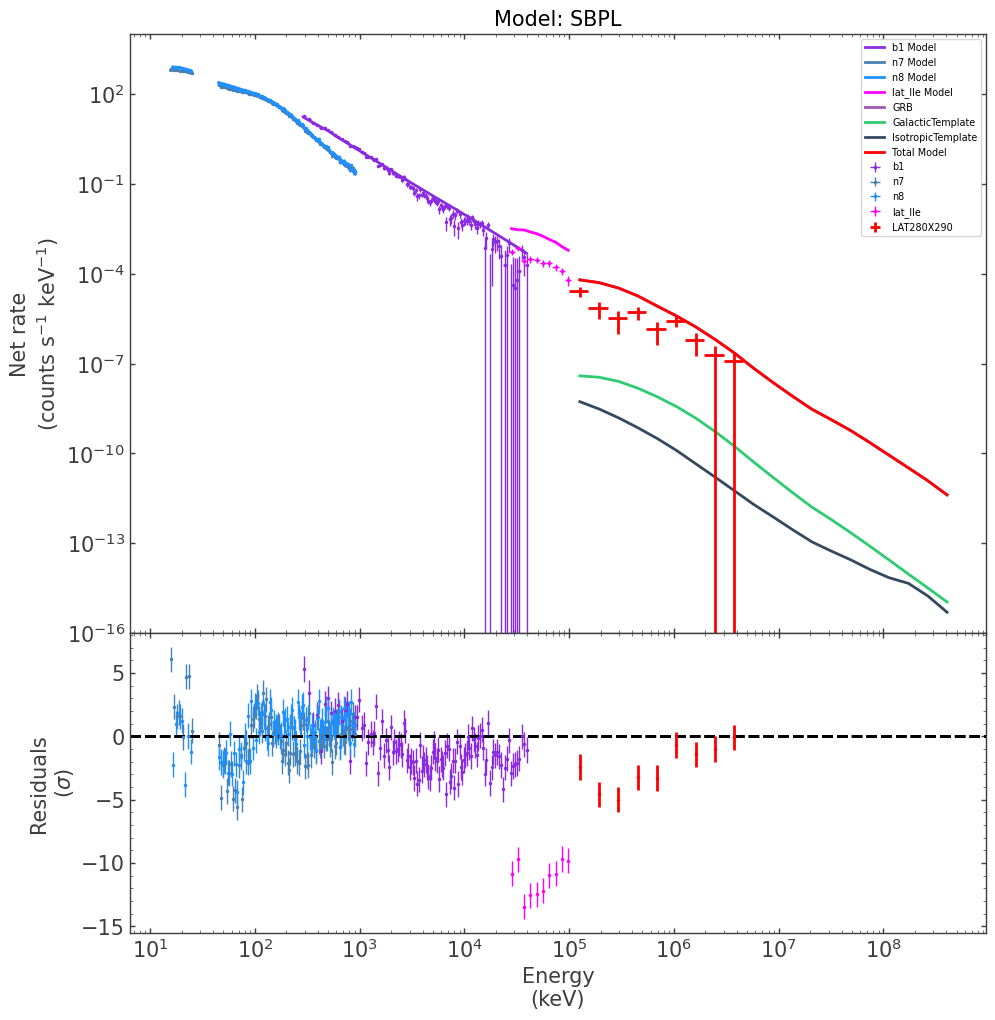}
    \includegraphics[width=0.4\linewidth]{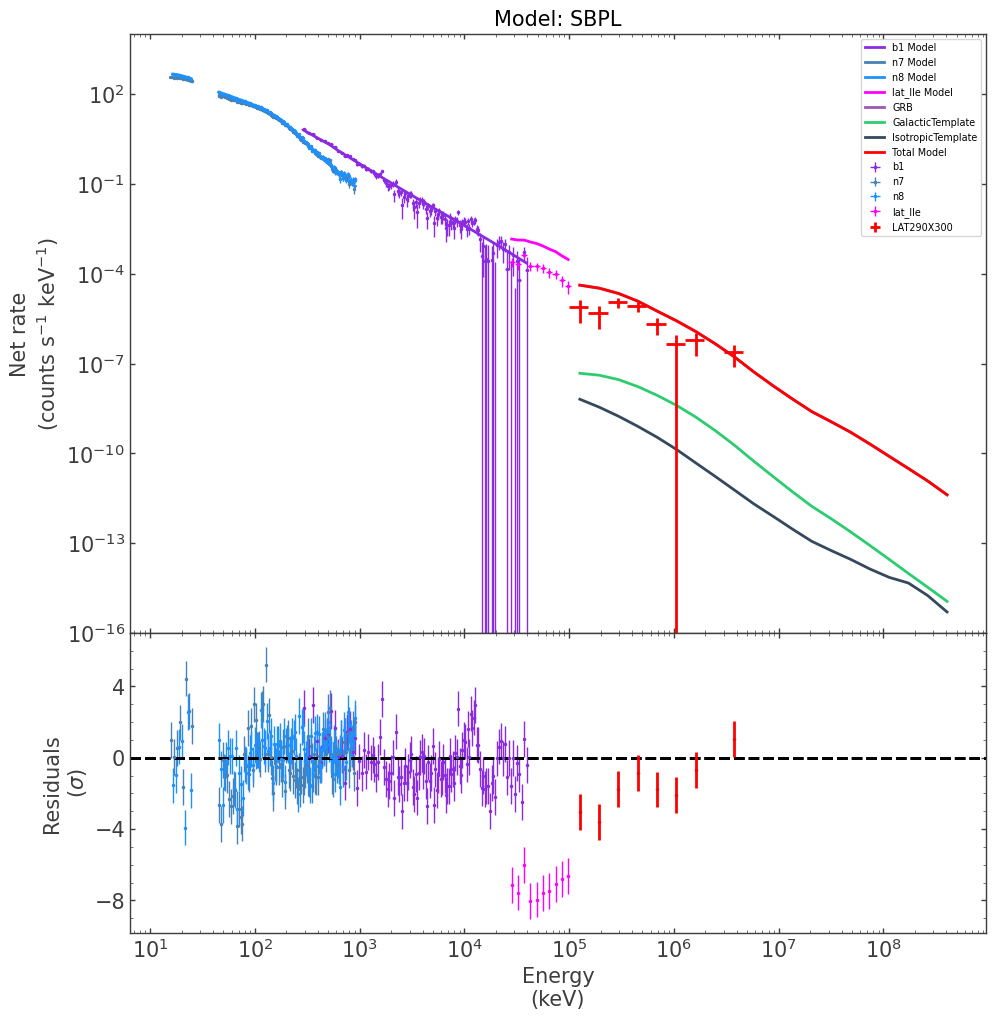}\\
    \includegraphics[width=0.4\linewidth]{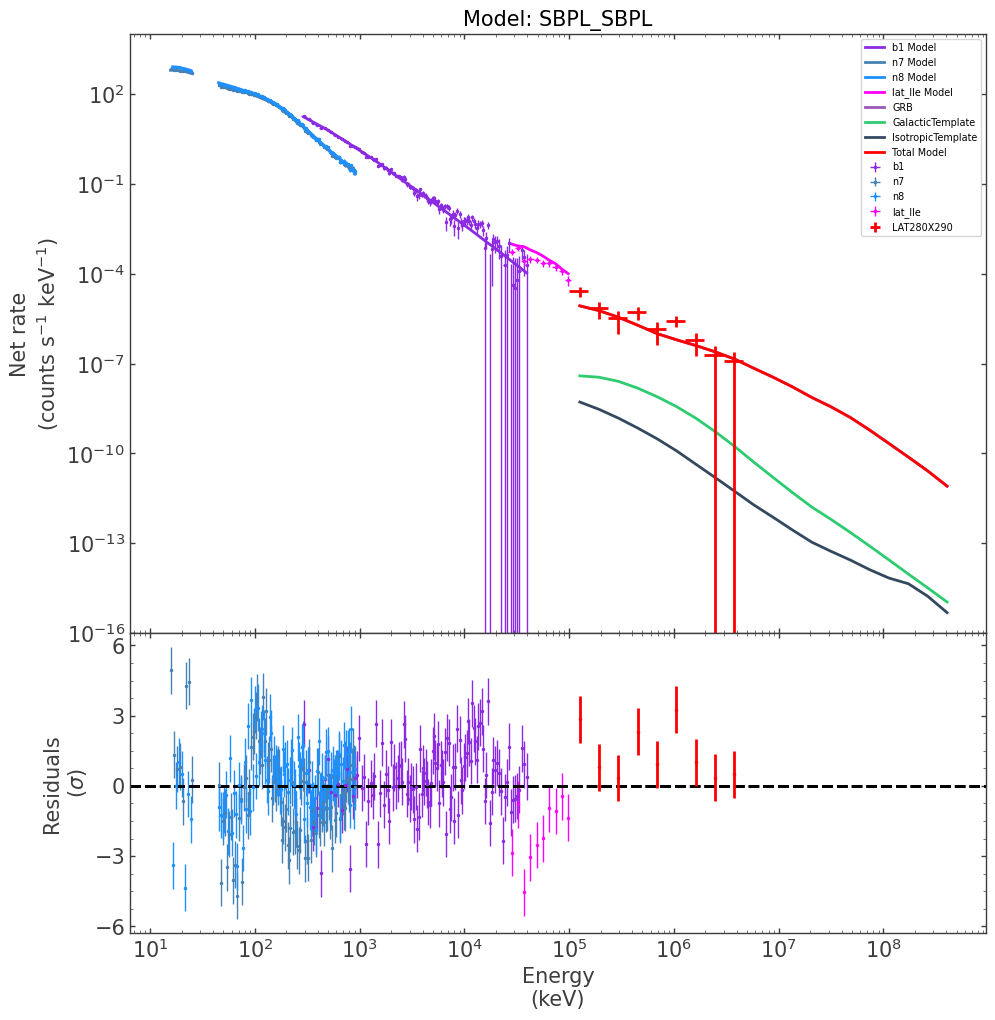}
    \includegraphics[width=0.4\linewidth]{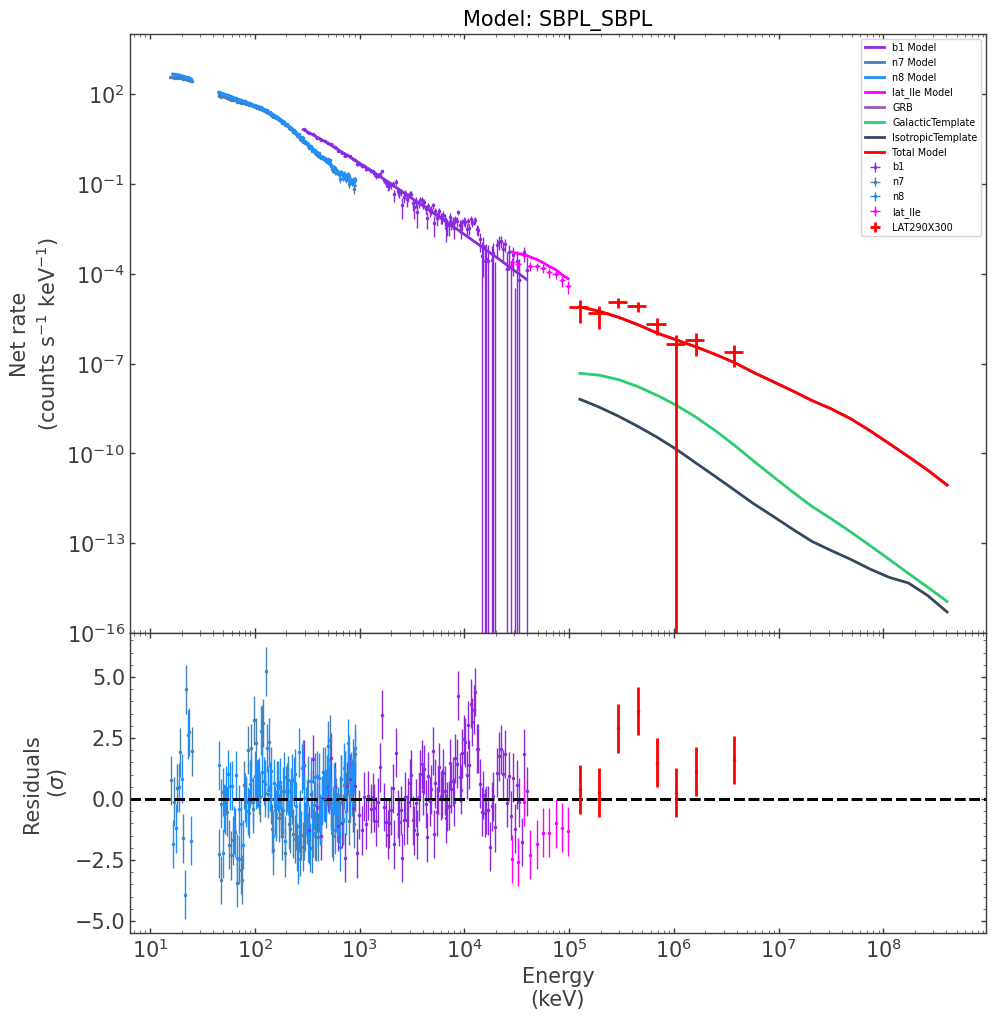}\\
    \includegraphics[width=0.4\linewidth]{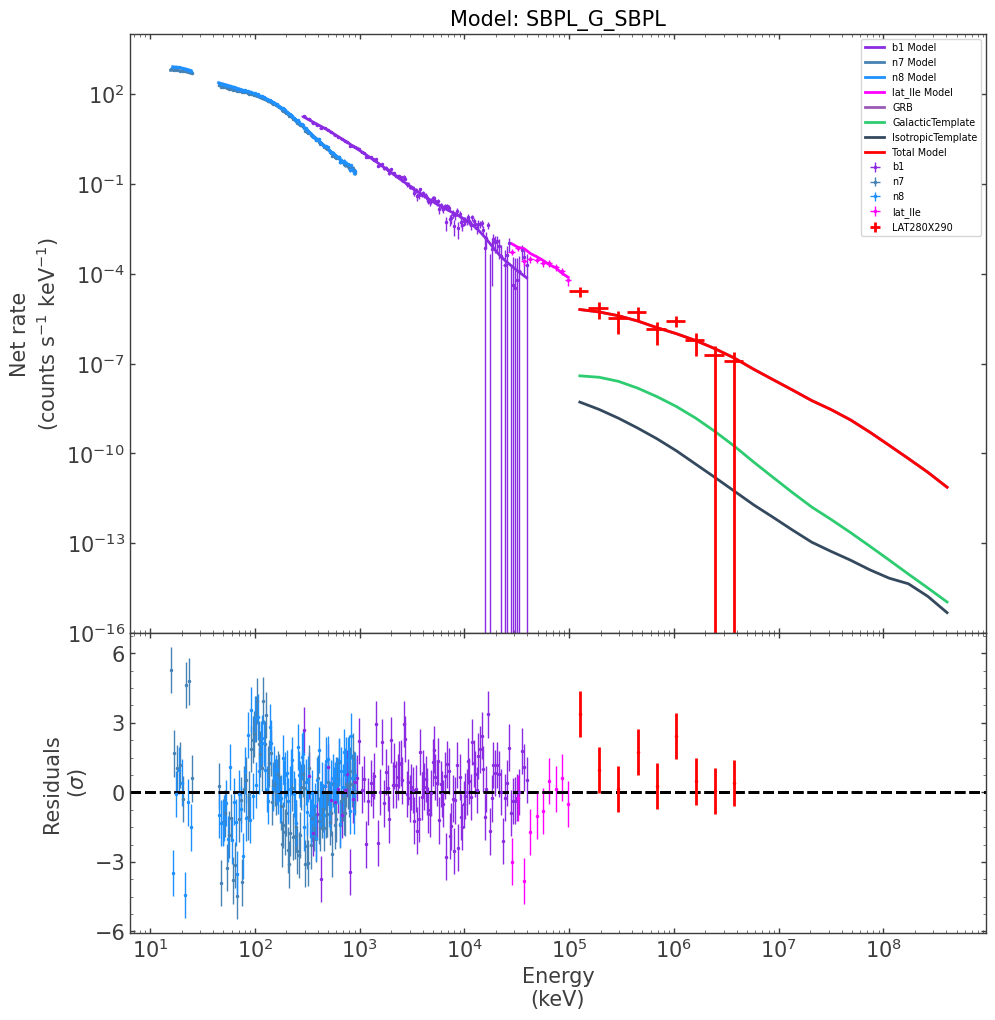}
    \includegraphics[width=0.4\linewidth]{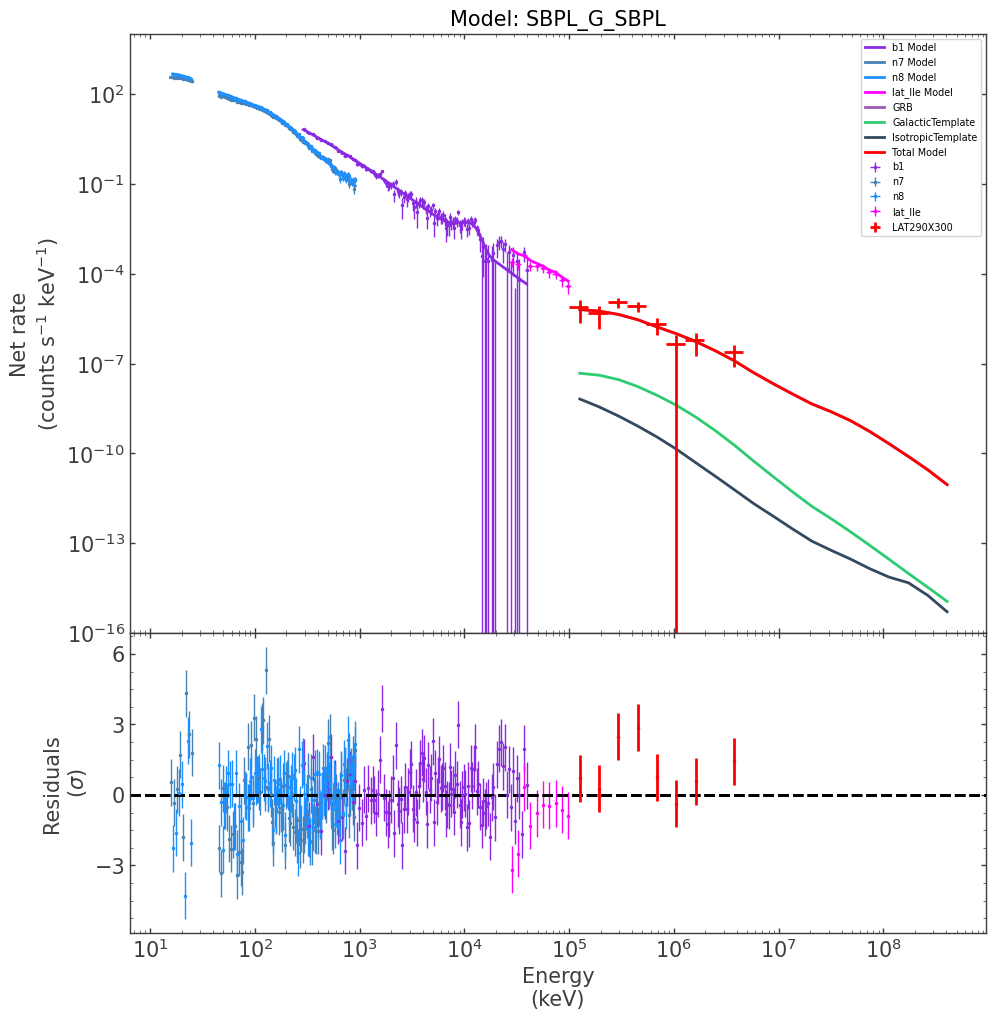}
\caption{Count spectra \nob{(in counts s$^{-1}$ keV$^{-1}$)} and residuals (in sigma units) for the [\trig+ 280.6,\,\trig+ 290.6] \nob{(left) and  [\trig+ 290.6,\,\trig+ 300.6]} \nob{(right)} time intervals. From \nob{top} to \nob{bottom}: the \texttt{SBPL}, \texttt{SBPL+SBPL} and \texttt{SBPL+G+SBPL} models.}
    \label{fig:interval12}
\end{figure*}

\begin{figure*}[t]
    \centering    
    \includegraphics[width=0.4\linewidth]{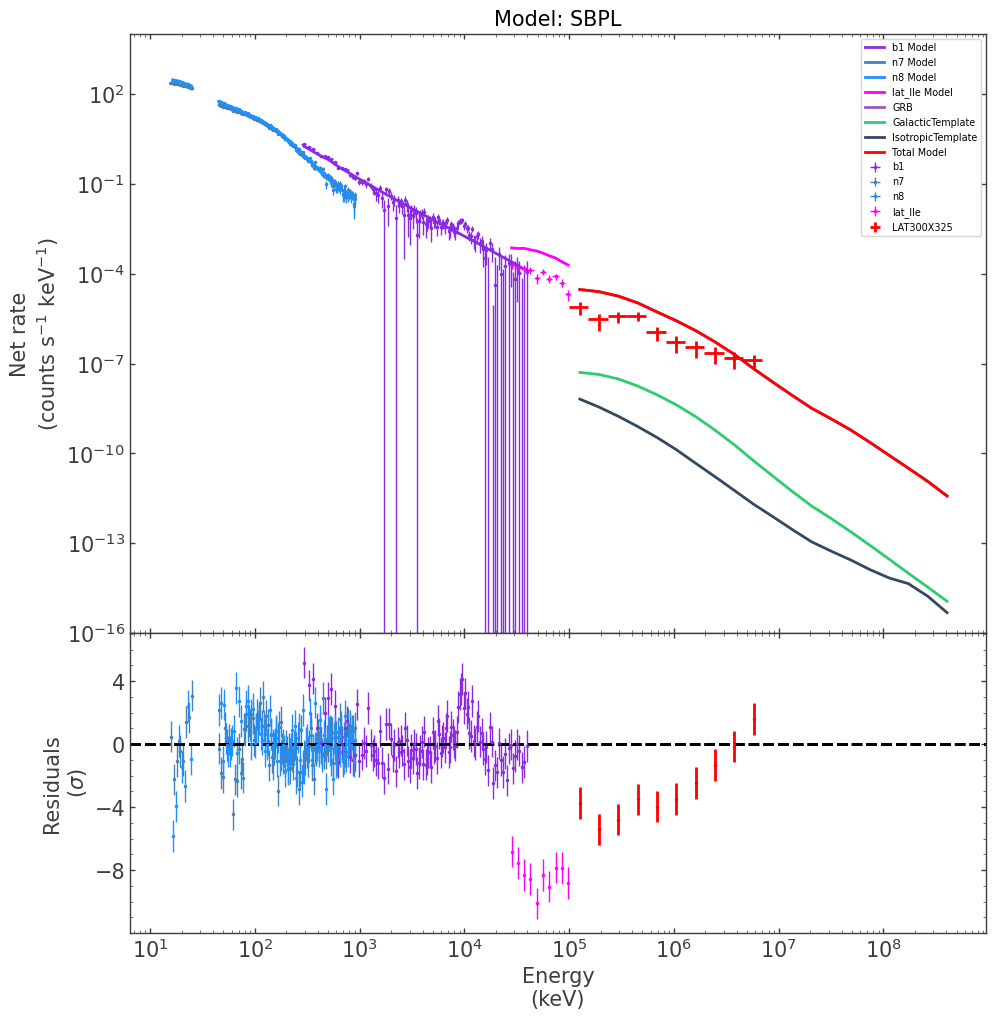}
    \includegraphics[width=0.4\linewidth]{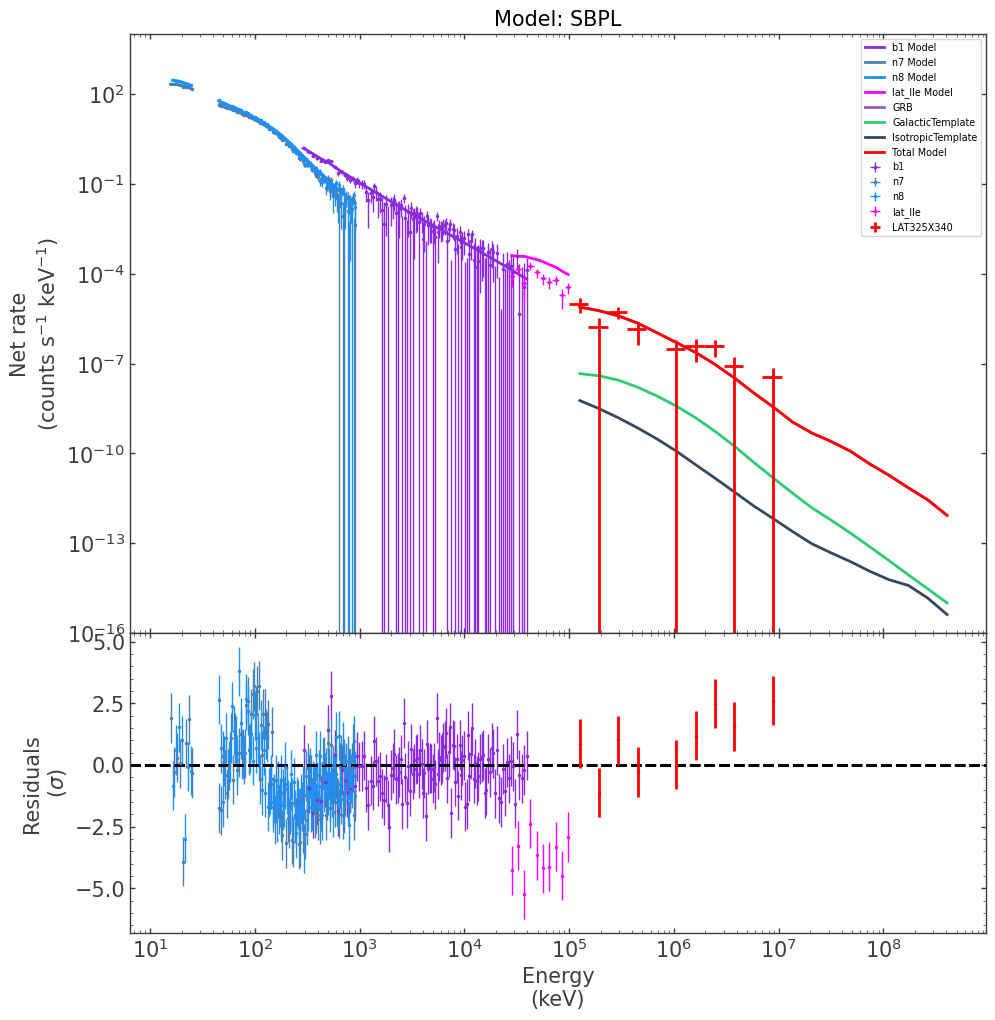}\\
    \includegraphics[width=0.4\linewidth]{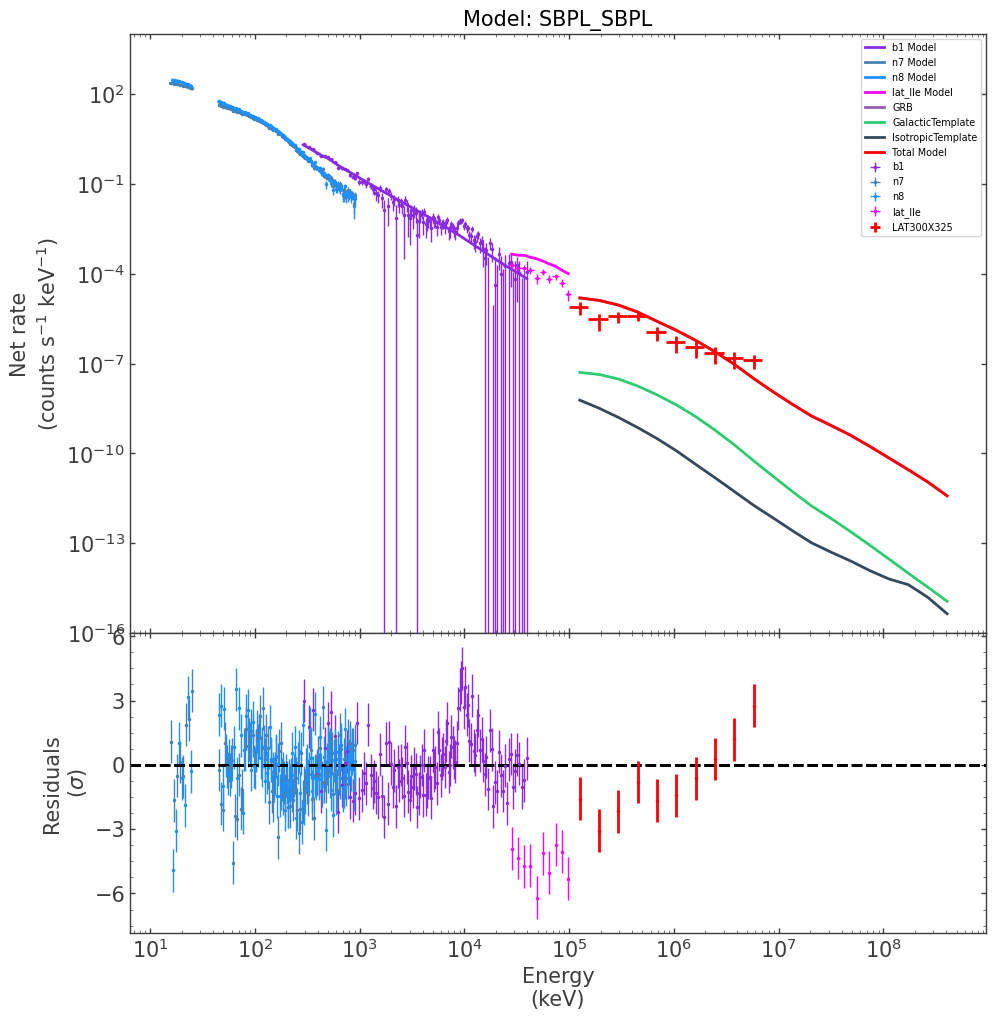}
    \includegraphics[width=0.4\linewidth]{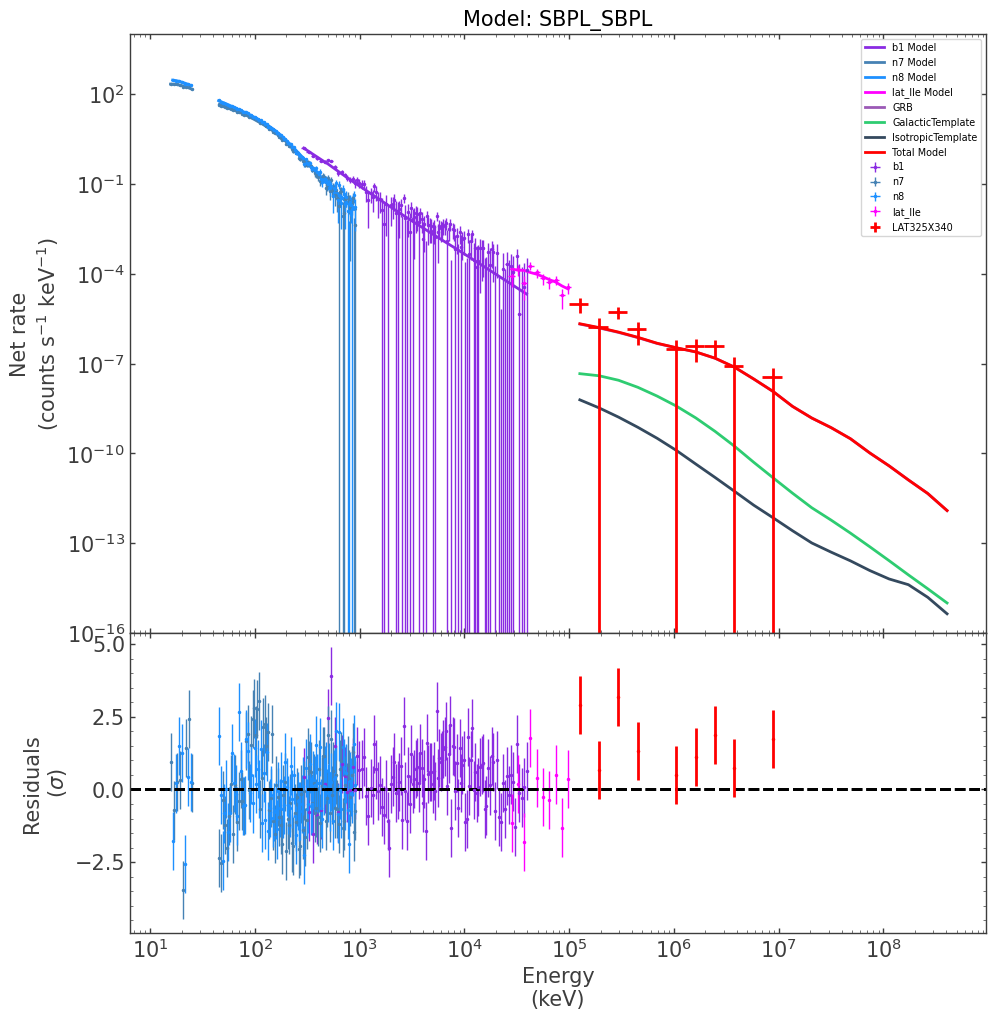}\\
    \includegraphics[width=0.4\linewidth]{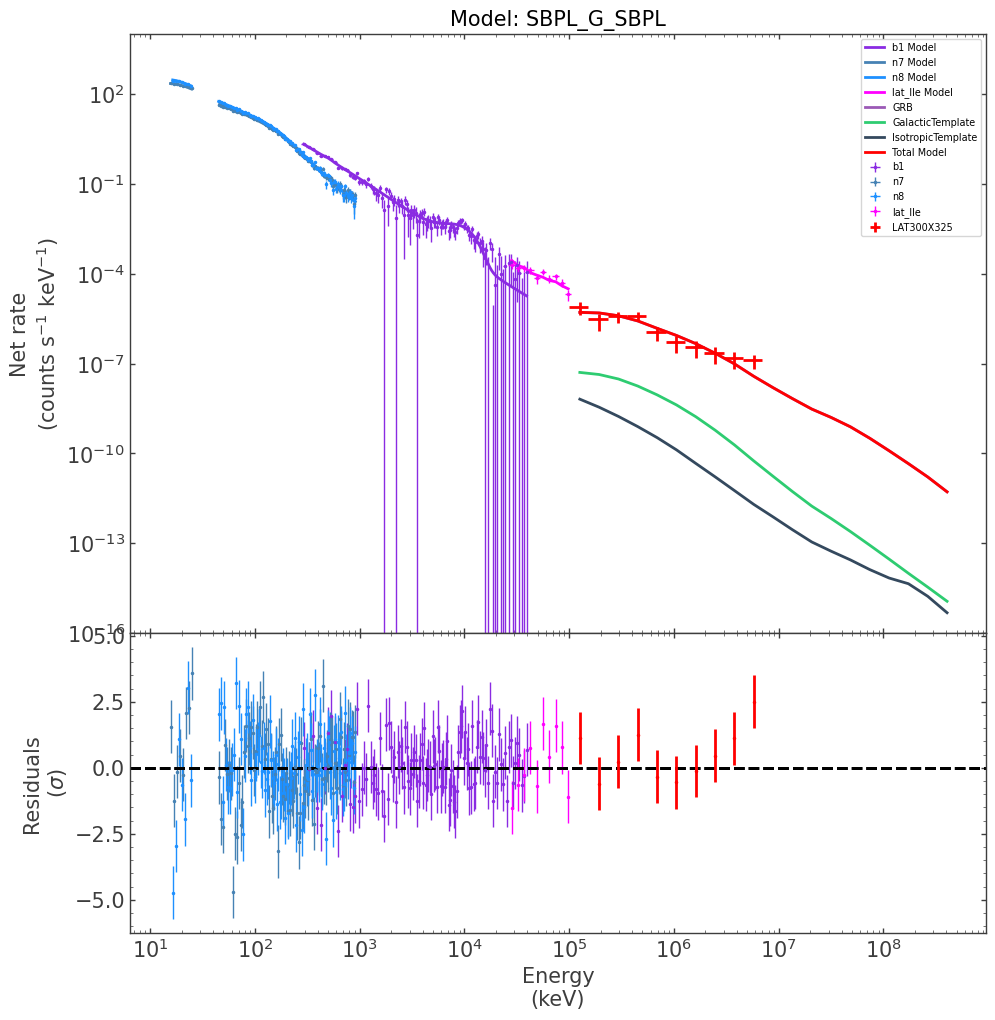}
    \includegraphics[width=0.4\linewidth]{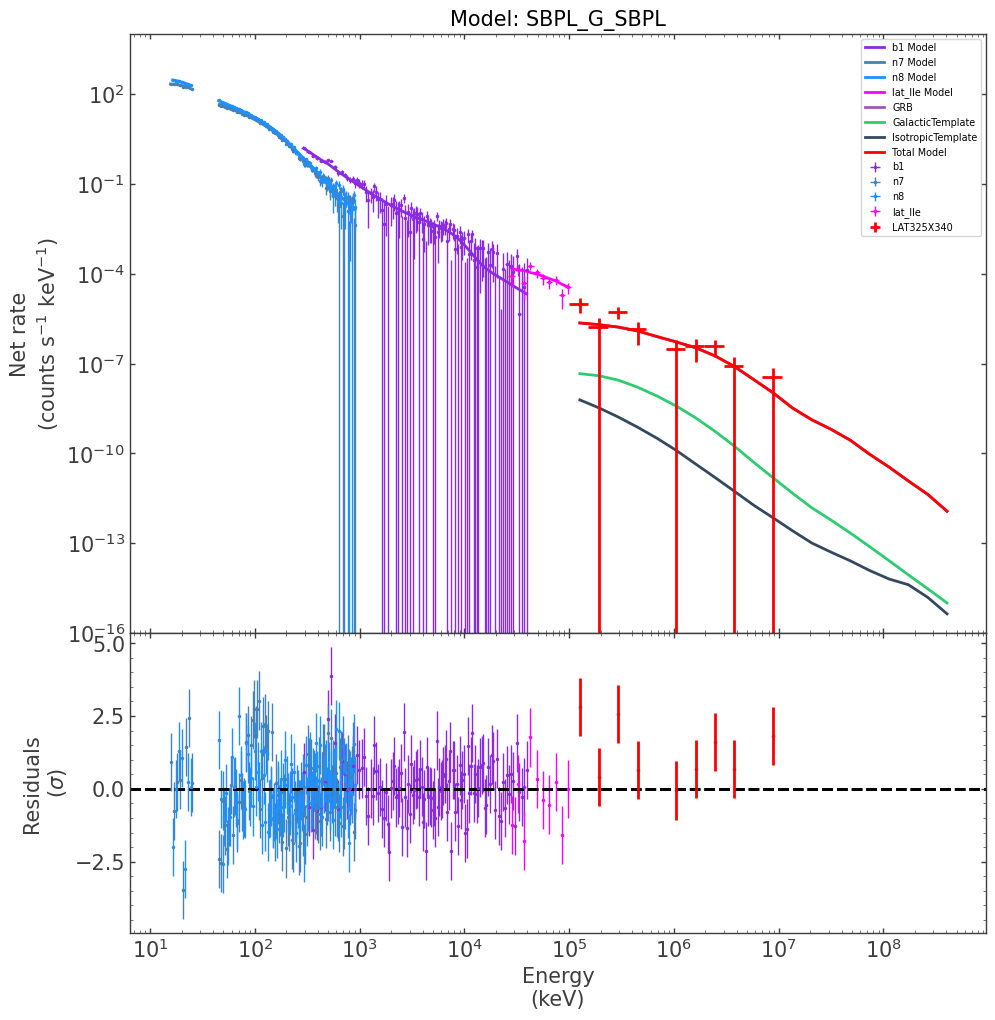}
\caption{Count spectra \nob{(in counts s$^{-1}$ keV$^{-1}$)} and residuals (in sigma units) for the [\trig+ 300.6,\,\trig+ 325.6] \nob{(left) and [\trig+ 325.6,\,\trig+ 340.6]} \nob{(right)} time intervals. From \nob{top} to \nob{bottom}: the \texttt{SBPL}, \texttt{SBPL+SBPL} and \texttt{SBPL+G+SBPL} models.}
    \label{fig:interval34}
\end{figure*}

\begin{figure*}[t]
    \centering    
    \includegraphics[width=0.4\linewidth]{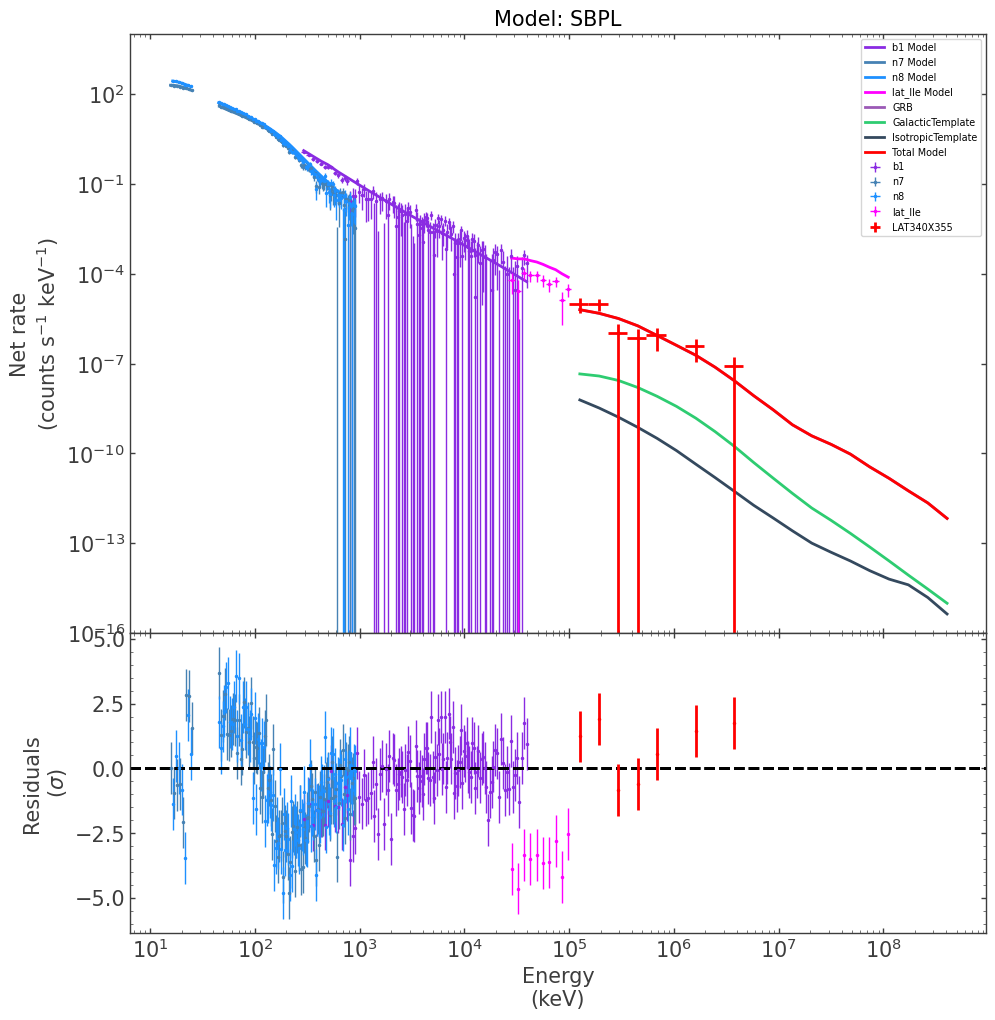}
    \includegraphics[width=0.4\linewidth]{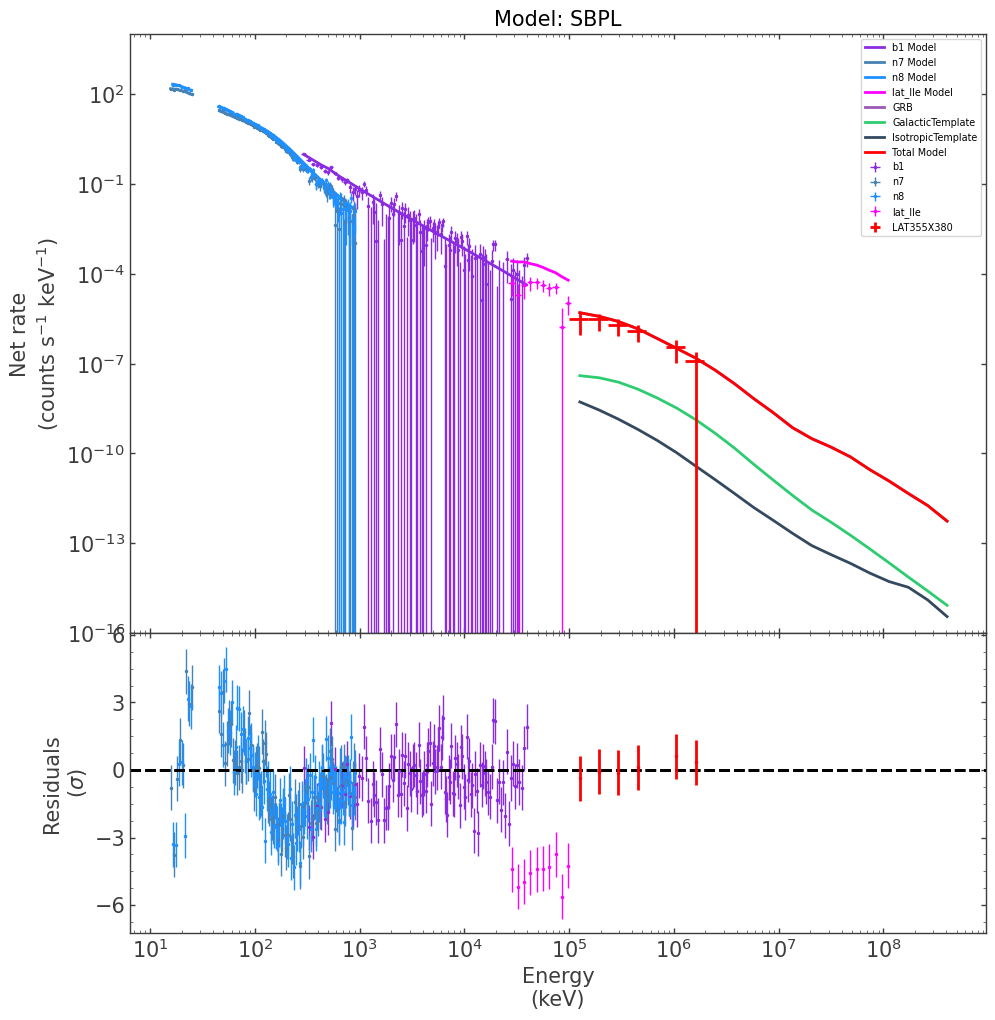}\\
    \includegraphics[width=0.4\linewidth]{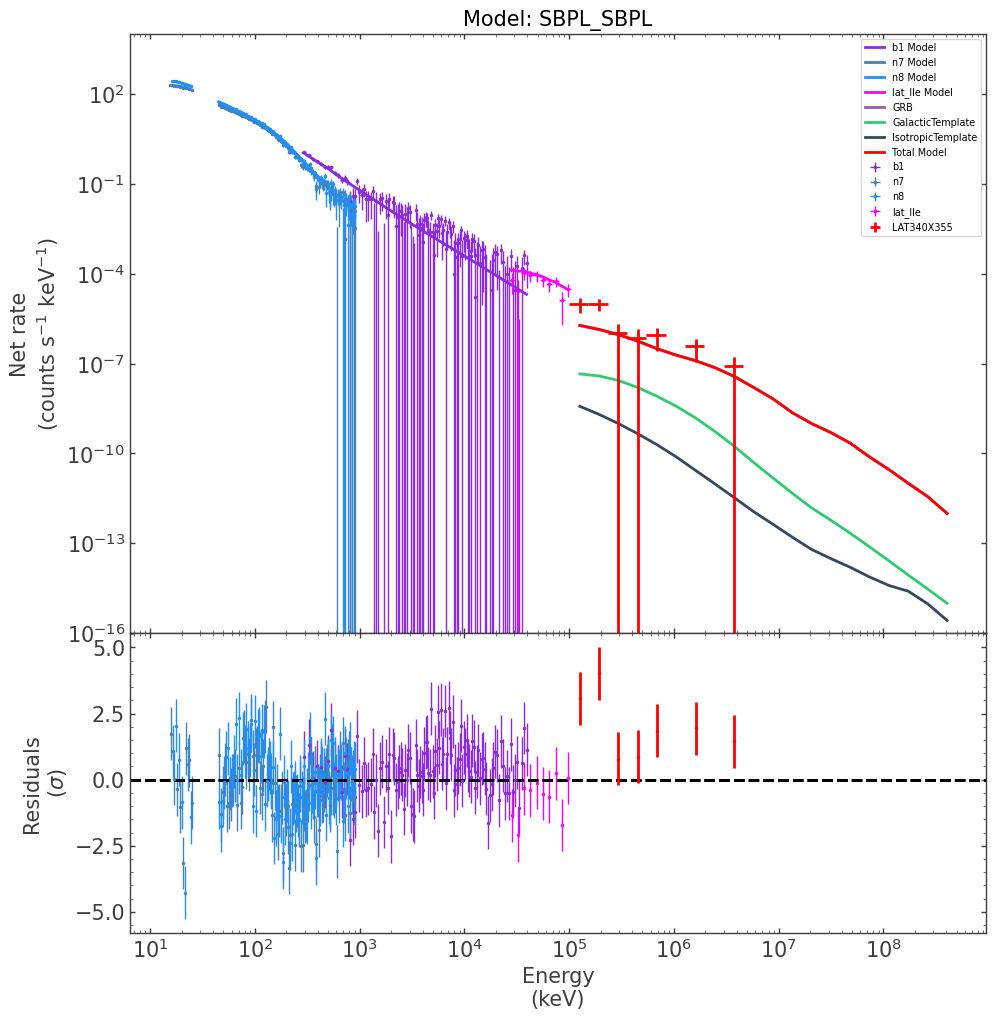}
    \includegraphics[width=0.4\linewidth]{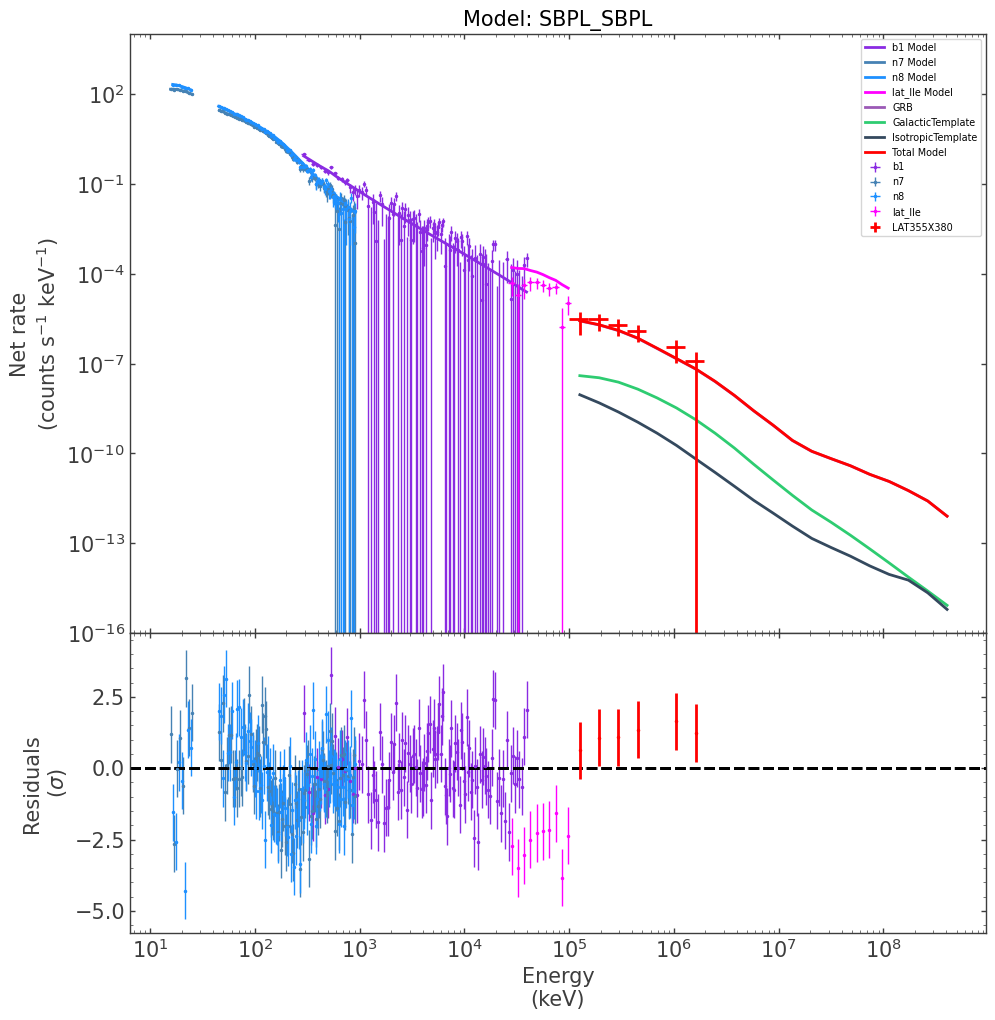}\\
    \includegraphics[width=0.4\linewidth]{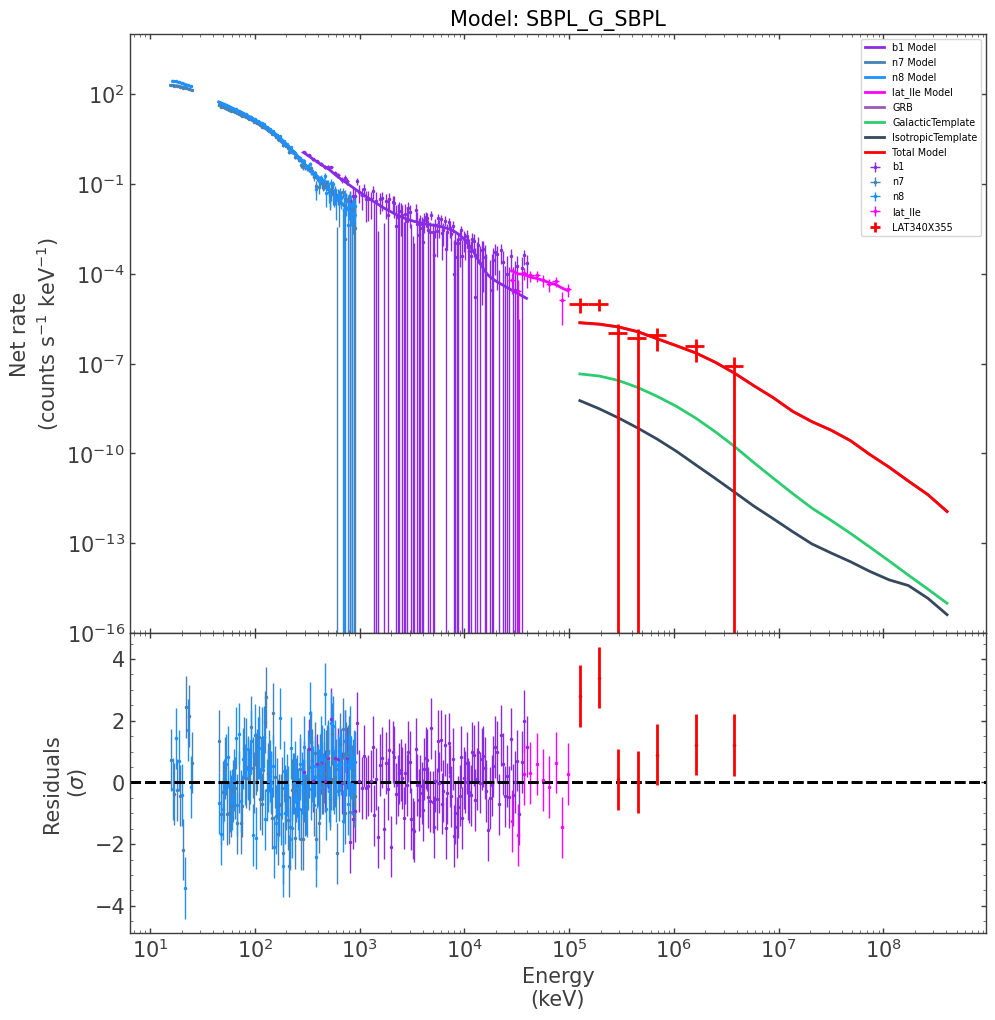}
    \includegraphics[width=0.4\linewidth]{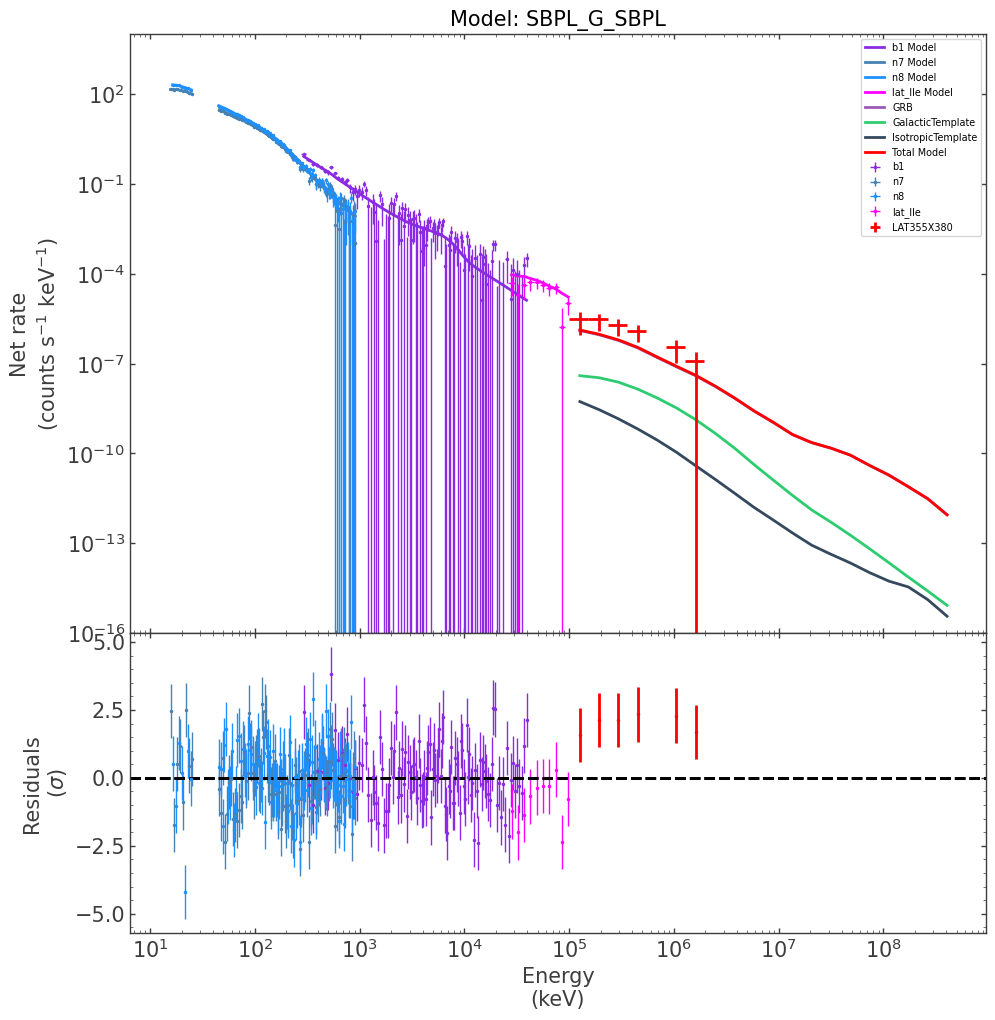}
\caption{Count spectra \nob{(in counts s$^{-1}$ keV$^{-1}$)} and residuals (in sigma units) for the [\trig+ 340.6,\,\trig+ 355.6] \nob{(left) and [\trig+ 355.6,\,\trig+ 380.6]} \nob{(right)} time intervals. From \nob{top} to \nob{bottom}: the \texttt{SBPL}, \texttt{SBPL+SBPL} and \texttt{SBPL+G+SBPL} models.}

    \label{fig:interval56}
\end{figure*}

\begin{figure*}[t]
    \centering    
    \includegraphics[width=0.4\linewidth]{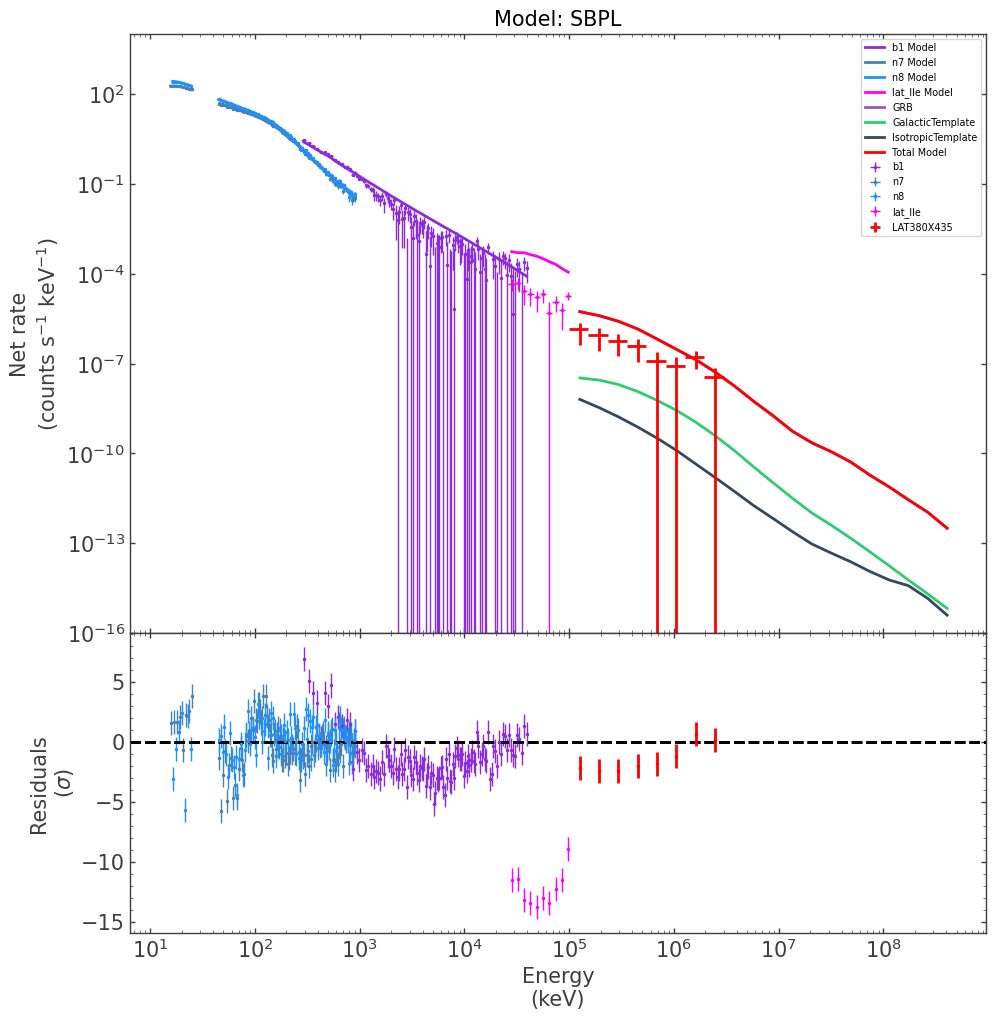}
    \includegraphics[width=0.4\linewidth]{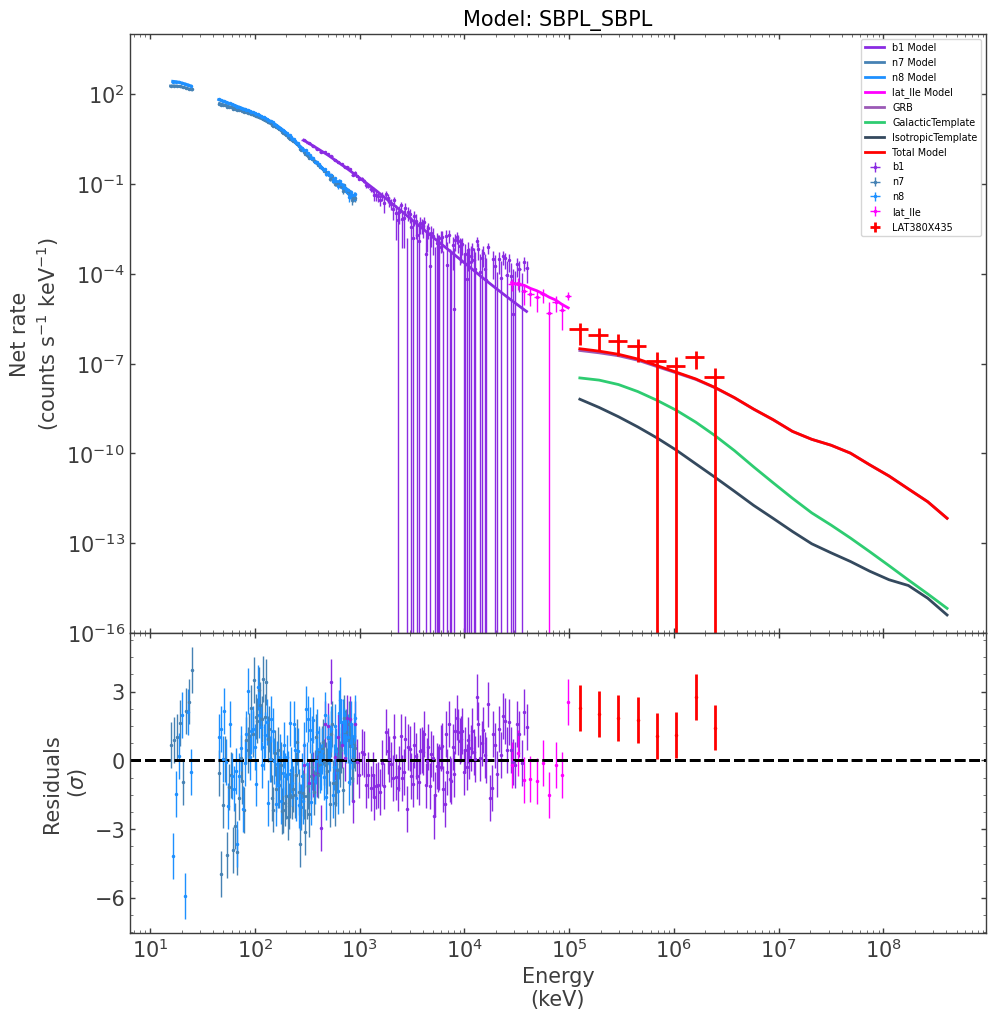}
    \includegraphics[width=0.4\linewidth]{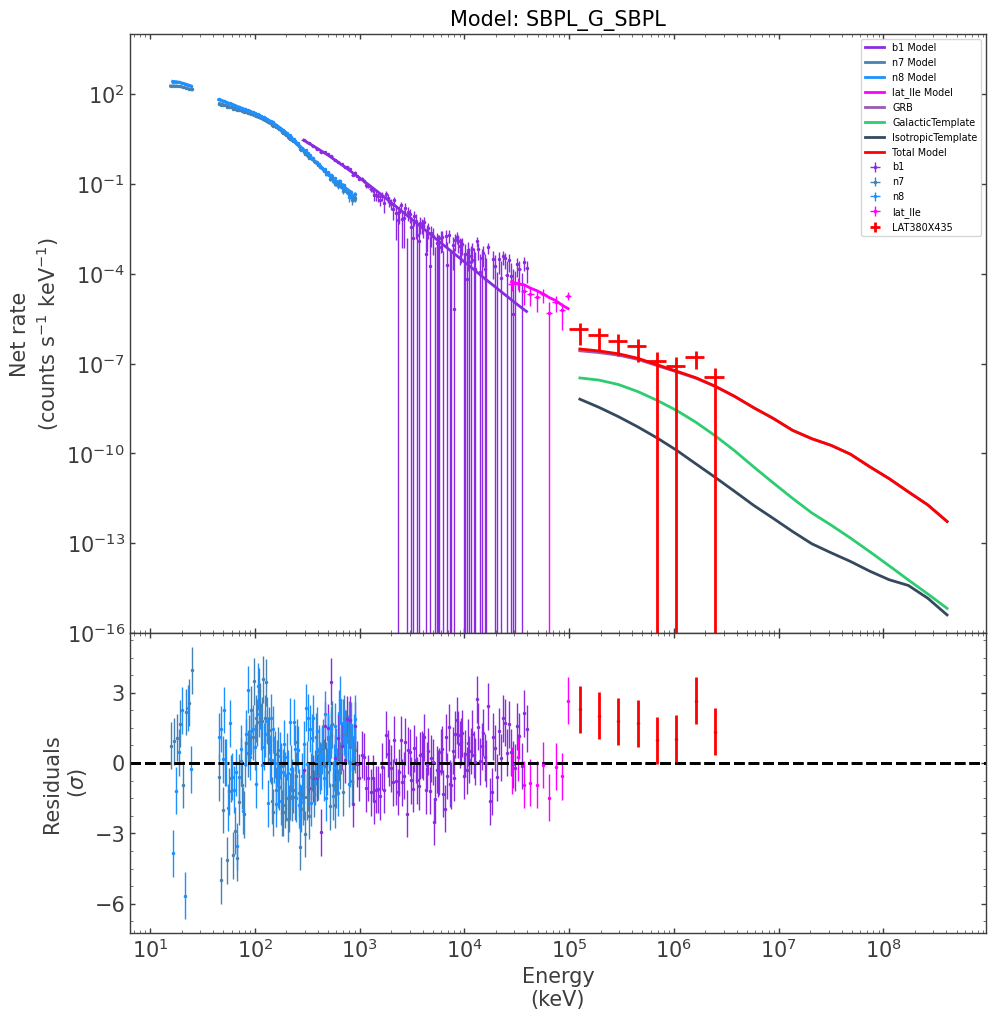}
\caption{Count spectra \nob{(in counts s$^{-1}$ keV$^{-1}$)} and residuals (in sigma units) for the [\trig+ 380.6,\,\trig+ 435.6] time interval: From \nob{top} to \nob{bottom}: the \texttt{SBPL}, \texttt{SBPL+SBPL}, \texttt{SBPL+G+SBPL} models.}
    \label{fig:interval7}
\end{figure*}

\begin{table}[h!]
    \centering
    \begin{tabular}{rr|rl}
        \hline
        Function & Parameter   & Prior & Bounds \\
        \hline
         \texttt{SBPL} & K$_1$      & Log uniform & ($10^{-6}$--$10^{6}$) [cm$^{-2}$ keV$^{-1}$ s$^{-1}$] \\
         & $\alpha_1$   & Uniform & ($-$2, 2)\\
         & E$_{b,1}$      & Log uniform & (1\,keV -- 10\,GeV) \\
         & $\beta_1$    & Uniform & ($-$10, 1)\\
         \hline
         \texttt{G} & F          & Log uniform & ($10^{-6}$--$10^{6}$) [cm$^{-2}$ s$^{-1}$ ] \\
         & $\mu$      & Uniform & (1\,MeV -- 20\,MeV)\\
         & $\sigma$   & Uniform & (1\,MeV -- 10\,MeV)\\
         \hline
         \texttt{PL} & K      & Log uniform & ($10^{-6}$--$10^{6}$) [cm$^{-2}$ keV$^{-1}$ s$^{-1}$] \\
         & $\Gamma_{\rm ph}$   & Uniform & ($-$2, 0)\\
        \hline
        \texttt{SBPL$_{\rm HE}$} & K$_2$      & Log uniform & ($10^{-15}$--$10^{-3}$) [cm$^{-2}$ keV$^{-1}$ s$^{-1}$] \\
         & $\alpha_2$   & Uniform & ($-$2, 2)\\
         & E$_{b,2}$      & Log uniform & (1\,GeV -- 100\,TeV) \\
         & $\beta_2$    & Uniform & ($-$10, 1)\\
         \hline

    \end{tabular}
    \caption{Priors used in the Bayesian fit of GBM and LAT data. We highlighted the second smoothly broken power law as \texttt{SBPL$_{\rm HE}$} to distinguish it from the \texttt{SBPL} at low energies. In both cases, the scale parameter is fixed to 0.5. The pivot energy for the \texttt{SBPL}  is fixed to its default value of 1~keV, while it is fixed to 1~GeV for the \texttt{SBPL$_{\rm HE}$}.}
    \label{tab:priors}
\end{table}
Numerical values of the parameters of the best-fit model (\texttt{SBPL+G+SBPL} and \texttt{SBPL+SBPL} in the last interval) are in Tab.~\ref{tab:joint_best}. 
In order to compare the residuals of the time-resolved spectral analysis, we show count spectra and residual for all the time intervals, to show the progression from \nob{top} to \nob{bottom} of \texttt{SBPL}, \texttt{SBPL+SBPL}, and \texttt{SBPL+G+SBPL} models in Figs.~\ref{fig:interval12}--\ref{fig:interval7}. 
Note that the LHASSO-WCDA spectrum is not displayed in the plots, but it is included in the fits.

\begin{sidewaystable}[h!]
\centering
\begin{tabular}{c | rrrrrrr}
\hline
Parameter &  \multicolumn{7}{c}{Interval from \trig (s)}\\
 & 280.6--290.6  & 290.6--300.6  & 300.6--325.6  & 325.6--340.6  & 340.6--355.6  & 355.6--380.6  & 380.6--435.6 \\
\hline
K$_1$ [cm$^{-2}$ keV$^{-1}$ s$^{-1}$] & 0.729$^{+0.006}_{-0.008}$ & 0.319$^{+0.002}_{-0.01}$ & 0.135$^{+0.001}_{-0.007}$ & 0.1390$^{+0.0002}_{-0.004}$ & 0.1186$^{+0.0004}_{-0.004}$ & (8.21$^{+0.02}_{-0.1}$)$\times$10$^{-2}$ & 0.179$^{+0.001}_{-0.003}$ \\
$\alpha_1$ & $-$1.499$^{+0.002}_{-0.004}$ & $-$1.682$^{+0.005}_{-0.002}$ & $-$1.926$^{+0.001}_{-0.006}$ &  $-$1.79 $\pm$ 0.02 & $-$1.75$^{+0.03}_{-0.02}$ & $-$1.24$^{+0.2}_{-0.07}$ & $-$1.689$^{+0.001}_{-0.004}$ \\
E$_{b,1}$ [MeV] & 0.90$^{+0.06}_{-0.02}$ & 1.16$^{+0.2}_{-0.04}$ & 2.121$^{+4}_{-0.005}$ & 0.10$^{+0.02}_{-0.01}$ & ( 8 $\pm$ 1)$\times$10$^{-2}$ & (2.0$^{+0.1}_{-0.4}$)$\times$10$^{-2}$ & 0.677$^{+0.1}_{-0.004}$ \\
$\beta_1$ & $-$2.84$^{+0.03}_{-0.09}$ & $-$2.78$^{+0.02}_{-0.3}$ & $-$3.02$^{+0.02}_{-5}$ & $-$2.36$^{+0.04}_{-0.03}$ & $-$2.52$^{+0.04}_{-0.05}$ & $-$2.273$^{+0.01}_{-0.006}$ & $-$2.99$^{+0.02}_{-0.2}$ \\
F [cm$^{-2}$ s$^{-1}$] & 0.52$^{+0.6}_{-0.06}$ & 0.198$^{+0.1}_{-0.003}$ & 0.254$^{+0.1}_{-0.003}$ & 0.29$^{+0.06}_{-0.1}$ & 0.23$^{+0.2}_{-0.01}$ & (6.9$^{+10.0}_{-0.9}$)$\times$10$^{-2}$ & -- \\
$\mu$ [MeV] & 6.5$^{+0.6}_{-5}$ & 11.5$^{+0.2}_{-2}$ & 8.8$^{+0.5}_{-1}$ & 3$^{+3}_{-1}$ & 5.8$^{+0.5}_{-4}$ & 5.13$^{+0.03}_{-3}$ & -- \\
$\sigma$ [MeV] & 5.3$^{+2}_{-0.1}$ & 2.0$^{+1}_{-0.1}$ & 3.2$^{+1}_{-0.1}$ & 4.0$^{+1}_{-0.9}$ & 3.5$^{+2}_{-0.2}$ & 2.2$^{+1}_{-0.2}$ & -- \\
K$_2$ [10$^{-11}\times$cm$^{-2}$ keV$^{-1}$ s$^{-1}$] & 3.1$^{+0.9}_{-1}$ & 2.0$^{+0.8}_{-0.9}$ & 1.84$^{+0.06}_{-0.8}$ & 2.12$^{+0.02}_{-0.3}$ & 1.4$^{+0.4}_{-0.3}$ & 0.28$^{+0.03}_{-0.2}$ & 0.5$^{+0.3}_{-0.2}$ \\
$\alpha_2$ &  $-$1.3 $\pm$ 0.2 &  $-$1.5 $\pm$ 0.2 & $-$1.51$^{+0.04}_{-0.2}$ & $-$1.2$^{+0.3}_{-0.2}$ &  $-$1.5 $\pm$ 0.1 &  $-$1.4 $\pm$ 0.1 &  $-$1.4 $\pm$ 0.2 \\
E$_{b,2}$ [MeV] & (1$^{+2}_{-0.7}$)$\times$10$^{4}$ & (2$^{+8}_{-1}$)$\times$10$^{4}$ & (2$^{+6}_{-0.1}$)$\times$10$^{4}$ & (4$^{+3}_{-2}$)$\times$10$^{3}$ & (1$^{+1}_{-0.8}$)$\times$10$^{4}$ & (7.0$^{+10.0}_{-2}$)$\times$10$^{4}$ & (2$^{+4}_{-1}$)$\times$10$^{4}$ \\
$\beta_2$ & $-$2.35$^{+0.02}_{-0.01}$ & $-$2.33$^{+0.01}_{-0.04}$ & $-$2.32$^{+0.02}_{-0.01}$ & $-$2.308$^{+0.02}_{-0.002}$ & $-$2.32$^{+0.01}_{-0.02}$ & $-$2.304$^{+0.008}_{-0.04}$ & $-$2.252$^{+0.002}_{-0.02}$ \\
\hline
\end{tabular}
\caption{Parameters for the best-fit model. For each time interval, we show the parameters for the \texttt{SBPL+G+SBPL} model, except for 380.6--435.6 where we show the ones for \texttt{SBPL+SBPL} as the additional \texttt{G} did not significantly improve the fit.}
\label{tab:joint_best}
\end{sidewaystable}

%
%
%
%
%
%
%
\bibliography{Fermi_GRB_2}

\begin{thebibliography}{}
\providecommand\natexlab[1]{#1}
\providecommand\JournalTitle[1]{#1}

\bibitem[{Abbasi {et~al.}(2023)Abbasi, Ackermann, Adams, Agarwalla, Aggarwal,
  Aguilar, Ahlers, Alameddine, Amin, Andeen, Anton, Argüelles, Ashida,
  Athanasiadou, Axani, Bai, V., Baricevic, Barwick, Basu, Bay, Beatty, Becker,
  Tjus, Beise, Bellenghi, BenZvi, Berley, Bernardini, Besson, Binder, Bindig,
  Blaufuss, Blot, Bontempo, Book, Borowka, Meneguolo, Böser, Botner,
  Böttcher, Bourbeau, Braun, Brinson, Brostean-Kaiser, Burley, Busse, Campana,
  Carloni, Carnie-Bronca, Chen, Chen, Chirkin, Choi, Clark, Classen, Coleman,
  Collin, Connolly, Conrad, Coppin, Correa, Countryman, Cowen, Dappen, Dave,
  Clercq, DeLaunay, López, Dembinski, Deoskar, Desai, Desiati, de~Vries,
  de~Wasseige, DeYoung, Diaz, Díaz-Vélez, Dittmer, Domi, Dujmovic, DuVernois,
  Ehrhardt, Eller, Engel, Erpenbeck, Evans, Evenson, Fan, Fazely, Fedynitch,
  Feigl, Fiedlschuster, Finley, Fischer, Fox, Franckowiak, Friedman, Fritz,
  Fürst, Gaisser, Gallagher, Ganster, Garcia, Garrappa, Gerhardt, Ghadimi,
  Glaser, Glauch, Glüsenkamp, Goehlke, Gonzalez, Goswami, Grant, Gray,
  Griffin, Griswold, Günther, Gutjahr, Haack, Hallgren, Halliday, Halve,
  Halzen, Hamdaoui, Minh, Hanson, Hardin, Harnisch, Hatch, Haungs, Helbing,
  Hellrung, Henningsen, Heuermann, Hickford, Hidvegi, Hill, Hill, Hoffman,
  Hoshina, Hou, Huber, Hultqvist, Hünnefeld, Hussain, Hymon, In, Iovine,
  Ishihara, Jansson, Japaridze, Jeong, Jin, Jones, Kang, Kang, Kang, Kappes,
  Kappesser, Kardum, Karg, Karl, Karle, Katz, Kauer, Kelley, Kheirandish, Kin,
  Kiryluk, Klein, Kochocki, Koirala, Kolanoski, Kontrimas, Köpke, Kopper,
  Koskinen, Koundal, Kovacevich, Kowalski, Kozynets, Kruiswijk, Krupczak,
  Kumar, Kun, Kurahashi, Lad, Gualda, Lamoureux, Larson, Lauber, Lazar, Lee,
  DeHolton, Leszczyńska, Lincetto, Liu, Liubarska, Lohfink, Love, Mariscal,
  Lu, Lucarelli, Ludwig, Luszczak, Lyu, Ma, Madsen, Mahn, Makino, Mancina,
  Sainte, Mariş, Marka, Marka, Marsee, Martinez-Soler, Maruyama, Mayhew,
  McElroy, McNally, Mead, Meagher, Mechbal, Medina, Meier, Meighen-Berger,
  Merckx, Merten, Micallef, Mockler, Montaruli, Moore, Morii, Morse, Moulai,
  Mukherjee, Naab, Nagai, Naumann, Necker, Neumann, Niederhausen, Nisa, Noell,
  Nowicki, Pollmann, Oehler, Oeyen, Olivas, Orsoe, Osborn, O’Sullivan,
  Pandya, Park, Parker, Paudel, Paul, de~los Heros, Peterson, Philippen,
  Pieper, Pizzuto, Plum, Popovych, Rodriguez, Pries, Procter-Murphy,
  Przybylski, Raab, Rack-Helleis, Rawlins, Rechav, Rehman, Reichherzer, Renzi,
  Resconi, Reusch, Rhode, Richman, Riedel, Roberts, Robertson, Rodan,
  Roellinghoff, Rongen, Rott, Ruhe, Ruohan, Ryckbosch, Athanasiadou, Safa,
  Saffer, Salazar-Gallegos, Sampathkumar, Herrera, Sandrock, Santander, Sarkar,
  Sarkar, Savelberg, Savina, Schaufel, Schieler, Schindler, Schlüter, Schmidt,
  Schneider, Schröder, Schumacher, Schwefer, Sclafani, Seckel, Seunarine,
  Sharma, Shefali, Shimizu, Silva, Skrzypek, Smithers, Snihur, Soedingrekso,
  Søgaard, Soldin, Sommani, Spannfellner, Spiczak, Spiering, Stamatikos,
  Stanev, Stein, Stezelberger, Stürwald, Stuttard, Sullivan, Taboada,
  Ter-Antonyan, Thompson, Thwaites, Tilav, Tollefson, Tönnis, Toscano, Tosi,
  Trettin, Tung, Turcotte, Twagirayezu, Ty, Elorrieta, Upshaw,
  Valtonen-Mattila, Vandenbroucke, van Eijndhoven, Vannerom, van Santen, Vara,
  Veitch-Michaelis, Venugopal, Verpoest, Veske, Walck, Watson, Weaver, Weigel,
  Weindl, Weldert, Wendt, Werthebach, Weyrauch, Whitehorn, Wiebusch, Willey,
  Williams, Wolf, Wrede, Wulff, Xu, Yanez, Yildizci, Yoshida, Yu, Yu, Yuan,
  Zhang, Zhelnin, \& Collaboration)}]{Abbasi_2023}
Abbasi, R., Ackermann, M., Adams, J., {et~al.} 2023,
  \href{http://dx.doi.org/10.3847/2041-8213/acc077}{\JournalTitle{\apjl}, 946,
  L26}

\bibitem[{{Abbasi} {et~al.}(2023){Abbasi}, {Ackermann}, {Adams}, {Agarwalla},
  {Aggarwal}, {Aguilar}, {Ahlers}, {Alameddine}, {Amin}, {Andeen}, {Anton},
  {Arg{\"u}elles}, {Ashida}, {Athanasiadou}, {Axani}, {Bai}, {Balagopal},
  {Baricevic}, {Barwick}, {Basu}, {Bay}, {Beatty}, {Becker}, {Becker Tjus},
  {Beise}, {Bellenghi}, {BenZvi}, {Berley}, {Bernardini}, {Besson}, {Binder},
  {Bindig}, {Blaufuss}, {Blot}, {Bontempo}, {Book}, {Borowka}, {Boscolo
  Meneguolo}, {B{\"o}ser}, {Botner}, {B{\"o}ttcher}, {Bourbeau}, {Braun},
  {Brinson}, {Brostean-Kaiser}, {Burley}, {Busse}, {Campana}, {Carloni},
  {Carnie-Bronca}, {Chen}, {Chen}, {Chirkin}, {Choi}, {Clark}, {Classen},
  {Coleman}, {Collin}, {Connolly}, {Conrad}, {Coppin}, {Correa}, {Countryman},
  {Cowen}, {Dappen}, {Dave}, {De Clercq}, {DeLaunay}, {Delgado L{\'o}pez},
  {Dembinski}, {Deoskar}, {Desai}, {Desiati}, {de Vries}, {de Wasseige},
  {DeYoung}, {Diaz}, {D{\'\i}az-V{\'e}lez}, {Dittmer}, {Domi}, {Dujmovic},
  {DuVernois}, {Ehrhardt}, {Eller}, {Engel}, {Erpenbeck}, {Evans}, {Evenson},
  {Fan}, {Fazely}, {Fedynitch}, {Feigl}, {Fiedlschuster}, {Finley}, {Fischer},
  {Fox}, {Franckowiak}, {Friedman}, {Fritz}, {F{\"u}rst}, {Gaisser},
  {Gallagher}, {Ganster}, {Garcia}, {Garrappa}, {Gerhardt}, {Ghadimi},
  {Glaser}, {Glauch}, {Gl{\"u}senkamp}, {Goehlke}, {Gonzalez}, {Goswami},
  {Grant}, {Gray}, {Griffin}, {Griswold}, {G{\"u}nther}, {Gutjahr}, {Haack},
  {Hallgren}, {Halliday}, {Halve}, {Halzen}, {Hamdaoui}, {Ha Minh}, {Hanson},
  {Hardin}, {Harnisch}, {Hatch}, {Haungs}, {Helbing}, {Hellrung}, {Henningsen},
  {Heuermann}, {Hickford}, {Hidvegi}, {Hill}, {Hill}, {Hoffman}, {Hoshina},
  {Hou}, {Huber}, {Hultqvist}, {H{\"u}nnefeld}, {Hussain}, {Hymon}, {In},
  {Iovine}, {Ishihara}, {Jansson}, {Japaridze}, {Jeong}, {Jin}, {Jones},
  {Kang}, {Kang}, {Kang}, {Kappes}, {Kappesser}, {Kardum}, {Karg}, {Karl},
  {Karle}, {Katz}, {Kauer}, {Kelley}, {Kheirandish}, {Kin}, {Kiryluk}, {Klein},
  {Kochocki}, {Koirala}, {Kolanoski}, {Kontrimas}, {K{\"o}pke}, {Kopper},
  {Koskinen}, {Koundal}, {Kovacevich}, {Kowalski}, {Kozynets}, {Kruiswijk},
  {Krupczak}, {Kumar}, {Kun}, {Kurahashi}, {Lad}, {Lagunas Gualda},
  {Lamoureux}, {Larson}, {Lauber}, {Lazar}, {Lee}, {Leonard DeHolton},
  {Leszczy{\'n}ska}, {Lincetto}, {Liu}, {Liubarska}, {Lohfink}, {Love}, {Lozano
  Mariscal}, {Lu}, {Lucarelli}, {Ludwig}, {Luszczak}, {Lyu}, {Ma}, {Madsen},
  {Mahn}, {Makino}, {Mancina}, {Marie Sainte}, {Mari{\c{s}}}, {Marka}, {Marka},
  {Marsee}, {Martinez-Soler}, {Maruyama}, {Mayhew}, {McElroy}, {McNally},
  {Mead}, {Meagher}, {Mechbal}, {Medina}, {Meier}, {Meighen-Berger}, {Merckx},
  {Merten}, {Micallef}, {Mockler}, {Montaruli}, {Moore}, {Morii}, {Morse},
  {Moulai}, {Mukherjee}, {Naab}, {Nagai}, {Naumann}, {Necker}, {Neumann},
  {Niederhausen}, {Nisa}, {Noell}, {Nowicki}, {Obertacke Pollmann}, {Oehler},
  {Oeyen}, {Olivas}, {Orsoe}, {Osborn}, {O'Sullivan}, {Pandya}, {Park},
  {Parker}, {Paudel}, {Paul}, {P{\'e}rez de los Heros}, {Peterson},
  {Philippen}, {Pieper}, {Pizzuto}, {Plum}, {Popovych}, {Prado Rodriguez},
  {Pries}, {Procter-Murphy}, {Przybylski}, {Raab}, {Rack-Helleis}, {Rawlins},
  {Rechav}, {Rehman}, {Reichherzer}, {Renzi}, {Resconi}, {Reusch}, {Rhode},
  {Richman}, {Riedel}, {Roberts}, {Robertson}, {Rodan}, {Roellinghoff},
  {Rongen}, {Rott}, {Ruhe}, {Ruohan}, {Ryckbosch}, {Athanasiadou}, {Safa},
  {Saffer}, {Salazar-Gallegos}, {Sampathkumar}, {Sanchez Herrera}, {Sandrock},
  {Santander}, {Sarkar}, {Sarkar}, {Savelberg}, {Savina}, {Schaufel},
  {Schieler}, {Schindler}, {Schl{\"u}ter}, {Schmidt}, {Schneider},
  {Schr{\"o}der}, {Schumacher}, {Schwefer}, {Sclafani}, {Seckel}, {Seunarine},
  {Sharma}, {Shefali}, {Shimizu}, {Silva}, {Skrzypek}, {Smithers}, {Snihur},
  {Soedingrekso}, {S{\o}gaard}, {Soldin}, {Sommani}, {Spannfellner}, {Spiczak},
  {Spiering}, {Stamatikos}, {Stanev}, {Stein}, {Stezelberger}, {St{\"u}rwald},
  {Stuttard}, {Sullivan}, {Taboada}, {Ter-Antonyan}, {Thompson}, {Thwaites},
  {Tilav}, {Tollefson}, {T{\"o}nnis}, {Toscano}, {Tosi}, {Trettin}, {Tung},
  {Turcotte}, {Twagirayezu}, {Ty}, {Unland Elorrieta}, {Upshaw},
  {Valtonen-Mattila}, {Vandenbroucke}, {van Eijndhoven}, {Vannerom}, {van
  Santen}, {Vara}, {Veitch-Michaelis}, {Venugopal}, {Verpoest}, {Veske},
  {Walck}, {Watson}, {Weaver}, {Weigel}, {Weindl}, {Weldert}, {Wendt},
  {Werthebach}, {Weyrauch}, {Whitehorn}, {Wiebusch}, {Willey}, {Williams},
  {Wolf}, {Wrede}, {Wulff}, {Xu}, {Yanez}, {Yildizci}, {Yoshida}, {Yu}, {Yu},
  {Yuan}, {Zhang}, {Zhelnin}, \& {IceCube Collaboration}}]{2023ApJ...946L..26A}
{Abbasi}, R., {Ackermann}, M., {Adams}, J., {et~al.} 2023,
  \href{http://dx.doi.org/10.3847/2041-8213/acc077}{\JournalTitle{\apjl}, 946,
  L26}

\bibitem[{{Abdalla} {et~al.}(2019){Abdalla}, {Adam}, {Aharonian}, {Ait
  Benkhali}, {Ang{\"u}ner}, {Arakawa}, {Arcaro}, {Armand}, {Ashkar}, {Backes},
  {Barbosa Martins}, {Barnard}, {Becherini}, {Berge}, {Bernl{\"o}hr},
  {Bissaldi}, {Blackwell}, {B{\"o}ttcher}, {Boisson}, {Bolmont}, {Bonnefoy},
  {Bregeon}, {Breuhaus}, {Brun}, {Brun}, {Bryan}, {B{\"u}chele}, {Bulik},
  {Bylund}, {Capasso}, {Caroff}, {Carosi}, {Casanova}, {Cerruti}, {Chand},
  {Chandra}, {Chen}, {Colafrancesco}, {Cury{\l}o}, {Davids}, {Deil}, {Devin},
  {deWilt}, {Dirson}, {Djannati-Ata{\"\i}}, {Dmytriiev}, {Donath},
  {Doroshenko}, {Dyks}, {Egberts}, {Emery}, {Ernenwein}, {Eschbach}, {Feijen},
  {Fegan}, {Fiasson}, {Fontaine}, {Funk}, {F{\"u}{\ss}ling}, {Gabici},
  {Gallant}, {Gat{\'e}}, {Giavitto}, {Giunti}, {Glawion}, {Glicenstein},
  {Gottschall}, {Grondin}, {Hahn}, {Haupt}, {Heinzelmann}, {Henri}, {Hermann},
  {Hinton}, {Hofmann}, {Hoischen}, {Holch}, {Holler}, {Horns}, {Huber},
  {Iwasaki}, {Jamrozy}, {Jankowsky}, {Jankowsky}, {Jardin-Blicq},
  {Jung-Richardt}, {Kastendieck}, {Katarzy{\'n}ski}, {Katsuragawa}, {Katz},
  {Khangulyan}, {Kh{\'e}lifi}, {King}, {Klepser}, {Klu{\'z}niak}, {Komin},
  {Kosack}, {Kostunin}, {Kreter}, {Lamanna}, {Lemi{\`e}re}, {Lemoine-Goumard},
  {Lenain}, {Leser}, {Levy}, {Lohse}, {Lypova}, {Mackey}, {Majumdar},
  {Malyshev}, {Marandon}, {Marcowith}, {Mares}, {Mariaud}, {Mart{\'\i}-Devesa},
  {Marx}, {Maurin}, {Meintjes}, {Mitchell}, {Moderski}, {Mohamed}, {Mohrmann},
  {Moore}, {Moulin}, {Muller}, {Murach}, {Nakashima}, {de Naurois},
  {Ndiyavala}, {Niederwanger}, {Niemiec}, {Oakes}, {O'Brien}, {Odaka}, {Ohm},
  {de Ona Wilhelmi}, {Ostrowski}, {Oya}, {Panter}, {Parsons}, {Perennes},
  {Petrucci}, {Peyaud}, {Piel}, {Pita}, {Poireau}, {Priyana Noel}, {Prokhorov},
  {Prokoph}, {P{\"u}hlhofer}, {Punch}, {Quirrenbach}, {Raab}, {Rauth},
  {Reimer}, {Reimer}, {Remy}, {Renaud}, {Rieger}, {Rinchiuso}, {Romoli},
  {Rowell}, {Rudak}, {Ruiz-Velasco}, {Sahakian}, {Sailer}, {Saito}, {Sanchez},
  {Santangelo}, {Sasaki}, {Schlickeiser}, {Sch{\"u}ssler}, {Schulz}, {Schutte},
  {Schwanke}, {Schwemmer}, {Seglar-Arroyo}, {Senniappan}, {Seyffert}, {Shafi},
  {Shiningayamwe}, {Simoni}, {Sinha}, {Sol}, {Specovius}, {Spir-Jacob},
  {Stawarz}, {Steenkamp}, {Stegmann}, {Steppa}, {Takahashi}, {Tavernier},
  {Taylor}, {Terrier}, {Tiziani}, {Tluczykont}, {Trichard}, {Tsirou}, {Tsuji},
  {Tuffs}, {Uchiyama}, {van der Walt}, {van Eldik}, {van Rensburg}, {van
  Soelen}, {Vasileiadis}, {Veh}, {Venter}, {Vincent}, {Vink}, {V{\"o}lk},
  {Vuillaume}, {Wadiasingh}, {Wagner}, {White}, {Wierzcholska}, {Yang},
  {Yoneda}, {Zacharias}, {Zanin}, {Zdziarski}, {Zech}, {Ziegler}, {Zorn},
  {{\.Z}ywucka}, {de Palma}, {Axelsson}, \& {Roberts}}]{2019Natur.575..464A}
{Abdalla}, H., {Adam}, R., {Aharonian}, F., {et~al.} 2019,
  \href{http://dx.doi.org/10.1038/s41586-019-1743-9}{\JournalTitle{\nat}, 575,
  464}

\bibitem[{Abdollahi {et~al.}(2020)Abdollahi, Acero, Ackermann, Ajello, Atwood,
  Axelsson, Baldini, Ballet, Barbiellini, Bastieri, Gonzalez, Bellazzini,
  Berretta, Bissaldi, Blandford, Bloom, Bonino, Bottacini, Brandt, Bregeon,
  Bruel, Buehler, Burnett, Buson, Cameron, Caputo, Caraveo, Casandjian, Castro,
  Cavazzuti, Charles, Chaty, Chen, Cheung, Chiaro, Ciprini, Cohen-Tanugi,
  Cominsky, Coronado-Bl√°zquez, Costantin, Cuoco, Cutini, D‚ÄôAmmando,
  DeKlotz, de~la Torre~Luque, de~Palma, Desai, Digel, Lalla, Mauro, Venere,
  Dom√≠nguez, Dumora, Dirirsa, Fegan, Ferrara, Franckowiak, Fukazawa, Funk,
  Fusco, Gargano, Gasparrini, Giglietto, Giommi, Giordano, Giroletti, Glanzman,
  Green, Grenier, Griffin, Grondin, Grove, Guiriec, Harding, Hayashi, Hays,
  Hewitt, Horan, J√≥hannesson, Johnson, Kamae, Kerr, Kocevski,
  Kovac‚Äôevic‚Äô, Kuss, Landriu, Larsson, Latronico, Lemoine-Goumard,
  Li, Liodakis, Longo, Loparco, Lott, Lovellette, Lubrano, Madejski, Maldera,
  Malyshev, Manfreda, Marchesini, Marcotulli, Mart√≠-Devesa, Martin,
  Massaro, Mazziotta, McEnery, Mereu, Meyer, Michelson, Mirabal, Mizuno,
  Monzani, Morselli, Moskalenko, Negro, Nuss, Ojha, Omodei, Orienti, Orlando,
  Ormes, Palatiello, Paliya, Paneque, Pei, Pe√±a-Herazo, Perkins, Persic,
  Pesce-Rollins, Petrosian, Petrov, Piron, Poon, Porter, Principe, Rain√≤,
  Rando, Razzano, Razzaque, Reimer, Reimer, Remy, Reposeur, Romani, Parkinson,
  Schinzel, Serini, Sgr√≤, Siskind, Smith, Spandre, Spinelli, Strong,
  Suson, Tajima, Takahashi, Tak, Thayer, Thompson, Tibaldo, Torres, Torresi,
  Valverde, Klaveren, van Zyl, Wood, Yassine, \& Zaharijas}]{4FGL}
Abdollahi, S., Acero, F., Ackermann, M., {et~al.} 2020,
  \href{http://dx.doi.org/10.3847/1538-4365/ab6bcb}{\JournalTitle{\apjs}, 247,
  33}

\bibitem[{{Abe} {et~al.}(2024){Abe}, {Abe}, {Acciari}, {Agudo}, {Aniello},
  {Ansoldi}, {Antonelli}, {Arbet Engels}, {Arcaro}, {Artero}, {Asano}, {Baack},
  {Babi{\'c}}, {Baquero}, {Barres de Almeida}, {Barrio}, {Batkovi{\'c}},
  {Baxter}, {Becerra Gonz{\'a}lez}, {Bednarek}, {Bernardini}, {Bernete},
  {Berti}, {Besenrieder}, {Bigongiari}, {Biland}, {Blanch}, {Bonnoli},
  {Bo{\v{s}}njak}, {Burelli}, {Busetto}, {Campoy-Ordaz}, {Carosi}, {Carosi},
  {Carretero-Castrillo}, {Castro-Tirado}, {Ceribella}, {Chai}, {Cifuentes},
  {Cikota}, {Colombo}, {Contreras}, {Cortina}, {Covino}, {D'Amico}, {D'Elia},
  {Da Vela}, {Dazzi}, {De Angelis}, {De Lotto}, {Del Popolo}, {Delfino},
  {Delgado}, {Mendez}, {Depaoli}, {Di Pierro}, {Di Venere}, {Prester},
  {Donini}, {Dorner}, {Doro}, {Elsaesser}, {Emery}, {Escudero}, {Fari{\~n}a},
  {Fattorini}, {Foffano}, {Font}, {Fukami}, {Fukazawa}, {L{\'o}pez},
  {Garczarczyk}, {Gasparyan}, {Gaug}, {Giesbrecht Paiva}, {Giglietto},
  {Giordano}, {Gliwny}, {Godinovi{\'c}}, {Grau}, {Green}, {Green}, {Hadasch},
  {Hahn}, {Hassan}, {Heckmann}, {Herrera}, {Hrupec}, {H{\"u}tten}, {Imazawa},
  {Inada}, {Iotov}, {Ishio}, {Jim{\'e}nez Mart{\'\i}nez}, {Jormanainen},
  {Kerszberg}, {Kluge}, {Kobayashi}, {Kouch}, {Kubo}, {Kushida}, {L{\'a}inez
  Lez{\'a}un}, {Lamastra}, {Leone}, {Lindfors}, {Linhoff}, {Lombardi}, {Longo},
  {L{\'o}pez-Coto}, {L{\'o}pez-Moya}, {L{\'o}pez-Oramas}, {Loporchio},
  {Lorini}, {Lyard}, {de Oliveira Fraga}, {Majumdar}, {Makariev}, {Maneva},
  {Mang}, {Manganaro}, {Mangano}, {Mannheim}, {Mariotti}, {Mart{\'\i}nez},
  {Mas-Aguilar}, {Mazin}, {Menchiari}, {Mender}, {Mi{\'c}anovi{\'c}}, {Miceli},
  {Miener}, {Miranda}, {Mirzoyan}, {Molero Gonz{\'a}lez}, {Molina}, {Mondal},
  {Moralejo}, {Morcuende}, {Nanci}, {Nava}, {Neustroev}, {Rosillo}, {Nigro},
  {Nikoli{\'c}}, {Nilsson}, {Nishijima}, {Ekoume}, {Noda}, {Nozaki}, {Ohtani},
  {Okumura}, {Otero-Santos}, {Paiano}, {Palatiello}, {Paneque}, {Paoletti},
  {Paredes}, {Pavleti{\'c}}, {Pavlovi{\'c}}, {Persic}, {Pihet}, {Pirola},
  {Podobnik}, {Moroni}, {Prandini}, {Principe}, {Priyadarshi}, {Rhode},
  {Rib{\'o}}, {Rico}, {Righi}, {Sahakyan}, {Saito}, {Satalecka}, {Saturni},
  {Schleicher}, {Schmidt}, {Schmuckermaier}, {Schubert}, {Schweizer},
  {Sciaccaluga}, {Sitarek}, {Sliusar}, {Sobczynska}, {Spolon}, {Stamerra},
  {Stri{\v{s}}kovi{\'c}}, {Strom}, {Strzys}, {Suda}, {Suutarinen}, {Tajima},
  {Takahashi}, {Takeishi}, {Tavecchio}, {Temnikov}, {Terauchi}, {Terzi{\'c}},
  {Teshima}, {Tosti}, {Truzzi}, {Tutone}, {Ubach}, {van Scherpenberg}, {Vazquez
  Acosta}, {Ventura}, {Verguilov}, {Viale}, {Vigorito}, {Vitale}, {Vovk},
  {Walter}, {Will}, {Yamamoto}, {Gomboc}, {Jordana-Mitjans}, {Melandri},
  {Mundell}, {Shrestha}, \& {Steele}}]{2024MNRAS.527.5856A}
{Abe}, H., {Abe}, S., {Acciari}, V.~A., {et~al.} 2024,
  \href{http://dx.doi.org/10.1093/mnras/stad2958}{\JournalTitle{MNRAS}, 527,
  5856}

\bibitem[{{Ackermann} {et~al.}(2010){Ackermann}, {Asano}, {Atwood}, {Axelsson},
  {Baldini}, {Ballet}, {Barbiellini}, {Baring}, {Bastieri}, {Bechtol},
  {Bellazzini}, {Berenji}, {Bhat}, {Bissaldi}, {Blandford}, {Bloom},
  {Bonamente}, {Borgland}, {Bouvier}, {Bregeon}, {Brez}, {Briggs}, {Brigida},
  {Bruel}, {Buson}, {Caliandro}, {Cameron}, {Caraveo}, {Carrigan},
  {Casandjian}, {Cecchi}, {{\c C}elik}, {Charles}, {Chiang}, {Ciprini},
  {Claus}, {Cohen-Tanugi}, {Connaughton}, {Conrad}, {Dermer}, {de Palma},
  {Dingus}, {Silva}, {Drell}, {Dubois}, {Dumora}, {Farnier}, {Favuzzi},
  {Fegan}, {Finke}, {Focke}, {Frailis}, {Fukazawa}, {Fusco}, {Gargano},
  {Gasparrini}, {Gehrels}, {Germani}, {Giglietto}, {Giordano}, {Glanzman},
  {Godfrey}, {Granot}, {Grenier}, {Grondin}, {Grove}, {Guiriec}, {Hadasch},
  {Harding}, {Hays}, {Horan}, {Hughes}, {J{\'o}hannesson}, {Johnson}, {Kamae},
  {Katagiri}, {Kataoka}, {Kawai}, {Kippen}, {Kn{\"o}dlseder}, {Kocevski},
  {Kouveliotou}, {Kuss}, {Lande}, {Latronico}, {Lemoine-Goumard}, {Llena
  Garde}, {Longo}, {Loparco}, {Lott}, {Lovellette}, {Lubrano}, {Makeev},
  {Mazziotta}, {McEnery}, {McGlynn}, {Meegan}, {M{\'e}sz{\'a}ros}, {Michelson},
  {Mitthumsiri}, {Mizuno}, {Moiseev}, {Monte}, {Monzani}, {Moretti},
  {Morselli}, {Moskalenko}, {Murgia}, {Nakajima}, {Nakamori}, {Nolan},
  {Norris}, {Nuss}, {Ohno}, {Ohsugi}, {Omodei}, {Orlando}, {Ormes}, {Ozaki},
  {Paciesas}, {Paneque}, {Panetta}, {Parent}, {Pelassa}, {Pepe},
  {Pesce-Rollins}, {Piron}, {Preece}, {Rain{\`o}}, {Rando}, {Razzano},
  {Razzaque}, {Reimer}, {Ritz}, {Rodriguez}, {Roth}, {Ryde}, {Sadrozinski},
  {Sander}, {Scargle}, {Schalk}, {Sgr{\`o}}, {Siskind}, {Smith}, {Spandre},
  {Spinelli}, {Stamatikos}, {Stecker}, {Strickman}, {Suson}, {Tajima},
  {Takahashi}, {Takahashi}, {Tanaka}, {Thayer}, {Thayer}, {Thompson},
  {Tibaldo}, {Toma}, {Torres}, {Tosti}, {Tramacere}, {Uchiyama}, {Uehara},
  {Usher}, {van der Horst}, {Vasileiou}, {Vilchez}, {Vitale}, {von Kienlin},
  {Waite}, {Wang}, {Wilson-Hodge}, {Winer}, {Wu}, {Yamazaki}, {Yang}, {Ylinen},
  \& {Ziegler}}]{2010ApJ...716.1178A}
{Ackermann}, M., {Asano}, K., {Atwood}, W.~B., {et~al.} 2010,
  \href{http://dx.doi.org/10.1088/0004-637X/716/2/1178}{\JournalTitle{\apj},
  716, 1178}

\bibitem[{{Ackermann} {et~al.}(2012){Ackermann}, {Ajello}, {Albert},
  {Allafort}, {Atwood}, {Axelsson}, {Baldini}, {Ballet}, {Barbiellini},
  {Bastieri}, {Bechtol}, {Bellazzini}, {Bissaldi}, {Blandford}, {Bloom},
  {Bogart}, {Bonamente}, {Borgland}, {Bottacini}, {Bouvier}, {Brandt},
  {Bregeon}, {Brigida}, {Bruel}, {Buehler}, {Burnett}, {Buson}, {Caliandro},
  {Cameron}, {Caraveo}, {Casandjian}, {Cavazzuti}, {Cecchi}, {{\c C}elik},
  {Charles}, {Chaves}, {Chekhtman}, {Cheung}, {Chiang}, {Ciprini}, {Claus},
  {Cohen-Tanugi}, {Conrad}, {Corbet}, {Cutini}, {D'Ammando}, {Davis}, {de
  Angelis}, {DeKlotz}, {de Palma}, {Dermer}, {Digel}, {Silva}, {Drell},
  {Drlica-Wagner}, {Dubois}, {Favuzzi}, {Fegan}, {Ferrara}, {Focke}, {Fortin},
  {Fukazawa}, {Funk}, {Fusco}, {Gargano}, {Gasparrini}, {Gehrels}, {Giebels},
  {Giglietto}, {Giordano}, {Giroletti}, {Glanzman}, {Godfrey}, {Grenier},
  {Grove}, {Guiriec}, {Hadasch}, {Hayashida}, {Hays}, {Horan}, {Hou}, {Hughes},
  {Jackson}, {Jogler}, {J{\'o}hannesson}, {Johnson}, {Johnson}, {Johnson},
  {Kamae}, {Katagiri}, {Kataoka}, {Kerr}, {Kn{\"o}dlseder}, {Kuss}, {Lande},
  {Larsson}, {Latronico}, {Lavalley}, {Lemoine-Goumard}, {Longo}, {Loparco},
  {Lott}, {Lovellette}, {Lubrano}, {Mazziotta}, {McConville}, {McEnery},
  {Mehault}, {Michelson}, {Mitthumsiri}, {Mizuno}, {Moiseev}, {Monte},
  {Monzani}, {Morselli}, {Moskalenko}, {Murgia}, {Naumann-Godo}, {Nemmen},
  {Nishino}, {Norris}, {Nuss}, {Ohno}, {Ohsugi}, {Okumura}, {Omodei},
  {Orienti}, {Orlando}, {Ormes}, {Paneque}, {Panetta}, {Perkins},
  {Pesce-Rollins}, {Pierbattista}, {Piron}, {Pivato}, {Porter}, {Racusin},
  {Rain{\`o}}, {Rando}, {Razzano}, {Razzaque}, {Reimer}, {Reimer}, {Reposeur},
  {Reyes}, {Ritz}, {Rochester}, {Romoli}, {Roth}, {Sadrozinski}, {Sanchez},
  {Saz Parkinson}, {Sbarra}, {Scargle}, {Sgr{\`o}}, {Siegal-Gaskins},
  {Siskind}, {Spandre}, {Spinelli}, {Stephens}, {Suson}, {Tajima}, {Takahashi},
  {Tanaka}, {Thayer}, {Thayer}, {Thompson}, {Tibaldo}, {Tinivella}, {Tosti},
  {Troja}, {Usher}, {Vandenbroucke}, {Van Klaveren}, {Vasileiou}, {Vianello},
  {Vitale}, {Waite}, {Wallace}, {Winer}, {Wood}, {Wood}, {Wood}, {Yang}, \&
  {Zimmer}}]{2012ApJS..203....4A}
{Ackermann}, M., {Ajello}, M., {Albert}, A., {et~al.} 2012,
  \JournalTitle{\apjs}, 203, 4

\bibitem[{{Ackermann} {et~al.}(2014){Ackermann}, {Ajello}, {Asano}, {Atwood},
  {Axelsson}, {Baldini}, {Ballet}, {Barbiellini}, {Baring}, {Bastieri},
  {Bechtol}, {Bellazzini}, {Bissaldi}, {Bonamente}, {Bregeon}, {Brigida},
  {Bruel}, {Buehler}, {Burgess}, {Buson}, {Caliandro}, {Cameron}, {Caraveo},
  {Cecchi}, {Chaplin}, {Charles}, {Chekhtman}, {Cheung}, {Chiang}, {Chiaro},
  {Ciprini}, {Claus}, {Cleveland}, {Cohen-Tanugi}, {Collazzi}, {Cominsky},
  {Connaughton}, {Conrad}, {Cutini}, {D'Ammando}, {de Angelis}, {DeKlotz}, {de
  Palma}, {Dermer}, {Desiante}, {Diekmann}, {Di Venere}, {Drell},
  {Drlica-Wagner}, {Favuzzi}, {Fegan}, {Ferrara}, {Finke}, {Fitzpatrick},
  {Focke}, {Franckowiak}, {Fukazawa}, {Funk}, {Fusco}, {Gargano}, {Gehrels},
  {Germani}, {Gibby}, {Giglietto}, {Giles}, {Giordano}, {Giroletti}, {Godfrey},
  {Granot}, {Grenier}, {Grove}, {Gruber}, {Guiriec}, {Hadasch}, {Hanabata},
  {Harding}, {Hayashida}, {Hays}, {Horan}, {Hughes}, {Inoue}, {Jogler},
  {J{\'o}hannesson}, {Johnson}, {Kawano}, {Kn{\"o}dlseder}, {Kocevski}, {Kuss},
  {Lande}, {Larsson}, {Latronico}, {Longo}, {Loparco}, {Lovellette}, {Lubrano},
  {Mayer}, {Mazziotta}, {McEnery}, {Michelson}, {Mizuno}, {Moiseev}, {Monzani},
  {Moretti}, {Morselli}, {Moskalenko}, {Murgia}, {Nemmen}, {Nuss}, {Ohno},
  {Ohsugi}, {Okumura}, {Omodei}, {Orienti}, {Paneque}, {Pelassa}, {Perkins},
  {Pesce-Rollins}, {Petrosian}, {Piron}, {Pivato}, {Porter}, {Racusin},
  {Rain{\`o}}, {Rando}, {Razzano}, {Razzaque}, {Reimer}, {Reimer}, {Ritz},
  {Roth}, {Ryde}, {Sartori}, {Parkinson}, {Scargle}, {Schulz}, {Sgr{\`o}},
  {Siskind}, {Sonbas}, {Spandre}, {Spinelli}, {Tajima}, {Takahashi}, {Thayer},
  {Thayer}, {Thompson}, {Tibaldo}, {Tinivella}, {Torres}, {Tosti}, {Troja},
  {Usher}, {Vandenbroucke}, {Vasileiou}, {Vianello}, {Vitale}, {Winer}, {Wood},
  {Yamazaki}, {Younes}, {Yu}, {Zhu}, {Bhat}, {Briggs}, {Byrne}, {Foley},
  {Goldstein}, {Jenke}, {Kippen}, {Kouveliotou}, {McBreen}, {Meegan},
  {Paciesas}, {Preece}, {Rau}, {Tierney}, {van der Horst}, {von Kienlin},
  {Wilson-Hodge}, {Xiong}, {Cusumano}, {La Parola}, \&
  {Cummings}}]{2014Sci...343...42A}
{Ackermann}, M., {Ajello}, M., {Asano}, K., {et~al.} 2014,
  \href{http://dx.doi.org/10.1126/science.1242353}{\JournalTitle{Science}, 343,
  42}

\bibitem[{{Ackermann} {et~al.}(2016){Ackermann}, {Ajello}, {Albert}, {Atwood},
  {Baldini}, {Ballet}, {Barbiellini}, {Bastieri}, {Bechtol}, {Bellazzini},
  {Bissaldi}, {Blandford}, {Bloom}, {Bonino}, {Bregeon}, {Britto}, {Bruel},
  {Buehler}, {Caliandro}, {Cameron}, {Caragiulo}, {Caraveo}, {Cavazzuti},
  {Cecchi}, {Charles}, {Chekhtman}, {Chiang}, {Chiaro}, {Ciprini},
  {Cohen-Tanugi}, {Cominsky}, {Costanza}, {Cutini}, {D'Ammando}, {de Angelis},
  {de Palma}, {Desiante}, {Digel}, {Di Mauro}, {Di Venere}, {Dom{\'\i}nguez},
  {Drell}, {Favuzzi}, {Fegan}, {Ferrara}, {Franckowiak}, {Fukazawa}, {Funk},
  {Fusco}, {Gargano}, {Gasparrini}, {Giglietto}, {Giommi}, {Giordano},
  {Giroletti}, {Godfrey}, {Green}, {Grenier}, {Guiriec}, {Hays}, {Horan},
  {Iafrate}, {Jogler}, {J{\'o}hannesson}, {Kuss}, {La Mura}, {Larsson},
  {Latronico}, {Li}, {Li}, {Longo}, {Loparco}, {Lott}, {Lovellette}, {Lubrano},
  {Madejski}, {Magill}, {Maldera}, {Manfreda}, {Mayer}, {Mazziotta},
  {Michelson}, {Mitthumsiri}, {Mizuno}, {Moiseev}, {Monzani}, {Morselli},
  {Moskalenko}, {Murgia}, {Negro}, {Nuss}, {Ohsugi}, {Okada}, {Omodei},
  {Orlando}, {Ormes}, {Paneque}, {Perkins}, {Pesce-Rollins}, {Petrosian},
  {Piron}, {Pivato}, {Porter}, {Rain{\`o}}, {Rando}, {Razzano}, {Razzaque},
  {Reimer}, {Reimer}, {Reposeur}, {Romani}, {S{\'a}nchez-Conde}, {Schmid},
  {Schulz}, {Sgr{\`o}}, {Simone}, {Siskind}, {Spada}, {Spandre}, {Spinelli},
  {Suson}, {Takahashi}, {Thayer}, {Tibaldo}, {Torres}, {Troja}, {Vianello},
  {Yassine}, \& {Zimmer}}]{FermiEBL}
{Ackermann}, M., {Ajello}, M., {Albert}, A., {et~al.} 2016,
  \href{http://dx.doi.org/10.1103/PhysRevLett.116.151105}{\JournalTitle{\prl},
  116, 151105}

\bibitem[{{Ackermann} {et~al.}(2018){Ackermann}, {Ajello}, {Baldini}, {Ballet},
  {Barbiellini}, {Bastieri}, {Bellazzini}, {Bissaldi}, {Blandford}, {Bloom},
  {Bonino}, {Bottacini}, {Brandt}, {Bregeon}, {Bruel}, {Buehler}, {Cameron},
  {Caputo}, {Caraveo}, {Castro}, {Cavazzuti}, {Charles}, {Cheung}, {Chiaro},
  {Ciprini}, {Cohen-Tanugi}, {Costantin}, {Cutini}, {D'Ammando}, {de Palma},
  {Desai}, {Di Lalla}, {Di Mauro}, {Di Venere}, {Favuzzi}, {Finke},
  {Franckowiak}, {Fukazawa}, {Funk}, {Fusco}, {Gargano}, {Gasparrini},
  {Giglietto}, {Giordano}, {Giroletti}, {Green}, {Grenier}, {Guillemot},
  {Guiriec}, {Hays}, {Hewitt}, {Horan}, {J{\'o}hannesson}, {Kensei}, {Kuss},
  {Larsson}, {Latronico}, {Lemoine-Goumard}, {Li}, {Longo}, {Loparco},
  {Lovellette}, {Lubrano}, {Magill}, {Maldera}, {Manfreda}, {Mazziotta},
  {McEnery}, {Meyer}, {Mizuno}, {Monzani}, {Morselli}, {Moskalenko}, {Negro},
  {Nuss}, {Omodei}, {Orienti}, {Orlando}, {Ormes}, {Palatiello}, {Paliya},
  {Paneque}, {Perkins}, {Persic}, {Pesce-Rollins}, {Piron}, {Porter},
  {Principe}, {Rain{\`o}}, {Rando}, {Rani}, {Razzaque}, {Reimer}, {Reimer},
  {Reposeur}, {Sgr{\`o}}, {Siskind}, {Spandre}, {Spinelli}, {Suson}, {Tajima},
  {Thayer}, {Tibaldo}, {Torres}, {Tosti}, {Valverde}, {Venters}, {Vogel},
  {Wood}, {Wood}, {Zaharijas}, {Fermi-LAT Collaboration}, \&
  {Biteau}}]{Fermi-LATextended18}
{Ackermann}, M., {Ajello}, M., {Baldini}, L., {et~al.} 2018,
  \href{http://dx.doi.org/10.3847/1538-4365/aacdf7}{\JournalTitle{\apjs}, 237,
  32}

\bibitem[{Agostinelli {et~al.}(2003)Agostinelli, Allison, Amako, Apostolakis,
  Araujo, Arce, Asai, Axen, Banerjee, Barrand, Behner, Bellagamba, Boudreau,
  Broglia, Brunengo, Burkhardt, Chauvie, Chuma, Chytracek, Cooperman, Cosmo,
  Degtyarenko, Dell'Acqua, Depaola, Dietrich, Enami, Feliciello, Ferguson,
  Fesefeldt, Folger, Foppiano, Forti, Garelli, Giani, Giannitrapani, Gibin,
  {Gómez Cadenas}, González, {Gracia Abril}, Greeniaus, Greiner, Grichine,
  Grossheim, Guatelli, Gumplinger, Hamatsu, Hashimoto, Hasui, Heikkinen,
  Howard, Ivanchenko, Johnson, Jones, Kallenbach, Kanaya, Kawabata, Kawabata,
  Kawaguti, Kelner, Kent, Kimura, Kodama, Kokoulin, Kossov, Kurashige, Lamanna,
  Lampén, Lara, Lefebure, Lei, Liendl, Lockman, Longo, Magni, Maire,
  Medernach, Minamimoto, {Mora de Freitas}, Morita, Murakami, Nagamatu,
  Nartallo, Nieminen, Nishimura, Ohtsubo, Okamura, O'Neale, Oohata, Paech,
  Perl, Pfeiffer, Pia, Ranjard, Rybin, Sadilov, {Di Salvo}, Santin, Sasaki,
  Savvas, Sawada, Scherer, Sei, Sirotenko, Smith, Starkov, Stoecker, Sulkimo,
  Takahata, Tanaka, Tcherniaev, {Safai Tehrani}, Tropeano, Truscott, Uno,
  Urban, Urban, Verderi, Walkden, Wander, Weber, Wellisch, Wenaus, Williams,
  Wright, Yamada, Yoshida, \& Zschiesche}]{AGOSTINELLI2003250}
Agostinelli, S., Allison, J., Amako, K., {et~al.} 2003,
  \href{http://dx.doi.org/https://doi.org/10.1016/S0168-9002(03)01368-8}{\JournalTitle{Nuclear
  Instruments and Methods in Physics Research Section A: Accelerators,
  Spectrometers, Detectors and Associated Equipment}, 506, 250}

\bibitem[{Aharonian {et~al.}(2023)Aharonian, Benkhali, Aschersleben, Ashkar,
  Backes, Baktash, Martins, Batzofin, Becherini, Berge, Bernlöhr, Bi,
  Böttcher, Boisson, Bolmont, de~Bony~de Lavergne, Borowska, Bouyahiaoui,
  Bradascio, Breuhaus, Brose, Brun, Bruno, Bulik, Burger-Scheidlin, Caroff,
  Casanova, Celic, Cerruti, Chand, Chandra, Chen, Chibueze, Chibueze, Cotter,
  Dai, Mbarubucyeye, Devin, Djannati-Ataï, Dmytriiev, Doroshenko, Egberts,
  Einecke, Ernenwein, Fegan, de~Clairfontaine, Filipovic, Fontaine, Füßling,
  Funk, Gabici, Ghafourizadeh, Giavitto, Glawion, Glicenstein, Goswami,
  Grolleron, Grondin, Hinton, Holch, Holler, Horns, Huang, Jamrozy, Jankowsky,
  Joshi, Jung-Richardt, Kasai, Katarzyński, Khatoon, Khélifi, Kluźniak,
  Komin, Konno, Kosack, Kostunin, Lang, Stum, Leitl, Lemière, Lemoine-Goumard,
  Lenain, Leuschner, Lohse, Lypova, Mackey, Malyshev, Malyshev, Marandon,
  Marchegiani, Marcowith, Martí-Devesa, Marx, Meyer, Mitchell, Mohrmann,
  Montanari, Moulin, Murach, Nakashima, de~Naurois, Niemiec, Noel, O’Brien,
  Ohm, Olivera-Nieto, de~Ona~Wilhelmi, Ostrowski, Panny, Panter, Parsons,
  Peron, Prokhorov, Prokoph, Pühlhofer, Punch, Quirrenbach, Reichherzer,
  Reimer, Reimer, Ren, Renaud, Reville, Rieger, Rowell, Rudak, Ruiz-Velasco,
  Sahakian, Salzmann, Santangelo, Sasaki, Schäfer, Schüssler, Schutte,
  Schwanke, Shapopi, Specovius, Spencer, Stawarz, Steenkamp, Steinmassl,
  Steppa, Sushch, Suzuki, Takahashi, Tanaka, Terrier, Tsuji, Uchiyama, Vecchi,
  Venter, Vink, Wagner, White, Wierzcholska, Wong, Zacharias, Zargaryan,
  Zdziarski, Zech, Zhu, Żywucka, \& Collaboration}]{Aharonian_2023}
Aharonian, F., Benkhali, F.~A., Aschersleben, J., {et~al.} 2023,
  \href{http://dx.doi.org/10.3847/2041-8213/acc405}{\JournalTitle{\apjl}, 946,
  L27}

\bibitem[{{Ai} \& {Gao}(2023)}]{2023ApJ...944..115A}
{Ai}, S., \& {Gao}, H. 2023,
  \href{http://dx.doi.org/10.3847/1538-4357/acb3bf}{\JournalTitle{\apj}, 944,
  115}

\bibitem[{{Ajello} {et~al.}(2014){Ajello}, {Albert}, {Allafort}, {Baldini},
  {Barbiellini}, {Bastieri}, {Bellazzini}, {Bissaldi}, {Bonamente}, {Brandt},
  {Bregeon}, {Brigida}, {Bruel}, {Buehler}, {Buson}, {Caliandro}, {Cameron},
  {Caraveo}, {Cecchi}, {Charles}, {Chekhtman}, {Chiang}, {Chiaro}, {Ciprini},
  {Claus}, {Cohen-Tanugi}, {Cominsky}, {Conrad}, {Cutini}, {D'Ammando}, {de
  Palma}, {Dermer}, {Desiante}, {Digel}, {Silva}, {Drell}, {Drlica-Wagner},
  {Favuzzi}, {Focke}, {Franckowiak}, {Fukazawa}, {Fusco}, {Gargano},
  {Gasparrini}, {Germani}, {Giglietto}, {Giommi}, {Giordano}, {Giroletti},
  {Glanzman}, {Godfrey}, {Grenier}, {Grove}, {Guiriec}, {Hadasch}, {Hayashida},
  {Hays}, {Horan}, {Hou}, {Hughes}, {Inoue}, {Jackson}, {Jogler},
  {J{\'o}hannesson}, {Johnson}, {Johnson}, {Kamae}, {Kn{\"o}dlseder},
  {Kocevski}, {Kuss}, {Lande}, {Larsson}, {Latronico}, {Longo}, {Loparco},
  {Lott}, {Lovellette}, {Lubrano}, {Mayer}, {Mazziotta}, {McEnery},
  {Michelson}, {Mizuno}, {Moiseev}, {Monte}, {Monzani}, {Morselli},
  {Moskalenko}, {Murgia}, {Murphy}, {Nakamori}, {Nemmen}, {Nuss}, {Ohno},
  {Ohsugi}, {Omodei}, {Orienti}, {Orlando}, {Ormes}, {Paneque}, {Panetta},
  {Perkins}, {Pesce-Rollins}, {Petrosian}, {Piron}, {Pivato}, {Porter},
  {Rain{\`o}}, {Rando}, {Razzano}, {Reimer}, {Reimer}, {Roth}, {Schulz},
  {Sgr{\`o}}, {Siskind}, {Spandre}, {Spinelli}, {Takahashi}, {Thayer},
  {Thayer}, {Thompson}, {Tibaldo}, {Tinivella}, {Tosti}, {Troja}, {Usher},
  {Vandenbroucke}, {Vasileiou}, {Vianello}, {Vitale}, {Werner}, {Winer},
  {Wood}, {Wood}, \& {Yang}}]{LLESolarFlare}
{Ajello}, M., {Albert}, A., {Allafort}, A., {et~al.} 2014,
  \href{http://dx.doi.org/10.1088/0004-637X/789/1/20}{\JournalTitle{\apj}, 789,
  20}

\bibitem[{{Ajello} {et~al.}(2019){Ajello}, {Arimoto}, {Axelsson}, {Baldini},
  {Barbiellini}, {Bastieri}, {Bellazzini}, {Bhat}, {Bissaldi}, {Blandford},
  {Bonino}, {Bonnell}, {Bottacini}, {Bregeon}, {Bruel}, {Buehler}, {Cameron},
  {Caputo}, {Caraveo}, {Cavazzuti}, {Chen}, {Cheung}, {Chiaro}, {Ciprini},
  {Costantin}, {Crnogorcevic}, {Cutini}, {Dainotti}, {D'Ammando}, {de la Torre
  Luque}, {de Palma}, {Desai}, {Desiante}, {Di Lalla}, {Di Venere}, {Fana
  Dirirsa}, {Fegan}, {Franckowiak}, {Fukazawa}, {Funk}, {Fusco}, {Gargano},
  {Gasparrini}, {Giglietto}, {Giordano}, {Giroletti}, {Green}, {Grenier},
  {Grove}, {Guiriec}, {Hays}, {Hewitt}, {Horan}, {J{\'o}hannesson}, {Kocevski},
  {Kuss}, {Latronico}, {Li}, {Longo}, {Loparco}, {Lovellette}, {Lubrano},
  {Maldera}, {Manfreda}, {Mart{\'\i}-Devesa}, {Mazziotta}, {Mereu}, {Meyer},
  {Michelson}, {Mirabal}, {Mitthumsiri}, {Mizuno}, {Monzani}, {Moretti},
  {Morselli}, {Moskalenko}, {Negro}, {Nuss}, {Ohno}, {Omodei}, {Orienti},
  {Orlando}, {Palatiello}, {Paliya}, {Paneque}, {Persic}, {Pesce-Rollins},
  {Petrosian}, {Piron}, {Poolakkil}, {Poon}, {Porter}, {Principe}, {Racusin},
  {Rain{\`o}}, {Rando}, {Razzano}, {Razzaque}, {Reimer}, {Reimer}, {Reposeur},
  {Ryde}, {Serini}, {Sgr{\`o}}, {Siskind}, {Sonbas}, {Spandre}, {Spinelli},
  {Suson}, {Tajima}, {Takahashi}, {Tak}, {Thayer}, {Torres}, {Troja},
  {Valverde}, {Veres}, {Vianello}, {von Kienlin}, {Wood}, {Yassine}, {Zhu}, \&
  {Zimmer}}]{2FLGC}
{Ajello}, M., {Arimoto}, M., {Axelsson}, M., {et~al.} 2019,
  \href{http://dx.doi.org/10.3847/1538-4357/ab1d4e}{\JournalTitle{\apj}, 878,
  52}

\bibitem[{{Ajello} {et~al.}(2020){Ajello}, {Arimoto}, {Axelsson}, {Baldini},
  {Barbiellini}, {Bastieri}, {Bellazzini}, {Berretta}, {Bissaldi}, {Blandford},
  {Bonino}, {Bottacini}, {Bregeon}, {Bruel}, {Buehler}, {Burns}, {Buson},
  {Cameron}, {Caputo}, {Caraveo}, {Cavazzuti}, {Chen}, {Chiaro}, {Ciprini},
  {Cohen-Tanugi}, {Costantin}, {Cutini}, {D'Ammando}, {DeKlotz}, {de la Torre
  Luque}, {de Palma}, {Desai}, {Di Lalla}, {Di Venere}, {Fana Dirirsa},
  {Fegan}, {Franckowiak}, {Fukazawa}, {Funk}, {Fusco}, {Gargano}, {Gasparrini},
  {Giglietto}, {Gill}, {Giordano}, {Giroletti}, {Granot}, {Green}, {Grenier},
  {Grondin}, {Guiriec}, {Hays}, {Horan}, {J{\'o}hannesson}, {Kocevski},
  {Kovac'evic'}, {Kuss}, {Larsson}, {Latronico}, {Lemoine-Goumard}, {Li},
  {Liodakis}, {Longo}, {Loparco}, {Lovellette}, {Lubrano}, {Maldera},
  {Malyshev}, {Manfreda}, {Mart{\'\i}-Devesa}, {Mazziotta}, {McEnery}, {Mereu},
  {Meyer}, {Michelson}, {Mitthumsiri}, {Mizuno}, {Monzani}, {Moretti},
  {Morselli}, {Moskalenko}, {Negro}, {Nuss}, {Omodei}, {Orienti}, {Orlando},
  {Palatiello}, {Paliya}, {Paneque}, {Pei}, {Persic}, {Pesce-Rollins},
  {Petrosian}, {Piron}, {Poon}, {Porter}, {Principe}, {Racusin}, {Rain{\`o}},
  {Rando}, {Rani}, {Razzano}, {Razzaque}, {Reimer}, {Reimer}, {Ryde}, {Saz
  Parkinson}, {Serini}, {Sgr{\`o}}, {Siskind}, {Spandre}, {Spinelli}, {Tajima},
  {Takagi}, {Takahashi}, {Tak}, {Thayer}, {Thompson}, {Torres}, {Troja},
  {Valverde}, {Van Klaveren}, {Wood}, {Yassine}, {Zaharijas}, {Mailyan},
  {Bhat}, {Briggs}, {Cleveland}, {Giles}, {Goldstein}, {Hui}, {Malacaria},
  {Preece}, {Roberts}, {Veres}, {Wilson-Hodge}, {Kienlin}, {Cenko}, {O'Brien},
  {Beardmore}, {Lien}, {Osborne}, {Tohuvavohu}, {D'Elia}, {D'A{\`\i}}, {Perri},
  {Gropp}, {Klingler}, {Capalbi}, {Tagliaferri}, {Stamatikos}, \& {De
  Pasquale}}]{2020ApJ...890....9A}
---. 2020,
  \href{http://dx.doi.org/10.3847/1538-4357/ab5b05}{\JournalTitle{\apj}, 890,
  9}

\bibitem[{{Aliu} {et~al.}(2014){Aliu}, {Aune}, {Barnacka}, {Beilicke},
  {Benbow}, {Berger}, {Biteau}, {Buckley}, {Bugaev}, {Byrum}, {Cardenzana},
  {Cerruti}, {Chen}, {Ciupik}, {Connaughton}, {Cui}, {Dickinson}, {Eisch},
  {Errando}, {Falcone}, {Federici}, {Feng}, {Finley}, {Fleischhack}, {Fortin},
  {Fortson}, {Furniss}, {Galante}, {Gillanders}, {Griffin}, {Griffiths},
  {Grube}, {Gyuk}, {H{\r{a}}kansson}, {Hanna}, {Holder}, {Hughes}, {Humensky},
  {Johnson}, {Kaaret}, {Kar}, {Kertzman}, {Khassen}, {Kieda}, {Krawczynski},
  {Krennrich}, {Lang}, {Madhavan}, {Maier}, {McArthur}, {McCann}, {Meagher},
  {Millis}, {Moriarty}, {Mukherjee}, {Nieto}, {O'Faol{\'a}in de Bhr{\'o}ithe},
  {Ong}, {Otte}, {Park}, {Pohl}, {Popkow}, {Prokoph}, {Pueschel}, {Quinn},
  {Ragan}, {Rajotte}, {Reyes}, {Reynolds}, {Richards}, {Roache}, {Sembroski},
  {Shahinyan}, {Smith}, {Staszak}, {Telezhinsky}, {Tucci}, {Tyler}, {Varlotta},
  {Vassiliev}, {Vincent}, {Wakely}, {Weiner}, {Weinstein}, {Welsing},
  {Wilhelm}, {Williams}, {Zitzer}, {McEnery}, {Perkins}, {Veres}, \&
  {Zhu}}]{2014ApJ...795L...3A}
{Aliu}, E., {Aune}, T., {Barnacka}, A., {et~al.} 2014,
  \href{http://dx.doi.org/10.1088/2041-8205/795/1/L3}{\JournalTitle{\apjl},
  795, L3}

\bibitem[{Allison {et~al.}(2006)Allison, Amako, Apostolakis, Araujo,
  Arce~Dubois, Asai, Barrand, Capra, Chauvie, Chytracek, Cirrone, Cooperman,
  Cosmo, Cuttone, Daquino, Donszelmann, Dressel, Folger, Foppiano, Generowicz,
  Grichine, Guatelli, Gumplinger, Heikkinen, Hrivnacova, Howard, Incerti,
  Ivanchenko, Johnson, Jones, Koi, Kokoulin, Kossov, Kurashige, Lara, Larsson,
  Lei, Link, Longo, Maire, Mantero, Mascialino, McLaren, Mendez~Lorenzo,
  Minamimoto, Murakami, Nieminen, Pandola, Parlati, Peralta, Perl, Pfeiffer,
  Pia, Ribon, Rodrigues, Russo, Sadilov, Santin, Sasaki, Smith, Starkov,
  Tanaka, Tcherniaev, Tome, Trindade, Truscott, Urban, Verderi, Walkden,
  Wellisch, Williams, Wright, \& Yoshida}]{1610988}
Allison, J., Amako, K., Apostolakis, J., {et~al.} 2006,
  \href{http://dx.doi.org/10.1109/TNS.2006.869826}{\JournalTitle{IEEE
  Transactions on Nuclear Science}, 53, 270}

\bibitem[{{Arimoto} {et~al.}(2020){Arimoto}, {Asano}, {Tachibana}, \&
  {Axelsson}}]{2020ApJ...891..106A}
{Arimoto}, M., {Asano}, K., {Tachibana}, Y., \& {Axelsson}, M. 2020,
  \href{http://dx.doi.org/10.3847/1538-4357/ab72f7}{\JournalTitle{\apj}, 891,
  106}

\bibitem[{{Atri} {et~al.}(2022){Atri}, {An}, {Giroletti}, {Zhang}, {Bright},
  {Farah}, {Fender}, {Geng}, {Ghirlanda}, {Giarratana}, {Li}, {Liu}, {Marcote},
  {Miller-Jones}, {Motta}, {Perez-Torres}, {Rhodes}, {Salafia}, {Wang}, {Wu},
  {Xu}, \& {Yang}}]{BOAT_RADIOPOSITION}
{Atri}, P., {An}, T., {Giroletti}, M., {et~al.} 2022, \JournalTitle{GRB
  Coordinates Network}, 32907, 1

\bibitem[{{Atwood} {et~al.}(2013){Atwood}, {Albert}, {Baldini}, {Tinivella},
  {Bregeon}, {Pesce-Rollins}, {Sgr{\`o}}, {Bruel}, {Charles}, {Drlica-Wagner},
  {Franckowiak}, {Jogler}, {Rochester}, {Usher}, {Wood}, {Cohen-Tanugi}, \& {S.
  Zimmer for the Fermi-LAT Collaboration}}]{2013arXiv1303.3514A}
{Atwood}, W., {Albert}, A., {Baldini}, L., {et~al.} 2013, \JournalTitle{ArXiv
  e-prints}, arXiv:1303.3514

\bibitem[{{Atwood} {et~al.}(2009){Atwood}, {Abdo}, {Ackermann}, {Althouse},
  {Anderson}, {Axelsson}, {Baldini}, {Ballet}, {Band}, {Barbiellini},
  {Bartelt}, {Bastieri}, {Baughman}, {Bechtol}, {B{\'e}d{\'e}r{\`e}de},
  {Bellardi}, {Bellazzini}, {Berenji}, {Bignami}, {Bisello}, {Bissaldi},
  {Blandford}, {Bloom}, {Bogart}, {Bonamente}, {Bonnell}, {Borgland},
  {Bouvier}, {Bregeon}, {Brez}, {Brigida}, {Bruel}, {Burnett}, {Busetto},
  {Caliandro}, {Cameron}, {Caraveo}, {Carius}, {Carlson}, {Casandjian},
  {Cavazzuti}, {Ceccanti}, {Cecchi}, {Charles}, {Chekhtman}, {Cheung},
  {Chiang}, {Chipaux}, {Cillis}, {Ciprini}, {Claus}, {Cohen-Tanugi},
  {Condamoor}, {Conrad}, {Corbet}, {Corucci}, {Costamante}, {Cutini}, {Davis},
  {Decotigny}, {DeKlotz}, {Dermer}, {de Angelis}, {Digel}, {do Couto e Silva},
  {Drell}, {Dubois}, {Dumora}, {Edmonds}, {Fabiani}, {Farnier}, {Favuzzi},
  {Flath}, {Fleury}, {Focke}, {Funk}, {Fusco}, {Gargano}, {Gasparrini},
  {Gehrels}, {Gentit}, {Germani}, {Giebels}, {Giglietto}, {Giommi}, {Giordano},
  {Glanzman}, {Godfrey}, {Grenier}, {Grondin}, {Grove}, {Guillemot}, {Guiriec},
  {Haller}, {Harding}, {Hart}, {Hays}, {Healey}, {Hirayama}, {Hjalmarsdotter},
  {Horn}, {Hughes}, {J{\'o}hannesson}, {Johansson}, {Johnson}, {Johnson},
  {Johnson}, {Johnson}, {Kamae}, {Katagiri}, {Kataoka}, {Kavelaars}, {Kawai},
  {Kelly}, {Kerr}, {Klamra}, {Kn{\"o}dlseder}, {Kocian}, {Komin}, {Kuehn},
  {Kuss}, {Landriu}, {Latronico}, {Lee}, {Lee}, {Lemoine-Goumard}, {Lionetto},
  {Longo}, {Loparco}, {Lott}, {Lovellette}, {Lubrano}, {Madejski}, {Makeev},
  {Marangelli}, {Massai}, {Mazziotta}, {McEnery}, {Menon}, {Meurer},
  {Michelson}, {Minuti}, {Mirizzi}, {Mitthumsiri}, {Mizuno}, {Moiseev},
  {Monte}, {Monzani}, {Moretti}, {Morselli}, {Moskalenko}, {Murgia},
  {Nakamori}, {Nishino}, {Nolan}, {Norris}, {Nuss}, {Ohno}, {Ohsugi}, {Omodei},
  {Orlando}, {Ormes}, {Paccagnella}, {Paneque}, {Panetta}, {Parent}, {Pearce},
  {Pepe}, {Perazzo}, {Pesce-Rollins}, {Picozza}, {Pieri}, {Pinchera}, {Piron},
  {Porter}, {Poupard}, {Rain{\`o}}, {Rando}, {Rapposelli}, {Razzano}, {Reimer},
  {Reimer}, {Reposeur}, {Reyes}, {Ritz}, {Rochester}, {Rodriguez}, {Romani},
  {Roth}, {Russell}, {Ryde}, {Sabatini}, {Sadrozinski}, {Sanchez}, {Sander},
  {Sapozhnikov}, {Parkinson}, {Scargle}, {Schalk}, {Scolieri}, {Sgr{\`o}},
  {Share}, {Shaw}, {Shimokawabe}, {Shrader}, {Sierpowska-Bartosik}, {Siskind},
  {Smith}, {Smith}, {Spandre}, {Spinelli}, {Starck}, {Stephens}, {Strickman},
  {Strong}, {Suson}, {Tajima}, {Takahashi}, {Takahashi}, {Tanaka}, {Tenze},
  {Tether}, {Thayer}, {Thayer}, {Thompson}, {Tibaldo}, {Tibolla}, {Torres},
  {Tosti}, {Tramacere}, {Turri}, {Usher}, {Vilchez}, {Vitale}, {Wang},
  {Watters}, {Winer}, {Wood}, {Ylinen}, \& {Ziegler}}]{2009ApJ...697.1071A}
{Atwood}, W.~B., {Abdo}, A.~A., {Ackermann}, M., {et~al.} 2009,
  \href{http://dx.doi.org/10.1088/0004-637X/697/2/1071}{\JournalTitle{\apj},
  697, 1071}

\bibitem[{{Ayala} \& {HAWC Collaboration}(2022)}]{2022GCN.32683....1A}
{Ayala}, H., \& {HAWC Collaboration}. 2022, \JournalTitle{GRB Coordinates
  Network}, 32683, 1

\bibitem[{{Band} {et~al.}(1993){Band}, {Matteson}, {Ford}, {Schaefer},
  {Palmer}, {Teegarden}, {Cline}, {Briggs}, {Paciesas}, {Pendleton}, {Fishman},
  {Kouveliotou}, {Meegan}, {Wilson}, \& {Lestrade}}]{1993ApJ...413..281B}
{Band}, D., {Matteson}, J., {Ford}, L., {et~al.} 1993,
  \href{http://dx.doi.org/10.1086/172995}{\JournalTitle{\apj}, 413, 281}

\bibitem[{{Bissaldi} {et~al.}(2022){Bissaldi}, {Omodei}, {Kerr}, \& {Fermi-LAT
  Team}}]{2022GCN.32637....1B}
{Bissaldi}, E., {Omodei}, N., {Kerr}, M., \& {Fermi-LAT Team}. 2022,
  \JournalTitle{GRB Coordinates Network}, 32637, 1

\bibitem[{{Broderick} {et~al.}(2012){Broderick}, {Chang}, \&
  {Pfrommer}}]{Broderick12}
{Broderick}, A.~E., {Chang}, P., \& {Pfrommer}, C. 2012,
  \href{http://dx.doi.org/10.1088/0004-637X/752/1/22}{\JournalTitle{\apj}, 752,
  22}

\bibitem[{{Burns} {et~al.}(2023{\natexlab{a}}){Burns}, {Svinkin}, {Fenimore},
  {Kann}, {Ag{\"u}{\'\i} Fern{\'a}ndez}, {Frederiks}, {Hamburg}, {Lesage},
  {Temiraev}, {Tsvetkova}, {Bissaldi}, {Briggs}, {Dalessi}, {Dunwoody},
  {Fletcher}, {Goldstein}, {Hui}, {Hristov}, {Kocevski}, {Lysenko}, {Mailyan},
  {Mangan}, {McBreen}, {Racusin}, {Ridnaia}, {Roberts}, {Ulanov}, {Veres},
  {Wilson-Hodge}, \& {Wood}}]{2023ApJ...946L..31B}
{Burns}, E., {Svinkin}, D., {Fenimore}, E., {et~al.} 2023{\natexlab{a}},
  \href{http://dx.doi.org/10.3847/2041-8213/acc39c}{\JournalTitle{\apjl}, 946,
  L31}

\bibitem[{{Burns} {et~al.}(2023{\natexlab{b}}){Burns}, {Svinkin}, {Fenimore},
  {Kann}, {Ag{\"u}{\'\i} Fern{\'a}ndez}, {Frederiks}, {Hamburg}, {Lesage},
  {Temiraev}, {Tsvetkova}, {Bissaldi}, {Briggs}, {Dalessi}, {Dunwoody},
  {Fletcher}, {Goldstein}, {Hui}, {Hristov}, {Kocevski}, {Lysenko}, {Mailyan},
  {Mangan}, {McBreen}, {Racusin}, {Ridnaia}, {Roberts}, {Ulanov}, {Veres},
  {Wilson-Hodge}, \& {Wood}}]{BOAT}
---. 2023{\natexlab{b}},
  \href{http://dx.doi.org/10.3847/2041-8213/acc39c}{\JournalTitle{\apjl}, 946,
  L31}

\bibitem[{Cao {et~al.}(2023)Cao, Aharonian, An, Axikegu, Bai, Bai, Bao,
  Bastieri, Bi, Bi, Cai, Cao, Cao, Cao, Chang, Chang, Chen, Chen, Chen, Chen,
  Chen, Chen, Chen, Chen, Chen, Chen, Chen, Cheng, Cheng, Cheng, Cui, Cui, Cui,
  Dai, Dai, Dai, Danzengluobu, {Della Volpe}, Dong, Duan, Fan, Fan, Fang, Fang,
  Feng, Feng, Feng, Feng, Feng, Gao, Gao, Gao, Gao, Gao, Gao, Ge, Geng, Gong,
  Gou, Gu, Guo, Guo, Guo, Guo, Han, He, He, He, He, He, Heller, Hor, Hou, Hou,
  Hou, Hu, Hu, Hu, Huang, Huang, Huang, Huang, Huang, Huang, Huang, Ji, Jia,
  Jia, Jiang, Jiang, Jiang, Jin, Kang, Ke, Kuleshov, Kurinov, Li, Li, Li, Li,
  Li, Li, Li, Li, Li, Li, Li, Li, Li, Li, Li, Li, Li, Li, Li, Liang, Liang,
  Lin, Liu, Liu, Liu, Liu, Liu, Liu, Liu, Liu, Liu, Liu, Liu, Liu, Liu, Liu,
  Liu, Liu, Long, Lu, Luo, Lv, Ma, Ma, Ma, Mao, Min, Mitthumsiri, Nan, Ou,
  Pang, Pattarakijwanich, Pei, Qi, Qi, Qiao, Qin, Ruffolo, S{\'{a}}iz, Shao,
  Shao, Shchegolev, Sheng, Song, Stenkin, Stepanov, Su, Sun, Sun, Sun, Tam,
  Tang, Tian, Wang, Wang, Wang, Wang, Wang, Wang, Wang, Wang, Wang, Wang, Wang,
  Wang, Wang, Wang, Wang, Wang, Wang, Wang, Wang, Wang, Wang, Wang, Wei, Wei,
  Wei, Wen, Wu, Wu, Wu, Wu, Wu, Xi, Xia, Xia, Xiang, Xiao, Xiao, Xin, Xin,
  Xing, Xiong, Xu, Xu, Xu, Xue, Yan, Yan, Yan, Yang, Yang, Yang, Yang, Yang,
  Yang, Yang, Yang, Yang, Yao, Yao, Ye, Yin, Yin, You, You, Yu, Yuan, Yue,
  Zeng, Zeng, Zeng, Zeng, Zha, Zhang, Zhang, Zhang, Zhang, Zhang, Zhang, Zhang,
  Zhang, Zhang, Zhang, Zhang, Zhang, Zhang, Zhang, Zhang, Zhang, Zhang, Zhang,
  Zhang, Zhao, Zhao, Zhao, Zhao, Zhao, Zheng, Zheng, Zhou, Zhou, Zhou, Zhou,
  Zhou, Zhou, Zhu, Zhu, Zhu, Zhu, \& Zuo}]{BOAT_LHAASO}
Cao, Z., Aharonian, F., An, Q., {et~al.} 2023,
  \href{http://dx.doi.org/10.1126/science.adg9328}{\JournalTitle{Science}, 380,
  1390}

\bibitem[{{Castro-Tirado} {et~al.}(2022){Castro-Tirado}, {Sanchez-Ramirez},
  {Hu}, {Caballero-Garcia}, {Castro Tirado}, {Fernandez-Garcia},
  {Perez-Garcia}, {Lombardi}, {Pandey}, {Yang}, \&
  {Zhang}}]{2022GCN.32686....1C}
{Castro-Tirado}, A.~J., {Sanchez-Ramirez}, R., {Hu}, Y.~D., {et~al.} 2022,
  \JournalTitle{GRB Coordinates Network}, 32686, 1

\bibitem[{{Corsi} {et~al.}(2010){Corsi}, {Guetta}, \&
  {Piro}}]{2010ApJ...720.1008C}
{Corsi}, A., {Guetta}, D., \& {Piro}, L. 2010,
  \href{http://dx.doi.org/10.1088/0004-637X/720/2/1008}{\JournalTitle{\apj},
  720, 1008}

\bibitem[{{Da Vela} {et~al.}(2023){Da Vela}, {Mart{\'\i}-Devesa}, {Saturni},
  {Veres}, {Stamerra}, \& {Longo}}]{DaVela23}
{Da Vela}, P., {Mart{\'\i}-Devesa}, G., {Saturni}, F.~G., {et~al.} 2023,
  \href{http://dx.doi.org/10.1103/PhysRevD.107.063030}{\JournalTitle{\prd},
  107, 063030}

\bibitem[{{de Ugarte Postigo} {et~al.}(2022){de Ugarte Postigo}, {Izzo},
  {Pugliese}, {Xu}, {Schneider}, {Fynbo}, {Tanvir}, {Malesani}, {Saccardi},
  {Kann}, {Wiersema}, {Gompertz}, {Thoene}, {Levan}, \& {Stargate
  Collaboration}}]{2022GCN.32648....1D}
{de Ugarte Postigo}, A., {Izzo}, L., {Pugliese}, G., {et~al.} 2022,
  \JournalTitle{GRB Coordinates Network}, 32648, 1

\bibitem[{{Dichiara} {et~al.}(2022){Dichiara}, {Gropp}, {Kennea}, {Kuin},
  {Lien}, {Marshall}, {Tohuvavohu}, {Williams}, \& {Neil Gehrels Swift
  Observatory Team}}]{2022GCN.32632....1D}
{Dichiara}, S., {Gropp}, J.~D., {Kennea}, J.~A., {et~al.} 2022,
  \JournalTitle{GRB Coordinates Network}, 32632, 1

\bibitem[{{Dzhappuev} {et~al.}(2022){Dzhappuev}, {Afashokov}, {Dzaparova},
  {Dzhatdoev}, {Gorbacheva}, {Karpikov}, {Khadzhiev}, {Klimenko}, {Kudzhaev},
  {Kurenya}, {Lidvansky}, {Mikhailova}, {Petkov}, {Podlesnyi}, {Pozdnukhov},
  {Romanenko}, {Rubtsov}, {Troitsky}, {Unatlokov}, {Vaiman}, {Yanin}, \&
  {Zhuravleva}}]{2022ATel15669....1D}
{Dzhappuev}, D.~D., {Afashokov}, Y.~Z., {Dzaparova}, I.~M., {et~al.} 2022,
  \JournalTitle{The Astronomer's Telegram}, 15669, 1

\bibitem[{{Feroz} {et~al.}(2009){Feroz}, {Hobson}, \& {Bridges}}]{Multinest}
{Feroz}, F., {Hobson}, M.~P., \& {Bridges}, M. 2009,
  \href{http://dx.doi.org/10.1111/j.1365-2966.2009.14548.x}{\JournalTitle{MNRAS},
  398, 1601}

\bibitem[{{Finke} {et~al.}(2010){Finke}, {Razzaque}, \&
  {Dermer}}]{2010ApJ...712..238F}
{Finke}, J.~D., {Razzaque}, S., \& {Dermer}, C.~D. 2010,
  \href{http://dx.doi.org/10.1088/0004-637X/712/1/238}{\JournalTitle{\apj},
  712, 238}

\bibitem[{{Frederiks} {et~al.}(2023){Frederiks}, {Svinkin}, {Lysenko},
  {Molkov}, {Tsvetkova}, {Ulanov}, {Ridnaia}, {Lutovinov}, {Lapshov},
  {Tkachenko}, \& {Levin}}]{BOAT_KW}
{Frederiks}, D., {Svinkin}, D., {Lysenko}, A.~L., {et~al.} 2023,
  \href{http://dx.doi.org/10.3847/2041-8213/acd1eb}{\JournalTitle{\apjl}, 949,
  L7}

\bibitem[{Fulton {et~al.}(2023)Fulton, Smartt, Rhodes, Huber, Villar, Moore,
  Srivastav, Schultz, Chambers, Izzo, Hjorth, Chen, Nicholl, Foley, Rest,
  Smith, Young, Sim, Bright, Zenati, de~Boer, Bulger, Fairlamb, Gao, Lin, Lowe,
  Magnier, Smith, Wainscoat, Coulter, Jones, Kilpatrick, McGill, Ramirez-Ruiz,
  Lee, Narayan, Ramakrishnan, Ridden-Harper, Singh, Wang, Kong, Ngeow, Pan,
  Yang, Davis, Piro, Rojas-Bravo, Sommer, \& Yadavalli}]{Fulton_2023}
Fulton, M.~D., Smartt, S.~J., Rhodes, L., {et~al.} 2023,
  \href{http://dx.doi.org/10.3847/2041-8213/acc101}{\JournalTitle{\apjl}, 946,
  L22}

\bibitem[{{Ghirlanda} {et~al.}(2018){Ghirlanda}, {Nappo}, {Ghisellini},
  {Melandri}, {Marcarini}, {Nava}, {Salafia}, {Campana}, \&
  {Salvaterra}}]{Ghirlanda+18Lorentz}
{Ghirlanda}, G., {Nappo}, F., {Ghisellini}, G., {et~al.} 2018,
  \href{http://dx.doi.org/10.1051/0004-6361/201731598}{\JournalTitle{\aap},
  609, A112}

\bibitem[{{Ghisellini} {et~al.}(2010){Ghisellini}, {Ghirlanda}, {Nava}, \&
  {Celotti}}]{2010MNRAS.403..926G}
{Ghisellini}, G., {Ghirlanda}, G., {Nava}, L., \& {Celotti}, A. 2010,
  \href{http://dx.doi.org/10.1111/j.1365-2966.2009.16171.x}{\JournalTitle{MNRAS},
  403, 926}

\bibitem[{{Ghisellini} {et~al.}(2005){Ghisellini}, {Tavecchio}, \&
  {Chiaberge}}]{2005A&A...432..401G}
{Ghisellini}, G., {Tavecchio}, F., \& {Chiaberge}, M. 2005,
  \href{http://dx.doi.org/10.1051/0004-6361:20041404}{\JournalTitle{\aap}, 432,
  401}

\bibitem[{{Giroletti} {et~al.}(2004){Giroletti}, {Giovannini}, {Feretti},
  {Cotton}, {Edwards}, {Lara}, {Marscher}, {Mattox}, {Piner}, \&
  {Venturi}}]{Giroletti-2004-ApJ}
{Giroletti}, M., {Giovannini}, G., {Feretti}, L., {et~al.} 2004,
  \href{http://dx.doi.org/10.1086/379663}{\JournalTitle{\apj}, 600, 127}

\bibitem[{{H.~E.~S.~S. Collaboration} {et~al.}(2021){H.~E.~S.~S.
  Collaboration}, {Abdalla}, {Aharonian}, {Ait Benkhali}, {Ang{\"u}ner},
  {Arcaro}, {Armand}, {Armstrong}, {Ashkar}, {Backes}, {Baghmanyan}, {Barbosa
  Martins}, {Barnacka}, {Barnard}, {Becherini}, {Berge}, {Bernl{\"o}hr}, {Bi},
  {Bissaldi}, {B{\"o}ttcher}, {Boisson}, {Bolmont}, {de Bony de Lavergne},
  {Breuhaus}, {Brun}, {Brun}, {Bryan}, {B{\"u}chele}, {Bulik}, {Bylund},
  {Caroff}, {Carosi}, {Casanova}, {Chand}, {Chandra}, {Chen}, {Cotter},
  {Cury{\l}o}, {Damascene Mbarubucyeye}, {Davids}, {Davies}, {Deil}, {Devin},
  {Dirson}, {Djannati-Ata{\"\i}}, {Dmytriiev}, {Donath}, {Doroshenko},
  {Dreyer}, {Duffy}, {Dyks}, {Egberts}, {Eichhorn}, {Einecke}, {Emery},
  {Ernenwein}, {Feijen}, {Fegan}, {Fiasson}, {Fichet de Clairfontaine},
  {Fontaine}, {Funk}, {F{\"u}{\ss}ling}, {Gabici}, {Gallant}, {Giavitto},
  {Giunti}, {Glawion}, {Glicenstein}, {Grondin}, {Hahn}, {Haupt}, {Hermann},
  {Hinton}, {Hofmann}, {Hoischen}, {Holch}, {Holler}, {H{\"o}rbe}, {Horns},
  {Huber}, {Jamrozy}, {Jankowsky}, {Jankowsky}, {Jardin-Blicq}, {Joshi},
  {Jung-Richardt}, {Kasai}, {Kastendieck}, {Katarzy{\'n}ski}, {Katz},
  {Khangulyan}, {Kh{\'e}lifi}, {Klepser}, {Klu{\'z}niak}, {Komin}, {Konno},
  {Kosack}, {Kostunin}, {Kreter}, {Lamanna}, {Lemi{\`e}re}, {Lemoine-Goumard},
  {Lenain}, {Leuschner}, {Levy}, {Lohse}, {Lypova}, {Mackey}, {Majumdar},
  {Malyshev}, {Malyshev}, {Marandon}, {Marchegiani}, {Marcowith}, {Mares},
  {Mart{\'\i}-Devesa}, {Marx}, {Maurin}, {Meintjes}, {Meyer}, {Mitchell},
  {Moderski}, {Mohrmann}, {Montanari}, {Moore}, {Morris}, {Moulin}, {Muller},
  {Murach}, {Nakashima}, {Nayerhoda}, {de Naurois}, {Ndiyavala}, {Niemiec},
  {Oakes}, {O'Brien}, {Odaka}, {Ohm}, {Olivera-Nieto}, {de Ona Wilhelmi},
  {Ostrowski}, {Panny}, {Panter}, {Parsons}, {Peron}, {Peyaud}, {Piel}, {Pita},
  {Poireau}, {Priyana Noel}, {Prokhorov}, {Prokoph}, {P{\"u}hlhofer}, {Punch},
  {Quirrenbach}, {Raab}, {Rauth}, {Reichherzer}, {Reimer}, {Reimer}, {Remy},
  {Renaud}, {Rieger}, {Rinchiuso}, {Romoli}, {Rowell}, {Rudak}, {Ruiz-Velasco},
  {Sahakian}, {Sailer}, {Salzmann}, {Sanchez}, {Santangelo}, {Sasaki},
  {Scalici}, {Sch{\"a}fer}, {Sch{\"u}ssler}, {Schutte}, {Schwanke},
  {Seglar-Arroyo}, {Senniappan}, {Seyffert}, {Shafi}, {Shapopi},
  {Shiningayamwe}, {Simoni}, {Sinha}, {Sol}, {Specovius}, {Spencer},
  {Spir-Jacob}, {Stawarz}, {Sun}, {Steenkamp}, {Stegmann}, {Steinmassl},
  {Steppa}, {Takahashi}, {Tam}, {Tavernier}, {Taylor}, {Terrier}, {Thiersen},
  {Tiziani}, {Tluczykont}, {Tomankova}, {Tsirou}, {Tuffs}, {Uchiyama}, {van der
  Walt}, {van Eldik}, {van Rensburg}, {van Soelen}, {Vasileiadis}, {Veh},
  {Venter}, {Vincent}, {Vink}, {V{\"o}lk}, {Wadiasingh}, {Wagner}, {Watson},
  {Werner}, {White}, {Wierzcholska}, {Wong}, {Yusafzai}, {Zacharias}, {Zanin},
  {Zargaryan}, {Zdziarski}, {Zech}, {Zhu}, {Zorn}, {Zouari}, {{\.Z}ywucka},
  {Evans}, \& {Page}}]{2021Sci...372.1081H}
{H.~E.~S.~S. Collaboration}, {Abdalla}, H., {Aharonian}, F., {et~al.} 2021,
  \href{http://dx.doi.org/10.1126/science.abe8560}{\JournalTitle{Science}, 372,
  1081}

\bibitem[{Hakkila \& Preece(2014)}]{Hakkila_2014}
Hakkila, J., \& Preece, R.~D. 2014,
  \href{http://dx.doi.org/10.1088/0004-637X/783/2/88}{\JournalTitle{\apj}, 783,
  88}

\bibitem[{{Huang} {et~al.}(2022){Huang}, {Hu}, {Chen}, {Zha}, {Liu}, {Yao},
  {Cao}, \& {Experiment}}]{2022GCN.32677....1H}
{Huang}, Y., {Hu}, S., {Chen}, S., {et~al.} 2022, \JournalTitle{GRB Coordinates
  Network}, 32677, 1

\bibitem[{{Jeffreys}(1939)}]{Jeffreys}
{Jeffreys}, H. 1939, {Theory of Probability}

\bibitem[{{KM3NeT Collaboration}(2022)}]{2022GCN.32741....1K}
{KM3NeT Collaboration}. 2022, \JournalTitle{GRB Coordinates Network}, 32741, 1

\bibitem[{{Kobayashi}(2000)}]{2000ApJ...545..807K}
{Kobayashi}, S. 2000,
  \href{http://dx.doi.org/10.1086/317869}{\JournalTitle{\apj}, 545, 807}

\bibitem[{{Kumar} \& {Barniol Duran}(2009)}]{2009MNRAS.400L..75K}
{Kumar}, P., \& {Barniol Duran}, R. 2009,
  \href{http://dx.doi.org/10.1111/j.1745-3933.2009.00766.x}{\JournalTitle{MNRAS},
  400, L75}

\bibitem[{{Kumar} \& {Panaitescu}(2000)}]{2000ApJ...541L..51K}
{Kumar}, P., \& {Panaitescu}, A. 2000,
  \href{http://dx.doi.org/10.1086/312905}{\JournalTitle{\apjl}, 541, L51}

\bibitem[{{Laing} \& {Bridle}(2002)}]{2002MNRAS.336..328L}
{Laing}, R.~A., \& {Bridle}, A.~H. 2002,
  \href{http://dx.doi.org/10.1046/j.1365-8711.2002.05756.x}{\JournalTitle{MNRAS},
  336, 328}

\bibitem[{{Laskar} {et~al.}(2023){Laskar}, {Alexander}, {Margutti},
  {Eftekhari}, {Chornock}, {Berger}, {Cendes}, {Duerr}, {Perley}, {Ravasio},
  {Yamazaki}, {Ayache}, {Barclay}, {Barniol Duran}, {Bhandari}, {Brethauer},
  {Christy}, {Coppejans}, {Duffell}, {Fong}, {Gomboc}, {Guidorzi}, {Kennea},
  {Kobayashi}, {Levan}, {Lobanov}, {Metzger}, {Ros}, {Schroeder}, \&
  {Williams}}]{BOAT_MW}
{Laskar}, T., {Alexander}, K.~D., {Margutti}, R., {et~al.} 2023,
  \href{http://dx.doi.org/10.3847/2041-8213/acbfad}{\JournalTitle{\apjl}, 946,
  L23}

\bibitem[{{Lesage} {et~al.}(2022){Lesage}, {Veres}, {Roberts}, {Burns},
  {Bissaldi}, \& {Fermi GBM Team}}]{2022GCN.32642....1L}
{Lesage}, S., {Veres}, P., {Roberts}, O.~J., {et~al.} 2022, \JournalTitle{GRB
  Coordinates Network}, 32642, 1

\bibitem[{{Lesage} {et~al.}(2023){Lesage}, {Veres}, {Briggs}, {Goldstein},
  {Kocevski}, {Burns}, {Wilson-Hodge}, {Bhat}, {Huppenkothen}, {Fryer},
  {Hamburg}, {Racusin}, {Bissaldi}, {Cleveland}, {Dalessi}, {Fletcher},
  {Giles}, {Hristov}, {Hui}, {Mailyan}, {Malacaria}, {Poolakkil}, {Roberts},
  {von Kienlin}, {Wood}, {Ajello}, {Arimoto}, {Baldini}, {Ballet}, {Baring},
  {Bastieri}, {Gonzalez}, {Bellazzini}, {Bissaldi}, {Blandford}, {Bonino},
  {Bruel}, {Buson}, {Cameron}, {Caputo}, {Caraveo}, {Cavazzuti}, {Chiaro},
  {Cibrario}, {Ciprini}, {Orestano}, {Crnogorcevic}, {Cuoco}, {Cutini},
  {D'Ammando}, {De Gaetano}, {Di Lalla}, {Di Venere}, {Dom{\'\i}nguez},
  {Fegan}, {Ferrara}, {Fleischhack}, {Fukazawa}, {Funk}, {Fusco}, {Galanti},
  {Gammaldi}, {Gargano}, {Gasbarra}, {Gasparrini}, {Germani}, {Giacchino},
  {Giglietto}, {Gill}, {Giroletti}, {Granot}, {Green}, {Grenier}, {Guiriec},
  {Gustafsson}, {Hays}, {Hewitt}, {Horan}, {Hou}, {Kuss}, {Latronico},
  {Laviron}, {Lemoine-Goumard}, {Li}, {Liodakis}, {Longo}, {Loparco},
  {Lorusso}, {Lovellette}, {Lubrano}, {Maldera}, {Manfreda},
  {Mart{\'\i}-Devesa}, {Mazziotta}, {McEnery}, {Mereu}, {Meyer}, {Michelson},
  {Mizuno}, {Monzani}, {Morselli}, {Moskalenko}, {Negro}, {Nuss}, {Omodei},
  {Orlando}, {Ormes}, {Paneque}, {Panzarini}, {Persic}, {Pesce-Rollins},
  {Pillera}, {Piron}, {Poon}, {Porter}, {Principe}, {Rain{\`o}}, {Rando},
  {Rani}, {Razzano}, {Razzaque}, {Reimer}, {Reimer}, {Ryde},
  {S{\'a}nchez-Conde}, {Parkinson}, {Scotton}, {Serini}, {Sgr{\`o}}, {Sharma},
  {Siskind}, {Spandre}, {Spinelli}, {Tajima}, {Torres}, {Valverde}, {Venters},
  {Wadiasingh}, {Wood}, \& {Zaharijas}}]{2023ApJ...952L..42L}
{Lesage}, S., {Veres}, P., {Briggs}, M.~S., {et~al.} 2023,
  \href{http://dx.doi.org/10.3847/2041-8213/ace5b4}{\JournalTitle{\apjl}, 952,
  L42}

\bibitem[{{Levan} {et~al.}(2013){Levan}, {Cenko}, {Perley}, \&
  {Tanvir}}]{2013GCN.14455....1L}
{Levan}, A.~J., {Cenko}, S.~B., {Perley}, D.~A., \& {Tanvir}, N.~R. 2013,
  \JournalTitle{GRB Coordinates Network}, 14455, 1

\bibitem[{{Li} \& {Ma}(1983)}]{LiMa}
{Li}, T.~P., \& {Ma}, Y.~Q. 1983,
  \href{http://dx.doi.org/10.1086/161295}{\JournalTitle{\apj}, 272, 317}

\bibitem[{{MAGIC Collaboration} {et~al.}(2019){MAGIC Collaboration}, {Acciari},
  {Ansoldi}, {Antonelli}, {Arbet Engels}, {Baack}, {Babi{\'c}}, {Banerjee},
  {Barres de Almeida}, {Barrio}, {Becerra Gonz{\'a}lez}, {Bednarek},
  {Bellizzi}, {Bernardini}, {Berti}, {Besenrieder}, {Bhattacharyya},
  {Bigongiari}, {Biland}, {Blanch}, {Bonnoli}, {Bo{\v{s}}njak}, {Busetto},
  {Carosi}, {Carosi}, {Ceribella}, {Chai}, {Chilingaryan}, {Cikota}, {Colak},
  {Colin}, {Colombo}, {Contreras}, {Cortina}, {Covino}, {D'Amico}, {D'Elia},
  {da Vela}, {Dazzi}, {de Angelis}, {de Lotto}, {Delfino}, {Delgado},
  {Depaoli}, {di Pierro}, {di Venere}, {Do Souto Espi{\~n}eira}, {Dominis
  Prester}, {Donini}, {Dorner}, {Doro}, {Elsaesser}, {Fallah Ramazani},
  {Fattorini}, {Fern{\'a}ndez-Barral}, {Ferrara}, {Fidalgo}, {Foffano},
  {Fonseca}, {Font}, {Fruck}, {Fukami}, {Gallozzi}, {Garc{\'\i}a L{\'o}pez},
  {Garczarczyk}, {Gasparyan}, {Gaug}, {Giglietto}, {Giordano}, {Godinovi{\'c}},
  {Green}, {Guberman}, {Hadasch}, {Hahn}, {Herrera}, {Hoang}, {Hrupec},
  {H{\"u}tten}, {Inada}, {Inoue}, {Ishio}, {Iwamura}, {Jouvin}, {Kerszberg},
  {Kubo}, {Kushida}, {Lamastra}, {Lelas}, {Leone}, {Lindfors}, {Lombardi},
  {Longo}, {L{\'o}pez}, {L{\'o}pez-Coto}, {L{\'o}pez-Oramas}, {Loporchio},
  {Machado de Oliveira Fraga}, {Maggio}, {Majumdar}, {Makariev}, {Mallamaci},
  {Maneva}, {Manganaro}, {Mannheim}, {Maraschi}, {Mariotti}, {Mart{\'\i}nez},
  {Masuda}, {Mazin}, {Mi{\'c}anovi{\'c}}, {Miceli}, {Minev}, {Miranda},
  {Mirzoyan}, {Molina}, {Moralejo}, {Morcuende}, {Moreno}, {Moretti},
  {Munar-Adrover}, {Neustroev}, {Nigro}, {Nilsson}, {Ninci}, {Nishijima},
  {Noda}, {Nogu{\'e}s}, {N{\"o}the}, {Nozaki}, {Paiano}, {Palacio},
  {Palatiello}, {Paneque}, {Paoletti}, {Paredes}, {Pe{\~n}il}, {Peresano},
  {Persic}, {Prada Moroni}, {Prandini}, {Puljak}, {Rhode}, {Rib{\'o}}, {Rico},
  {Righi}, {Rugliancich}, {Saha}, {Sahakyan}, {Saito}, {Sakurai}, {Satalecka},
  {Schmidt}, {Schweizer}, {Sitarek}, {{\v{S}}nidari{\'c}}, {Sobczynska},
  {Somero}, {Stamerra}, {Strom}, {Strzys}, {Suda}, {Suri{\'c}}, {Takahashi},
  {Tavecchio}, {Temnikov}, {Terzi{\'c}}, {Teshima}, {Torres-Alb{\`a}}, {Tosti},
  {Tsujimoto}, {Vagelli}, {van Scherpenberg}, {Vanzo}, {Vazquez Acosta},
  {Vigorito}, {Vitale}, {Vovk}, {Will}, {Zari{\'c}}, \&
  {Nava}}]{2019Natur.575..455M}
{MAGIC Collaboration}, {Acciari}, V.~A., {Ansoldi}, S., {et~al.} 2019,
  \href{http://dx.doi.org/10.1038/s41586-019-1750-x}{\JournalTitle{\nat}, 575,
  455}

\bibitem[{{Meegan} {et~al.}(2009){Meegan}, {Lichti}, {Bhat}, {Bissaldi},
  {Briggs}, {Connaughton}, {Diehl}, {Fishman}, {Greiner}, {Hoover}, {van der
  Horst}, {von Kienlin}, {Kippen}, {Kouveliotou}, {McBreen}, {Paciesas},
  {Preece}, {Steinle}, {Wallace}, {Wilson}, \&
  {Wilson-Hodge}}]{2009arXiv0908.0450M}
{Meegan}, C., {Lichti}, G., {Bhat}, P.~N., {et~al.} 2009,
  \href{http://dx.doi.org/10.1088/0004-637X/702/1/791}{\JournalTitle{\apj},
  702, 791}

\bibitem[{{Meliani} {et~al.}(2008){Meliani}, {Keppens}, \&
  {Giacomazzo}}]{Meliani-2008-AandA}
{Meliani}, Z., {Keppens}, R., \& {Giacomazzo}, B. 2008,
  \href{http://dx.doi.org/10.1051/0004-6361:20079185}{\JournalTitle{\aap}, 491,
  321}

\bibitem[{{Nakar} \& {Piran}(2004)}]{2004MNRAS.353..647N}
{Nakar}, E., \& {Piran}, T. 2004,
  \href{http://dx.doi.org/10.1111/j.1365-2966.2004.08099.x}{\JournalTitle{MNRAS},
  353, 647}

\bibitem[{{Nappo} {et~al.}(2014){Nappo}, {Ghisellini}, {Ghirlanda}, {Melandri},
  {Nava}, \& {Burlon}}]{GammaNappo}
{Nappo}, F., {Ghisellini}, G., {Ghirlanda}, G., {et~al.} 2014,
  \href{http://dx.doi.org/10.1093/mnras/stu1832}{\JournalTitle{MNRAS}, 445,
  1625}

\bibitem[{{Negro} {et~al.}(2023){Negro}, {Di Lalla}, {Omodei}, {Veres},
  {Silvestri}, {Manfreda}, {Burns}, {Baldini}, {Costa}, {Ehlert}, {Kennea},
  {Liodakis}, {Marshall}, {Mereghetti}, {Middei}, {Muleri}, {O'Dell},
  {Roberts}, {Romani}, {Sgr{\'o}}, {Terashima}, {Tiengo}, {Viscolo}, {Di
  Marco}, {La Monaca}, {Latronico}, {Matt}, {Perri}, {Puccetti}, {Poutanen},
  {Ratheesh}, {Rogantini}, {Slane}, {Soffitta}, {Lindfors}, {Nilsson},
  {Kasikov}, {Marscher}, {Tavecchio}, {Cibrario}, {Gunji}, {Malacaria},
  {Paggi}, {Yang}, {Zane}, {Weisskopf}, {Agudo}, {Antonelli}, {Bachetti},
  {Baumgartner}, {Bellazzini}, {Bianchi}, {Bongiorno}, {Bonino}, {Brez},
  {Bucciantini}, {Capitanio}, {Castellano}, {Cavazzuti}, {Chen}, {Ciprini}, {De
  Rosa}, {Del Monte}, {Di Gesu}, {Donnarumma}, {Doroshenko}, {Dovc̆iak},
  {Enoto}, {Evangelista}, {Fabiani}, {Ferrazzoli}, {Garcia}, {Hayashida},
  {Heyl}, {Iwakiri}, {Jorstad}, {Kaaret}, {Karas}, {Kislat}, {Kitaguchi},
  {Kolodziejczak}, {Krawczynski}, {Maldera}, {Marin}, {Marinucci}, {Mitsuishi},
  {Mizuno}, {Ng}, {Oppedisano}, {Papitto}, {Pavlov}, {Peirson},
  {Pesce-Rollins}, {Petrucci}, {Pilia}, {Possenti}, {Ramsey}, {Rankin},
  {Spandre}, {Swartz}, {Tamagawa}, {Taverna}, {Tawara}, {Tennant}, {Thomas},
  {Tombesi}, {Trois}, {Tsygankov}, {Turolla}, {Vink}, {Wu}, \&
  {Xie}}]{IXPEpapaer}
{Negro}, M., {Di Lalla}, N., {Omodei}, N., {et~al.} 2023,
  \href{http://dx.doi.org/10.3847/2041-8213/acba17}{\JournalTitle{\apjl}, 946,
  L21}

\bibitem[{{Neronov} \& {Semikoz}(2009)}]{Neronov09}
{Neronov}, A., \& {Semikoz}, D.~V. 2009,
  \href{http://dx.doi.org/10.1103/PhysRevD.80.123012}{\JournalTitle{\prd}, 80,
  123012}

\bibitem[{{Norris} {et~al.}(2005){Norris}, {Bonnell}, {Kazanas}, {Scargle},
  {Hakkila}, \& {Giblin}}]{norris:2005}
{Norris}, J.~P., {Bonnell}, J.~T., {Kazanas}, D., {et~al.} 2005,
  \href{http://dx.doi.org/10.1086/430294}{\JournalTitle{\apj}, 627, 324}

\bibitem[{{Omodei} {et~al.}(2022){Omodei}, {Bruel}, {Bregeon}, {Pesce-Rollins},
  {Horan}, E., {Pillera}, \& {Fermi-LAT Team}}]{2022GCN.32760....1O}
{Omodei}, N., {Bruel}, P., {Bregeon}, J., {et~al.} 2022, \JournalTitle{GRB
  Coordinates Network}, 32760, 1

\bibitem[{{Pelassa} {et~al.}(2010){Pelassa}, {Preece}, {Piron}, {Omodei},
  {Guiriec}, {Fermi LAT}, \& {GBM collaborations}}]{PelassaLLE}
{Pelassa}, V., {Preece}, R., {Piron}, F., {et~al.} 2010, \JournalTitle{ArXiv
  e-prints}, \href{http://arxiv.org/abs/1002.2617}{{\sffamily arXiv:1002.2617
  [astro-ph.HE]}}

\bibitem[{{Pillera} {et~al.}(2022){Pillera}, {Bissaldi}, {Omodei}, {La Mura},
  {Longo}, \& {Fermi-LAT team}}]{2022GCN.32658....1P}
{Pillera}, R., {Bissaldi}, E., {Omodei}, N., {et~al.} 2022, \JournalTitle{GRB
  Coordinates Network}, 32658, 1

\bibitem[{{Piron} {et~al.}(2019){Piron}, {Longo}, {Axelsson}, {Arimoto},
  {Racusin}, {Bissaldi}, \& {Fermi-LAT Collaboration}}]{2019GCN.25574....1P}
{Piron}, F., {Longo}, F., {Axelsson}, M., {et~al.} 2019, \JournalTitle{GRB
  Coordinates Network}, 25574, 1

\bibitem[{Poolakkil {et~al.}(2021)Poolakkil, Preece, Fletcher, Goldstein, Bhat,
  Bissaldi, Briggs, Burns, Cleveland, Giles, Hui, Kocevski, Lesage, Mailyan,
  Malacaria, Paciesas, Roberts, Veres, von Kienlin, \&
  Wilson-Hodge}]{Poolakkil_2021}
Poolakkil, S., Preece, R., Fletcher, C., {et~al.} 2021,
  \href{http://dx.doi.org/10.3847/1538-4357/abf24d}{\JournalTitle{\apj}, 913,
  60}

\bibitem[{Racusin {et~al.}(2009)Racusin, Liang, Burrows, Falcone, Sakamoto,
  Zhang, Zhang, Evans, \& Osborne}]{Racusin_2009}
Racusin, J.~L., Liang, E.~W., Burrows, D.~N., {et~al.} 2009,
  \href{http://dx.doi.org/10.1088/0004-637X/698/1/43}{\JournalTitle{\apj}, 698,
  43}

\bibitem[{{Ravasio} {et~al.}(2018){Ravasio}, {Oganesyan}, {Ghirlanda}, {Nava},
  {Ghisellini}, {Pescalli}, \& {Celotti}}]{2018A&A...613A..16R}
{Ravasio}, M.~E., {Oganesyan}, G., {Ghirlanda}, G., {et~al.} 2018,
  \href{http://dx.doi.org/10.1051/0004-6361/201732245}{\JournalTitle{\aap},
  613, A16}

\bibitem[{Ravasio {et~al.}(2024)Ravasio, Salafia, Oganesyan, Mei, Ghirlanda,
  Ascenzi, Banerjee, Macera, Branchesi, Jonker, Levan, Malesani, Mulrey,
  Giuliani, Celotti, \& Ghisellini}]{BOAT_LINE}
Ravasio, M.~E., Salafia, O.~S., Oganesyan, G., {et~al.} 2024,
  \href{http://dx.doi.org/10.1126/science.adj3638}{\JournalTitle{Science}, 385,
  452}

\bibitem[{{Razzaque}(2010)}]{2010ApJ...724L.109R}
{Razzaque}, S. 2010,
  \href{http://dx.doi.org/10.1088/2041-8205/724/1/L109}{\JournalTitle{\apjl},
  724, L109}

\bibitem[{{Sari} \& {Piran}(1999)}]{SP99}
{Sari}, R., \& {Piran}, T. 1999,
  \href{http://dx.doi.org/10.1086/307508}{\JournalTitle{\apj}, 520, 641}

\bibitem[{{Scargle} {et~al.}(2013){Scargle}, {Norris}, {Jackson}, \&
  {Chiang}}]{2013ApJ...764..167S}
{Scargle}, J.~D., {Norris}, J.~P., {Jackson}, B., \& {Chiang}, J. 2013,
  \href{http://dx.doi.org/10.1088/0004-637X/764/2/167}{\JournalTitle{\apj},
  764, 167}

\bibitem[{{Schnoor} {et~al.}(2022){Schnoor}, {Nicholson}, \&
  {Welch}}]{2022GCN.32744....1S}
{Schnoor}, P.~W., {Nicholson}, P., \& {Welch}, D.~L. 2022, \JournalTitle{GRB
  Coordinates Network}, 32744, 1

\bibitem[{{Tak} {et~al.}(2019){Tak}, {Omodei}, {Uhm}, {Racusin}, {Asano}, \&
  {McEnery}}]{CLOSURE_RELATIONS_TAK}
{Tak}, D., {Omodei}, N., {Uhm}, Z.~L., {et~al.} 2019,
  \href{http://dx.doi.org/10.3847/1538-4357/ab3982}{\JournalTitle{\apj}, 883,
  134}

\bibitem[{Tavani {et~al.}(2023)Tavani, Piano, Bulgarelli, Foffano, Ursi,
  Verrecchia, Pittori, Casentini, Giuliani, Longo, Panebianco, Piano,
  Baroncelli, Fioretti, Parmiggiani, Argan, Trois, Vercellone, Cardillo,
  Antonelli, Barbiellini, Caraveo, Cattaneo, Chen, Costa, Monte, Cocco,
  Donnarumma, Evangelista, Feroci, Gianotti, Labanti, Lazzarotto, Lipari,
  Lucarelli, Marisaldi, Mereghetti, Morselli, Pacciani, Pellizzoni, Perotti,
  Picozza, Pilia, Rapisarda, Rappoldi, Rubini, Soffitta, Trifoglio, Vittorini,
  \& D‚ÄôAmico}]{BOAT_AGILE}
Tavani, M., Piano, G., Bulgarelli, A., {et~al.} 2023,
  \href{http://dx.doi.org/10.3847/2041-8213/acfaff}{\JournalTitle{\apjl}, 956,
  L23}

\bibitem[{{Tiengo} {et~al.}(2023{\natexlab{a}}){Tiengo}, {Pintore}, {Vaia},
  {Filippi}, {Sacchi}, {Esposito}, {Rigoselli}, {Mereghetti}, {Salvaterra},
  {{\v{S}}iljeg}, {Bracco}, {Bo{\v{s}}njak}, {Jeli{\'c}}, \&
  {Campana}}]{2023ApJ...946L..30T}
{Tiengo}, A., {Pintore}, F., {Vaia}, B., {et~al.} 2023{\natexlab{a}},
  \href{http://dx.doi.org/10.3847/2041-8213/acc1dc}{\JournalTitle{\apjl}, 946,
  L30}

\bibitem[{{Tiengo} {et~al.}(2023{\natexlab{b}}){Tiengo}, {Pintore}, {Vaia},
  {Filippi}, {Sacchi}, {Esposito}, {Rigoselli}, {Mereghetti}, {Salvaterra},
  {{\v{S}}iljeg}, {Bracco}, {Bo{\v{s}}njak}, {Jeli{\'c}}, \&
  {Campana}}]{BOAT_XMM}
---. 2023{\natexlab{b}},
  \href{http://dx.doi.org/10.3847/2041-8213/acc1dc}{\JournalTitle{\apjl}, 946,
  L30}

\bibitem[{{Vasilopoulos} {et~al.}(2023){Vasilopoulos}, {Karavola},
  {Stathopoulos}, \& {Petropoulou}}]{2023MNRAS.521.1590V}
{Vasilopoulos}, G., {Karavola}, D., {Stathopoulos}, S.~I., \& {Petropoulou}, M.
  2023, \href{http://dx.doi.org/10.1093/mnras/stad375}{\JournalTitle{MNRAS},
  521, 1590}

\bibitem[{{Vianello} {et~al.}(2018){Vianello}, {Gill}, {Granot}, {Omodei},
  {Cohen-Tanugi}, \& {Longo}}]{2018ApJ...864..163V}
{Vianello}, G., {Gill}, R., {Granot}, J., {et~al.} 2018,
  \href{http://dx.doi.org/10.3847/1538-4357/aad6ea}{\JournalTitle{\apj}, 864,
  163}

\bibitem[{{Vianello} {et~al.}(2015{\natexlab{a}}){Vianello}, {Omodei}, \&
  {Fermi/LAT collaboration}}]{2015arXiv150203122V}
{Vianello}, G., {Omodei}, N., \& {Fermi/LAT collaboration}. 2015{\natexlab{a}},
  \JournalTitle{2014 Fermi Symposium proceedings - eConf C141020.1},
  \href{http://arxiv.org/abs/1502.03122}{{\sffamily arXiv:1502.03122
  [astro-ph.HE]}}

\bibitem[{{Vianello} {et~al.}(2015{\natexlab{b}}){Vianello}, {Lauer}, {Younk},
  {Tibaldo}, {Burgess}, {Ayala}, {Harding}, {Hui}, {Omodei}, \& {Zhou}}]{3ML}
{Vianello}, G., {Lauer}, R.~J., {Younk}, P., {et~al.} 2015{\natexlab{b}},
  \href{http://dx.doi.org/10.48550/arXiv.1507.08343}{\JournalTitle{The 34th
  International Cosmic Ray Conference (ICRC), 30 July-6 August, 2015, The
  Hague, The Netherlands}, arXiv:1507.08343}

\bibitem[{{Williams} {et~al.}(2023){Williams}, {Kennea}, {Dichiara},
  {Kobayashi}, {Iwakiri}, {Beardmore}, {Evans}, {Heinz}, {Lien}, {Oates},
  {Negoro}, {Cenko}, {Buisson}, {Hartmann}, {Jaisawal}, {Kuin}, {Lesage},
  {Page}, {Parsotan}, {Pasham}, {Sbarufatti}, {Siegel}, {Sugita}, {Younes},
  {Ambrosi}, {Arzoumanian}, {Bernardini}, {Campana}, {Capalbi}, {Caputo},
  {D'A{\`\i}}, {D'Avanzo}, {D'Elia}, {De Pasquale}, {Eyles-Ferris}, {Ferrara},
  {Gendreau}, {Gropp}, {Kawai}, {Klingler}, {Laha}, {Melandri}, {Mihara},
  {Moss}, {O'Brien}, {Osborne}, {Palmer}, {Perri}, {Serino}, {Sonbas},
  {Stamatikos}, {Starling}, {Tagliaferri}, {Tohuvavohu}, {Zane}, \&
  {Ziaeepour}}]{BOAT_SWIFT}
{Williams}, M.~A., {Kennea}, J.~A., {Dichiara}, S., {et~al.} 2023,
  \href{http://dx.doi.org/10.3847/2041-8213/acbcd1}{\JournalTitle{\apjl}, 946,
  L24}

\bibitem[{{Wood} {et~al.}(2017){Wood}, {Caputo}, {Charles}, {Di Mauro},
  {Magill}, {Perkins}, \& {Fermi-LAT Collaboration}}]{2017ICRC...35..824W}
{Wood}, M., {Caputo}, R., {Charles}, E., {et~al.} 2017,
  \href{http://dx.doi.org/10.22323/1.301.0824}{in International Cosmic Ray
  Conference, Vol. 301, 35th International Cosmic Ray Conference (ICRC2017)},
  824

\bibitem[{{Xia} {et~al.}(2022){Xia}, {Wang}, {Yuan}, \&
  {Fan}}]{2022GCN.32748....1X}
{Xia}, Z.-Q., {Wang}, Y., {Yuan}, Q., \& {Fan}, Y.-Z. 2022, \JournalTitle{GRB
  Coordinates Network}, 32748, 1

\bibitem[{{Xia} {et~al.}(2024){Xia}, {Wang}, {Yuan}, \&
  {Fan}}]{2024NatCo..15.4280X}
---. 2024,
  \href{http://dx.doi.org/10.1038/s41467-024-48668-5}{\JournalTitle{Nature
  Communications}, 15, 4280}

\bibitem[{{Zhang} {et~al.}(2024){Zhang}, {Xiong}, {Mao}, {Zhang}, {Xue},
  {Zheng}, {Liu}, {Zhang}, {Wang}, {Ge}, {Yi}, {Song}, {An}, {Cai}, {Li},
  {Peng}, {Tan}, {Wang}, {Wen}, {Wang}, {Xiao}, {Zhang}, {Zhang}, \&
  {Zheng}}]{2024arXiv240312851Z}
{Zhang}, Y.-Q., {Xiong}, S.-L., {Mao}, J.-R., {et~al.} 2024,
  \href{http://dx.doi.org/10.48550/arXiv.2403.12851}{\JournalTitle{arXiv
  e-prints}, arXiv:2403.12851}

\end{thebibliography}
\bibliographystyle{yahapj}

\end{document}